# A new approach to the theory of Brownian coagulation and diffusion-limited reactions

*M.S. Veshchunov*

Nuclear Safety Institute (IBRAE), Russian Academy of Sciences
52, B. Tulskaya, Moscow, 115191, Russia

Moscow Institute of Physics and Technology (MIPT) (State University)

9, Institutskii per., Dolgoprudny, Moscow Region, 141700, Russia

phone: +7(495) 955 2218, fax: +7(495)9580040, e-mail: msvesh@gmail.com

## Abstract

An overview of the author's papers on the new approach to the Brownian coagulation theory and its generalization to the diffusion-limited reaction rate theory is presented. The traditional diffusion approach of the Smoluchowski theory for coagulation of colloids is critically analysed and shown to be valid only in the particular case of coalescence of small particles with large ones, $R_1 << R_2$. It is shown that, owing to rapid diffusion mixing, coalescence of comparable size particles occurs in the kinetic regime, realized under condition of homogeneous spatial distribution of particles, in the two modes, continuum and free molecular. However, the expression for the collision frequency function in the continuum mode of the kinetic regime formally coincides with the standard expression derived in the diffusion regime for the particular case of large and small particles. Transition from the continuum to the free molecular mode can be described by the interpolation expression derived within the new analytical approach with fitting parameters that can be specified numerically, avoiding semi-empirical assumptions of the traditional models. A similar restriction arises in the traditional approach to the diffusion-limited reaction rate theory, based on generalization of the Smoluchowski theory for coagulation of colloids. In particular, it is shown that the traditional approach is applicable only to the special case of reactions (A + B $\rightarrow$ C) with a large reaction radius, $\bar{r}_A << R_{AB} << \bar{r}_B$ (where $\bar{r}_A$, $\bar{r}_B$ are the mean inter-particle distances), and becomes inappropriate to calculation of the reaction rate in the case of a relatively small reaction radius, $R_{AB} << \bar{r}_A, \bar{r}_B$. In the latter, more general case particles collisions occur mainly in the kinetic regime (rather than in the diffusion one) characterized by homogeneous (at random) spatial distribution of particles. The calculated reaction rate for a small reaction radius in 3-d formally coincides with the expression derived in the traditional approach for reactions with a large reaction radius, however, notably deviates at large times from the traditional result in the plane (2-d) geometry, that has wide applications also in the membrane biology as well as in some other important areas.

## Table of Contents

# Part 1. Brownian coagulation theory

## 1.1. Introduction

Brownian motion refers to the continuous random movement (or diffusion) of particles suspended in a fluid. Brownian agglomeration occurs when, as a result of their random motion, particles collide and stick together. The theoretical treatment of agglomeration consists on keeping count of the number of particles as a result of collisions and determining the collision frequency function which depends on the particle sizes, concentrations and transport mechanisms in the system. Coagulation is regarded as a special case of agglomeration where there is instantaneous coalescence of particles after collision.

Brownian coagulation was first calculated basing on the Brownian diffusion theory by Smoluchowski [1] and further developed by Chandrasekhar [2]. The theory was essentially based on assumption that the local coagulation rate should be equal to the diffusive current of particles. Namely, the expression for the collision frequency was obtained by solving the diffusion equation for particles around one particle that is assumed to be fixed using the relative diffusion coefficient $D_1 + D_2$ for moving particles.

In application to the case of suspending gas, it was outlined that this expression is valid only for particles that are large enough that they experience the surrounding gas as a continuum (so called "continuum regime"), whereas for particles with radius $R$ much smaller than the mean free path $\lambda_m$ of the surrounding molecules the "free molecular regime" (corresponding to large Knudsen numbers, $\mathrm{Kn} = \lambda_m / R >> 1$) was considered (see, e.g. [3]). An expression for the collision frequency function in the free molecular regime was derived in the gas-kinetic approach assuming rigid elastic spheres.

Fuchs proposed a semi-empirical interpolation formula for the whole particle diameter range [4]. In fact this formula interpolates two regimes corresponding to large and small particle sizes in comparison with the "mean free path" $a$ (also termed as the "mean drift distance" or "persistence distance") of the Brownian particles (rather than $\lambda_m$ !). The formula is reduced to the standard diffusion expression [1, 2] for relatively large particles, $R >> a$, and to the free molecular regime collision frequency function in the opposite limiting case, $R << a$ (rather than the above mentioned and widely used inequality, $\mathrm{Kn} = \lambda_m / R >> 1$).

The classical problem of Brownian coagulation was reconsidered in the author's papers [5-7] with the main conclusion that the usual approach to the calculation of the local coagulation rate via the particle diffusive current (in the continuum regime) is valid only in the particular case of collisions between large and small Brownian particles, $R_1 >> \bar{r} >> R_2$ (where $\bar{r} \approx n^{-1/3}$ is the mean inter-particle distance), and cannot be applied to a more general case of particles of comparable sizes, $R_1, R_2 << \bar{r}$. In particular, the significance and the necessity to introduce a new length scale, $\bar{r} \approx n^{-1/3}$, that is not frequently used in dispersed flows, however, plays an important role in the new approach, was revealed.

In order to expose the main inconsistency of the traditional approach, the diffusion equation for the ensemble of Brownian particles is re-derived in the first order approximation for the small concentration $n$ of comparable size ($\approx R$) particles, $n^{1/3}R << 1$, with a special attention to restrictions on the system parameters that provide applicability of the diffusion approach (see Section 1.2).

On this basis, in the second order approximation of $n^{1/3}R << 1$ it is shown (Section 1.3) that coalescence of comparable size particles, $R_1, R_2 << \bar{r}$, occurs in the kinetic regime (rather than in the diffusion regime) characterized by homogeneous spatial distribution function of the colliding particles (rather than by their concentration profiles) practically in the whole considered range $n^{1/3}R << 1$, owing to rapid diffusion mixing of particles. Such a mixing takes place on the scale of the mean inter-particle distance, $l \approx \bar{r}$, with the characteristic diffusion mixing time $\tau_d$ that is generally small in comparison with the characteristic reaction time $\tau_c$ (≈ the mean time between two subsequent collisions of a particle), i.e. $\tau_d << \tau_c$ (Section 1.3.2). This implies that a random distribution of particles is attained during a time step $\tau_d << \delta t << \tau_c$, chosen for calculation of the collision rate, which thus can be searched in the kinetic approach as the collision frequency of two particles randomly located in unit volume. The latter value can be equally calculated as the rate of volume sweeping $\delta\langle V_{12}\rangle / \delta t$ by the effective particle of radius $R_{12} = R_1 + R_2$ migrating with the diffusivity $D_{12} = D_1 + D_2$ [5].

The volume swept by a Brownian particle is known as the Wiener sausage [8]; in particular, this quantity equals the probability that a diffusing point-wise particle is absorbed by a large trap for time $t$ [9]. For this reason, for small particles (of radius $R_1$), sinking in a large trap (of radius $R_2 >> R_1$), the volume sweeping rate coincides with the condensation rate constant (in the steady-state approximation). For comparable size particles, $R_1, R_2 << \bar{r}$, it eventually determines the collision rate, as justified in the kinetic approach [5]. This explains why the formal expression for the collision frequency of particles, derived by Smoluchowski [1] and Chandrasekhar [2] for the diffusion regime and being relevant only in the particular case of coalescence of small particles with large ones, correctly fits to numerous experimental measurements of the coagulation rate of comparable size Brownian particle in the continuum mode, cf. [10].

The whole range of the kinetic regime can be subdivided into two intervals of the model parameters, $a << R$ and $a >> R$, corresponding to different modes ("continuum" and "free molecular") of the kinetic regime, and the "transition" interval, $a \approx R$, amid the two limiting cases (Section 1.4).

In the continuum mode of the kinetic regime, $a << R$ (Section 1.4.1), the formal expression for the collision frequency of particles (of comparable sizes) coincides (in fact, fortuitously) with that derived in [1, 2] for the diffusion regime (being relevant only in the particular case of coalescence of small particles with large ones, $R_1 << \bar{r} << R_2$). This formal coincidence apparently explains why the traditional approach correctly describes numerous experimental measurements of the Brownian particles coalescence rate.

In the opposite case $a >> R$ (Section 1.4.2), the standard free molecular expression for the collision frequency function is valid. It is shown that, despite the free molecular expression can be rigorously derived (Section 1.4) only in the case of very high collision frequency (when two subsequent collisions of a particle with other ones occur within one drift period), it can be properly extended to the whole range $a/R >> 1$.

Since the transition interval $a \approx R$ (Section 1.4.3) also belongs to the kinetic regime characterized by homogeneous spatial distribution of particles, the collision frequency can be numerically calculated in the same approach, generalizing the analytical method applied in the limiting cases $a << R$ and $a >> R$. On this base, new interpolation expressions were derived in the first approximation [5-7], based on the *simple* random walk theory (with the fixed elementary drift, or persistence distance, of migrating particles) that allowed a relatively simple derivation of the new interpolation formulas by fitting to the calculated points (Section 1.4.4), avoiding semi-empirical assumptions of the existing models.

This approach was further improved in the author's paper [10] by a more realistic consideration of random walks with stochastically distributed lengths (Section 1.5). The new set of calculation points obtained in the next approximation of the random walk theory (with stochastic lengths) reliably confirmed (within the calculation accuracy) the interpolation expression derived in the first (simple random walk) approximation.

The subsequent improvement of the sweeping rate calculations can be obtained by application of the Langevin equation for calculation of particle trajectories. The Langevin equation [11], derived under condition that forces can be split up into a systematic part (friction term) and a statistical part (stochastic term, or the Langevin force) that is grounded at times $t \geq \tau_0$ (where $\tau_0$ is the particle relaxation time), can be equally reduced at large times $t >> \tau_0$ to the Einstein diffusion equation for the Brownian particle motion [12], which can be properly described by the random walk theory. For this reason, one should expect also for the collision rate a rather good coincidence of the semi-analytical predictions of the random walk models for Brownian particles migration with results of numerical methods based on the Langevin equation. This conclusion was well confirmed by calculations of the two-particle collision rate in the Langevin-equation-based mean first passage time calculations of Gopalakrishnan and Hogan [13].

In the author's paper [14] the kinetic approach, originally applied to calculation of the sweeping rate in frames of the simplified random walk models, was further advanced by application of the Langevin theory for Brownian particles migration. The new results allowed an additional justification and further improvement of the interpolation expressions for the coagulation kernel obtained in the random walk theory, also in comparison with the other approaches (Section 1.6).

In the transition mode the semi-empirical flux matching theory, proposed by Fuchs [4], with various definitions of the absorption sphere radius in subsequent models, is traditionally applied to consideration of hard sphere collisions. Again, the theory is also well grounded in the case of collisions of small particles with a large trap. However, for coagulation of comparable size particles this theory inherits the main deficiency of the traditional approach, since in the transition mode the diffusion theory cannot be used near the outside surface of the absorbing sphere, where the external and internal fluxes are matched. As shown in the author's paper [15], a relatively good agreement with the more rigorous results of the kinetic approach in the case of comparable size rigid particles is also fortuitous and promptly disappears in a more general case of a finite sticking probability, $P_{12} < 1$, resulting in erroneous predictions of the traditional flux matching theory for the transition regime (Section 1.7).

The kinetic approach can be also extended to consideration of vapour molecule condensation, following the author's papers [10] for heavy vapour molecules and [16] for light vapour molecules (Section 1.8).

Extension of the Smoluchowski theory to transitions from dilute (low number concentration of particles) to dense regime (high number concentration of particles) of Brownian coagulation by consideration (following in the author's paper [17]) of triple collisions in the kinetic approach, is presented in Section 1.9.

Discussion on validity of the earlier approaches and frameworks of their justification is presented in Section 1.10.

Generalization of the new kinetic approach to the diffusion-limited reaction rate theory will be presented in Part 2.

### 1.2. Diffusion relaxation in ensemble of Brownian particles

Let us consider a continuous spatial distribution of centres of Brownian particles with the mean radius $R$ and mass $m$ that migrate throughout a fluid sample with the heat velocity and randomly change direction of the velocity for the relaxation time $\tau_0$. This consideration can be justified under an assumption $\tau_m << \tau_0$, where $\tau_m$ is the mean time between stochastic collisions of a particle with the surrounding fluid molecules.

The suspending fluid can be considered as a continuous medium, if the size $L_m$ of the elementary volume $\delta V_m = L_m^3$ (over which the averaging is carried out) is large enough in comparison with the mean inter-molecular distance $\bar{r}_m \approx n_m^{-1/3}$, where $n_m$ is the fluid molecules concentration, $L_m >> n_m^{-1/3}$. For this reason the minimum distance (or the length scale) $dr$ between two possible positions of a particle centre, $\mathbf{r}$ and $\mathbf{r} + d\mathbf{r}$, that can be considered in this approach, corresponds to the size $L_m$ of the elementary volume and thus $dr >> n_m^{-1/3}$.

On the other hand, the local number concentration of particles $n(\mathbf{r})$, defined as the number of particles in the unit volume, becomes a strongly fluctuating value, if this unit volume is comparable with (or smaller than) the local inter-particle distance $n^{-1/3}(\mathbf{r})$. In order to consider $n(\mathbf{r})$ as a macroscopic value (i.e. when its thermodynamic fluctuations are small in comparison with its value, $\sqrt{\langle (\delta n)^2 \rangle} \approx \sqrt{n} << n$), the size of the elementary volume $\delta \widetilde{V} = L^3$, with respect to which $n(\mathbf{r})$ is defined, has to be large enough in comparison with the local inter-particle distance, $L >> n^{-1/3}(\mathbf{r})$. For this reason, the local concentration of particles $n(\mathbf{r})$ can be related to the probability of having the centre of one particle in the elementary volume $d^3 r$ at $\mathbf{r}$, $p(\mathbf{r},t) = P(\mathbf{r},t) d^3 r$, where $P(\mathbf{r},t)$ is the probability density, after averaging over the elementary volume $\delta\widetilde{V} = L^3$, $n(\mathbf{r},t) = \langle p(\mathbf{r},t)\rangle_{\delta\widetilde{V}} \equiv (\delta\widetilde{V})^{-1}\int P(\mathbf{r},t)\delta\widetilde{V}$.

In the first order of approximation $n^{1/3}R << 1$, collisions (and coagulations) of migrating particles can be neglected. Let us designate $w(\mathbf{r},\xi)d^3\xi$ the probability that the centre of a particle located at $\mathbf{r}$ will relocate to $\mathbf{r}-\xi$ in $\Delta t = \tau_0$. On this relaxation time scale, $\tau_0$, the particle relocates to some distance and randomly change the direction of its velocity, i.e. the particle elementary drifts are uncorrelated, realizing a Markov process (which is not the case, if the variation time $\Delta t$ is chosen less than $\tau_0$). For this reason, the probability to find this centre in the position $\mathbf{r}-\xi$ in $\tau_0$ is $P(\mathbf{r},t)w(\mathbf{r},\xi)d^3rd^3\xi$. Correspondingly, variation of the probability density at $\mathbf{r}$ in $\tau_0$ is $P(\mathbf{r},t+\tau_0) - P(\mathbf{r},t) = \int [w(\mathbf{r}+\xi,\xi)P(\mathbf{r}+\xi,t) - w(\mathbf{r},\xi)P(\mathbf{r},t)]\, d^3\xi$, where the first term on the right hand side represents the increase of the concentration at $\mathbf{r}$ owing to drifts from the neighbour positions, and the second term represents the reduction of the concentration owing to drifts from the position $\mathbf{r}$. For large times, $t >> \tau_0$, the left hand side can be decomposed as $P(\mathbf{r},t+\tau_0) - P(\mathbf{r},t) \approx \tau_0 \dfrac{\partial P(\mathbf{r},t)}{\partial t}$, and the rate equation takes the form

$$\frac{\partial P(\mathbf{r},t)}{\partial t} = \frac{1}{\tau_0}\int [w(\mathbf{r}+\xi,\xi)P(\mathbf{r}+\xi,t) - w(\mathbf{r},\xi)P(\mathbf{r},t)]\, d^3\xi . \quad (1.1)$$

In the continuum approach it is assumed that the length scale $l$ of $P(\mathbf{r})$ variation is large in comparison with the characteristic drift (or jump) distance $a$ of a particle in the time interval $\tau_0$, i.e. $a << l$ . Since $w(\mathbf{r},\xi)$ rapidly decreases to zero with increase of $\xi$, varying at small distances, the sub-integral term $w(\mathbf{r}+\xi,\xi)P(\mathbf{r}+\xi)$ can be decomposed in the Fokker-Planck approximation as $w(\mathbf{r}+\xi,\xi)P(\mathbf{r}+\xi,t) \approx w(\mathbf{r},\xi)P(\mathbf{r},t) + \xi\dfrac{\partial}{\partial \mathbf{r}}(w(\mathbf{r},\xi)P(\mathbf{r},t)) + \dfrac{1}{2}\xi_\alpha\xi_\beta \dfrac{\partial^2}{\partial r_\alpha \partial r_\beta}(w(\mathbf{r},\xi)P(\mathbf{r},t))$. In the lack of external fields, $w(\mathbf{r},\xi)$ is a spatially homogeneous function, $w(\mathbf{r}+\xi,\xi) = w(\mathbf{r},\xi) = w(0,\xi) \equiv w(\xi)$, and the kinetic Eq. (1.1) takes form

$$\frac{\partial P}{\partial t} = \frac{\partial}{\partial r_\alpha}\left\{A_\alpha P + \frac{\partial}{\partial r_\beta}(B_{\alpha\beta}P)\right\}, \quad (1.2)$$

where $A_\alpha = \dfrac{1}{\tau_0}\int \xi_\alpha w(\xi)d^3\xi$, $B_{\alpha\beta} = \dfrac{1}{2\tau_0}\int \xi_\alpha\xi_\beta w(\xi)d^3\xi$. Since the spatially homogeneous function $w(\mathbf{r},\xi)$ is also independent of the direction of the relocation vector $\xi$, then $B_{\alpha\beta} = B\delta_{\alpha\beta}$, $A_\alpha = \tau_0^{-1}\int \xi_\alpha w(\xi)\, d^3\xi = \tau_0^{-1}\langle \xi_\alpha \rangle = 0$, resulting in

$$\frac{\partial P}{\partial t} = D\nabla^2 P, \quad (1.3)$$

where

$$D = \frac{1}{6\tau_0}\int \xi^2 w(\boldsymbol{\xi}) d^3\xi = \frac{a^2}{6\tau_0}, \tag{1.4}$$

and $a^2 = \langle \xi^2 \rangle = \int \xi^2 w(\boldsymbol{\xi})\, d^3\xi$ , i.e. $a$ is the mean-square relocation distance in $\tau_0$.

Since Eq. (1.3) was derived neglecting collisions (and coagulations) of particles (in the first order of approximation $n^{1/3}R << 1$), the particles migrate independently from each other, and thus the probability density $P(\mathbf{r},t)$ is a smooth (on the scale $l >> a$) function, which can be considered as a linear superposition, $P(\mathbf{r},t) = \sum_i P_i(\mathbf{r},t)$, of the probability densities (at $\mathbf{r}$ in $t$) of all particles, $i = 1,...,N$, with some initial distribution in space at $t = 0$. Each of these probabilities is described by the Einstein-Fokker equation, which is formally equivalent to Eq. (1.3), but derived for a single particle self-diffusion, cf. [4].

After averaging over the scale of the mean interparticle distance, $l >> n^{-1/3}$, Eq. (1.3) can be transformed into the diffusion equation for the particle concentration,

$$\frac{\partial n}{\partial t} = D\nabla^2 n. \tag{1.5}$$

If particle collisions (and coagulations) are taken into consideration (in the next approximation of $n^{1/3}R << 1$), $P(\mathbf{r},t)$ cannot be considered anymore as a superposition of the probability densities of independently moving particles (described by the Einstein-Fokker equation), and thus generally does not obey Eq. (1.3).

However, in-between two subsequent collisions in a local ensemble of particles the system evolution can be described by Eq. (1.3), starting from the moment of the former collision. During this period a local heterogeneity in the particles distribution density, emerged after the former collision on the length scale of the mean inter-particle distance, $l \approx \bar{r} \approx n^{-1/3}$, tends to relax, in accordance Eq. (1.3), within the characteristic diffusion time $\tau_d \approx n^{-2/3}/6D$ (if $a << l \approx \bar{r}$, which is obviously valid at $a \le R << \bar{r}$). Relaxation to the homogeneous distribution will be completed, if two-particle collisions (with the characteristic time $\tau_c$) are relatively infrequent, $\tau_d << \tau_c$, that is valid practically in the whole range of $n^{1/3}R << 1$, as will be shown in the next Section 1.3.

On the contrary, the coagulation rate equation is defined in terms of the particle concentrations $n(\mathbf{r},t)$ (see the next Section 1.3), and for this reason Eq. (1.5), defined on a larger scale $l >> \bar{r} \approx n^{-1/3}$, should be used in the coagulation theory (e.g., for calculation of the particle fluxes).

The Brownian particles move with the root mean-square heat velocity (in accordance with the equipartition theorem)

$$u_T = \sqrt{\langle \mathbf{u}^2 \rangle} = (3kT/m)^{1/2} = 1.5(kT/\pi\rho R^3)^{1/2}, \tag{1.6}$$

related to the mean thermal speed

$$\bar{c} = (8kT/\pi m)^{1/2} = u_T (3\pi/8)^{-1/2}, \tag{1.6a}$$

where $\rho$ is the mass density of the particles, each particle changing a random direction of its drift for the relaxation time $\tau_0$, which can be estimated from the Langevin theory of Brownian movement [11] as

$$\tau_0 \approx mb = \frac{mD}{kT}, \tag{1.7}$$

where $b$ is the particle mobility, calculated in the Stokes regime ($\mathrm{Re} << 1$) as

$$b = \frac{C_c}{6\pi\eta R}, \tag{1.8}$$

$\eta$ is the liquid viscosity, $C_c$ is the Cunningham slip correction factor for spherical particles, depending on the Knudsen number, $\mathrm{Kn} = \lambda_m / R$, in the form [18]

$$C_c = 1 + \mathrm{Kn}\left( A_1 + A_2 \exp\left( -\frac{A_3}{\mathrm{Kn}} \right) \right), \tag{1.9}$$

with $A_1$ = 1.257, $A_2$ = 0.40 and $A_3$ =1.1 [19], or $A_1$ = 1.165, $A_2$ = 0.483 and $A_3$ = 0.997 obtained in recent experiments [20].

After substitution of Eqs. (1.4) and (1.6) in Eq. (1.7), the mean-square drift distance $a$ can be roughly evaluated as

$$a \approx (6kT/m)^{1/2} \tau_0 = u_T \tau_0 \sqrt{2}. \tag{1.10}$$

Validity of the current approach for description of Brownian particles can be justified, as mentioned above, under the assumption $\tau_m << \tau_0$, where $\tau_m$ is the mean time between stochastic collisions of a particle with the surrounding fluid molecules. This condition can be violated only for very small particles in the same size range as the (gas) molecules. Indeed, from Eq. (1.10) $\tau_0$ is estimated as $\tau_0 = a/\sqrt{2}u_T = a(m/6kT)^{1/2}$, whereas $\tau_m$ is estimated from the gas-kinetic theory as $\tau_m \approx (n_m R^2 u_m)^{-1} \approx (n_m R^2)^{-1} (m_m \pi / 8kT)^{1/2} \approx (n_m R^2)^{-1} (\rho_m \pi^2 R_m^3 / 6kT)^{1/2}$, where $m_m$, $\rho_m$ and $u_m$ are the mass, mass density and thermal velocity of the gas molecules, respectively. Therefore, $\tau_m << \tau_0$ is valid when $aR^2 n_m >> (6\pi/8)^{1/2} (m_m/m)^{1/2}$. Substituting an estimation of the gas molecules mean free path, $\lambda_m \approx (\sqrt{2}\pi R_m^2 n_m)^{-1}$, where $R_m$ is the molecule effective radius, one eventually obtains the limitation on the Knudsen number, $\mathrm{Kn} = \lambda_m / R << (2/3\pi^3)^{1/2} (m/m_m)^{1/2} aR/R_m^2$, which is really not very strong (since the r.h.s. of the inequality is generally very large for Brownian particles) and becomes essential only in the limit when the diffusing particles are in the same size range as the gas molecules (i.e. $\mathrm{Kn} >> 1$). Actually, in this range $C_c \approx 1.2 \cdot \mathrm{Kn}$, and the gas viscosity can be evaluated from the gas-kinetic theory as $\eta \approx (\pi m_m kT)^{1/2} / 4\pi R_m^2$, therefore, $a \approx 3\sqrt{2} D/u_T = (C_c/\pi\eta R)(mkT/6)^{1/2} \approx \mathrm{Kn}(m/m_m)^{1/2} R_m^2/R$, and the above derived restriction takes a simple form, $m/m_m >> 1$, or $R >> R_m$ (if $\rho \approx \rho_m$). Therefore, a conclusion can be derived that the theory of Brownian particle migration can be used in a wide range of the particle sizes, $R_m << R << n^{-1/3}$, in accordance with a more strict justification of the Langevin equation by Mazur and Oppenheim [21].

### 1.3. Coagulation rate equation

In the second order of approximation $n^{1/3} R << 1$, pair-wise collisions of particles during their Brownian migration can be taken into consideration. In this approximation, collisions which occur among any combination consisting of more than two particles, can be ignored (and will be further considered in Section 1.9).

For a continuous size distribution of particles $N(R,t)dR$, the number of particles of radius $R$ to $R + dR$ per unit volume, under an assumption that collided particles are randomly distributed in space and, upon collision, immediately coalesce to form a new particle, the Smoluchowski coagulation equation takes the form

$$\frac{\partial N(R,t)}{\partial t} = \frac{1}{2}\int_0^\infty\int_0^\infty N(R_1,t)N(R_2,t)\delta\left[R - \left(R_1^3 + R_2^3\right)^{1/3}\right]\beta(R_1,R_2)dR_1 dR_2 - N(R,t)\int_0^\infty N(R_1,t)\beta(R,R_1)dR_1, \quad (1.11)$$

where $\beta(R_1, R_2)$ is the collision frequency function. The first term on the right hand side of Eq. (1.11) represents the increase in number density at *R* due to pair-wise coalescence between all particles, and the second term represents the removal of particles of radius *R* due to pair-wise coalescence between particles of radius *R* with particles of all radii.

Under the basic condition of the Smoluchowski theory on spatial homogeneity of the particle distribution, $N(R,\vec{r},t) = N(R,t)$, the kernel $\beta(R_1, R_2)$ can be defined as a number of collisions in unit time per unit volume between two particles of radii $R_1$ and $R_2$ randomly located in space.

*1.3.1. Applicability of the diffusion approach to particles coagulation*

In the case when all particles, located in a medium of infinite extent, are of comparable sizes $\approx R_1$ and their mean concentration $n_1$ obeys the condition $n_1 R_1^3 << 1$, the particles can be considered as point objects ($R_1 << \bar{r}_1$, where $\bar{r}_1 \approx n_1^{-1/3}$ is the mean inter-particles distance), which in accordance with the diffusion equation, Eq. (1.3), tend to relax to a homogeneous spatial distribution.

The situation critically changes in the case when a relatively large particle (with the characteristic size $R_2 >> R_1$) emerges in the ensemble of small particles. The large particle cannot be considered as a point object, if $n_1 R_2^3 \geq 1$. In this case the large particle should be considered as macroscopic with respect to small ones, since its size $R_2$ is much larger than the mean inter-particle distance $\bar{r}_1 \approx n_1^{-1/3}$ in the ensemble of small particles (cf. Section 1.2) and, therefore, an additional (absorbing) boundary condition for diffusion of small particles emerges on the large particle surface. For this reason, the induced by this boundary condition heterogeneity in the spatial distribution of small particles does not tend to disappear with time, as it was in the previous case of comparable size particles, and the steady state concentration profile of small particles around the large particle centre, $n_1(r) = n_1(R_{12}) + (\bar{n}_1 - n_1(R_{12})) \cdot (1 - R_{12}/r)$, is formed at $t >> R_{12}^2 / \pi D_1$, where $R_{12} = R_1 + R_2 \approx R_2$ is the "influence sphere" radius [2]. The diffusion flux of small particles in this concentration profile, $J_{dif}(R_{12}) = 4\pi D_1 R_{12}(\bar{n}_1 - n_1(R_{12})) \approx 4\pi D_1 R_2 \bar{n}_1$, if $n_1(R_{12}) << \bar{n}_1$, determines the condensation rate of small particles in the large particle trap, and, following analysis of [1, 2], the collision frequency function, taking into consideration migration of the traps, eventually takes the form

$$\beta_{dif}(R_1, R_2) = 4\pi(D_1 + D_2)(R_1 + R_2). \tag{1.12}$$

For determination of the applicability range of this result, it should be noted that the characteristic size $l$ of the zone around the large particle in which the small particles concentration varies (and where the diffusion flux is calculated, $J_{dif}(R_{12}) \propto \partial n_1 / \partial r\big|_{r=R_{12}} \approx (n_1(R_{12} + \Delta r) - n_1(R_{12}))/\Delta r$, $\Delta r << R_{12}$), is comparable with $R_{12} \approx R_2$, i.e. $l \approx R_2$. In order to maintain the concentration profile of small particles around the large one, this value must naturally exceed the mean distance $n_1^{-1/3}$ between small particles in the vicinity of the large particle surface (which in its turn is much larger than that far outside the absorbing boundary $\bar{n}_1^{-1/3}$), $R_2 \approx l >> n_1^{-1/3}(R_{12}) >> \bar{n}_1^{-1/3}$, or $\bar{n}_1 R_2^3 >> 1$. This condition naturally coincides with the general requirement to applicability of the diffusion equation, Eq. (1.5), $l >> \bar{r}_1 \approx \bar{n}_1^{-1/3}$.

Therefore, the standard diffusion approach [2] is valid only for coalescence of large particles with small ones (with sufficiently high concentration), $R_1 << \bar{n}_1^{-1/3} << R_2$, and thus cannot be applied to consideration of particles of comparable sizes.

It is also important to note that Eq. (1.12) was derived in neglect of mutual coalescences between small particles, i.e. under an (implicit) assumption that the rate of small particles mutual collisions, $\nu_c(R_1, R_1) = 4\pi(D_1 + D_1)(R_1 + R_1)\bar{n}_1/2 = 8\pi D_1 R_1 \bar{n}_1$ is negligible in comparison with the rate of their condensation in the trap, $\nu_{cd}(R_1, R_2) = 4\pi(D_1 + D_2)(R_1 + R_2)\bar{n}_1 \approx 4\pi D_1 R_2 \bar{n}_1$. It is straightforward to see that this assumption, $\nu_c(R_1, R_1) << \nu_{cd}(R_1, R_2)$, is valid if $R_2 >> R_1$; this additionally confirms the important conclusion concerning applicability of the diffusion approach.

In the opposite case $R_2 \approx R_1 \approx R$ the limit of the point-wise particles restores, which is characterized by the particles tendency to a homogeneous spatial distribution (with the mean concentration $n$). However, after coalescence of two particles (taken into account in the second order of approximation $nR^3 << 1$), the local number of particles alters step-wise (from two to one) that induces a local heterogeneity in the probability density $P(\mathbf{r},t)$ on the length scale of the mean inter-particle distance $\bar{r} \approx n^{-1/3}$. Consequently, a spatially homogeneous distribution of coalescing particles (characterized by a homogeneous probability density) can be assumed only under condition that heterogeneities of such scale, $n^{-1/3}$, rapidly relax by the particles self-diffusion, Eq. (1.3), in-between two subsequent collisions of a particle, and thus do not evolve in heterogeneous distribution of the particles concentration $n(\mathbf{r},t)$ on a larger time scale.

*1.3.2. Diffusion mixing condition*

Therefore, the condition of the particles rapid relaxation, or "mixing", takes the form $\tau_c >> \tau_d$, where $\tau_d \approx n^{-2/3}/6D$ is the characteristic time of the particles (diffusion) redistribution on the length scale of the induced heterogeneities (i.e. of the mean inter-particle distance $n^{-1/3}$) and $\tau_c$ is the mean time between two subsequent collisions of a particle. In this case it may be assumed in calculations that the coalescing particles are randomly distributed in the space, and for this reason, the collision frequency can be correctly calculated in the kinetic approach.

It is shown in [5-7] that in fact coalescence of comparable size (i.e. point-wise) particles occurs in the kinetic regime practically in the whole considered range $n^{1/3}R << 1$. Indeed, the mixing condition, $\tau_c >> \tau_d \approx n^{-2/3}/6D$, can be equally represented in the form $\lambda n^{1/3} >> 1$, where $\lambda \approx (6D\tau_c)^{1/2}$ is the characteristic distance between two subsequent collisions of a particle with other ones, that is generalization of the mean free path definition to the case of the Brownian particles (do not confuse with the mean drift distance $a$, which can be very small in comparison with $\lambda$!). Therefore, the mixing condition has a clear physical sense that $\lambda$ cannot be smaller than the mean inter-particle distance $\bar{r} \approx n^{-1/3}$; this can be also confirmed by direct evaluation of $\tau_c/\tau_d$ in both limiting cases $a >> R$ and $a << R$ (see Section 1.5.5).

Correspondingly, it will be further assumed that (comparable size) particles are randomly distributed in space and, in-between two subsequent collisions of a particle, the random distribution quickly reinstates owing to the particles diffusion relaxation (or mixing). In this case (corresponding to the kinetic regime) the spatial distribution of particle centres $n(\mathbf{r},t)$ can be considered as a homogeneous function characterized by their mean concentration $n(t)$, i.e. $n(\mathbf{r},t) = n(t)$.

In this kinetic regime (characterized by a homogeneous particle distribution, rather than by their concentration profiles) the original multi-particle problem is rigorously reduced to consideration of two-particle collisions. This significantly simplifies the coagulation problem and justifies the phenomenological form of the pair-wise kernel $\beta(R_1,R_2)$ in the Smoluchowski kinetic equation, Eq. (1.11), derived for spatially homogeneous systems.

In fact, after some initial time period during which the self-preserving size distribution of particles is attained (see, e.g. [22]), the majority of particles are concentrated in a relatively narrow size-band (within one order of magnitude, $0.2\bar{R} \le R \le 2\bar{R}$ ) around the mean size $\bar{R}(t)$, where the particle concentration decreases from the maximum value by $\approx 3-4$ orders of magnitude (see Fig. 1.1). This allows excluding from consideration, with a good accuracy, the particle sizes outside this narrow band. On the other hand, the remaining sizes (located within this band), being distributed within one order of magnitude, can be considered as comparable. Therefore, only collisions among comparable size particles (distributed around the mean size) may be taken into consideration, despite formally integration over the particle sizes in the coagulation equation, Eq. (1.11), is extended from $-\infty$ to $+\infty$, inducing some restrictions on applicability of the coagulation equation.

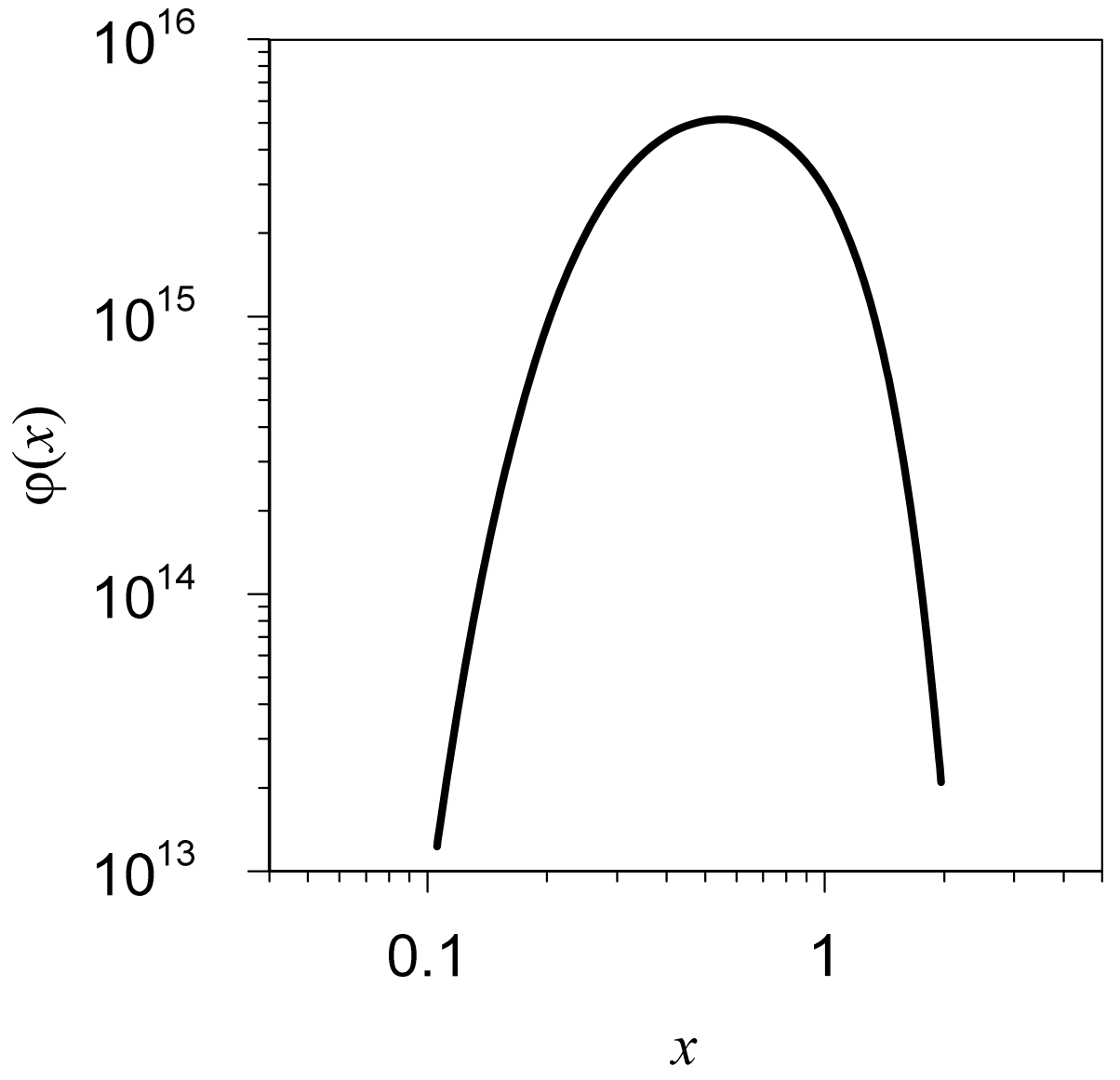

Fig. 1.1. Example of calculated self-preserving size distribution function $\varphi(R/\overline{R})$ for Brownian coagulation for the particle volume fraction $f$ =0.3% in the continuum mode (from ([17]).

Indeed, for a particle of radius $R$ with the concentration $n(R)$ the collision rate has the maximum value for collisions with particles from the central size-band (around the mean size $\overline{R}$ ), e.g., in the continuum regime, $\tau_c \geq \left(\beta(R,\overline{R})n(\overline{R})\right)^{-1} \approx \left(4\pi D\overline{R}n(\overline{R})\right)^{-1}$. Hence the diffusion mixing condition, $\tau_c >> \tau_d \approx n^{-2/3}(R)/6D$, takes the form $n(R)/n(\overline{R}) >> \overline{R}n^{1/3}(R)$, or $n^{2/3}(R)/n^{2/3}(\overline{R}) >> \overline{R}n^{1/3}(\overline{R})$, which coincides with the basic dilution condition, $n^{1/3}R << 1$, for particles from the central size-band, $R \approx \overline{R}$ . For smaller and larger particles the mixing condition fails, owing to a steep dependence of the similarity function $\psi_d(\eta = R/\overline{R}) = n(R)/n(\overline{R})$ outside the central size-band, and thus determines the applicability range of the coagulation equation, Eq. (1.11), justified for mixed, homogeneous systems. This applicability range broadens with decrease of the mean particle concentration, i.e. Eq. (1.11) is valid in a wide size range only for highly diluted systems, $\overline{R}n^{1/3}(\overline{R}) \to 0$.

## 1.4. Coagulation kernel

*1.4.1. High collision frequency ($\tau_c << \tau_0$)*

At first, the case of high collision frequency $\tau_c^{-1}$, corresponding to an inequality $\tau_c << \tau_0$, will be considered. In this case two subsequent collisions of a particle occur within the drift time $\tau_0$, that can be considered in the traditional free molecular (or "ballistic") approximation. In this approximation it is assumed that the particles move straight with their heat velocities $\vec{u}_i$ and randomly change direction of the velocity with the frequency $\tau_0^{-1}$.

Let us consider two particles of radii $R_1$ and $R_2$ migrating in a sample of unit volume. The first ("parent") particle of radius $R_1$ can be surrounded by a sphere with the radius $R_1 + R_2$. If the second particle centre falls into this exclusion zone, coalescence would occur.

During the time step $\delta t$, where $\tau_m << \delta t << \tau_c$, the two particles relocate with respect to each other with the relative velocity vector $\mathbf{u}_{12} = \mathbf{u}_1 - \mathbf{u}_2$. As a result, the exclusion zone also relocates to a distance $|\mathbf{u}_1 - \mathbf{u}_2|\delta t$ and the particle of radius $R_2$ may be swept out by the parent particle of radius $R_1$ with the probability, $\delta V = \pi(R_1 + R_2)^2 |\mathbf{u}_1 - \mathbf{u}_2| \delta t$. Correspondingly, the probability of the two particles coalescence in $\delta t$ is equal to $\langle \delta V \rangle = \pi(R_1 + R_2)^2 \langle |\mathbf{u}_1 - \mathbf{u}_2| \rangle \delta t$, where averaging of the swept volume is carried out over the Maxwell distribution of the two independent heat velocity vectors $\mathbf{u}_1$ and $\mathbf{u}_2$, resulting in $\langle |\mathbf{u}_1 - \mathbf{u}_2| \rangle = \sqrt{(8kT/\pi)(m_1^{-1} + m_2^{-1})}$ $= \sqrt{(6kT/\pi^2 \rho)(R_1^{-3} + R_2^{-3})}$.

If there are $N_2$ particles of radius $R_2$ in a sample of unit volume randomly distributed in the space (as above assumed), the probability of coalescence of the parent particle with a particle of radius $R_2$ reduces to $N_2 \langle \delta V \rangle$. In the first order of approximation $\delta t / \tau_c << 1$, one can neglect variation of $N_2$ in the time step $\delta t$ (occurred owing to coalescences of particles of radius $R_2$ with other particles). Besides, $N_2$ is considered to be small enough that one can ignore collisions which occur in $\delta t$ with more than one particles of radius $R_2$. This implies that only one particle can coalesce with the parent particle with the probability $N_2 \langle \delta V \rangle = N_2 \pi (R_1 + R_2)^2 \langle |\mathbf{u}_1 - \mathbf{u}_2| \rangle \delta t$ $= N_2 (R_1 + R_2)^2 \sqrt{(6kT/\rho)(R_1^{-3} + R_2^{-3})} \delta t$.

Furthermore, the number of collisions between particles of radii $R_1$ and $R_2$ in $\delta t$, if there are $N_1$ particles of radius $R_1$ randomly distributed in the sample of unit volume, is $N_{12} = \delta t N_1 N_2 (R_1 + R_2)^2 \sqrt{(6kT/\rho)(R_1^{-3} + R_2^{-3})}$. If there are more than two size groups of particles in the sample, then the number of collisions between each pair of groups $p$ and $q$ in $\delta t$ is $N_{pq} = \delta t N_p N_q (R_p + R_q)^2 \sqrt{(6kT/\rho)(R_p^{-3} + R_q^{-3})}$. Since all $N_i$ are small enough, we can ignore collisions which occur among any combination consisting of more than two particles.

Therefore, for a continuous size distribution of particles, $n(R)dR$, the collision frequency function in this kinetic regime takes the standard free molecular form [4]

$$\beta_{fm}(R_1,R_2)=\left(\frac{6kT}{\rho}\right)^{1/2}(R_1+R_2)^2\left(\frac{1}{R_1^3}+\frac{1}{R_2^3}\right)^{1/2}. \qquad (1.13)$$

In the mean field approximation taking into consideration monodisperse particles size distribution, the probability of the two particles coalescence in $\delta t$ becomes equal to $(3kT/\rho)^{1/2}8NR^{1/2}\delta t$. Considering each particle in turn, the total rate of loss of particles by coalescence is given by:

$$\frac{dn}{dt}=-4\left(\frac{3kT}{\rho}\right)^{1/2}n^2R^{1/2}=-\frac{n}{\tau_c}, \qquad (1.14)$$

where the additional factor of ½ is introduced to avoid counting each interaction twice, and $\tau_c=\left((3kT/\rho)^{1/2}4nR^{1/2}\right)^{-1}$ is the coalescence time corresponding to the characteristic period between two subsequent collisions of a particle. This confirms that variation of the particles concentration can be neglected in the first order of approximation $\delta t/\tau_c<<1$ during the time step $\delta t$, as above assumed.

Therefore, the applicability range of Eq. (1.13) derived under condition $\tau_c<<\tau_0$, where $\tau_0\approx a/\sqrt{2}u_T=aR^{3/2}(2\pi\rho/9kT)^{1/2}$, takes the form $a/R>>(3/2\pi)^{1/2}(4nR^3)^{-1}$, or $(R/R_m)^{1/2}<<\left(16\sqrt{\pi}/3\right)(\rho/\rho_m)^{1/2}nR^3C_c$, which should additionally obey the relationship $m/m_m>>1$ or $R/R_m>>(\rho_m/\rho)^{1/3}$, derived in Section 1.2 (from the condition $\tau_m<<\tau_0$). This can take place under condition of very high Knudsen numbers, $\mathrm{Kn}>>(\rho_m/\rho)^{5/6}(16nR^3)^{-1}$.

However, as will be shown in Section 1.5.2, in a more general case $a>>R$ when a particle makes many jumps in-between its two subsequent collisions with other particles ($\tau_c>>\tau_0$), the swept volume per unit time is nearly a constant (in time) value, and for this reason it coincides (in the first order approximation of $R/a<<1$) with the above calculated value $\delta V/\delta t=\pi(R_1+R_2)^2\langle|\vec{u}_1-\vec{u}_2|\rangle$, representing the ratio of the mean swept volume during one jump (designated below in Section 1.5 as $\delta V_0$) to the jump period $\tau_0$. Therefore, the applicability range of Eq. (1.13) can be extended to this case, $a>>R$, which corresponds to a more realistic condition, $R/R_m<<0.5(\rho/\rho_m)^{1/3}(n_m^{1/3}R_m)^{-2}$, or $\mathrm{Kn}>>0.5(\rho_m/\rho)^{1/3}(n_m^{1/3}R_m)^{-1}$.

### *1.4.2. Low collision frequency ($\tau_c>>\tau_0$)*

In the case $\tau_c>>\tau_0$ a particle makes many diffusion drifts (or jumps) between its two subsequent collisions with other particles. In order to calculate the value of the collision frequency in this regime, a relatively large time step $\delta t>>\tau_0$ should be chosen. On the other hand, it should be sufficiently small in comparison with $\tau_c$, in order to ignore variation in $\delta t$ of the mean number concentration of surrounding particles (owing to their mutual coalescences) during the time step, $\tau_c >> \delta t >> \tau_0$. Besides, this time step $\delta t$ should be large enough in comparison with the diffusion relaxation (or mixing) time $\tau_d \approx n^{-2/3}/6D$, in order to sustain the main assumption of the kinetic regime on random (homogeneous) distribution of coalescing particles, $\tau_c >> \delta t >> \tau_d$.

Again, let us consider two particles of radii $R_1$ and $R_2$ randomly located in a sample of unit volume. The first ("parent") particle of radius $R_1$ can be surrounded by a sphere with the radius $R_1 + R_2$. If the second particle centre is located in this exclusion zone, coalescence would occur.

As shown in [1, 2], the relative displacements between two particles describing Brownian motions independently of each other and with the diffusion coefficients $D_1$ and $D_2$ also follow the law of Brownian motion with the diffusion coefficient $D_1 + D_2$. Indeed, since the two particles are not correlated in motion, $\langle(\mathbf{r}_1 \cdot \mathbf{r}_2)\rangle = 0$, the Einstein equation for the relative displacement of two particles gives

$$D_{12} = \frac{\langle(\mathbf{r}_1 - \mathbf{r}_2)^2\rangle}{6t} = \frac{\langle(\mathbf{r}_1)^2\rangle}{6t} - \frac{\langle(\mathbf{r}_1 \cdot \mathbf{r}_2)\rangle}{3t} + \frac{\langle(\mathbf{r}_2)^2\rangle}{6t} = \frac{\langle(\mathbf{r}_1)^2\rangle}{6t} + \frac{\langle(\mathbf{r}_2)^2\rangle}{6t} = D_1 + D_2. \quad (1.15)$$

Therefore, in order to calculate the probability of collisions between the two particles, one can equivalently consider the second particle as immobile whereas the first one migrating with the effective diffusion coefficient, $D_{12} = D_1 + D_2$.

In the first approximation (corresponding to the *simple* random walk theory) it is assumed that the effective (mobile) particle of radius $R_1 + R_2$ drifts to the fixed distance $a_{12}$ in random directions with the frequency $\nu_{12} = \tau_{12}^{-1}$, those are unknown (searched) values obeying the relationship for the particle diffusivity, $D_{12} = a_{12}^2/6\tau_{12}$.

As a result of a jump, the exclusion zone also relocates to the distance $a_{12}$ and opens the possibility that the second (immobile) particle with its centre located in a zone with the volume

$$\delta V_0 = \pi a_{12}(R_1 + R_2)^2, \quad (1.16)$$

may be swept out by the mobile particle (dashed zone in Fig. 1.2).

It is important to note that this result does not depend on the ratio between $a$ and $R$, and therefore is valid in the whole considered range $R_m << R << n^{-1/3}$.

The model parameters $a_{12}$ and $\tau_{12}$ will be self-consistently determined (below in Sections 1.5.2-1.5.4) by comparison of the collision frequency calculated in the simple random walk approach at high Knudsen numbers, $\mathrm{Kn} \to \infty$, with that calculated in the free molecular approach (considering the original particles moving straight with their heat velocities).

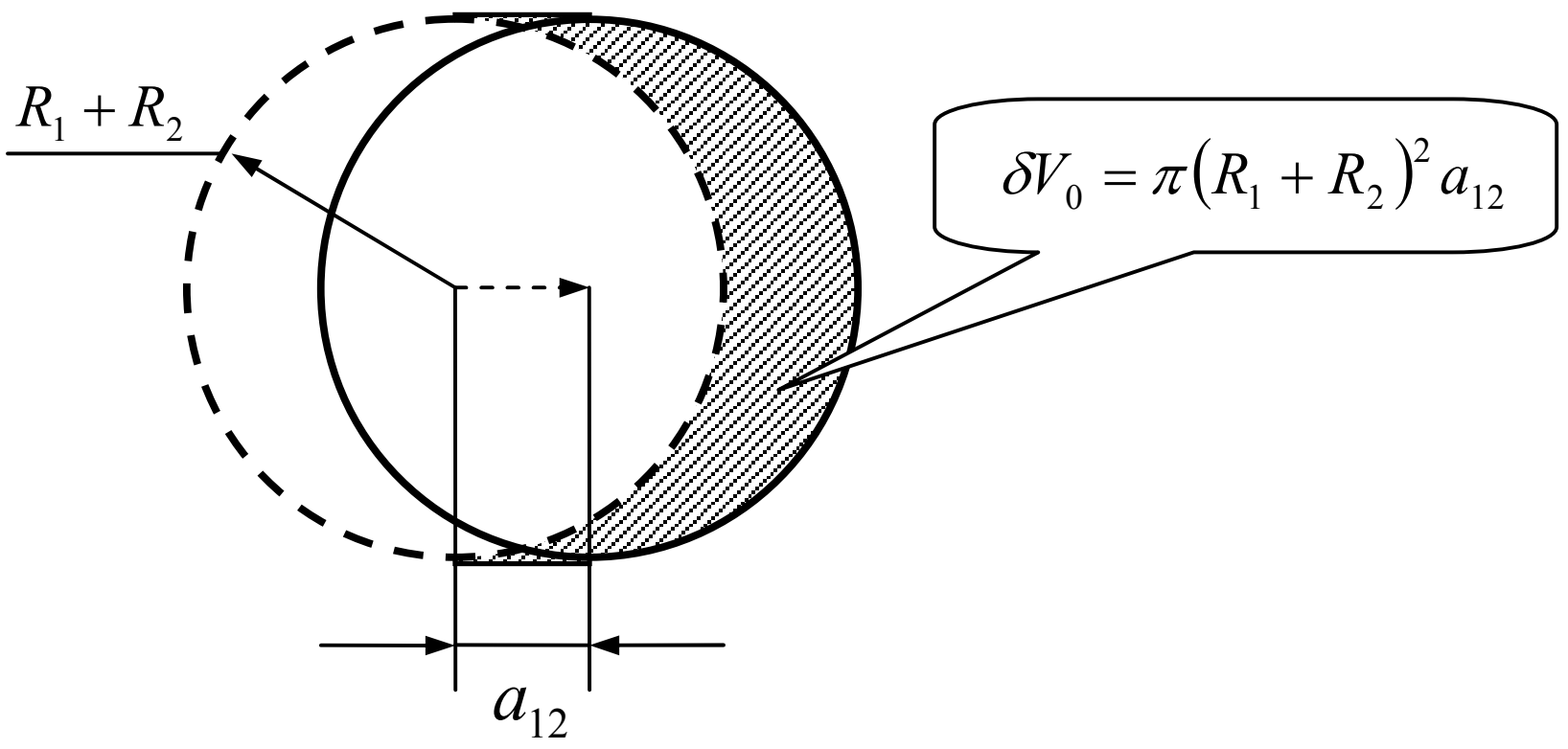

Fig. 1.2. Schematic representation of the swept zone for colliding particles of radii $R_1$ and $R_2$.

*1.4.3. Continuum mode ($a << R$)*

During the time step $\delta t >> \tau_{12}$ the effective mobile particle (of radius $R_1 + R_2$) makes many jumps, $k = \delta t/\tau_{12} >> 1$, in random directions, however, the total swept zone volume $\delta V$ that determines the probability of the two particles coalescence in $\delta t$, will be smaller than $k\,\delta V_0 = \delta V_0 \delta t/\tau_{12}$, owing to strong overlapping of the swept zone segments at $a_{12} << R_{12}$. This limit corresponds to the continuum mode of the kinetic regime, realized at low Knudsen numbers and characterized by a random spatial distribution of particles, quickly reinstated in-between two subsequent collisions of a particle (the validity of this latter condition will be checked below in Section 1.5.5).

If the particle concentrations are sufficiently low, the only events which occur (during the chosen time step $\delta t$) with non-negligible probability are those corresponding to the case where there is no particle centre contained in the swept volume $\delta V_{12}$ and the case where there is one particle centre contained in $\delta V_{12}$. The Poisson relation giving the probability of having the centre of one particle of radius $R_2$ in the swept volume $\delta V_{12}$ then reduces to $N_2 \delta V_{12}$, where $N_2$ is the number of particles of radius $R_2$ per unit volume. Therefore, the number of coalescences between particles of radii $R_1$ and $R_2$ in unit time (per unit volume), if there are $N_1$ particles of radius $R_1$ per unit volume, is $(\delta V_{12}/\delta t)N_1 N_2$ (which will be apparently smaller than $\delta V_0 N_1 N_2/\tau_{12}$). If there are more than two size groups of particles in the sample, then the rate of collisions between each pair of groups $p$ and $q$ is $(\delta V_{pq}/\delta t)N_p N_q$. If $N_p$ and $N_q$ are small enough, collisions which occur during the chosen time step $\delta t$ among any combination consisting of more than two particles can be ignored.

In order to calculate the volume $\delta V_{12}$ swept in $\delta t$, let us uniformly (in random) fill up the space with auxiliary (fictitious) point immobile particles ("markers") of radius $R_* \to 0$ with a relatively high concentration, $n_* >> R_{12}^{-3}$. In order to adequately resolve a fine structure (with the characteristic length of $a_{12} << R_{12}$) of the swept zone, the markers concentration $n_*$ should additionally obey the condition that the number of swept markers $\delta N_*^{(0)}$ during one jump must be large, $\delta N_*^{(0)} = \pi R_{12}^2 a_{12} n_* >> 1$, or $n_* >> \left(\pi R_{12}^2 a_{12}\right)^{-1}$. In this case the swept volume can be calculated as the total number $\delta N_*$ of the swept markers divided by their concentration, $\delta V_{12} = \delta N_* / n_*$.

In its turn, for the same reasons (concerning relative displacements of diffusing particles), calculation of the sweeping rate of randomly distributed immobile markers by a large particle of radius $R_{12}$ migrating with the diffusivity $D_{12}$ is equivalent to calculation of the condensation rate of the mobile markers migrating with the diffusivity $D_{12}$ in the immobile trap of radius $R_{12}$ (see Appendix A).

Owing to $n_* R_{12}^3 >> 1$, this problem of the (point-wise) markers condensation in the large (macroscopic) trap can be adequately solved in the continuum approach of [2], as above explained in Section 1.3.1. In this approach the total number of swept markers in $\delta t$ is equal to $\delta N_* = 4\pi D_{12} R_{12} n_* \left(\delta t + 4R_{12}\sqrt{\delta t/\pi D_{12}}\right)$, and the swept volume per unit time in this case is equal to $\left(\delta V_{12}/\delta t\right) = n_*^{-1}\left(\delta N_*/\delta t\right) = 4\pi\left(D_1 + D_2\right)\left(R_1 + R_2\right) = 4\pi D_{12} R_{12}$, if the time step is sufficiently large, $\delta t >> 16 R_{12}^2 / \pi D_{12}$. In particular, this implies that the ratio of $\left(\delta V_{12}/\delta t\right)\tau_{12}$ to $\delta V_0 = \pi a_{12} R_{12}^2$ is equal to $\left(\delta V_{12}/\delta t\right)/\left(\delta V_0/\tau_{12}\right) = 2a_{12}/3R_{12}$.

The spatial variation of the markers concentration occurs on the length scale $l$ which is comparable with $R_{12}$ (see Section 1.3.1), i.e. $l \approx R_{12}$. In accordance with the condition of the diffusion equation applicability, $a << l$ (see Section 1.2), this result is valid only in the case $a_{12} << R_{12}$.

It is straightforward to see that the necessary condition $\delta t >> 16 R_{12}^2/\pi D_{12}$ is valid for the chosen time step $\delta t >> \tau_d \approx n^{-2/3}/6D$ under an assumption $\tau_d \ge 16 R_{12}^2/\pi D_{12}$ or $n^{1/3} R \le \left(\pi/96\right)^{1/2} \approx 0.2$, that is in agreement with the basic requirement $n^{1/3} R << 1$. However, the other necessary condition $16 R_{12}^2/\pi D_{12} << \delta t << \tau_c \approx \left(8\pi D R n\right)^{-1}$ induces a more strong restriction on the concentration, $n^{1/3} R << 0.2$. In this case, the number of coalescences $\left(\delta V_{12}/\delta t\right) N_1 N_2$ between particles of radii $R_1$ and $R_2$ in the unit time becomes equal to $4\pi\left(D_1 + D_2\right)\left(R_1 + R_2\right) N_1 N_2$.

Therefore, the collision frequency function in the kinetic equation, Eq. (1.11), takes the form,

$$\beta_{kin}^{(con)}\left(R_1, R_2\right) = 4\pi\left(D_1 + D_2\right)\left(R_1 + R_2\right), \qquad (1.17)$$

or, in the limit of small Knudsen numbers, $\mathrm{Kn} << 1$ (see Eqs. (1.7) – (1.9)), when $C_c \to 1$,

$$\beta_{kin}^{(con)}(R_1, R_2) = \frac{2kT}{3\eta}\left(\frac{1}{R_1} + \frac{1}{R_2}\right)(R_1 + R_2). \qquad (1.17a)$$

More accurately, one should split the integrals in Eq. (1.11) into two parts, $\int_0^\infty\int_0^\infty dR_1 dR_2 = \int_0^{n^{-1/3}}\int_0^{n^{-1/3}} dR_1 dR_2 + \int_{n^{-1/3}}^{\infty}\int_0^{n^{-1/3}} dR_1 dR_2$, and to use the kernel $\beta_{kin}^{(con)}$ from Eq. (1.17) (derived for particles of comparable sizes under condition $n^{1/3}R << 1$) in the first part and the kernel $\beta_{dif}$ from Eq. (1.12) (correct for coalescence of small and large particles) in the second part (and neglecting coalescences among large, and thus very slow, particles with $R >> n^{-1/3}$). However, since the two kernels $\beta_{kin}^{(con)}$ and $\beta_{dif}$ formally coincide, Eq. (1.11) can be used with a good accuracy.

It should be outlined that the coincidence of Eqs. (1.12) and (1.17) is fortuitous, since Eq. (1.12) was derived in the diffusion regime (by consideration of concentration profiles of small particles around large ones), whereas Eq. (1.17) was derived in the kinetic regime (by consideration of uniform spatial distribution of Brownian particles). The nature of this internal symmetry will be revealed below in Section 1.6.2.

### *1.4.4. Free molecular mode ( $a >> R$ )*

In the opposite limit $a_{12} >> R_{12}$ one can neglect the mean relative volume of the swept zone segments intersections $\propto R_{12}^3/\delta V_0 \propto R_{12}/a_{12} << 1$. In this approximation the mean volume swept by the exclusion zone of radius $R_{12}$ per unit time, $\delta V/\delta t$, is a constant value equal to the sweeping rate during the jump period, $\delta V_0/\tau_{12}$. For this reason, the above applied requirement to the time step $\delta t >> \tau_0 \approx \tau_{12}$ for calculation of the collision rate (see Section 1.5) is not anymore necessary, i.e. it can be correctly calculated within one jump period.

In accordance with the free molecular (or ballistic) approach traditionally applied to consideration of particles movement and collisions on the time scale of one jump period (see Section 1.4), it is assumed that the original particles move straight during their jump periods. For this reason, the probability $\delta P_{12}^*$ of a collision in $\delta t^* << \tau_{12}$ of two original particles of radii $R_1$ and $R_2$, migrating in a sample of unit volume, can be considered as a constant value, equal to the sweeping rate of the effective mobile particle (of radius $R_{12}$), $\delta P_{12}^* = \delta V_0/\tau_{12} = \pi R_{12}^2 a_{12}/\tau_{12}$. On the other hand, this probability can be calculated in the free molecular approach (Section 1.4) as $\delta P_{12}^* = \pi R_{12}^2\sqrt{(8kT/\pi)(m_1^{-1} + m_2^{-1})} = \pi R_{12}^2\sqrt{(\bar{c}_1^2 + \bar{c}_2^2)}$, i.e. $\bar{c}_{12} = a_{12}/\tau_{12} = \sqrt{(\bar{c}_1^2 + \bar{c}_2^2)}$, where $\bar{c}_1$ and $\bar{c}_2$ are the mean thermal speeds of the two original particles defined in Eq. (1.6a). Correspondingly, for the effective diffusivity $D_{12}$, defined in Section 1.5, one obtains $D_{12} = a_{12}^2/6\tau_{12} = a_{12}\bar{c}_{12}/6$, or $a_{12} = 6D_{12}/\bar{c}_{12}$.

Therefore, the total swept volume $\delta V$ (after $k = \delta t/\tau_{12} >> 1$ jumps) is equal to $k\,\delta V_0 = \delta V_0 \delta t/\tau_{12}$, and the number of coalescences $(\delta V_{12}/\delta t)N_1 N_2$ between particles of radii $R_1$ and $R_2$ (with the number densities $N_1$ and $N_2$, respectively) in the unit time is equal to $N_1 N_2\, \delta V_0 / \tau_{12} = N_1 N_2 (R_1 + R_2)^2 \sqrt{(6kT/\rho)(R_1^{-3} + R_2^{-3})}$. The kernel of Eq. (1.11), $\beta_{kin}^{(fm)}$, in this case is equal to $\delta V_0 / \tau_{12}$ that coincides with the free molecular expression for $\beta_{fm}$, Eq. (1.13), i.e.

$$\beta_{kin}^{(fm)} = \delta V_0 / \tau_{12} = (R_1 + R_2)^2 \sqrt{(6kT/\rho)(R_1^{-3} + R_2^{-3})}. \qquad (1.18)$$

Hence, this case corresponds to the free molecular mode of the kinetic regime and is realized at high Knudsen numbers, $\mathrm{Kn} >> 0.5 \cdot (\rho_m/\rho)^{1/3} (n_m^{1/3} R_m)^{-1}$.

*1.4.5. Transition mode ($a \approx R$)*

In the transition interval $a \approx R$, which also belongs to the kinetic regime (characterized by homogeneous spatial distribution of particles), the collision rate can be calculated similarly to the two above considered limiting cases by evaluation of the mean swept volume per unit time, $\beta_{kin} = (\delta V / \delta t)$, however, in the numerical approach. In this approach the exact values of $\beta_{kin}$ can be calculated at different $a_{12}/R_{12}$ and then approximated with an analytical expression [7].

For the numerical evaluation of the mean swept volume per unit time, $\delta V / \delta t$, a random migration of a particle of the radius $R_{12}$ with the fixed jump distance $a_{12}$ and jump frequency $\nu_{12} = \tau_{12}^{-1}$ is numerically generated. The randomly generated data describe the subsequent positions of the particle centre trajectory, which can be further used for calculation of the swept volume. For this calculation a similar to the above described procedure of random spatial distribution of auxiliary point immobile particles (markers) with a relatively high concentration, $n_* >> (\pi R_{12}^2 a_{12})^{-1}$, is numerically realized using the Monte Carlo method. Each marker found in the swept volume is counted only once.

For each number of jumps $k$, several (up to 15) random trajectories and for each trajectory, several (up to 5) realizations of markers random spatial distribution are generated. Numbers of trajectories and markers distributions are increased until the calculated swept volume, averaged over these realizations, become insensitive to further increase of these numbers.

The number of jumps $k = \delta t / \tau_{12}$ is increased until the ratio of $(\delta V / \delta t)\tau_{12}$ to $\delta V_0$ attains a steady-state value, which in accordance with the above presented analytical calculations has to be equal to $2a_{12}/3R_{12}$ in the limit $a_{12} << R_{12}$ and to 1 in the limit $a_{12} >> R_{12}$. The minimum number of jumps necessary for attainment of the steady-state regime smoothly decreases from $k_{\min} >> (96/\pi)(R_{12}/a_{12})^2$ (corresponding to $\delta t >> 16 R_{12}^2 / \pi D_{12}$) at large $R_{12}/a_{12} >> 1$ (see Section 1.5.1) to $k_{\min} = 1$ at small $R_{12}/a_{12} << 1$ (see Section 1.5.2). Therefore, in the transition interval $a_{12}/R_{12} \approx 1$ the inequality $k >> 30 (R_{12}/a_{12})^2$ is a conservative requirement for the time step, that is confidently valid under the necessary condition $\delta t >> \tau_d \approx n^{-2/3}/6D$, or $k >> (n^{-1/3}/a_{12})^2$ (imposed in Section 1.5 and confirmed below in Section 1.5.5) along with the basic assumption $n^{1/3} R << 1$, since in this case $k >> (n^{-1/3}/a_{12})^2 \geq 30 (R_{12}/a_{12})^2$.

From an obvious geometry ("scaling") consideration it is clear that the steady-state value of the mean swept volume per unit time depends only on the ratio $a_{12}/R_{12}$ (rather than on $a_{12}$ and $R_{12}$ separately), that is confirmed by numerical calculations.

Examples of numerical calculations for dependence of the inverse (for convenience of graphical representation) ratio $(\delta V_0/\tau_{12})/(\delta V_{12}/\delta t)$, that is equal to $\beta_{kin}^{(fm)}/\beta_{kin}$, on the number of jumps $k$ for several values of the parameter $R_{12}/a_{12}$ are presented in Fig. 1.3. For large $R_{12}/a_{12} \geq 10$, the calculated steady-state values are in a satisfactory agreement (with the error < 1%) with the analytically predicted (in the limit $R_{12}/a_{12} >> 1$) values, $1.5R_{12}/a_{12}$.

The steady-state values of the ratio $(\delta V_0/\tau_{12})/(\delta V_{12}/\delta t)$ calculated in a wide range of the parameter values, $0.1 \leq a_{12}/R_{12} \leq 50$, are collected in Table 1.1 and presented below in Fig. 1.4 (by centers of circles). Using these data, an interpolation curve was searched by fitting to the calculated points and following the general requirements derived from the analytical consideration.

Since the transition regime belongs to the kinetic regime characterized by homogeneous spatial distribution of particles, in the case of a finite sticking probability for two-particle collisions, $P_{12} = \exp(-E_a/kT) \leq 1$, the coagulation rate should be calculated as the collision rate multiplied by this probability factor,

$$\beta'_{kin} = P_{12}(\delta V_{12}/\delta t). \qquad (1.19)$$

Table 1.1. Steady-state values of $(\delta V_0/\tau_{12})/(\delta V_{12}/\delta t) = \left(\beta_{kin}/\beta_{kin}^{(fm)}\right)^{-1}$ calculated in the *simple* random walk approach at different values of $a_{12}/R_{12} = 6D_{12}/\bar{c}_{12}R_{12}$.

| $a_{12}/R_{12}$ | 0.05 | 0.1 | 0.2 | 0.3 | 0.4 | 0.5 | 2/3 | 5/6 | 1.0 | 1.25 |
|---|---|---|---|---|---|---|---|---|---|---|
| $\frac{\delta V_0/\tau_{12}}{\delta V_{12}/\delta t}$ | 30.00 | 15.00 | 7.72 | 5.38 | 4.22 | 3.49 | 2.82 | 2.39 | 2.13 | 1.81 |
| $a_{12}/R_{12}$ | 5/3 | 2.0 | 10/3 | 5.0 | 10.0 | 15.0 | 20.0 | 30.0 | 40.0 | 50.0 |
| $\frac{\delta V_0/\tau_{12}}{\delta V_{12}/\delta t}$ | 1.59 | 1.47 | 1.21 | 1.12 | 1.05 | 1.03 | 1.01 | 1.01 | 1.00 | 1.00 |

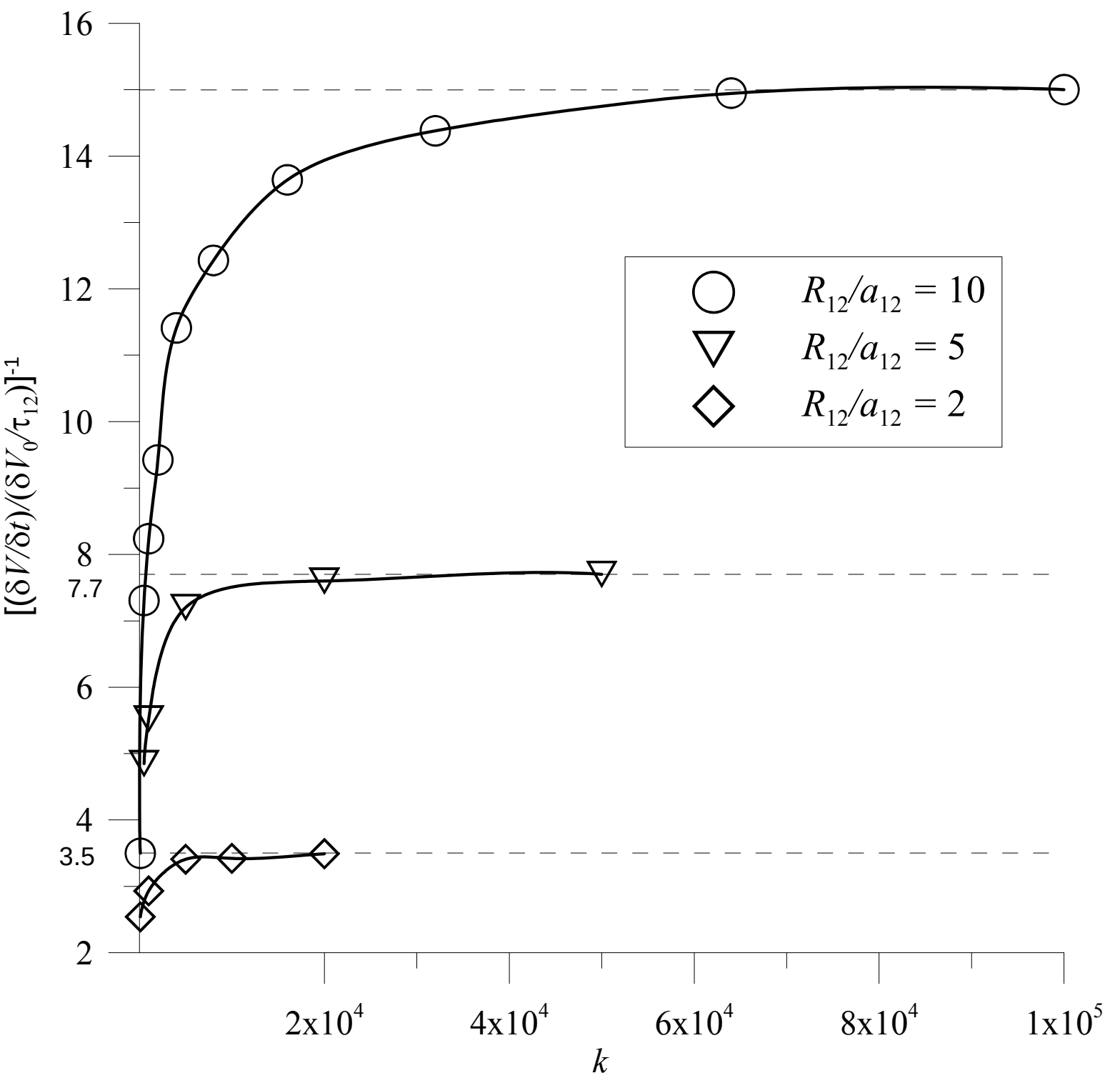

Fig. 1.3. Calculated dependence of the inverse relative mean sweeping rate $\left[\left(\delta\langle V_{12}\rangle/\delta t\right)/\left(\delta V_0/\tau_{12}\right)\right]^{-1}$ on the number of jumps $k$.

*1.4.6. Interpolation formulas*

Namely, taking into account (as above explained) that $\left(\delta V_{12}/\delta t\right)/\left(\delta V_0/\tau_{12}\right)$ depends only on the ratio $a_{12}/R_{12}$ and varies from $2a_{12}/3R_{12}$ (at $a_{12} << R_{12}$) to 1 (at $a_{12} >> R_{12}$), the calculated value should be approximated by an analytical expression as a function of the unique parameter $a_{12}/R_{12}$. Since $\beta_{kin}^{(con)} = \left(\delta V_0/\tau_{12}\right)\left(2a_{12}/3R_{12}\right)$ and $\beta_{kin}^{(fm)} = \left(\delta V_0/\tau_{12}\right)$, this parameter is equal to $a_{12}/R_{12} = 1.5\left(\beta_{kin}^{(con)}/\beta_{kin}^{(fm)}\right)$, therefore, the interpolation expression for $\beta_{kin}$ must be searched in the form

$$\beta_{kin} = \beta_{kin}^{(con)} f_1\left(a_{12}/R_{12}\right) = \beta_{kin}^{(con)} \tilde{f}_1\left(\beta_{kin}^{(con)}/\beta_{kin}^{(fm)}\right), \quad (1.20)$$

or in the equivalent form

$$\beta_{kin} = \beta_{kin}^{(fm)} f_2\left(a_{12}/R_{12}\right) = \beta_{kin}^{(fm)} \tilde{f}_2\left(\beta_{kin}^{(con)}/\beta_{kin}^{(fm)}\right), \quad (1.20a)$$

which can be also represented in terms of the independent set of the three parameters $R_{12} = R_1 + R_1$, $D_{12} = D_1 + D_1$ and $\bar{c}_{12} = \sqrt{\left(\bar{c}_1^2 + \bar{c}_1^2\right)} = \sqrt{\left(8kT/\pi\right)\left(m_1^{-1} + m_2^{-1}\right)}$ as

$$\beta_{kin} = \pi R_{12}^2 \bar{c}_{12} \tilde{f}_2 \left( 4D_{12} / R_{12} \bar{c}_{12} \right), \qquad (1.20b)$$

since $\beta_{kin}^{(fm)} = \pi R_{12}^2 \bar{c}_{12}$ and $\beta_{kin}^{(con)} = 4\pi D_{12} R_{12}$.

On the other hand, these forms, Eqs. (1.20) and (1.20a), can be specified more definitely by analytical consideration of the markers condensation in the trap of radius $R_{12}$ (in the reformulated problem of the immobile particle and point markers migrating by random walks with the diffusivity $D_{12}$). Indeed, in the limiting case $a_{12} << R_{12}$, the markers diffusion is the rate determining step, whereas in the opposite case $a_{12} >> R_{12}$, when the jump distance of the markers significantly exceeds the length scale of their concentration heterogeneity $l \approx R_{12}$ induced by the trap of radius $R_{12}$, the markers migrates in the kinetic ("free molecular") regime. Similarly to the classical problem of vapour molecules condensation in a large immobile trap [4], the transition regime for the markers with $a_{12} \approx R_{12}$ can be described (however, only qualitatively, see Section 1.6.2 below) by flux matching at the adsorbing sphere radius, $R_{abs} = R_{12} + \Delta_{12}$, separating zones of the different regimes of the markers migration, kinetic (inside the absorbing sphere) and diffusion (outside the sphere).

The general solution of the flux matching problem can be searched in the form $\beta_{kin} = \beta_{kin}^{(con)} \varphi(\tilde{\Delta}, \Gamma)$, where $\Gamma = a_{12}/R_{12}$ and $\tilde{\Delta} = \Delta_{12}/R_{12}$, which should additionally obey the general restriction of Eq. (1.20), thus $\tilde{\Delta}$ has to be searched as a function of $\Gamma$.

Owing to an uncertainty in determination of $\Delta_{12}$, various semi-empirical models for condensation of point particles in a trap were proposed (see, e.g., [23]), which in the simplest approach can be reduced to the general form (with various sets of parameters ($B_1$ and $B_2$) [24]

$$\beta_{kin} = \beta_{kin}^{(con)} \frac{1 + B_1 \Gamma}{1 + B_2 \Gamma + (2/3) B_1 \Gamma^2}, \qquad (1.21)$$

where $\Gamma = a_{12}/R_{12}$.

In order to extrapolate the results of the point particles condensation problem to consideration of a polydisperse system of large particles, the traditional models use additional semi-empirical assumptions connected with an uncertainty in determination of the "mean free path" $a_{12}$.

Besides, the flux matching theory of Fuchs [4] cannot be justified for consideration of comparable size particles. Indeed, applicability of the continuum diffusion theory outside the absorbing sphere can be grounded, following the above presented analysis in Section 1.3.1, only in the case when the absorbing sphere radius is much larger than the mean inter-particle distance, $\bar{r} << R_{abs}$. Therefore, in the transition regime, $R_{12} \approx \Delta_{12} \approx a_{12}$, the traditional semi-empirical approach to particles coagulation based on the flux matching theory, cannot be applied to consideration of comparable size particles, $R_1 \approx R_2 \approx 0.5 R_{12} << \bar{r}$, since in this case the opposite condition, $\bar{r} >> R_{abs}$, is realized, and a good coincidence with the more rigorous results of the kinetic approach in the case of comparable size particles (presented below) is also fortuitous (as in the continuum mode).

In the new approach these problems are consistently resolved using for the parameter $a_{12}/R_{12} = 1.5\left(\beta_{kin}^{(con)}/\beta_{kin}^{(fm)}\right)$ the analytical expressions, Eqs. (1.17) and (1.18), which allow explicit calculation of $a_{12}$:

$$a_{12} = \frac{\pi(D_1 + D_2)}{\sqrt{(kT/6\rho)(R_1^{-3} + R_2^{-3})}} = \frac{6(D_1 + D_2)}{\sqrt{(\bar{c}_1^2 + \bar{c}_2^2)}} = \frac{6D_{12}}{\bar{c}_{12}}, \tag{1.22}$$

where $\bar{c}_i = (8kT/\pi m_i)^{1/2} = (6kT/\pi^2 \rho R_i^3)^{1/2}$, in accordance with consideration in Section 1.5.2.

This value of $a_{12}$ is different from that obtained in the traditional approach by formal extension of the mean free path expression for vapour molecules in the condensation problem [25], $\lambda = 3D/\bar{c}$, to the considered problem of polydisperse particles coagulation (see, e.g. [23])

$$\lambda_{12} = 3D_{12}/\bar{c}_{12}\,. \tag{1.23}$$

In terms of the three basic parameters $R_{12}$, $D_{12}$ and $\bar{c}_{12}$, Eq. (1.21) can be equivalently presented in the form

$$\beta_{kin} = 4\pi D_{12} R_{12} \frac{(2/3) + B_1\Gamma_1}{(2/3) + B_2\Gamma_1 + B_1\Gamma_1^2} = \pi R_{12}^2 \bar{c}_{12} \frac{(2/3)\Gamma_1 + B_1\Gamma_1^2}{(2/3) + B_2\Gamma_1 + B_1\Gamma_1^2}, \tag{1.24}$$

where $\Gamma_1 = \beta_{kin}^{(con)}/\beta_{kin}^{(fm)} = 4D_{12}/R_{12}\bar{c}_{12}$, or $\Gamma_1 = (2/3)\Gamma$.

The key problem of determination of $\tilde{\Delta}$ (resolved semi-empirically in the traditional approach) can be resolved more accurately using the new numerical approach to the swept volume calculation. In this approach the exact values of $\beta_{kin}$ are calculated at different $a_{12}/R_{12}$ (Table 1.1) and then approximated using Eq. (1.21) or a more sophisticated expression with an extended set of fitting parameters, e.g. in the form

$$\beta_{kin} = \beta_{kin}^{(con)} \frac{1 + B_1'\Gamma + B_4'\Gamma^2}{1 + B_2'\Gamma + B_3'\Gamma^2 + (2/3)B_4'\Gamma^3}, \tag{1.25}$$

that is consistent with Eq. (1.20).

The values of the parameters in Eq. (1.25) can be determined using the least-squares method, that gives $B_1' = 14.24$, $B_2' = 13.61$, $B_3' = 9.52$, $B_4' = 3.52$ and provides a relatively high accuracy with the maximum error of ≤ 1% in comparison with the calculation points (from Table 1.1), as presented in Fig. 1.4 (solid line).

A somewhat reduced accuracy with the maximum error of ≤ 3% can be attained using the simplified model, Eq. (1.21), with parameters $B_1 = 0.51$, $B_2 = 0.78$, determined using the least-squares method (dashed line in Fig. 1.4).

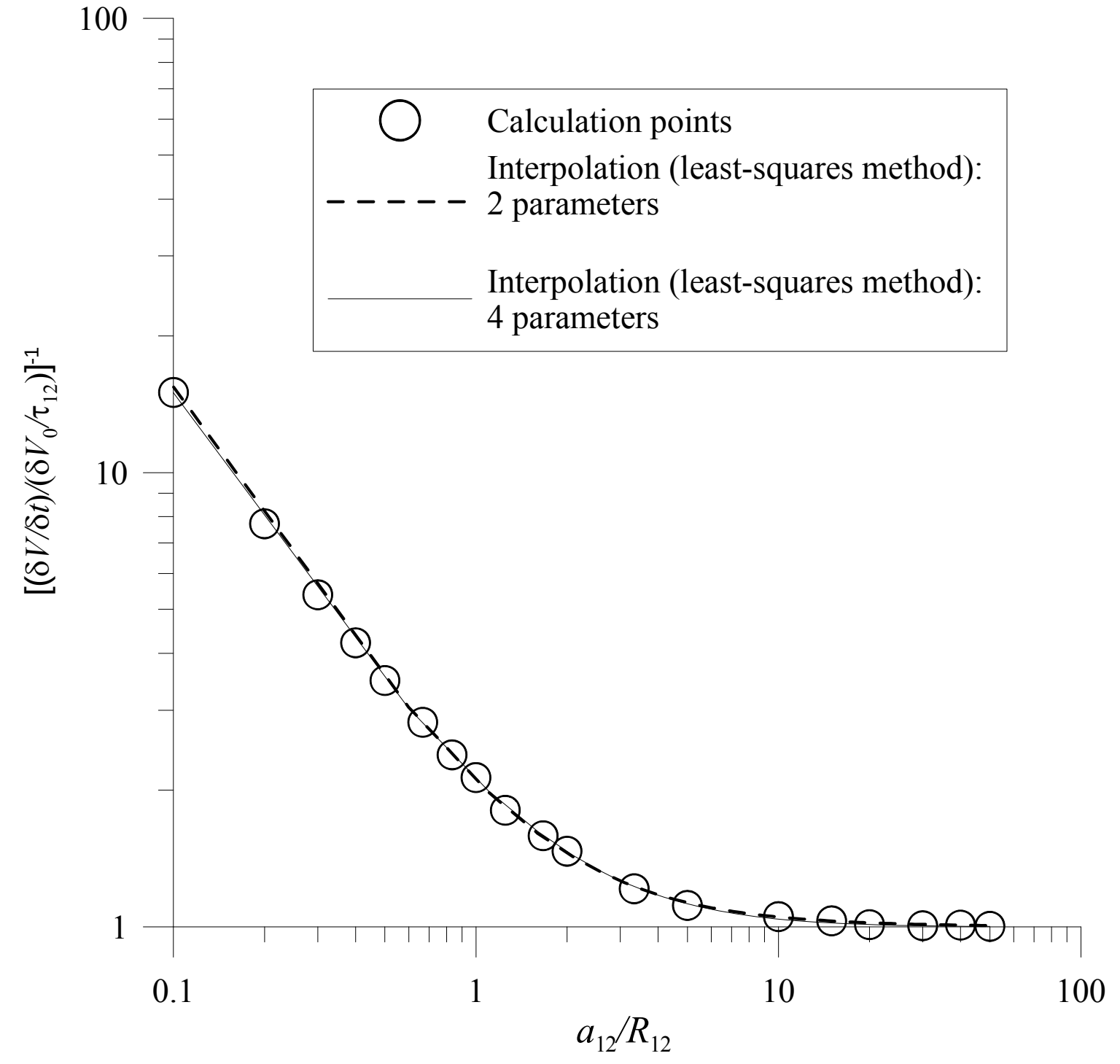

Fig. 1.4. Dependence of the steady-state values of the inverse relative mean sweeping rate $[(\delta V_{12}/\delta t)/(\delta V_0/\tau_{12})]^{-1} = \left(\beta_{kin}/\beta_{kin}^{(fm)}\right)^{-1}$ on the parameter $a_{12}/R_{12} = 6D_{12}/R_{12}\bar{c}_{12}$, using interpolation curves, Eq. (1.25) with $B_1' = 14.24$, $B_2' = 13.61$, $B_3' = 9.52$, $B_4' = 3.52$ (solid line), and Eq. (1.21) with $B_1 = 0.51$, $B_2 = 0.78$ (dashed line), in comparison with the calculated points (centres of circles).

This optimal set of parameters for Eq. (1.21) is somewhat different from $B_1 = 0.5$, $B_2 = 0.855$, obtained by formal application of Fuchs – Sutugin's interpolation formula for condensation problem [25] with $\tilde{\Gamma} = \lambda_{12}/R_{12}$ and Eq. (1.23) (reformulated in terms of $\Gamma = a_{12}/R_{12} = 2\tilde{\Gamma}$) that provides an accuracy with the maximum error of ≈ 4% in comparison with the calculation points.

A similar result can be obtained using Eq. (1.21) with $B_1 = 1/3$, $B_2 = 2/3$, proposed in [24] by formal application of Dahneke's formula [26] with $\widetilde{\Gamma} = \lambda_{12}/R_{12}$ and Eq. (1.23), that eventually takes the form (in terms of $\Gamma = a_{12}/R_{12}$)

$$\beta_{kin} = \beta_{kin}^{(con)} \frac{1+\widetilde{\Delta}}{1+(2/3)\Gamma\left(1+\widetilde{\Delta}\right)}, \tag{1.26}$$

with

$$\widetilde{\Delta} = B_1\Gamma = \Gamma/3, \tag{1.27}$$

and provides an accuracy with ≈ 3.5% maximum error for this set of parameters.

Eq. (1.26) formally coincides with the other semi-empirical interpolation formulas from the literature [4, 27], however, with different expressions for $\widetilde{\Delta}$. Namely, the expression for $\widetilde{\Delta}$ proposed by Fuchs [4] has the form

$$\widetilde{\Delta} = \frac{\Delta_{12}}{R_{12}} = \frac{\left(\Delta_1^2 + \Delta_2^2\right)^{1/2}}{R_{12}}, \tag{1.28}$$

with

$$\frac{\Delta_i}{2R_i} = \frac{1}{3}\left(\frac{1}{\Psi_i}\right)\left(\left(1+\Psi_i\right)^3 - \left(1+\Psi_i^2\right)^{3/2}\right) - 1, \tag{1.28a}$$

or, following Wright [27], with

$$\frac{\Delta_i}{2R_i} = \left(\frac{1}{\Psi_i^2}\right)\left(\frac{1}{5}\left(1+\Psi_i\right)^5 - \frac{1}{3}\left(1+\Psi_i\right)^3\left(1+\Psi_i^2\right) + \frac{2}{15}\left(1-\Psi_i^2\right)^{5/2}\right) - 1, \tag{1.28b}$$

where $\Psi_i = l_i/2R_i$ and $l_i = 8D_i/\pi\bar{c}_i$.

Both these expressions, Eqs. (1.28a) and (1.28b), however, do not match with the above-derived general requirement of the theory, Eq. (1.20). For this reason, a more adequate expression for $\widetilde{\Delta}$ can be applied using corrected Fuchs' and Wright's formulas proposed in [27]

$$\widetilde{\Delta} = \frac{\Delta_{12}}{R_{12}} = \frac{1}{3}\left(\frac{1}{\Psi_{12}}\right)\left(\left(1+\Psi_{12}\right)^3 - \left(1+\Psi_{12}^2\right)^{3/2}\right) - 1, \tag{1.29a}$$

and

$$\widetilde{\Delta} = \frac{\Delta_{12}}{R_{12}} = \left(\frac{1}{\Psi_{12}^2}\right)\left(\frac{1}{5}\left(1+\Psi_{12}\right)^5 - \frac{1}{3}\left(1+\Psi_{12}\right)^3\left(1+\Psi_{12}^2\right) + \frac{2}{15}\left(1-\Psi_{12}^2\right)^{5/2}\right) - 1, \tag{1.29b}$$

where $\Psi_{12} = l_{12}/R_{12}$ and $l_{12} = 8(D_1 + D_2)/\pi(\bar{c}_1^2 + \bar{c}_2^2)^{1/2}$.

Results of approximation with these semi-empirical expressions, Eqs. (1.29a) and (1.29b), being rather similar to each other (within < 1%), notably deviate from the calculation points (by 4–6%) in a wide range of the transition interval, $0.1 \le a_{12}/R_{12} \le 0.9$. Hence, in order to compare behaviour of various interpolation expressions in this area, they are represented in the traditional form, $\beta_{kin}/\beta_{kin}^{(con)}$ (as a function of $a_{12}/R_{12}$) in Fig. 1.5, demonstrating a relatively high error (up to $\approx 6\%$ at $a_{12}/R_{12} \approx 0.3$) in predictions of these semi-empirical models in comparison with the new formula and the calculation points.

Therefore, the new approach allows calculation of the collision frequency function in the transition regime, without use of the semi-empirical assumptions (from the literature) in derivation of the interpolation expression and with a higher accuracy (with respect to the calculation points), which can be further improved in the next approximation of the random walk theory (see Section 1.6).

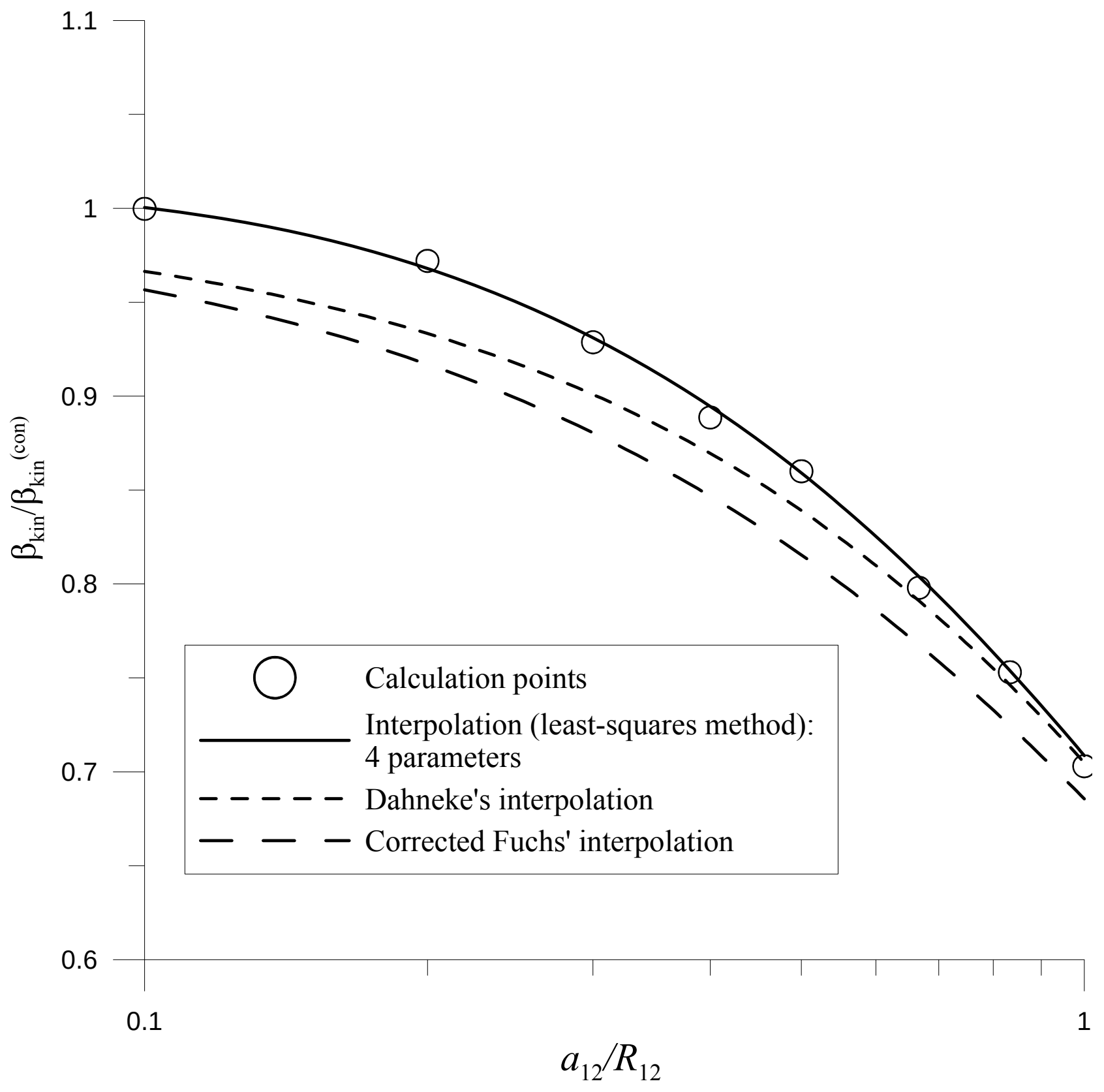

Fig. 1.5. Verification of reformulated Dahneke's and corrected (in [24]) Fuchs' semi-empirical interpolation formulas for $\beta_{kin}/\beta_{kin}^{(con)}$ as a function of $a_{12}/R_{12}$ against the newly derived four-parameter formula and the calculated points (from Table 1.1).

*1.4.7. Applicability range of the kinetic approach*

In the mean field approximation considering monodisperse particles size distribution (which delivers a reasonable asymptotic solution for the coagulation problem at large times), the probability of two neighbour particles coalescence in $\delta t$ can be evaluated from Eq. (1.17) for the case $a << R$. Indeed, the total rate of loss of particles by coalescence is given by

$$\frac{dn}{dt} = -8\pi DRn^2 = -\frac{n}{\tau_c} \qquad (1.30)$$

where $\tau_c = (8\pi DRn)^{-1}$ is the coalescence time corresponding to the characteristic time between two subsequent collisions of a particle, during which the mean inter-particle distance $\bar{r} \approx n^{-1/3}$ increases by a factor of $\approx 1.3$. This confirms that variation of the mean particles concentration can be neglected in the first order of approximation $\delta t / \tau_c << 1$ during the time step $\delta t$, as assumed in Section 1.5.

Eq. (1.30) should be solved simultaneously with the condition of the total mass conservation

$$R^3 n = const. = R_0^3 n_0 . \qquad (1.31)$$

As already outlined in Section 1.3.1, owing to coalescence of two particles from the ensemble of point particles, the local number of particles alters step-wise (from two to one) that induces a local heterogeneity of the particles spatial distribution function on the length scale of the mean inter-particle distance $\approx n^{-1/3}$. Therefore, the uniform spatial distribution of coalescing particles can be considered under condition that the heterogeneities of such scale, $n^{-1/3}$, can rapidly relax by the particles diffusion during the characteristic time between two subsequent collisions of a particle, $\tau_c$. This condition of the particles relaxation, or mixing, takes the form $\tau_c >> \tau_d$, where $\tau_d \approx n^{-2/3}/6D$ is the characteristic time of the particles (diffusion) redistribution on the length scale of the induced heterogeneities. Only in this case the probability to find a new particle in the vicinity of the coalesced particle at the moment of their subsequent collision will be still determined by the mean concentration of particles, as assumed in derivation of Eqs. (1.17), (1.18) and (1.24).

Substitution of $\tau_c = (8\pi DRn)^{-1}$ and $\tau_d \approx n^{-2/3}/6D$ in the relationship $\tau_c >> \tau_d$ yields the mixing condition in the form $n^{1/3}R << 3/4\pi$, which is practically indistinguishable from the basic condition $n^{1/3}R << 1$, within the accuracy of the characteristic times evaluation. The other condition, $\tau_c >> \tau_0$, required in the current approach for choosing a relatively large time step $\delta t >> \tau_0$, takes the form $8\pi DRn\tau_0 = (4/3)\pi Ra^2 n = (4/3)\pi (a/R)^2 nR^3 << 1$, that is valid owing to $n^{1/3}R << 1$ and $a/R \le 1$.

In the transition regime, when $a \le R$ (or $\Gamma \le 1$), the value of $\tau_c$ derived from the first part of Eq. (1.24) is comparable with the continuum mode value, $\tau_c \approx (8\pi DRn)^{-1}$, and thus also obeys the mixing condition $\tau_c >> \tau_d$.

It is straightforward to see that the mixing condition $\tau_c >> \tau_d$ can be equivalently represented in the form $\lambda >> \bar{r} \approx n^{-1/3}$, where $\lambda \approx (6D\tau_c)^{1/2}$ is the characteristic distance between two subsequent collisions of a particle with other particles (or the mean free path of Brownian particles). From the above presented consideration it is seen that the limit $\lambda n^{1/3} \approx 1$, corresponding to $\tau_c \approx \tau_d$, can be attained only in the case of very high densities, $n^{1/3}R \approx 1$, that is beyond the applicability range of the basic approximation, $n^{1/3}R << 1$ (or $R/\bar{r} << 1$).

This situation has a clear physical sense that the mean free path of particles in the diluted system cannot be less than the mean inter-particle distance and is qualitatively similar to behaviour of ordinary molecular gases. Indeed, the mean free path $\lambda_m$ of the gas molecules is estimated as $\lambda_m \approx (n_m\sigma^2)^{-1}$, where $n_m \approx \bar{r}_m^{-3}$ is the gas density and $\sigma \approx d^2$ is the molecular cross-section, therefore, $\lambda_m \approx \bar{r}_m(\bar{r}_m/d)^2 \approx d(\bar{r}_m/d)^3$, which results in the relationship $\lambda_m >> \bar{r}_m >> d$ under the ideal (diluted) gas condition, $d/\bar{r}_m << 1$ (analogous to the basic condition of the Brownian particles theory, $R/\bar{r} << 1$). The mixing condition implies in this case that after each collision of gas molecules their random distribution quickly reinstates owing to their relocations on the length scale of the mean inter-molecular distance, $\bar{r}_m$. Since the gas molecules move straight with the mean thermal speed $u_m$ in-between their collisions, the characteristic collision and mixing times can be evaluated as $\tau_c \approx \lambda_m/u_m$ and $\tau_{mix} \approx \bar{r}_m/u_m$, respectively. Therefore, the mixing condition, $\tau_c >> \tau_{mix}$, directly corresponds to the above justified relationship, $\lambda_m >> \bar{r}_m$.

The diffusion mixing condition, $\tau_c >> \tau_d$, is valid also in the opposite case $R << a << \bar{r}$, corresponding to the free molecular mode. Indeed, substituting $\tau_c \approx (\pi R^2\bar{c}n)^{-1}$ from Eq. (1.14) in this inequality, one obtains the mixing condition in the form $a/R >> (4\pi/3)(6\pi)^{1/2}n^{1/3}R \approx 18n^{1/3}R$, which is valid practically in the whole range of the free molecular mode applicability, $a/R >> 1$ (taking into account that $n^{1/3}R << 1$).

In the other range $a \ge \bar{r}$ of the free molecular mode ($R << a$), the particle mixing (on the scale $l \approx \bar{r}$) occurs at $\tau_{mix} \approx \bar{r}/\bar{c} \le \tau_0$, whereas $\tau_c \approx (\pi R^2\bar{c}n)^{-1}$, therefore the mixing condition, $\tau_c >> \tau_{mix}$, is valid owing to $(\bar{r}/R)^2 >> 1$.

In the transition regime, when $a \ge R$ (or $\Gamma \ge 1$), the value of $\tau_c$ derived from the second part of Eq. (1.24) is comparable with the free molecular mode value, $\tau_c \approx (\pi R^2\bar{c}n)^{-1}$, and thus also obeys the mixing condition $\tau_c >> \tau_d$ (in the case $R << a << \bar{r}$) or $\tau_c >> \tau_{mix}$ (in the case $a \ge \bar{r}$).

Therefore, the kinetic regime is realized for comparable size particles practically in the whole range of the considered approximation $n^{1/3}R \ll 1$, however, with various expressions for the collision frequency in the different intervals of the parameter $a/R$ (corresponding to the different modes of the kinetic regime).

## 1.5. Next approximation of the random walk theory

The above presented approach based on the *simple* random walk theory (with the fixed jump distance of migrating particle) allowed a relatively simple and rapid calculation of the new interpolation formulas for the coalescence rate, which correctly reduce to the analytical expressions in the two limiting cases, $a \ll R$ and $a \gg R$, avoiding semi-empirical assumptions of the traditional models, however, not in completely self-consistent manner. Namely, one of the model parameters, $a_{12}$, describing the elementary drift (or jump) distance of the effective migrating particle, was derived in Section 1.5.2 by comparison of the collision frequency calculated in the simple random walk approach (neglecting variation of the jump distance) at high Knudsen numbers, $\mathrm{Kn} \to \infty$, with that calculated in the free molecular approach, which takes into consideration the Maxwell distribution of the particles velocities (and thus resulting in a finite distribution of the jump distances).

Nevertheless, since this parameter $a_{12}$ enters in the calculated collision frequency implicitly (via $D_{12}$), the collision kernel, Eq. (1.21) or Eq. (1.25), represented in terms of the independent set of the three basic parameters $R_{12}$, $D_{12}$ and $\bar{c}_{12}$, Eq. (1.24), were apparently calculated with a satisfactory accuracy (considered as the first approximation of the random walk theory).

In order to overcome this inconsistency and to verify the simplified approach, the next approximation of the random walk theory taking into consideration variable length of the jumps will be studied in the current Section (following [10]).

*1.5.1. Brownian particles coagulation*

The general expression for the mean-square relocation distance, $\hat{r} = \sqrt{\langle \vec{r}^2 \rangle}$, of a Brownian particle (of mass $m$), derived by Uhlenbeck and Ornstein [12]

$$\hat{r}^2 = 2u_T^2\tilde{\tau}\left[t - \tilde{\tau}\left(1 - \exp\left(-t/\tilde{\tau}\right)\right)\right], \quad (1.32)$$

where $u_T = \sqrt{\langle \vec{u}^2 \rangle} = \left(3kT/m\right)^{1/2}$ is the root mean-square heat velocity, $\tilde{\tau} = mb = mD/kT$ is the characteristic relaxation time, $b$ is the particle mobility, and thus taking the form

$$\hat{r}^2 = 6D\left[t - \tilde{\tau}\left(1 - \exp\left(-t/\tilde{\tau}\right)\right)\right], \quad (1.33)$$

can be correctly reduced at $t \gg \tilde{\tau}$ to the Einstein relationship

$$\hat{r}^2 = 6Dt, \quad (1.34)$$

and at $t \ll \tilde{\tau}$ to

$$\hat{r} = u_T t, \tag{1.35}$$

or equivalently, in terms of the mean relocation distance, to

$$\bar{r} = \bar{c}t, \tag{1.36}$$

where $\bar{c} = (8kT/\pi m)^{1/2}$ is the mean thermal speed.

This limiting expression, Eq. (1.36), substantiates the free molecular (or ballistic) approach traditionally applied to consideration of particles movement (and collisions) at short times, $t \le \tilde{\tau}$. Namely, in derivation of the collision rate of Brownian particles with large ratio $a >> R$, it is assumed [4] that particles obey the simple kinetic theory of gases, i.e. move straight with their heat velocities and randomly change direction of the velocity for the relaxation time (or the drift period), $\tau \approx \tilde{\tau}$. On the other hand, at large times $t >> \tilde{\tau}$ (corresponding to the calculation time step chosen in the coagulation rate equation, $\delta t >> \tilde{\tau}$) particle migration can be considered as the Markov process obeying the diffusion equation [4], and thus can be adequately described by the random walk theory.

The simplest approximation of this approach considering elementary drifts (or jumps) with the fixed length (so called *simple* random walks) was applied to the problem of Brownian coagulation in the Sections 1.2-1.5. As explained above, this approach is not completely self-consistent, since it neglects the finite distribution of particles speeds during their elementary drifts (under assumption of the fixed drift length), which is taken in consideration, however, in calculation of the collision probability within the elementary drift period $\tau$.

This inconsistency can be removed in the next approximation of the random walk theory. Indeed, distribution for the velocity of a Brownian particle (of mass $m$ and radius $R$ at temperature $T$) has a finite statistical dispersion (near the mean heat velocity) described by the Maxwell law. On the other hand, variation in time of the velocity components of the particle, moving under external stochastic forces $\vec{F}(t)$ exerted by the carrier gas molecules, calculated, following [12], as

$$u_i(t) = u_i(0)\exp(-t/\tilde{\tau}) + f(F_i, t), \tag{1.37}$$

where $f(F_i,t) = \int_0^t e^{-(t-s)/\tilde{\tau}} F(s)ds$, $\langle f(F_i,t)\rangle = 0$, $i = x, y, z$, is determined by the characteristic relaxation time, $\tilde{\tau} = mb = mD/kT$, which is independent of the initial velocity $u_i(0)$. For instance, taking the mean over an ensemble of particles, which have started at $t = 0$ with the same velocity $u_i(0)$, one gets $\langle u_i(t)\rangle = u_i(0)\exp(-t/\tilde{\tau})$, whereas $\langle u_i^2\rangle = (kT/m) + [u_i^2(0) - (kT/m)]\exp(-2t/\tilde{\tau})$. From these expressions it is seen that the "memory" of the initial velocity is persistent within the characteristic relaxation time $\tilde{\tau}$, which is independent of the initial velocity $u_i(0)$, and $\langle u_i^2\rangle \to kT/m$ (at $t >> \tilde{\tau}$) in accordance with the equipartition theorem. These important conclusions substantiate consideration of elementary drifts (or jumps) with the fixed elementary drift time, $\tau_0 \approx \tilde{\tau}$.

In higher approximations of the random walk theory a finite statistical dispersion of the elementary drift times $\tau$ near the mean value $\tau_0$ could be taken into consideration; however, this dispersion apparently reduces with increase of the particle mass, owing to statistically large number of stochastic hits by the carrier gas molecules during the elementary drift period (proportionally to $\tilde{\tau}/\tau_m = mb/\tau_m$, where $\tau_m^{-1}$ is the mean collision frequency of the particle with the carrier gas molecules).

As a result, the distribution for the elementary drift distance of a particle in the applied approximation of the random walk theory (with $\tau = \tau_0$) obeys a law deduced from the Maxwell distribution law for its speed. For instance, this essentially simplifies the random walk theory for Brownian particles in comparison with that for the Boltzmann gas, where both the drift distance and drift time between two subsequent collisions of a gas molecule are stochastically varying values.

In this approximation for Brownian particles, the probability $w(\mathbf{r},\boldsymbol{\xi})d^3\xi$ of a particle (from an ensemble of identical particles) at $\mathbf{r}$ to relocate to $\mathbf{r}-\boldsymbol{\xi}$ in $\tau_0$ is equal to the probability for this particle to have the velocity $\mathbf{u} = \boldsymbol{\xi}/\tau_0$, which is determined by the Maxwell distribution law, $\tilde{w}(\mathbf{u})d^3u = \left(\frac{m}{2\pi kT}\right)^{3/2} \exp\left(-\frac{mu^2}{2kT}\right)d^3u$, or $w(\mathbf{r},\boldsymbol{\xi})d^3\xi = \tilde{w}(\mathbf{u} = \boldsymbol{\xi}/\tau_0)d^3u = \tilde{w}(\boldsymbol{\xi}/\tau_0)d^3\xi/\tau_0^3$. Therefore, the mean-square drift distance can be calculated as $\hat{a}^2 = \langle \xi^2 \rangle = \int_0^\infty \xi^2 w(\mathbf{r},\boldsymbol{\xi})d^3\xi = \int_0^\infty \frac{4\pi}{\tau_0^3}\left(\frac{m}{2\pi kT}\right)^{3/2} \xi^4 \exp\left(-\frac{m\xi^2}{2kT\tau_0^2}\right)d\xi = \frac{3kT}{m}\tau_0^2$ (cf. Section 1.2), or

$$\hat{a} = \left(\frac{3kT}{m}\right)^{1/2} \tau_0 = u_T \tau_0, \tag{1.38}$$

whereas the mean relocation distance $\bar{a}$ during one walk can be calculated as

$$\bar{a} = \langle |\bar{u}| \rangle \tau_0 = \bar{c}\,\tau_0. \tag{1.39}$$

For the diffusion coefficient in the ensemble of particles, $D = \frac{1}{6\tau_0}\int \xi^2 w(\boldsymbol{\xi})d^3\xi$ (cf. Section 1.2), one obtains

$$D = \hat{a}^2/6\tau_0 = u_T^2\tau_0/6, \tag{1.40}$$

or, using Eqs. (1.38) and (1.39),

$$D = \hat{a}^2/6\tau_0 = \left(\bar{a}^2/6\tau_0\right)\left(u_T/\bar{c}\right)^2 = \pi\,\bar{a}^2/16\tau_0 \approx \bar{a}^2/5\tau_0, \tag{1.41}$$

where

$$\hat{a}/\bar{a} = u_T/\bar{c} = (3\pi/8)^{1/2}. \tag{1.42}$$

As shown in Section 1.4, evaluation of the two-particle collision frequency (required for calculation of the coagulation kernel in the ensemble of comparable size particles under the mixing condition) can be reduced to calculation of the mean value of the volume sweeping rate $(\delta V/\delta t)$ by the effective particle of radius $R_{12} = R_1 + R_1$ migrating by random walks with diffusivity $D_{12} = D_1 + D_1$. In accordance with the above presented consideration (for walks of variable length), the diffusivity is determined by the root mean-square length $\hat{a}_{12}$ of the particle jumps as $D_{12} = \hat{a}_{12}^2/6\tau_{12}$ (with a fixed value of $\tau_{12}$). On the other hand, the mean swept volume $\delta V_0$ during one jump is determined by the mean jump distance $\overline{a}_{12}$ and is calculated as $\delta V_0 = \pi\overline{a}_{12}(R_1 + R_2)^2$.

In the continuum mode ($a << R$) the mean swept volume per unit time (averaged over relatively long time step $\delta t$ including many jump periods, $\tau_c >> \delta t >> \tau_{12}$), which determines the collision frequency $\beta_{kin}^{(con)} = \delta\langle V\rangle/\delta t$, is smaller in comparison with the mean swept volume per one jump period, $(\delta V_0/\tau_{12})$, owing to strong overlapping of the swept zone segments, and is equal to $\langle \delta V_{12}/\delta t\rangle = 4\pi(D_1 + D_2)(R_1 + R_2) = 4\pi D_{12} R_{12}$ (as shown in Section 1.5.1). In particular, this implies (using Eq. (1.41)) that the ratio of $\langle \delta V_{12}/\delta t\rangle$ to $\delta V_0/\tau_{12}$ is equal to $\langle \delta V_{12}/\delta t\rangle/(\delta V_0/\tau_{12}) = 2\hat{a}_{12}^2/3\overline{a}_{12}R_{12}$.

In the opposite, free molecular limit ($a >> R$) one can neglect the relative volume of the swept zone segments intersections. For this reason, the mean volume swept per unit time, $\langle \delta V_{12}/\delta t\rangle$, is a constant value equal to the mean volume swept per one jump period, $\delta V_0/\tau_{12}$. In its turn, the latter can be considered as a constant value, $\delta V_0/\tau_{12} = \pi R_{12}^2\,\overline{a}_{12}/\tau_{12}$, equal to the mean sweeping rate $\delta V^*/\delta t^* = \pi R_{12}^2 c_{12}$ during a short time step $\delta t^* << \tau_{12}$. On the other hand, the mean swept volume $\delta V^*$ represents the probability of a collision in $\delta t^*$ of two original particles of radii $R_1$ and $R_2$, migrating in a sample of unit volume, that can be calculated in the free molecular approach as $\langle \delta V_{12}^*/\delta t^*\rangle = \pi R_{12}^2\sqrt{(8kT/\pi)(m_1^{-1} + m_2^{-1})} = \pi R_{12}^2\sqrt{(\overline{c}_1^2 + \overline{c}_2^2)}$, i.e. $\overline{c}_{12} = \overline{a}_{12}/\tau_{12} = \sqrt{(\overline{c}_1^2 + \overline{c}_2^2)}$, where $\overline{c}_1$ and $\overline{c}_2$ are the mean thermal speeds of the two original particles, or $\tau_{12} = \overline{a}_{12}/\overline{c}_{12}$. Correspondingly, for the effective particle diffusivity $D_{12}$ one obtains $D_{12} = \hat{a}_{12}^2/6\tau_{12} = \hat{a}_{12}^2\overline{c}_{12}/6\overline{a}_{12} = (\pi/16)\overline{a}_{12}\overline{c}_{12}$, or $\overline{a}_{12} = (16/\pi)D_{12}\overline{c}_{12}^{-1}$, instead of Eq. (1.22) calculated in the simple random walk approach of Section 1.4 (with the fixed jump distance, $a_{12} = const.$).

Nevertheless, the collision kernel, $\beta_{kin}^{(fm)}$, which is independent of $\overline{a}_{12}$, reduces to the same equation, $\beta_{kin}^{(fm)} = \delta V_0/\tau_{12} = (R_1 + R_2)^2\sqrt{(6kT/\rho)(R_1^{-3} + R_2^{-3})}$. Besides, the ratio $\beta_{kin}^{(con)}/\beta_{kin}^{(fm)} = (\pi/4)\overline{a}_{12}/R_{12}$, being represented in terms of the independent parameters $R_{12}$, $D_{12}$ and $\overline{c}_{12}$, is reduced to the same value $\beta_{kin}^{(con)}/\beta_{kin}^{(fm)} = 4D_{12}/R_{12}\overline{c}_{12}$, calculated in the simple random walk approach.

In the transition interval $a \approx R$, the collision rate can be calculated similarly to the two above considered limiting cases by evaluation of the mean swept volume per unit time, $\beta_{kin} = \langle \delta V_{12}/\delta t\rangle$, in the numerical approach. However, in contrast to the simple random walk approximation of Section 1.4.3, the jump distance of the effective particle (of radius $R_{12}$ and mass $m_{12}$) in random directions is calculated as $a_{12} = u_{12}\tau_{12}$, where the particle speed $u_{12}$ is generated as a random value with the probability density function, $f(u_{12}) = \sqrt{\frac{2}{\pi}\left(\frac{m_{12}}{kT}\right)^3} u_{12}^2 \exp\left(-\frac{m_{12}u_{12}^2}{2kT}\right)$, where $m_{12}^{-1} = \left(m_1^{-1} + m_2^{-1}\right)$, providing self-consistency of the above derived expression for the thermal speed, $\bar{c}_{12} = \left(8kT/\pi m_{12}\right)^{1/2} = \sqrt{\left(\bar{c}_1^2 + \bar{c}_1^2\right)} = \sqrt{\left(8kT/\pi\right)\left(m_1^{-1} + m_2^{-1}\right)}$. The generated data describe the successive positions of the particle centre trajectory, which can be further used for calculation of the swept volume. For this calculation the same procedure of random spatial distribution of auxiliary (fictitious) point immobile particles (markers) with a relatively high concentration, $n_* >> \left(\pi R_{12}^2 a_{12}\right)^{-1}$, described in Section 1.4.3, is numerically realized using the Monte Carlo method. Each marker found in the swept volume is counted only once.

*1.5.2. Numerical calculations*

The number of jumps $k = \delta t/\tau_{12}$ is increased until the ratio of $\delta\langle V_{12}\rangle/\delta t$ to $\delta V_0/\tau_{12}$ attains a steady-state value, which in accordance with the above presented analytical calculations has to be equal to $(\pi/4)\bar{a}_{12}/R_{12}$ in the limit $a_{12} << R_{12}$ and to 1 in the limit $a_{12} >> R_{12}$. In order to diminish statistical dispersion of the calculation results, a relatively large value of $k \approx 10^5$ was chosen for each trajectory.

Nevertheless, owing to variation of the jump distance, dispersion of the calculated value $\left(\delta\langle V_{12}\rangle/\delta t\right)/\left(\delta V_0/\tau_{12}\right)$ is notably larger in comparison with that calculated for jumps with the fixed length (in Section 1.4.3). For this reason, 150-200 calculations for each value of $R_{12}/a_{12}$ have been carried out, forming a smooth distribution plot near the mean values, adequately approximated by the normal distribution function, Fig. 1.6.

Results of these calculations for the mean values $\langle\delta V/\delta t\rangle = \beta_{kin}^{(con)}$ are summarized in Table 1.2 and plotted in Fig. 1.7 in dependence on the parameter $a_{12}/R_{12}$, used in the previous calculations in Section 1.4 and formulated in terms of the independent variables as $a_{12}/R_{12} = 6D_{12}/\bar{c}_{12}R_{12}$. This parameter is related to the currently used parameter $\bar{a}_{12}/R_{12}$ as $a_{12}/R_{12} = \left(\bar{a}_{12}/R_{12}\right)3\pi/8$. As seen from Fig. 1.7, the new set of calculation points is credibly described (within the calculation accuracy) by the interpolation expression derived in the simple random walk approach [7] and represented in terms of the independent variables $D_{12}$, $R_{12}$ and $\bar{c}_{12}$ as

$$\frac{\beta_{kin}}{\beta_{kin}^{(fm)}} = \frac{(2/3)\Gamma_1 + B_1\Gamma_1^2}{(2/3) + B_2\Gamma_1 + B_1\Gamma_1^2}, \qquad (1.43)$$

where $\Gamma_1 = 4D_{12}/R_{12}\bar{c}_{12}$ and $B_1 = 0.51$, $B_2 = 0.78$.

A similar result can be confirmed for the advanced four-parameter interpolation expression, Eq. (1.25), represented in terms of the independent variables as

$$\frac{\beta_{kin}}{\beta_{kin}^{(fm)}} = \Gamma_1 \frac{(2/3) + B_1'\Gamma_1 + (3/2)B_4'\Gamma_1^2}{(2/3) + B_2'\Gamma_1 + (3/2)B_3'\Gamma_1^2 + (3/2)B_4'\Gamma_1^3}, \tag{1.44}$$

where $\Gamma_1 = 4D_{12}/R_{12}\bar{c}_{12}$ and $B_1' = 14.24$, $B_2' = 13.61$, $B_3' = 9.52$, $B_4' = 3.52$.

From these results one can conclude that the new interpolation expression derived in the simple random walk approach is reliably confirmed by calculations in the next approximation of the random walk theory (with stochastically varying jump distance) and thus can be used instead of the formulas derived semi-empirically.

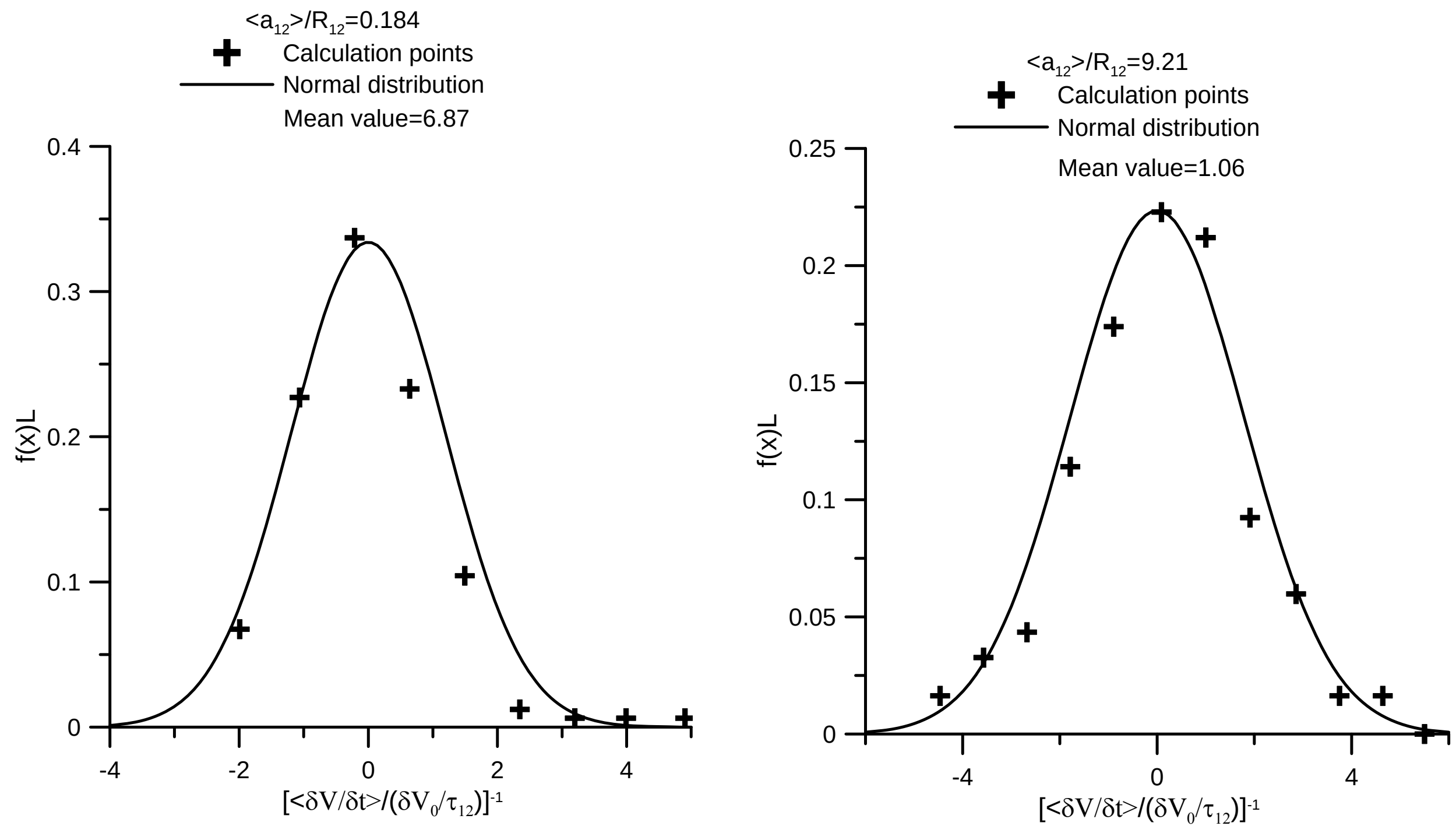

Fig. 1.6. Examples of calculations of the probability density (in arbitrary units) for $\left[\left(\delta\langle V_{12}\rangle/\delta t\right)/\left(\delta V_0/\tau_{12}\right)\right]^{-1}$ at two different values of the mean drift distance $\bar{a}_{12}/R_{12}$; 150-200 calculation points for each plot are grouped in intervals of equal width (≈ 10% of the whole distribution width), from [10].

Table 1.2. Steady-state values of $(\delta V_0/\tau_{12})/(\delta V_{12}/\delta t) = \left(\beta_{kin}/\beta_{kin}^{(fm)}\right)^{-1}$ calculated in the next approximation of the random walk theory at different values of $\bar{a}_{12}/R_{12} = 16D_{12}/\pi\bar{c}_{12}R_{12}$, which is related to $a_{12}/R_{12} = 6D_{12}/\bar{c}_{12}R_{12}$ from Table 1.1 as $a_{12}/R_{12} = (\bar{a}_{12}/R_{12})3\pi/8$.

| | | | | | | | | | |
|---|---|---|---|---|---|---|---|---|---|
| $\bar{a}_{12}/R_{12}$ | 0.046 | 0.057 | 0.092 | 0.169 | 0.184 | 0.23 | 0.47 | 0.594 | 0.92 |
| $a_{12}/R_{12}$ | 0.054 | 0.067 | 0.109 | 0.2 | 0.217 | 0.271 | 0.553 | 0.7 | 1.085 |
| $\frac{\delta V_0/\tau_{12}}{\langle \delta V_{12}/\delta t \rangle}$ | 27.40 | 22.38 | 13.69 | 7.965 | 6.87 | 6.12 | 3.29 | 2.765 | 2.04 |
| $\bar{a}_{12}/R_{12}$ | 1.57 | 3.395 | 9.21 | 21.93 | | | | | |
| $a_{12}/R_{12}$ | 1.843 | 4.0 | 10.85 | 25.84 | | | | | |
| $\frac{\delta V_0/\tau_{12}}{\langle \delta V_{12}/\delta t \rangle}$ | 1.54 | 1.196 | 1.06 | 1.02 | | | | | |

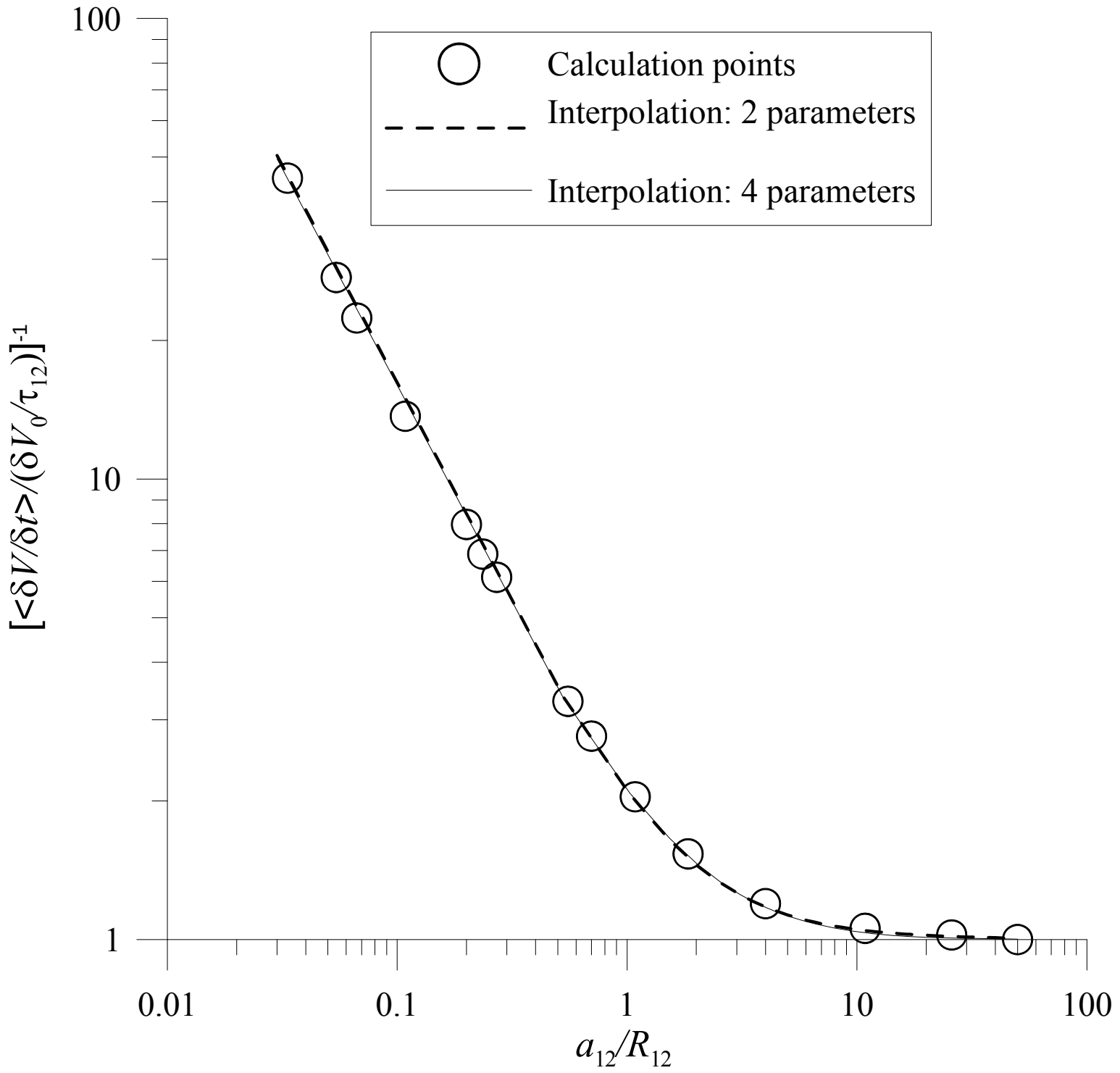

Fig. 1.7. Verification of the interpolation expressions for $\left[(\delta\langle V_{12}\rangle/\delta t)/(\delta V_0/\tau_{12})\right]^{-1} = \left(\beta_{kin}/\beta_{kin}^{(fm)}\right)^{-1}$ as a function of $a_{12}/R_{12} = 6D_{12}/R_{12}\bar{c}_{12}$ obtained in the first approximation of the random walk theory with fixed jump distance (from Fig. 1.4) against the new set of calculation points obtained in the next approximation with variable jump distance (from Table 1.2), from [10].

## 1.6. Advancement of the kinetic approach to Brownian coagulation on the base of the Langevin theory

In the current Section the kinetic approach, originally applied to calculation of the sweeping rate in frames of the simplified random walk models (Sections 1.4 and 1.5), is further advanced by application of the Langevin theory for Brownian particles migration, following the author's paper [14]. The new results allow an additional justification and further improvement of the interpolation expressions for the coagulation kernel obtained in the random walk theory, also in comparison with the other approache.1. Brownian particles trajectories

Single-particle trajectories can be described by the Langevin differential equation (cf. [12])

$$m\frac{d\mathbf{v}}{dt} = -f\mathbf{v} + \mathbf{X}, \qquad (1.45)$$

where $m$ and $f$ are the particle mass and friction coefficient, $\mathbf{v}$ is the particle velocity vector, $\mathbf{X}$ is the stochastic, rapidly fluctuating force on the particles from the bombardment of fluid molecules, which obey the relationships

$$\langle \mathbf{X}(t) \rangle = 0, \ \langle \mathbf{X}(t)\mathbf{X}(t') \rangle = 6fkT\delta(t-t'). \qquad (1.46)$$

Partial integration of Eq. (1.45), following Ermak and Buckholz [28], and introduction of dimensionless variables, $\tau = \frac{f}{m}t$, $\mathbf{v}^* = \frac{m\mathbf{v}}{fR}$, $\mathbf{r}^* = \frac{\mathbf{r}}{R}$, following Isella and Drossinos [29] and Gopalakrishnan and Hogan [13], result in

$$\mathbf{v}^* = \mathbf{v}_0^* \exp(-\tau) + \mathbf{B}_1^*, \qquad (1.47)$$

$$\mathbf{r}^* = \mathbf{r}_0^* + (\mathbf{v}^* + \mathbf{v}_0^*)(1-\exp(-\tau))(1+\exp(-\tau))^{-1} + \mathbf{B}_2^*, \qquad (1.48)$$

where $\mathbf{B}_1^*$ and $\mathbf{B}_2^*$ represent Gaussian-distributed independent random numbers with zero mean and a variance of unity,

$$\langle \mathbf{B}_1^{*2} \rangle = 3Kn_D^2(1-\exp(-2\tau)), \qquad (1.49)$$

$$\langle \mathbf{B}_2^{*2} \rangle = 6Kn_D^2\left(\tau - 2(1-\exp(-\tau))(1+\exp(-\tau))^{-1}\right), \qquad (1.50)$$

where $Kn_D = \frac{\sqrt{kTm}}{fR}$.

### *1.6.1. Collisions of equal size particle*

As above explained, owing to rapid diffusion mixing in the ensemble of equal size Brownian particles (of mass $m$ and diffusivity $D = kT/f$), the coagulation problem for hard-sphere entities can be properly reduced to calculation of the collision rate of two entities, or equally, of the collision rate of the effective particle of radius $R_{12} = R_1 + R_2 = 2R$ moving with the effective diffusivity $D_{12} = D_1 + D_2 = 2D$ with an immobile point-wise particle, or equally, of the collision rate of the immobile effective particle of radius $R_{12} = 2R$ with a point-wise particle, moving with the effective diffusivity $D_{12} = 2D$. Indeed, the relative motion of the two original particles, $\mathbf{r}_{12}^*(t) = \mathbf{r}_1^*(t) - \mathbf{r}_2^*(t)$, exactly reduces to the Langevin equation for a single particle, as shown by Narsimhan and Ruckenstein [30],

$$\frac{m}{2}\frac{d\mathbf{v}}{dt} = -\frac{f}{2}\mathbf{v} + \mathbf{X}(t), \tag{1.51}$$

where $\mathbf{v} = \mathbf{v}_1 - \mathbf{v}_2$ and

$$\mathbf{X}(t) = \frac{1}{2}(\mathbf{X}_1(t) - \mathbf{X}_2(t)), \qquad \langle \mathbf{X}(t)\mathbf{X}(t')\rangle = \left(\frac{f}{2}\right)\cdot 6kT\delta(t-t'), \tag{1.52}$$

corresponding to the motion of the effective particle of mass $m_{12} = m/2$ with the friction coefficient $f_{12} = f/2$ (or with the diffusivity $D_{12} = kT/f_{12} = 2kT/f = 2D$), characterized by the diffusion Knudsen number $Kn_{\mathrm{D}} = \frac{\sqrt{kTm_{12}}}{f_{12}R_{12}} = \frac{\sqrt{kT\,m/2}}{fR}$ [13]. The latter is related to the dimensionless parameter $\Gamma = 6D_{12}/R_{12}\bar{c}_{12} = a_{12}/R_{12}$, considered in the simple random walk theory approach [5], where $a_{12}$ is the elementary walk (or drift) distance, $\bar{c}_{12} = (8kT/\pi m)^{1/2}$ is the mean thermal speed and $R_{12}$ is the radius of the effective particle, as $Kn_{\mathrm{D}} = \Gamma\sqrt{2}/3\sqrt{\pi} \approx 0.266\Gamma$.

In the next approximation of the random walk theory with (linear) walks in random directions with the heat velocity for the relaxation time $\tau_0$ [10], the effective particle migrates with the diffusivity $D_{12} = D_1 + D_2$, or equivalently with the friction coefficient $f_{12}^{-1} = (f_1^{-1} + f_2^{-1})$, and its velocity distribution obeys the Maxwell law, $\varphi(u_{12}) = \sqrt{\frac{2}{\pi}\left(\frac{m_{12}}{kT}\right)^3} u_{12}^2 \exp\left(-\frac{m_{12}u_{12}^2}{2kT}\right)$, with the effective mass $m_{12}^{-1} = (m_1^{-1} + m_2^{-1})$, the thermal speed $\bar{c}_{12} = \sqrt{(\bar{c}_1^2 + \bar{c}_2^2)} = (8kT/\pi m_{12})^{1/2}$ and the mean persistence distance $\bar{a}_{12} = (16/\pi)D_{12}/\bar{c}_{12}$ (which is therefore related to $\Gamma$ as $\bar{a}_{12}/R_{12} = (8/3\pi)\Gamma$). As above explained, particle trajectories are correctly described in this approximation at a large time scale, $t >> \tau_0$. At a small time scale, $t \le \tau_0$, the trajectories become smooth and non-linear, and are more adequately described by the Langevin equation. In order to take this effect into consideration, the time step in numerical calculations should be chosen small in comparison with the relaxation time, $\delta t << \tau_0$ (despite the Langevin approach is not justified on this time scale; this determines the uncertainty of the advanced approach).

Following the general procedure of the kinetic approach, the collision rate can be evaluated as the mean volume swept per unit time, $\beta = \delta\langle V\rangle/\delta t$, by the effective particle moving in accordance with Eq. (8), similarly to the previous calculations in frames of the simplified random walk models. In the free molecular limit, $\Gamma \to \infty$, one can neglect the volume of the trajectory intersections in comparison with the total swept volume, the latter coincides in this case with the straightened trajectory volume, $V_0 = \pi R_{12}^2 \bar{v}_{12} \tau = \pi (2R)^2 \tau \sqrt{8kT/\pi(m/2)} = 16\pi R^2 \tau \sqrt{kT/\pi m}$, and thus $\beta_{fm} = \delta\langle V_0 \rangle / \delta t = V_0/\tau = 16\pi R^2 \sqrt{kT/\pi m}$. In the continuum limit, $\Gamma \to 0$, the collision rate should converge to the value $\beta_c = 4\pi D_{12} R_{12} = \beta_{fm}\left(\frac{2}{3}\Gamma\right)$, justified analytically in [5].

### *1.6.2. Numerical calculations*

For the swept volume calculation in the transition mode with intermediate Knudsen numbers, the same procedure, described by Azarov and Veshchunov [7], of random spatial distribution of auxiliary (fictitious) point immobile particles (markers) with a relatively high concentration in a simulation box (completely covering the trajectory), was numerically realized using the Monte Carlo method. Each marker found in the swept volume is counted only once.

In order to diminish the dependence of the calculation results on the time step, $\delta t$, its value should be chosen small in comparison with the characteristic relaxation time $\tau_0 = f/m$, i.e. $\delta t \ll \tau_0$, or $\delta\tau = \delta t/\tau_0 \ll 1$. Numerical calculations, presented in Fig. 1.8, show that this dependence really becomes sufficiently weak, if $\delta\tau$, characterized in this Figure by the number of microsteps $N = \tau_0/\delta t = (\delta\tau)^{-1} \gg 1$ in one macrostep $\tau_0$, varies from 0.05 (at $\Gamma = 10$) to 0.1 (at $\Gamma = 0.2$).

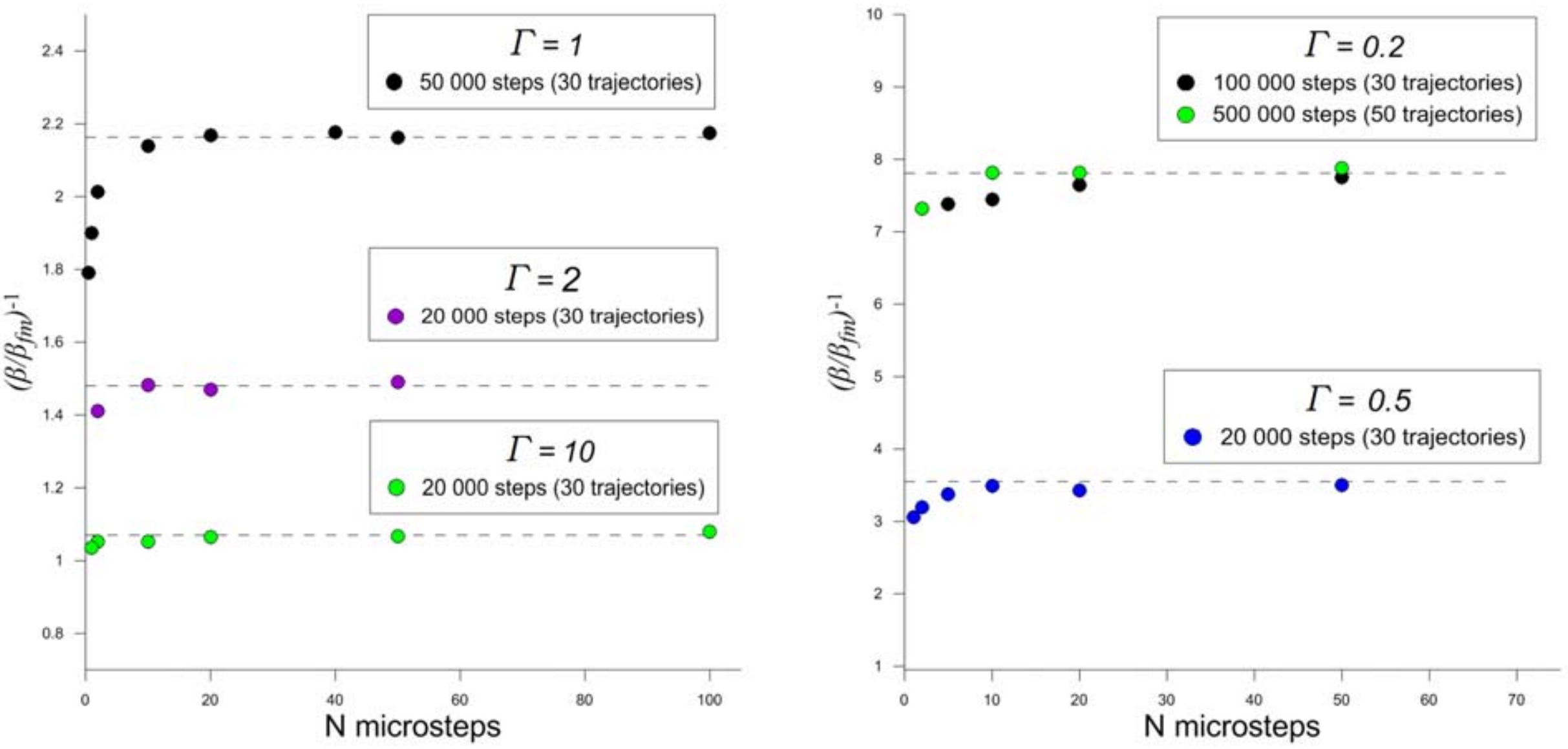

Fig. 1.8. Dependence of the inverse sweeping rate $\left((\delta\langle V \rangle/\delta t)/(V_0/\tau)\right)^{-1} = (\beta/\beta_{fm})^{-1}$ on the number of time microsteps $N = \tau_0/\delta t = (\delta\tau)^{-1}$ at different values of $\Gamma$.

The length of each calculated trajectory, characterized by the number of macrosteps, $k = t/\tau_0$ , was increased until the ratio of $\beta = \delta\langle V\rangle/\delta t$ to $\beta_{fm} = \delta\langle V_0\rangle/\delta t$ attained a steady-state value. As in the previous calculations (within the random walk approximation), the necessary number of steps notably increases with the decrease of the Knudsen number; namely, for $Kn = 10$ the steady state is attained at $k \approx 2\cdot 10^4$, whereas for $Kn = 0.2$ the minimum number is close to $k \approx 5\cdot 10^5$.

In order to provide a relatively high accuracy of the mean value determination with the standard error of the mean $SEM_x = \sqrt{\frac{1}{N\cdot(N-1)}\sum_{i=1}^{N}(x_i - \bar{x})^2} < 1\%$, from 60 to 150 calculations for each value of $Kn$ were carried out, forming a smooth distribution plot near the mean value, adequately approximated by the normal distribution function, Fig. 1.9.

The number of markers (Monte Carlo points) used in calculations for each trajectory was increased until the calculated mean value of the sweeping rate became invariant with respect to a further increase of this number, Fig. 1.10. It was shown that for the sufficiently large number of Monte Carlo points in a simulation box covering a trajectory (normally $10^5$ - $10^6$ points, depending on the length of a trajectory) a further increase of this number affects only the width of the distribution, rather than the searched mean value.\

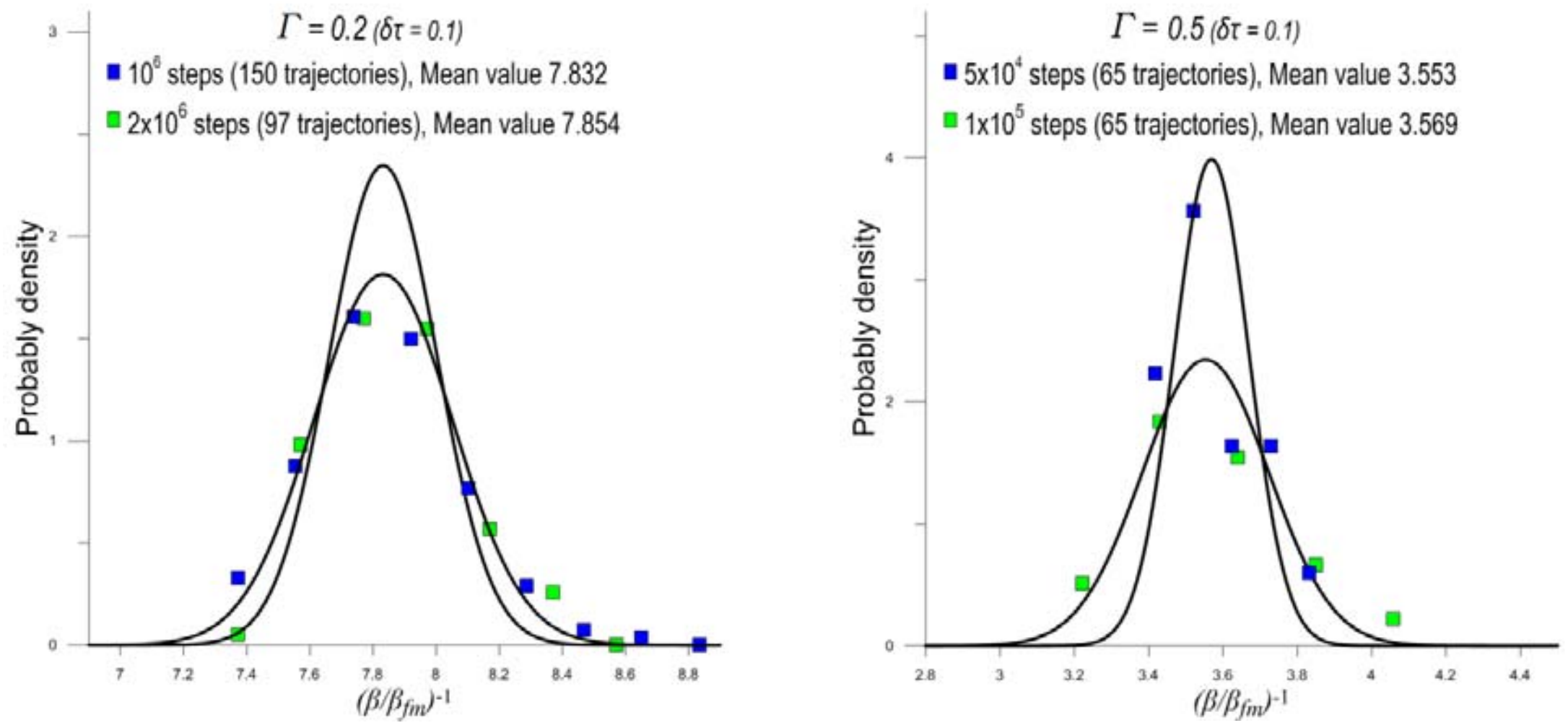

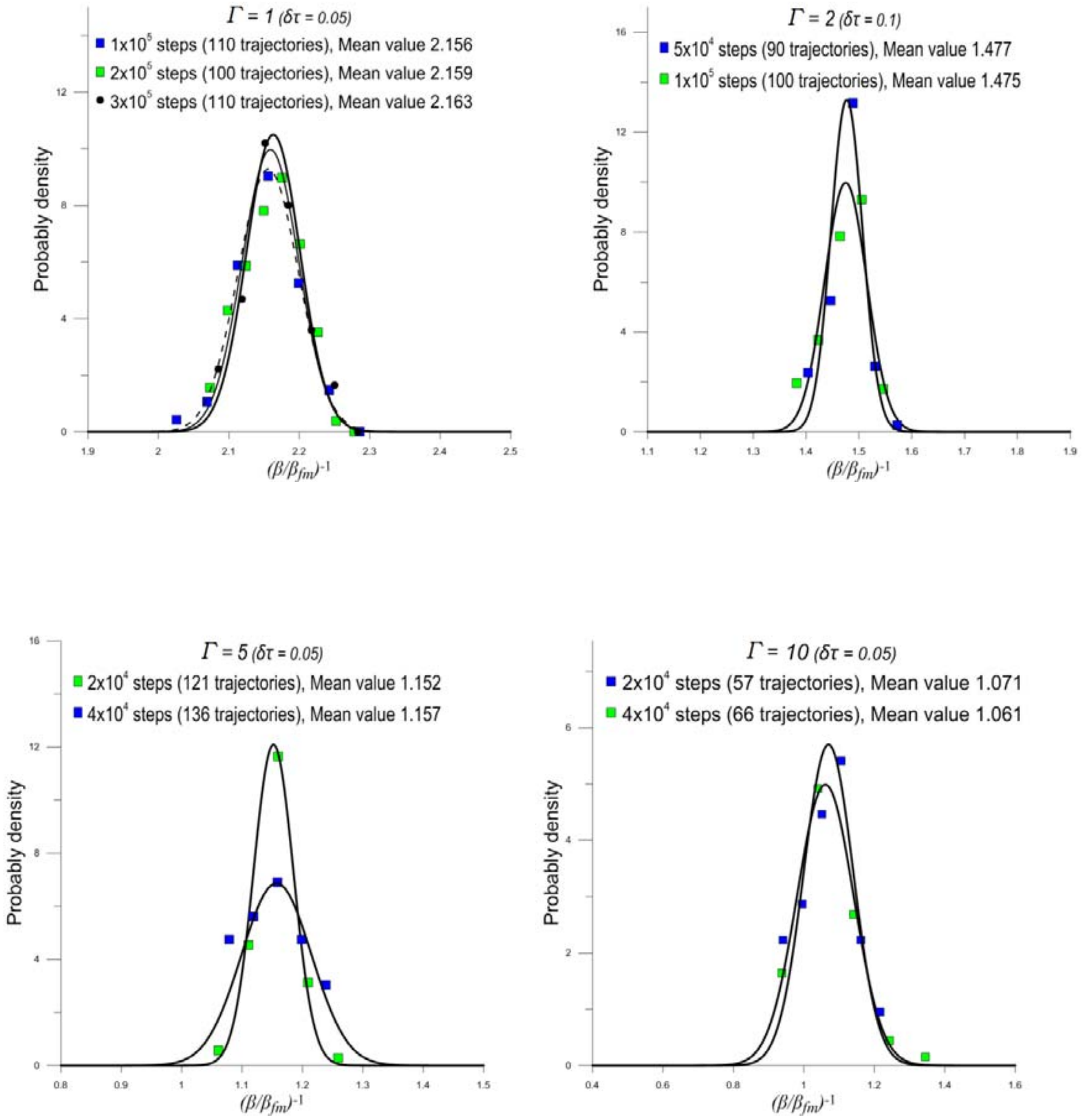

Fig. 1.9. Examples of calculation of the probability density $f(x)$ for $x = (\beta/\beta_{fm})^{-1} = (d\langle V\rangle/dt / d\langle V_0\rangle/dt)^{-1}$ at different $\Gamma$; for each number of macro-steps $k$, 60-150 trajectories with the microstep $\delta\tau = \delta t/\tau_0$ are generated resulting in calculation points, which are grouped in intervals of equal width $L$ (≈ 10% of the whole distribution width) and form normal distributions (lines) around the mean values (at given $k$).

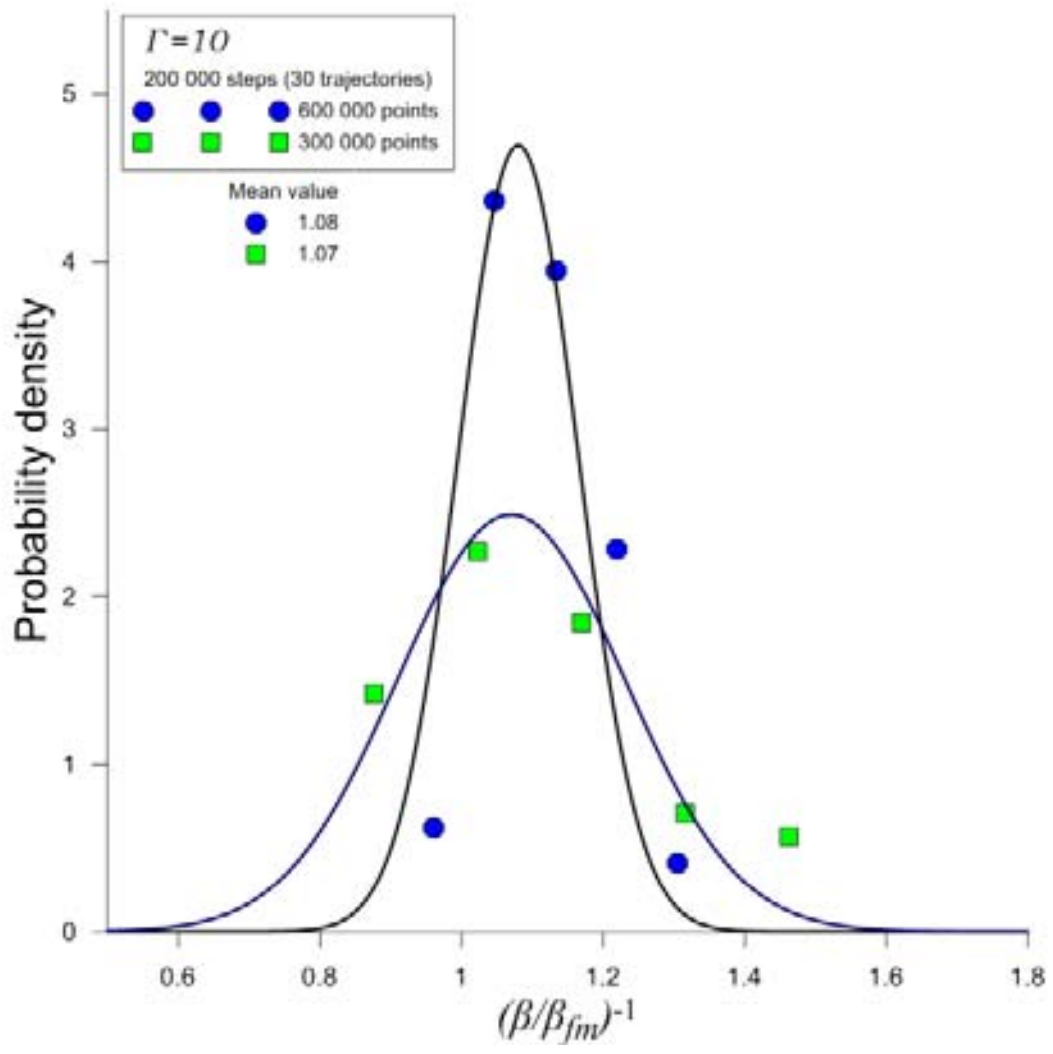

Fig. 1.10. Example of calculated dependence of the distribution function on the number of Monte Carlo points used in calclulations ( $\Gamma = 10$, $k = 200\,000$).

Results of calculations of the mean values $[(\delta V/\delta t)/(V_0/\tau)]^{-1} = \beta_{fm}/\beta$ are summarized in Table 1.3 and plotted in Fig. 1.11 in dependence on $\Gamma = 6D_{12}/R_{12}\bar{c}_{12} = 3\sqrt{\pi/2}Kn_D$ along with the interpolation curve

$$\frac{\beta}{\beta_{fm}} = \frac{\Gamma + A\cdot\Gamma^2 + B\cdot\Gamma^3}{1.5 + C\cdot\Gamma + D\cdot\Gamma^2 + B\cdot\Gamma^3}, \tag{1.53}$$

with four parameters found by least squares, $A = 9.55\cdot 10^5$, $B = 0.345\cdot A$, $C = 0.145\cdot A$, $D = 1.11\cdot A$, which correctly reduces to $2\Gamma/3$ in the limit $\Gamma << 1$ and to 1 in the limit $\Gamma >> 1$ and provides accuracy within ≈ 1%. This accuracy is comparable with that attained in experiments, e.g. [31], and thus Eq. (1.53) can be directly applied to the detailed analysis of experimental observations using the Smoluchowski rate equation, Eq. (1.11), with the refined kernel, Eq. (1.53) (which is beyond the scope of the current paper).

Somewhat reduced accuracy with the maximum error of ≈2% can be attained with the 2-parameter interpolation expression

$$\frac{\beta}{\beta_{fm}} = \frac{\Gamma(1+0.94\Gamma)}{1.5 + 0.7\Gamma + \Gamma(1+0.94\Gamma)}, \tag{1.54}$$

which is comparable with the maximum deviation from the new calculation points of the expressions derived in the simple random walk approximation of the kinetic approach [10], and is notably higher than accuracy provided by the semi-empirical expressions of Fuchs [4] and Dahneke [26] with the maximum error of ≈4-7% in the area of large diffusion Knudsen numbers, $2 \le \Gamma \le 10$, as demonstrated in Fig. 1.11. In this area a better agreement is attained with the interpolation expression of Gopalakrishnan and Hogan [13], which was obtained using the mean first passage time algorithm (with the standard deviation of 3−5%). In the area of smaller diffusion Knudsen numbers, $0.1 \le \Gamma \le 0.2$, the latter expression deviates from the calculation points more markedly, $\approx$3%, that exceeds the maximum calculation error ($\le$ 1−2%) of the new interpolation expressions, Eqs. (1.53) and (1.54), but is still within its own calculation error (3−5%).

Table 1.3. Calculation points of the mean values $\beta_{fm}/\beta$ with the standard error of the mean (SEM) and deviations (in %) of the interpolation curves

| Γ | Calculation points | SEM | 4-parameter fit | 2-parameter fit |
|---|---|---|---|---|
| 0.2 | 7.870 | 0.7% | 0.02% | 0.4% |
| 0.5 | 3.569 | 0.44% | -0.2% | -1.21% |
| 1 | 2.160 | 0.93% | -0.1% | -1.21% |
| 2 | 1.475 | 0.3% | 1% | 1.92% |
| 5 | 1.157 | 0.4% | -0.9% | 2.03% |
| 10 | 1.061 | 1% | -0.36% | 2.04% |

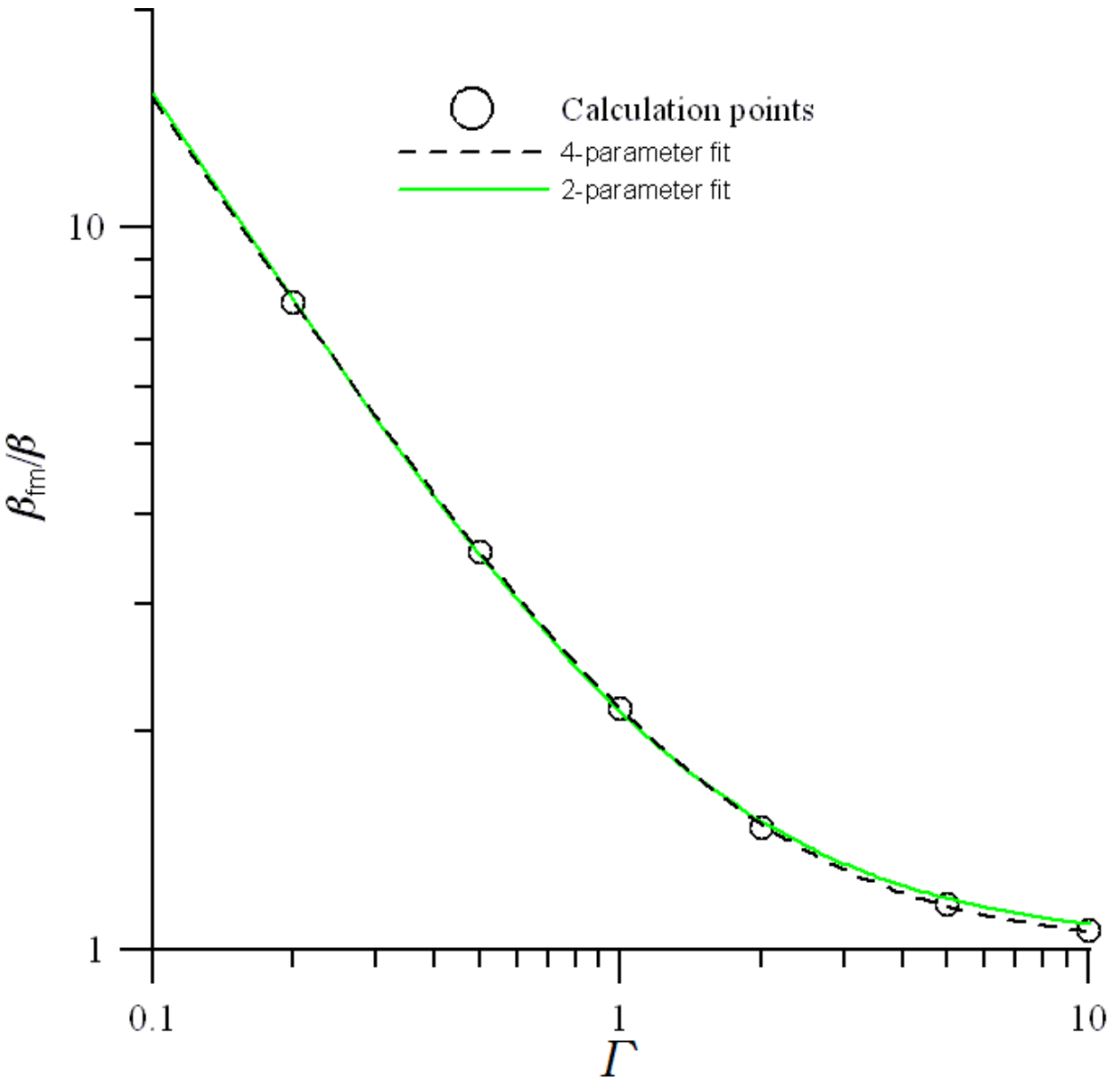

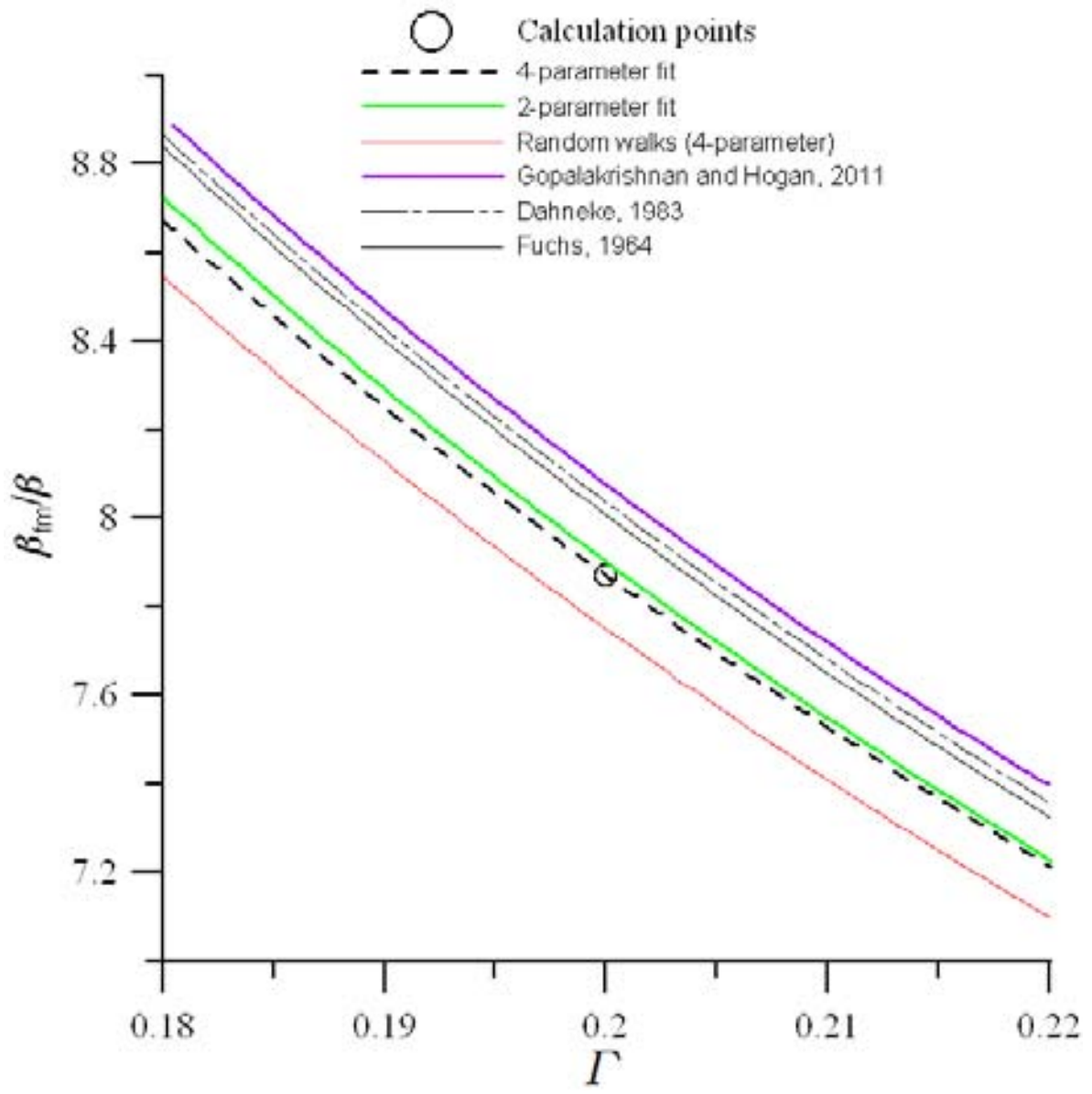

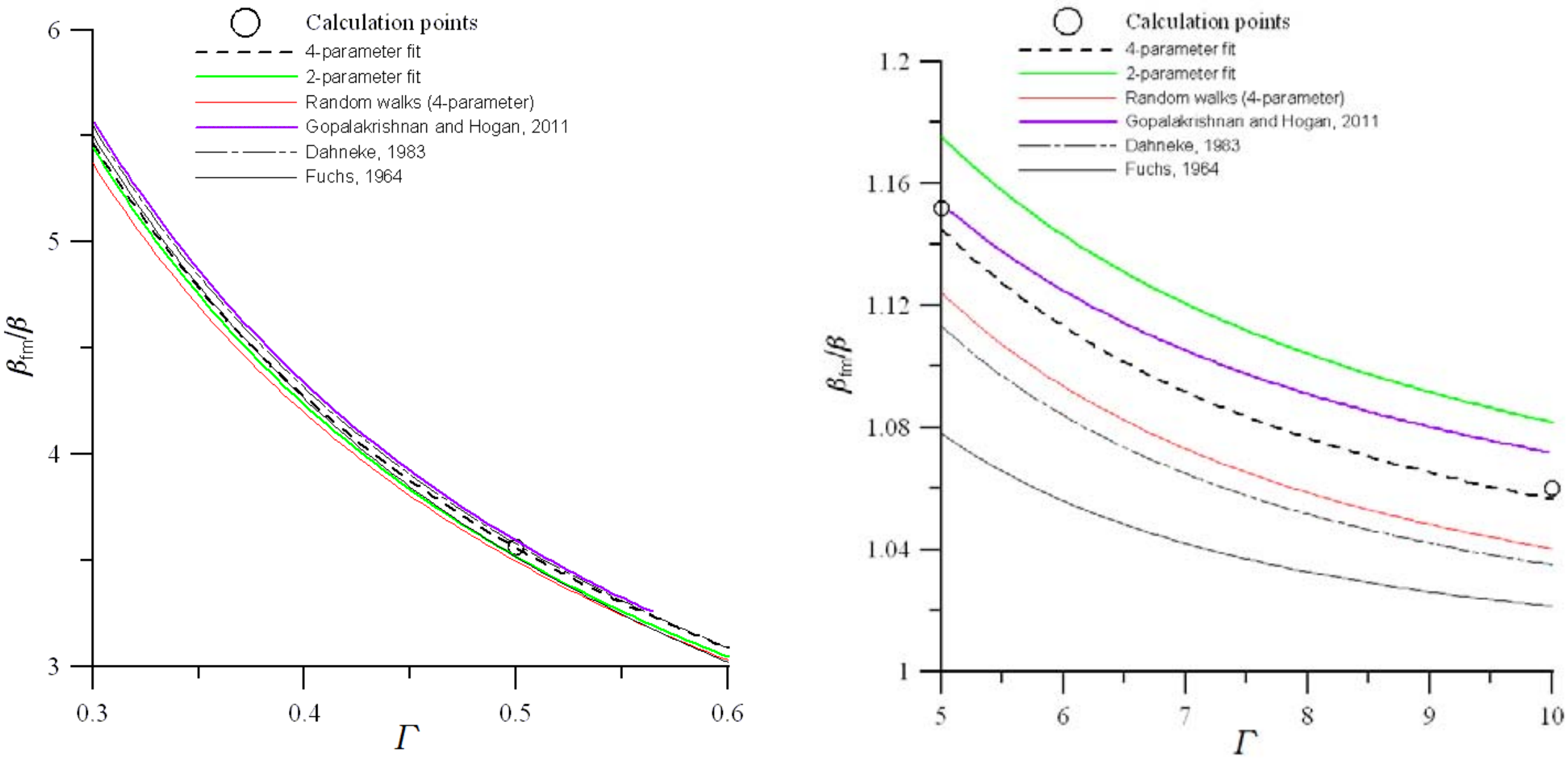

Fig. 1.11. Comparison of the calculated steady-state values of the inverse relative mean sweeping rate $((\delta V/\delta t)/(V_0/\tau))^{-1} = (\beta/\beta_{fm})^{-1}$ on the parameter $\Gamma$ with various interpolation curves.

The new interpolation expressions, Eqs. (1.53) and (1.54), are directly applicable to consideration of monodisperse systems (manageable in the initial stage of aerosol coagulation experiments), diffusion-limited reactions among identical species (A + A → C, where C does not affect the reaction, cf. Part 2), and also the condensation problem in the particular case of heavy vapour molecules (suspended in a light molecule gas), as explained in [13] and [10]. Applicability of the interpolation expressions to polydisperse systems will be discussed in the next Section.

*1.6.3. Collision rate of different size particles*

As explained in Section 1.3.2, the original multi-particle problem can be reduced to consideration of two-particle collisions owing to the diffusion mixing of comparable size particles, $R_1, R_2 << \bar{r}$. This significantly simplifies the coagulation problem and justifies the phenomenological form of the pair-wise kernel $\beta(R_1, R_2)$ in the Smoluchowski kinetic equation, Eq. (1.11), derived for spatially homogeneous systems. Besides, it allows determination of the applicability range of Eq. (1.11).

In the case of collisions between two particles of different sizes $R_i$, $i$ =1,2, the motion of each particle is described by Eq. (1.45) with the corresponding parameters $m_i$ and $f_i$. Calculating the trajectory of each particle $\mathbf{r}_i^*(t)$ as described in Section 1.6.1, one can find the trajectory $\mathbf{r}_{12}^*(t) = \mathbf{r}_1^*(t) - \mathbf{r}_2^*(t)$ of the effective particle and then calculate the rate of the volume sweeping by this particle of the assigned radius $R_{12} = R_1 + R_2$. After introduction of the dimensionless variables, the system of the Langevin equations for the two particles is completely determined by the three values, $m_1/m_2$ , $f_1/f_2$ and $Kn_D$ , as demonstrated by Gopalakrishnan and Hogan [13]. Furthermore, those authors reasonably asserted that the most realistic case of collisions between entities of similar density should be considered, noting that this can lead to a (non-specified in their paper) correlation between the examined $m_1/m_2$ and $f_1/f_2$ values.

Indeed, in the case of two particles of similar density in the suspending gas (with the mean free path of the gas molecules $\lambda_m$), all the three values $m_1/m_2 = (R_1/R_2)^3$, $f_1/f_2 = R_1 C(Kn_2)/R_2 C(Kn_1)$ and $Kn_D = \sqrt{kTm_{12}}/f_{12}R_{12}$ , are unambiguously determined by the two independent parameters $R_1$ and $R_2$ (or equally, by $R_1$ and $R_2/R_1$), where $C(Kn_i) = 1 + Kn_i(A_1 + A_2 \exp(-A_3/Kn_i))$ is the sleep correction factor for a spherical particle of radius $R_i$, depending on the Knudsen number, $Kn_i = \lambda_m/R_i$, with $A_1$ = 1.257, $A_2$ = 0.40 and $A_3$ = 1.1 [19], or $A_1$ = 1.165, $A_2$ = 0.483 and $A_3$ = 0.997 [20]. Therefore, only two of them (say, $m_1/m_2$ and $Kn_D$) can be considered as independent, whereas the value of the third one ( $f_1/f_2$ ) is determined by the values of the first two ones. This may explain a strong functional dependency on $\theta_m = m_1/m_2$ and $\theta_f = f_1/f_2$ with the maximum deviation between calculation points up to ≈ 13−14%, revealed in calculations of Gopalakrishnan and Hogan (2011) in the transition regime for $Kn_D$ = 0.5 and 1, considering $\theta_m$ and $\theta_f$ as independent values (Table S3 of Supplemental Information).

Nevertheless, basing on the results of the analysis of (Veshchunov and Azarov, 2012) in the random walk approximation, where the sweeping rate was derived as a function of the single parameter, $\bar{a}_{12}/R_{12} = (8/3\pi)\Gamma = (8/\sqrt{2\pi})Kn_D$, one may assume that, with the adequate choice of the third parameter $\theta_f$, this dependency on $\theta_m$ becomes weak and thus can be neglected also in the more accurate Langevin approach (despite the equation for the effective particle relocation does not exactly reduce to the Langevin equation for a single particle, in contrast to the above considered case of equal size particles).

In order to verify this assumption, additional calculations were carried out for the transition mode with an arbitrary choice of the parameters $Kn_D$ and $m_1/m_2$ (for comparable size particles), which unambiguously determine the value of the third parameter $f_1/f_2$ . As above explained, in fact one can choose two arbitrary values of the radii $R_1$ and $R_2$ (within one order of magnitude for comparable size particles) and then determine all other parameters. For instance, for two particles of radii $R_1$ = 10 nm and $R_2$ = 5 nm in the typical suspending gas with the molecular radius $R_m$ = 0.3 nm and the mean free path $\lambda_m$ = 68 nm one can determine the model parameters $Kn_D = 0.841$, $m_1/m_2 = 8$, $f_1/f_2 = 1.44$, whereas for another choice of the radii, $R_1$ = 20 nm and $R_2$ = 5 nm, one obtains $Kn_D = 0.451$, $m_1/m_2 = 64$ , $f_1/f_2 = 14.77$ . Results of calculations with these two sets of the parameters performed with sufficiently small microsteps, $\delta\tau = \delta t/\tau_0 = 0.05$ and sufficiently large number of persistent macrosteps, $k = t/\tau_0$ = $10^5$, are presented in in Fig. 1.12. The calculated mean values for the collision rate match to the interpolation expression, Eq. (1.53), with the maximum deviation less than 1%, i.e. within the accuracy limits of the interpolation curve determination.

These two examples illustrate that the interpolation expression, Eq. (1.53), obtained for equal size particles, might be valid also for particles of different (comparable) sizes. A more reliable confirmation requires an extended set of calculations for different combinations of $Kn_{\mathrm{D}}$ and $m_1/m_2$ , which are beyond the scope of the current paper and, being particularly time consuming, are planned in the near future.

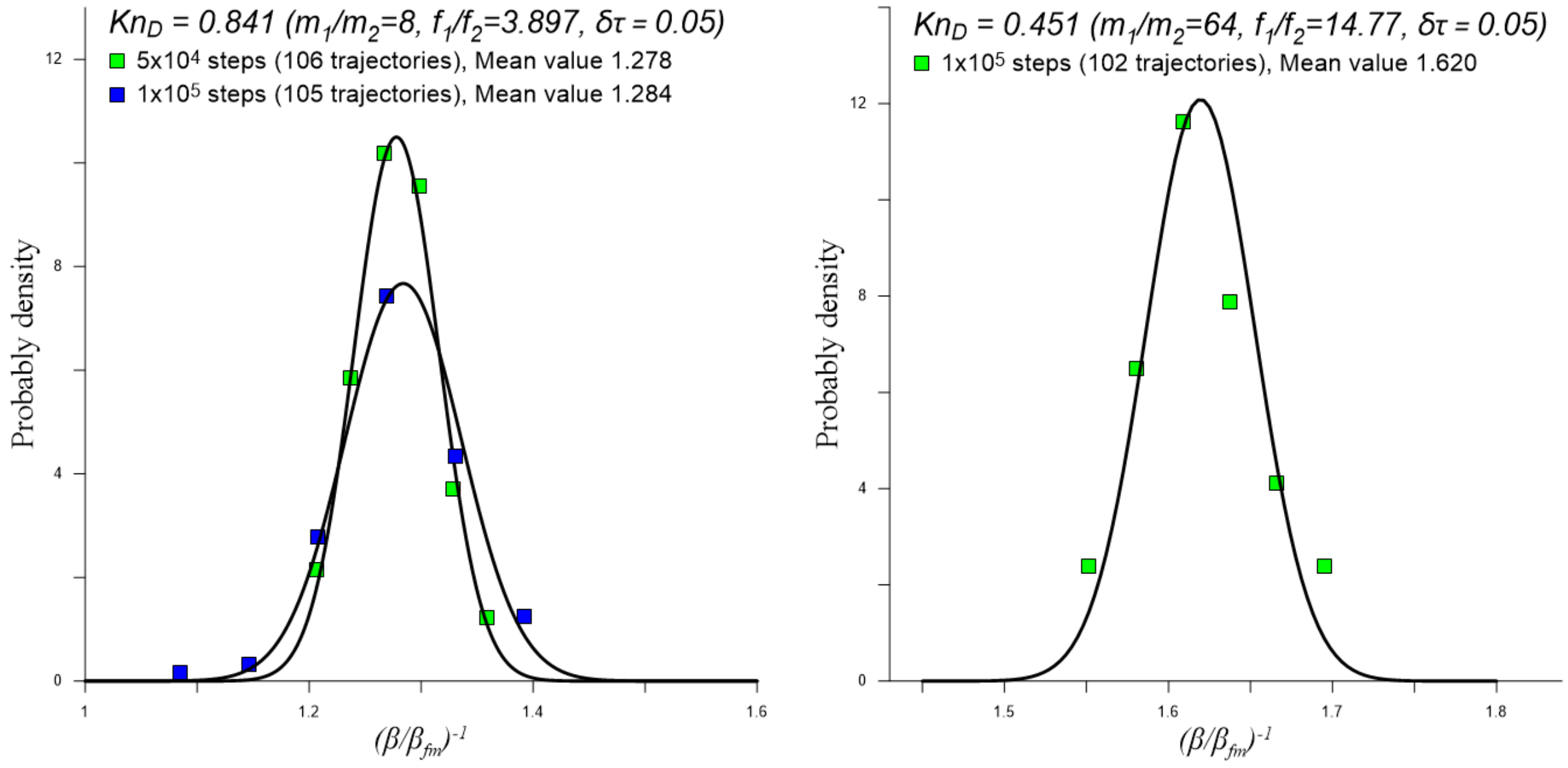

Fig. 1.12. Calculation of the probability density $f(x)$ for $x = (\beta/\beta_{fm})^{-1}$ in the case of different size particles with $Kn_D = 0.841$ ($\Gamma = 3.16$), $m_1/m_2 = 8$, $f_1/f_2 = 1.44$ (*left*) and $Kn_D = 0.451$ ($\Gamma = 1.7$), $m_1/m_2 = 64$, $f_1/f_2 = 14.77$ (*right*).

### *1.6.4. Conclusions*

The kinetic approach to calculation of the coagulation kernel, derived in the two subsequent approximations of the random walk theory in the authors' previous papers avoiding semi-empirical assumptions of the traditional models, is further advanced by application of the Langevin theory for Brownian particle migration. In this approach the original multi-particle problem is rigorously reduced to consideration of two-particle collisions, and the collision rate can be evaluated as the rate of the volume sweeping by the effective Brownian particle of radius $R_{12} = R_1 + R_2$ migrating with the relative velocity of two original particles, $\mathbf{v}_{12}(t) = \mathbf{v}_1(t) - \mathbf{v}_2(t)$. Correspondingly, in the random walk approximation the effective particle migration was described by the Einstein diffusion equation (to which the Langevin equation properly reduces at a large time scale, $t >> \tau_0$, where $\tau_0$ is the particle persistent time) with the diffusivity $D_{12} = D_1 + D_2$. In the advanced approach the trajectories of the original particles are more adequately described at a shorter time scale, $t \approx \tau_0$, by the Langevin equation, and thus the rate of the volume sweeping by the effective particle can be calculated more accurately.

Numerical calculations of the sweeping rate in the transition mode were realized using the Monte Carlo method. Calculation points for equal size particles were interpolated by the new analytical expression (with the maximum deviation from the calculation points $\approx 1\%$), which has somewhat higher accuracy in comparison with the interpolation expressions derived in the random walk approximation of the kinetic approach (with the maximum deviation from the new calculation points $\approx 2\%$) and notably improves predictions of the traditional semi-empirical correlations (with the maximum deviation $\approx 4-7\%$).

It is outlined that the new interpolation expressions are properly applicable to monodisperse systems (manageable in the initial stage of aerosol coagulation experiments), diffusion-limited reactions among identical species and the condensation problem in the particular case of heavy vapour molecules.

The applicability of the new interpolation expressions to the collisions of different (but comparable) size particles, earlier substantiated by the authors in the random walk approximation, was verified by the two sets of calculations with an arbitrary choice of two particles radii $R_1$ and $R_2$ (within one order of magnitude), which determined two different combinations of the model parameters $m_1/m_2$ , $Kn_\mathrm{D}$ and $f_1/f_2$ (only two of them being independent). The calculated collision rates accurately matched to the new interpolation expression with the maximum deviation less than 1% (i.e. within the accuracy of this expression determination $\approx 1\%$), thus demonstrating that the interpolation expression, derived for equal size particles, might be also valid for particles of different sizes. Recent calculations of the two-particle coagulation rate in the Langevin approach using the mean first passage time method by Gopalakrishnan and Hogan [13] also confirmed this assumption, however with a lower accuracy. For more precise confirmation of this assumption an extended set of calculations for different combinations $m_1/m_2$ and $Kn_\mathrm{D}$ (for comparable size particles) are planned in the near future.

From the obtained calculation results one can generally conclude that the interpolation expressions for the collision rate derived in the random walk approximation of the kinetic approach were additionally justified with a reasonable accuracy and further improved in the higher approximation of the kinetic approach using the more precise Langevin equation for the particle trajectories.

## 1.7. Extension of the new approach to slow Brownian coagulation with finite sticking probability

Collision and subsequent coagulation of aero-colloidal particles in a stationary medium is a result of their Brownian motion and of the mutual forces of interaction between them. The interactions between particles due to the dispersion forces are mainly responsible for the coagulation of two colliding particles. The effect of these short-range forces on coagulation is usually accounted for through a phenomenological sticking probability, i.e., the probability of coagulation upon collision. Since the dispersion forces of large particles are usually strong, it is generally assumed that, when a particle collides with another particle or with a filter element, this sticking probability $P_{12}$ is equal to unity, i.e., every collision leads to coagulation.

### *1.7.1. Collisions of compact spherical Brownian particles*

For compact solid spherical particles an approach was developed by Narsimhan and Ruckenstein [30], who proposed a model for the Brownian coagulation of electrically neutral aerosol particles taking into consideration the inter-particle forces due to the van der Waals attraction and Born repulsion. On this basis, the coagulation coefficient was calculated through the Monte Carlo simulation for the entire range of Knudsen numbers. As a result, the reflected particles which have a kinetic energy less than the depth of the potential well are assumed to be captured in the potential well. Nevertheless, for small particles the potential well is not deep, and the probability of escape of particles from the potential well becomes higher, i.e., the sticking probability becomes less than unity. For particles of very small sizes $\le 10$ nm the model predicts vanishingly small sticking probabilities.

This conclusion was in a good agreement with that of Dahneke [32] and Loeffler [33], who have developed a simple model for predicting the critical normal impact velocity which would allow colliding particles to escape (rebound) from an inter-particle potential well of a finite depth. Their estimations showed that the critical velocities become comparable with the thermal velocities for very small particles of diameter $\le 10$ nm. A similar conclusion was derived by Wang and Kasper [34] who identified efficiency of “thermal rebound” from the filter surface for small particles of the same size range $\le 10$ nm, which is of practical importance for aerosols generated by radioactive decay.

The semi-empirical flux matching theory, proposed by Fuchs [4] for the transition mode, with an appropriate definition of the absorption sphere radius is traditionally applied to consideration of hard sphere collisions. The theory is well grounded in the case of collisions of small particles with a large trap. However, for coalescence of comparable size particles this theory inherits the main deficiency of the traditional diffusion approach (for the continuum mode), since in the transition mode the diffusion theory cannot be used near the outside surface of the absorbing sphere, where the external and internal fluxes are matched.

Application of Fuchs’ flux matching theory to analysis of Brownian coagulation with a finite sticking probability [4] resulted in conclusion that the coagulation rate is a linear function of the sticking probability only in the free molecular mode, whereas in the transition mode the dependence weakens and vanishes at very small Knudsen numbers. In particular, this result, being in contradiction with consideration of “slow coagulation” by Smoluchowski [1], led Fuchs to a qualitative conclusion that coincidence of the measured and calculated coagulation rate constants cannot be considered as an argument for highly sticking of particles collisions. To overcome this contradiction and inconsistency of the flux matching theory in application to coagulation, a more adequate approach should be applied.

The revealed in [10] (see also Section 1.8 below) equivalence between the coagulation and condensation problems vanishes in the case of finite sticking probability, $P_{12} \le 1$, i.e. beyond the hard sphere approximation. Indeed, in the problem of comparable size particles coagulation, which is properly reduced to the kinetic consideration of spatially homogeneous system, the random distribution of particles quickly reinstates after each collision (independently, whether it was effective or not), owing to the particles rapid diffusion relaxation (or mixing). Therefore, in this kinetic regime the coagulation rate should be calculated as the collision rate multiplied by this probability factor,

$$\beta'_{12} = \beta_{12} P_{12} . \quad (1.55)$$

In the limit $a_{12} << R_{12}$ the general interpolation formula for the coagulation kernel $\beta_{12}$ calculated in Section 1.5.4 can be reduced with a good accuracy to

$$\beta_{12} \approx \left( \frac{1}{\beta_{fm}} + \frac{1}{\beta_{con}} \right)^{-1} = \frac{4\pi D_{12} R_{12} R_{12}^2 \sqrt{8\pi kT\left(m_1^{-1} + m_2^{-1}\right)}}{4\pi D_{12} R_{12} + R_{12}^2 \sqrt{8\pi kT\left(m_1^{-1} + m_2^{-1}\right)}}, \tag{1.56}$$

corresponding to the so called harmonic mean approximation of the Fuchs theory (cf., e.g., [24]), with $\beta_{con} = 4\pi D_{12} R_{12}$ and $\beta_{fm} = R_{12}^2 \sqrt{8\pi kT\left(m_1^{-1} + m_2^{-1}\right)} = R_{12}^2 \bar{c}_{12}$.

On the other hand, in the problem of a dense gas of Brownian particles condensation in a large trap, to which the flux matching theory of Fuchs can be applied with a reasonable accuracy, a small particle after each ineffective collision with the trap drifts apart within its mean drift distance, e.g. remains in the absorbing sphere, separating zones of the different modes of the particle migration, free-molecular (i.e. kinetic), inside the absorbing sphere, and diffusion, outside the sphere. Thus, in the flux matches the free-molecular flux in the internal zone (where particles are well mixed) should be multiplied by the sticking probability factor,

$$\beta'_{fm} = P_{12} \beta_{fm}. \tag{1.57}$$

As a result, in the case of a small thickness of the shell, $a_{12} << R_{12}$, the condensation rate can be calculated as

$$\beta''_{12} \approx \left( \frac{1}{P_{12}\beta_{fm}} + \frac{1}{\beta_{con}} \right)^{-1} = \frac{4\pi D_{12} R_{12} R_{12}^2 \sqrt{8\pi kT\left(m_1^{-1} + m_2^{-1}\right)} P_{12}}{4\pi D_{12} R_{12} + R_{12}^2 \sqrt{8\pi kT\left(m_1^{-1} + m_1^{-1}\right)} P_{12}} , \tag{1.58}$$

This expression (correct for the condensation problem) was extended by Fuchs also to consideration of comparable size particles (and further used by other authors, see, e.g. [3]), resulting in erroneous predictions for the slow coagulation process.

Indeed, comparison of Eqs. (1.56) and (1.58) in the transition mode, $a_{12} \approx R_{12}$ (or $\beta_{fm} \approx \beta_{con}$), shows that

$$\frac{\beta'_{12}}{\beta''_{12}} = \frac{4\pi D_{12} R_{12} + R_{12}^2 \sqrt{8\pi kT\left(m_1^{-1} + m_1^{-1}\right)} P_{12}}{4\pi D_{12} R_{12} + R_{12}^2 \sqrt{8\pi kT\left(m_1^{-1} + m_2^{-1}\right)}} \approx \frac{\mathrm{Kn_D} + P_{12}}{\mathrm{Kn_D} + 1} < 1, \tag{1.59}$$

where

$$\frac{4\pi D_{12} R_{12}}{R_{12}^2 \sqrt{8\pi kT\left(m_1^{-1} + m_2^{-1}\right)}} = \frac{4 D_{12}}{R_{12} \bar{c}_{12}} = \frac{\pi a_{12}}{4 R_{12}} \approx \mathrm{Kn_D} \approx 1. \tag{1.60}$$

Therefore, inconsistency of the traditional approach in application to comparable size particles in the transition mode can be experimentally measured, if $P_{12} < 1$.

Indeed, as above explained, the sticking probability becomes notably small for very small aerosol particles ≤ 10 nm [30]. This conclusion is in a good agreement with that of Dahneke [32] and Loeffler [33], who have developed a simple model for predicting the critical normal impact velocity which would allow colliding particles to escape (rebound) from an inter-particle potential well of a finite depth *E*. When elastic particles collide, the incoming kinetic energy of the particles is converted into elastic strain energy as they deform in the vicinity of contact. The kinetic energy (other than that dissipated by the internal friction of the solids or remaining as elastic vibrations) is restored as the particles rebound. The sticking probability is essentially zero when the energy dissipated in the solids and fluid is negligible. Their estimations showed that the critical velocities which are generally very high, become comparable with the thermal velocities for small particles ≤ 10 nm. This size range can be of practical importance for aerosols generated by radioactive decay.

It is straightforward to see that coagulation of such small particles occurs in the transition mode, where Eq. (1.59) is valid. Indeed, in the atmosphere under normal conditions, where the carrier gas (e.g. nitrogen) molecules are characterized by the size $2R_m \approx 0.3$ nm and the mean free path $\lambda_m \approx 3.3$ nm, the diffusion Knudsen number can be evaluated as $\mathrm{Kn_D} = a/R \approx C_c (m/m_m)^{1/2} (R_m/R)^2 \approx C_c (R_m/R)^{1/2}$ (see Section 1.2), where $C_c \approx 1 + \mathrm{Kn}(1.257 + 0.4\exp(-1.1/\mathrm{Kn}))$ is the slip correction factor for spherical particles, $\mathrm{Kn} = \lambda_m/R$, and thus for 1 nm≤ $R$ ≤ 10 nm varies in the range $0.2 \le \mathrm{Kn_D} \le 2$.

In this range the ratio in Eq. (1.59) becomes really smaller than 1, demonstrating inconsistency of flux matching theory; however, effect is not very strong. For this reason, to demonstrate more clearly the deficiency of the traditional flux matching theory, extension of the coagulation theory to the Brownian particles aggregation and reaction kinetics, where the sticking probability can be really very small, will be presented in the next sections. In these particular cases of the Brownian collision theory the considered effect becomes much more pronounced and thus can be more easily detected experimentally.

*1.7.2. Collisions of non-spherical Brownian particles and aggregates*

Collisional growth of non-spherical particles plays an important role in many aerosol systems [35, 36] and aqueous colloids [37, 38], particularly those in which aggregates form and evolve. In actual experimental situations various factors influence the aggregation process. One important physical parameter affecting the aggregation phenomena is the cluster mobility. In the diffusion-limited regime, clusters approach each other by Brownian motion. The static properties, such as the fractal dimension, are rather insensitive to the cluster mobility, but the dynamic properties, such as the cluster size distribution, are critically affected.

Mountain et al. [39] and Mulholland et al. [40] modified the collision kernels to describe not only spherical, but also fractal particles

$$\beta_{con} = \frac{2kT}{3\mu}\left(\frac{1}{V_1^{1/d_f}} + \frac{1}{V_2^{1/d_f}}\right)\left(V_1^{1/d_f} + V_2^{1/d_f}\right), \tag{1.61}$$

and

$$\beta_{fm} = \left(\frac{3}{4\pi}\right)^{2/d_f - 1/2} r_0^{2-6/d_f} \left(\frac{6kT}{\rho_0}\right)^{2/d_f - 1/2} \left(\frac{1}{V_1} + \frac{1}{V_2}\right)^{1/2} \left(V_1^{1/D_f} + V_2^{1/D_f}\right)^2, \tag{1.62}$$

where $d_f \leq 3$ is the fractal dimension of aggregates, $V_1$ and $V_2$ are volumes of aggregates, $r_0$ and $\rho_0$ are the primary particle radius and density, respectively.

Based on these modified collision kernels, Oh and Sorensen [41] compared predicted agglomeration rates to light scattering measurements made as a function of height in a flame. They reported good agreement considering the uncertainty inherent in the value of the soot index of refraction. For the transition modes in experiments with sooting flames, the flux matching theory of Fuchs was applied by Maricq [42].

Other factors such as chemical reactivity, kinetic energy, mass, etc. of the clusters also influence the aggregation process. Depending on these parameters the coagulation of two clusters may or may not take place during a collision. In the simulations these factors can be taken into account through the sticking probability $P_{12}$ of two clusters. The effects of sticking probability on the fractal dimension of cluster-cluster aggregates has been investigated numerically by Meakin [43], and by Kolb and Jullien [44]. Lowering the sticking probability tends to make the clusters more compact, particularly on short length scales. This process, known as reaction-limited aggregation, leads to more compact structures than the diffusion limited or ballistic models.

The reaction rate and other factors influencing the formation of permanent bonds between clusters can generally be controlled experimentally. Therefore, the understanding of the effects of chemical bonding on the cluster-size distribution is an important problem because it might provide an explanation for the variety of results observed in the experiments with combustion aerosols.

In the case of aqueous colloid aggregation, static and dynamic light scattering are also predominant tools used to probe the structure and kinetics of aggregation. There is a host of experimental studies on colloidal particle aggregation using prototype materials such as colloidal gold, silica, polystyrene latex, as well as other kinds of particles, over a sufficiently broad range of physico-chemical conditions by varying the concentration of a simple indifferent electrolyte, which regulates the electrostatic interactions between particles. The aggregation is initiated by addition of a small amount of electrolyte which displaces the charged ions from the surface of the colloid. The rate of aggregation is directly controlled by the amount of electrolyte added and can be varied over many orders of magnitude, see, e.g., [45].

The origin of the kinetic behavior can be understood by consideration of the nature of the short-range interaction energy between two approaching particles. This can be discussed within the Derjaguin-Landau-Verwey-Overbeek model, which consists of a screened Coulomb repulsive barrier, with a height $E_b$ determined by the surface charge, and a screening length, $l_g$, determined by the ion concentration in solution [41]. The probability $P_{12}$ of two particles sticking upon approach to within $l_g$, is proportional to $\exp(-E_b/kT)$. Initially, $E_b >> kT$ and $l_g \approx 10$ nm and the colloid does not aggregate. Electrolyte displaces the charged ions adsorbed on the colloid surface, without significantly changing the ionic concentration in solution. This reduces $E_b$ without affecting $l_g$, and therefore directly affects the sticking probability. Addition of sufficient electrolyte displaces nearly all the charge so that $E_b << kT$. However, addition of less electrolyte causes only a small amount of the charge to be displaced, so that $E_b \geq kT$, and $P_{12} << 1$. In the slow regime, a very sensitive dependence of the initial rate of aggregation on the concentration of electrolyte added is observed. If the surface coverage of electrolyte is maintained constant, a temperature dependence consistent with the rate being proportional to $\exp(-E_b/kT)$ is observed. Therefore, the conclusion derived in Section 1.7.1 on inconsistency of the traditional flux matching approach in application to comparable size particles in the transition mode, if $P_{12} < 1$, becomes even more important in the case of the aggregation process.

The effects of chemical bonding (e.g. in combustion aerosols) and of screened electrostatic interactions between particles (in aqueous colloids) on the cluster-cluster aggregation kinetics can be considered in terms of chemical reactions between Brownian particles, presented in Part 2 (Section 2.3.2).

## 1.8. Vapour condensation

The steady state flow of vapour molecules to a spherical particle in a sample of volume $V \to \infty$, when the particle radius $R$ is sufficiently large compared to the mean free path $\lambda_v$ of the diffusing (with the diffusivity $D$) vapour molecules (or $\mathrm{Kn} = \lambda_v/R << 1$, where $\mathrm{Kn}$ is the Knudsen number), is given by Maxwell's solution of the continuum transport equation,

$$F_c = 4\pi D_v R n_v, \tag{1.63}$$

where $n_v = N_v/V$ is the mean concentration of vapour molecules and $n_s$ is their saturation concentration (i.e. at vapour-solid equilibrium). This expression can be applied to calculation of the spherical particle growth kinetics

$$\frac{1}{\Omega_p}\frac{dV_p}{dt} = F_c, \tag{1.64a}$$

where $V_p$ is the particle volume, $\Omega_p$ is the volume of a condensed vapour molecule in the particle, or

$$\frac{dR}{dt} = \frac{D\Omega_p}{R}\left(n_v - n_s\right), \tag{1.64b}$$

under the steady state growth condition, $R^{-1}\, dR/dt << \tau_{ss}^{-1}$, where $\tau_{ss} \approx R^2/\pi D$ is the characteristic time for attainment of the steady state solution of the vapour diffusion problem with the fixed value of $R$ (i.e. with the immobile boundary), which can be deduced from analysis of the general, non-stationary solution of the diffusion problem, $F_c = 4\pi D_v R\left(n_v - n_s\right)\left(1 + R/\sqrt{\pi D t}\right)$ (cf., e.g. [4]). This provides an insignificant restriction on the applicability of the steady state approximation, $\left(n_v - n_s\right)\Omega_p << \pi$, and thus Eqs. (1.64) can be used with a good accuracy.

At the other extreme, $\mathrm{Kn} >> 1$, the expression for the vapour flow based on the kinetic theory of gases can be used

$$F_{fm} = \pi R^2 \bar{c}_v n_v = \frac{R\bar{c}_v}{4D_v} F_c, \tag{1.65}$$

where $\bar{u}_v = (8kT/\pi m_v)^{1/2}$ is the mean thermal speed of vapour molecules, $\alpha$ is the molecular accommodation coefficient. Both expressions are no longer valid in the transition mode, when the mean free path of the diffusive vapour molecules becomes comparable with the particle radius, $\mathrm{Kn} \approx 1$.

Early investigations of Knudsen aerosol condensation used the flux matching theory of Fuchs [4], that is, by considering the non-continuum effects to be limited to a region $R \le r \le \Delta$ beyond the droplet surface and assuming that for $r \ge \Delta$ continuum theory applies. The absorbing sphere radius $\Delta$, then, is of the order of the mean free path $\lambda_v$ and within this inner region the simple kinetic theory of gases is assumed to apply. Fuchs, by matching the fluxes for the two domains at $r = \Delta$, obtained the flux ratio $F/F_{fm}$, as follows

$$\frac{F}{F_{fm}} = \frac{1+\Delta/R}{1+\Delta/R+R\bar{c}_v/4D_v}, \tag{1.66}$$

The value $\Delta$ was not specified in the original model and must be adjusted empirically or estimated by independent theory. Several choices for $\Delta$ have been proposed; the simplest, due to Fuchs, is $\Delta = 0$. Dahneke [26] suggested $\Delta = \lambda_v$, and using $\lambda_v = 2D_v/\bar{u}_v$ in definition of the Knudsen number (designated here as $\mathrm{Kn}_{\mathrm{Da}}$), obtained

$$\frac{F}{F_c} = \frac{1+\mathrm{Kn}_{\mathrm{Da}}}{1+2\mathrm{Kn}_{\mathrm{Da}}(1+\mathrm{Kn}_{\mathrm{Da}})\alpha^{-1}}, \tag{1.67}$$

but numerous other possibilities exist, as reviewed by Davis [23].

Besides, the basic equations used in the flux matching theory cannot be strictly justified and thus the obtained expressions for the condensation flux can be used only for qualitative consideration. Indeed, the diffusion equation for the vapour molecules concentration, which is applied in the flux matching theory outside the absorbing sphere, is valid under the general condition that the length scale $L$ of the concentration variation is large in comparison with the mean free path, $L >> \lambda_v$. Taking into account that the diffusion concentration profile substantially varies outside the absorbing sphere (of radius $\Delta \approx R + \lambda_v$) on the scale of $L \approx \Delta$, the condition of the diffusion equation validity in the vicinity of the absorbing sphere takes the form, $\Delta >> \lambda_v$, or $R >> \lambda_v$. This condition significantly restricts applicability of the flux matching theory to small Knudsen numbers, $\mathrm{Kn} = \lambda_v/R << 1$, i.e. only small corrections to the flux, Eq. (1), in the near-continuum regime can be properly evaluated in the flux matching approach.The concentration distributions of the diffusing species and background gas in the transition mode are governed rigorously by the Boltzmann equation. However, there does not exist a general solution to the Boltzmann equation valid over the full range of Knudsen numbers for arbitrary masses of the diffusing species, $m_v$, and the background gas, $m_g$; nonetheless, the problem can be studied more rigorously in the two limiting cases of heavy and light vapour molecules, when the partial vapour pressure is much less than the gas pressure and thus the vapour - vapour collisions can be neglected.

*1.8.1. Heavy vapour molecules condensation*

The new approach can be applied to consideration of the condensation problem in the particular case of heavy vapour molecules (suspended in the light molecule gas) [10], since, as shown in Section 1.2, a diffusing species can be considered as a Brownian particle, if its mass $m_v$ is large in comparison with the mass $m_m$ of the carrier gas molecules, $z = m_v / m_m >> 1$.

The new approach becomes especially important in the transition mode, $a_v \approx R_p$, where $a_v$ is the mean free path of the vapour molecules and $R_p$ is the radius of the central particle, since in this case the traditionally used flux matching theory is not valid (either for heavy vapour molecules or for light ones).

Indeed, the diffusion equation for the vapour molecules concentration $n_v(\vec{r},t)$, which is applied in the flux matching theory outside the absorbing sphere, is valid under the general condition that the length scale $L$ of the concentration variation is large in comparison with the mean free path, $L >> a_v$ (see Section 1.2). Taking into account that the diffusion concentration profile substantially varies outside the absorbing sphere (of radius $\Delta \approx R_p + a_v$) on the scale of $L \approx \Delta$ (cf. Section 1.3.1), the condition of the diffusion equation validity in the vicinity of the absorbing sphere takes the form, $\Delta >> a_v$, or $R_p >> a_v$. This condition significantly restricts applicability of the flux matching theory to relatively small Knudsen numbers $\mathrm{Kn} = a_v / R_p << 1$, i.e. only small corrections to the flux in the near-continuum regime can be properly calculated in the flux matching approach. Besides, the Maxwellian distribution near the particle surface (resulting in zero flux inside the absorbing sphere) can be justified only at $\mathrm{Kn} \to \infty$, and for this reason the free molecular expression for the surface flux in the flux matching equation cannot be strictly applied at finite $\mathrm{Kn}$.

Moreover, as explained in Section 1.5.4, the additional uncertainty of the traditional semi-empirical approach to the classical problem of condensation in a large immobile trap of small Brownian particles (migrating by random walks), which is connected with determination of the absorbing sphere radius, does exist, which can be rigorously resolved in the new approach.

The volume swept by a Brownian particle is known as the Wiener sausage [8]. In particular, this quantity equals the probability that a diffusing Brownian point particle is absorbed by the only trap of radius $R_p$ for time $t$ (see, e.g., [9]). For this reason, the rate of volume sweeping, which determines the Smoluchowski constant for comparable size particles, or $R_{12} << \bar{r}_1, \bar{r}_2$ (as justified in the new kinetic approach), coincides with the condensation rate constant for small particles sinking in a large trap, $\bar{r}_v << R_p$.

Indeed, vapour molecules of radius $R_v$, small in comparison with the radius $R_p$ of the trap particle, $R_v << R_p$, can be considered as point-wise particles, i.e. their mutual collisions can be neglected and thus the condensation rate can be properly calculated as the collision probability of one molecule (randomly located in space with the probability equal to their mean concentration $n$) with the trap. In fact, the collision frequency of vapour molecules with the particle can be conservatively estimated as $\tau_{vp}^{-1} \approx 4\pi(\Delta + R_v)(D_v + D_p)n_v \approx 4\pi(R_p + a_v)D_v n_v$, whereas the collision frequency between heavy vapour molecules can be evaluated with the same accuracy as $\tau_{vv}^{-1} \approx 4\pi R_v^2 \bar{u}_v n_v \approx 4\pi R_v^2 n_v D_v / a_v$. Therefore, the collisions between vapour molecules can be neglected, if $\tau_{vp}^{-1} >> \tau_{vv}^{-1}$, or $(R_p/R_v) + (a_v/R_v) >> (R_v/a_v)$, which is always valid owing to $R_v/a_v << 1$.

Since the Brownian particles coagulation problem is reduced under the mixing condition to the similar two-particle collision problem, this allows rigorous consideration of the condensation problem using the above calculated collision kernel, Eq. (1.43) or Eq. (1.44), with the particular values of the parameters $D_{12} = D_{vp} = D_v + D_p \approx D_v$ and $R_{12} = R_{vp} = R_v + R_p \approx R_p$.

In the whole range of the continuum mode, $a_v << R_p$ (or equally $a_{vp} << R_{vp}$, where $a_{vp} \approx a_v$ is the elementary drift distance, or the mean free path of the vapour molecules), the condensation kernel calculated from consideration of the two-particle problem (in neglect of mutual collisions between point-wise vapour molecules, as explained above), is equal to $\beta_{vp} = 4\pi D_{vp} R_{vp}$ (cf. Section 1.5.1). On the other hand, in the specific range of the continuum mode, $\bar{r} \approx n^{-1/3} << R_p$ (or $\bar{r} << R_{vp}$), corresponding to a dense gas of heavy vapour molecules, the condensation kernel can be equally calculated (as $\beta_{vp} = 4\pi D_{vp} R_{vp}$) from consideration of the diffusion flux (into the trap) in the multi-particle system of the vapour molecules (i.e. in the diffusion approach), cf. Section 1.3.1. Therefore, the condensation kernel derived in the diffusion approach formally coincides with that in the whole range of the continuum mode, including also the range $\bar{r} >> R_{vp} >> a_{vp}$ (if $\bar{r} >> a_{vp}$), corresponding to a rarefied gas of vapour molecules, where diffusion approach is not anymore valid, however, the two-particle collision problem has the same solution, see Fig. 1.13. This explains why the solution obtained in the diffusion regime can be correctly extended beyond the applicability range of this regime (to the whole range of the continuum mode).

In application of this kernel to consideration of the coagulation problem in the continuum mode, $a_{12} << R_{12}$, this feature apparently elucidates the reason for fortuitous coincidence (revealed in Section 1.5.1) of the formal expressions for the coagulation kernel, derived either in the diffusion regime, that is valid for collisions between large and small particles (i.e. in the case $R_1 << \bar{r} << R_2$, or $\bar{r} << R_{12}$), or in the continuum mode of the kinetic regime, that is valid in the range $\bar{r} >> R_{12} >> a_{12}$.

For the same reason, in the transition mode semi-empirical predictions of the flux matching theory based on the Fuchs approach [4] with an appropriate definition of the absorption sphere radius are (fortuitously) in a reasonable agreement with more rigorous theoretical analysis of the current approach.

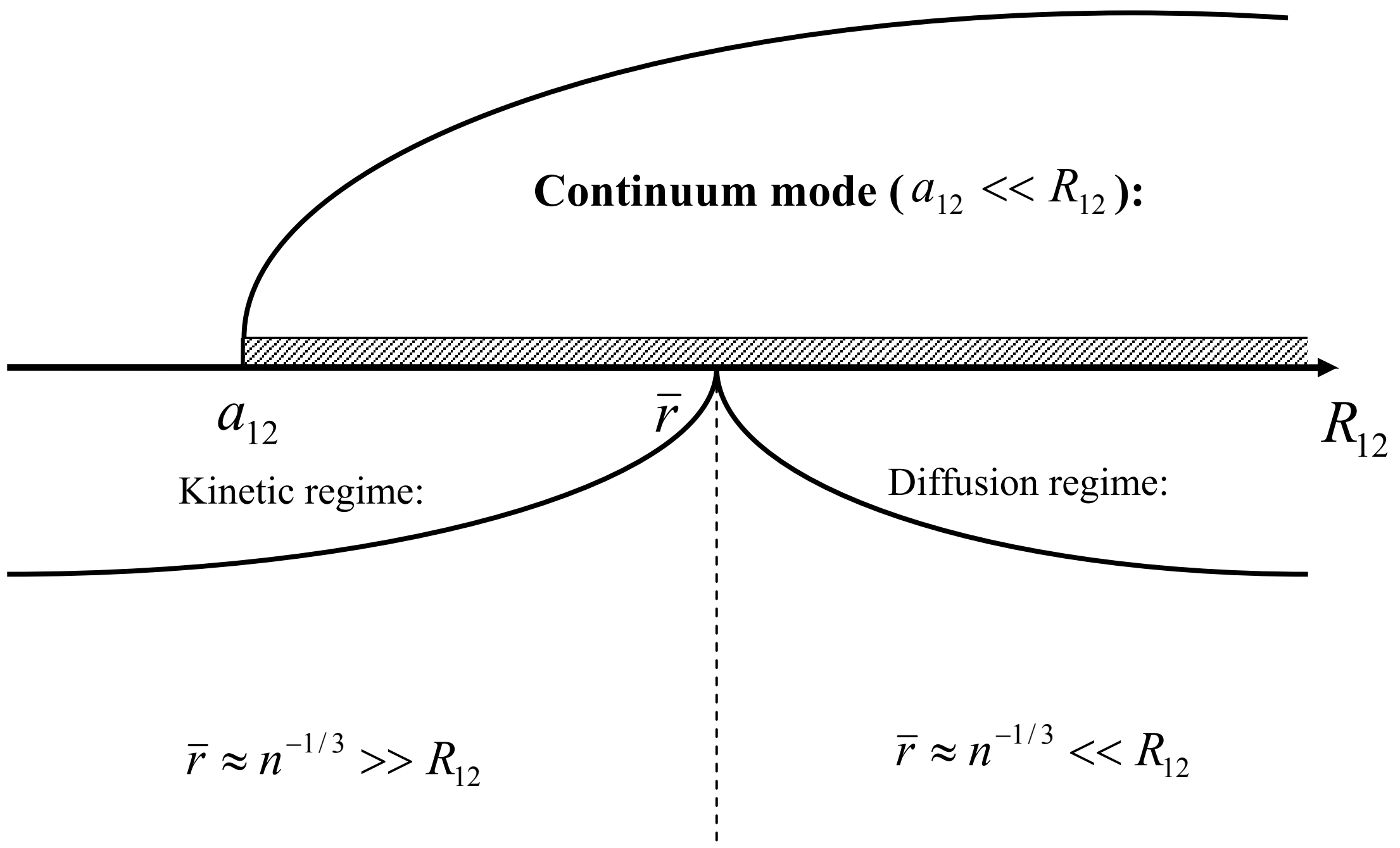

Fig. 1.13. Schematic representation of various ranges of the parameter $R_{12}$: dashed zone corresponds to the range, where the collision kernel is described by the unique analytical expression, $\beta_{12} = 4\pi D_{12} R_{12}$, derived under the general condition $a_{12} << R_{12}$, corresponding to the continuum mode either in the diffusion regime or in the kinetic regime.

Furthermore, in the considered limit of heavy vapour molecules, $m_v >> m_g$, which are in the same size range as the gas molecules (corresponding to $a_v / R_v >> 1$), diffusivity of Brownian particles, $D = bkT$, where $b$ is the particle mobility, calculated in the Stokes regime ($\mathrm{Re} << 1$) as $b = C_c / 6\pi\eta R$, consistently reduces, as noticed in [47], to diffusivity of the heavy vapour molecules calculated from the Boltzmann kinetic equation (cf. [48])

$$D_v = \frac{3}{2}\sqrt{\frac{(kT)^3}{2\pi m_g}}\frac{1}{P d_{gv}^2} = \frac{3}{2}\sqrt{\frac{kT}{2\pi m_g}}\frac{1}{n_g d_{vg}^2}, \qquad (1.68)$$

where $P = (n_g + n_v)kT \approx n_g kT$ is the gas pressure, $\sigma_t = \pi d_{gv}^2 / 4$ is the transport cross-section of vapour-gas interactions, $d_{gv} \approx d_g + d_v = 2(R_g + R_v)$ is the collision diameter in the hard spheres approximation, applied in the current approach.

This allows direct calculation of the key microscopic parameters, $\tau_0$ and $\hat{a}$, by comparison of Eqs. (1.40) and (1.68),

$$\tau_0 = \frac{3\sqrt{3}m_v}{\sqrt{2\pi}n_g m_g^{1/2}(R_v + R_g)^2 (kT)^{1/2}} \approx \frac{3\sqrt{3}m_v}{\sqrt{2\pi}n_g m_g^{1/2} R_v^2 (kT)^{1/2}}, \tag{1.69}$$

$$\hat{a} = \left(\frac{3kT}{m_v}\right)^{1/2} \tau_0 = \frac{9m_v^{1/2}}{\sqrt{2\pi}n_g m_g^{1/2}(R_v + R_g)^2} \approx \frac{9m_v^{1/2}}{\sqrt{2\pi}n_g m_g^{1/2} R_v^2}, \tag{1.70}$$

which are in a sound agreement with the qualitative estimations, presented above in the end of Section 1.2.

*1.8.2. Light vapour molecules condensation*

The problem can be simplified also in the opposite case of small mass ratio, $z = m_v / m_g \to 0$ [16]. This problem assumes considerable similarity to the problems encountered in the neutron transport studies, under assumptions that the concentration and velocity distribution of the gas molecules are only slightly perturbed by the evaporation process and that once a vapour molecule encounters the surface of a particle its probability of sticking is unity, $\alpha = 1$. Namely, it was noticed that the problem of a flux, to a black sphere, of neutrons, diffused isotropically by heavy atoms, is completely equivalent to the examined problem of droplet growth for this case, $z = m_v / m_g \to 0$. The former problem has received considerable attention from the workers in neutron transport theory and some more accurate solutions have been obtained. In this approach, Fuchs and Sutugin [25] fitted Sahni's [49] theoretical solution to the Boltzmann equation by means of the expression

$$\frac{F}{F_c} = \frac{1 + \mathrm{Kn}_{\mathrm{FS}}}{1 + 1.7104\mathrm{Kn}_{\mathrm{FS}} + (4/3)\mathrm{Kn}_{\mathrm{FS}}^2}, \tag{1.71}$$

which correctly represents asymptotic solutions for large and small Knudsen numbers; however, becomes approximate in the transition mode, where series expansion of the Bessel function (with a finite argument) in the integral equation analysed by Sahni [49] cannot be truncated after the first terms. Besides, some additional deviation of values calculated according to this interpolation formula, Eq. (1.71), from numerical results of Sahni attained 2 to 6%. The mean free path included in the definition of the Knudsen number (designated here as $\mathrm{Kn}_{\mathrm{FS}}$) was used from the zeroth order kinetic theory (the Meyer formula)

$$\lambda_v = \frac{3D_v}{\bar{c}_v}, \tag{1.72}$$

which was not justified for the higher order kinetic theory, used in derivation of Eq. (1.71), and was in disagreement with other options, e.g. assumed by Loyalka (see below Eq. (1.74)) or derived by Davis [23] from the Chapman-Enskog theory, $\lambda_v = 32D_v / 3\pi\bar{u}_v(1 + z)$.

Loyalka [50] used another approximation to the problem by linearization of the Boltzmann equation for the vapour molecules distribution function (so called BGK model [50]) and searching the solution using variation technique. Numerical results of these calculations were fitted by an interpolation formula, proposed by Loyalka [52], as

$$\frac{F}{F_c} = \frac{(1+1.333\mathrm{Kn_L})}{1+1.333\mathrm{Kn_L} + \left(1.333\sqrt{\pi}\mathrm{Kn_L} + 1.0161\right)\mathrm{Kn_L}}, \tag{1.73}$$

with the mean free path included in definition of the Knudsen number (designated here as $\mathrm{Kn_L}$) given by

$$\lambda_v = \frac{4D_v}{\sqrt{\pi}\bar{c}_v}. \tag{1.74}$$

However, Williams and Loyalka [53] pointed out that Eq. (1.73) does not have the correct shape near the free molecular limit and should be used with caution.

In order to resolve discrepancies between various interpolation formulas in the transition mode, a new approach, similar to that recently proposed by the author [10] to consideration of the heavy vapour molecules condensation, $z = m_v/m_g \to \infty$, was developed in [16] in application to light vapours.

*1.8.2.1. Model formulation*

The problem of condensation (neglecting evaporation and assuming $\alpha = 1$) on a large immobile particle of radius $R$ of monoatomic vapour molecules (of mass $m_v$ and radius $R_v$) suspended in the ideal gas of heavy gas molecules (of mass $m_g >> m_v$ and radius $R_g$), is considered. It is additionally assumed that the concentration of light molecules in the gas mixture is small, $n_v << 1$. In this case one can neglect mutual collisions of vapour atoms and take into consideration only collisions between light and heavy molecules. Since the mean heat energies of the translational motion of all species are the same (at fixed temperature), the mean thermal speed of heavy molecules is small in comparison with that of light ones and in the limit $z = m_v/m_g \to 0$ they can be considered as immobile (the so called Lorentz gas).

After a collision of a light molecule with a heavy one the latter remains immobile, whereas the light molecule velocity changes its direction and keeps its absolute value, cf. [54]. Therefore, collisions of light monoatomic vapour molecules with heavy gas molecules are elastic and thus the scattering cross-section of such collisions does not depend on the scattering direction, i.e. the light atom velocity distribution after a collision is completely isotropic.

Under these conditions the probability $wdt$ of a collision between time $t$ and $t + dt$ of each vapour molecule with immobile heavy gas molecules, randomly distributed in space, depends only on its speed $u$, which is invariant in time, and thus $w(u)$ does not depend on time. Therefore, the exponential distribution of time intervals $\tilde{t}$ between successive collisions of a vapour molecule can be justified in this case,

$$\omega(\tilde{t}) = w(u)\exp(-w(u)\tilde{t}) = \tau_u^{-1}\exp(-\tilde{t}/\tau_u), \tag{1.75}$$

where $\tau_u = w^{-1}(u)$ is the mean time between collisions. Indeed, knowing the collision probability $w$, it is possible to calculate the survival probability $P(t)$, that is the probability that an atom survives a time $t$ without suffering a collision, from the equation, $P(t+dt) = P(t)(1 - wdt)$, which has a solution, $P(t) = \exp(-wt)$ obeying $P(0) = 1$. Therefore, the probability that an atom, after surviving without collisions for a time $t$, suffers a collision in the time interval between $t$ and $t + dt$, is $\varphi(t)dt = P(t)wdt = \exp(-wt)wdt$, which obeys $\int_0^\infty \varphi(t)dt = 1$, and thus $\tau_u = \int_0^\infty \varphi(t)tdt = w^{-1}(u)$.

If $z$ is not small, the speed of a vapour molecule may vary after each collision and $w$ is not anymore constant in time. Therefore, in the general case Eq. (1.75) can be used only for qualitative estimations of the average collision time $\bar{\tau}(\bar{u})$ of the molecules travelling with the mean thermal speed $\bar{u}$ [55].

Since the speed of a light vapour molecule does not change after collisions, the diffusivity $D_u$ of a vapour atom, migrating with the speed $u$ in a random direction after each collision, can be calculated as

$$D_u = \lim_{k\to\infty} \frac{1}{6k\tau_u} \sum_{i=1}^{k} \rho_i^2 = \frac{1}{6\tau_u} \left\langle \rho^2 \right\rangle, \tag{1.76}$$

where $\rho_i$ is the length of $i$-th drift (or jump) between two consecutive collisions, and $\left\langle \rho^2 \right\rangle = \lim_{k\to\infty} \frac{1}{k} \sum_{i=1}^{k} \rho_i^2$ is the mean value of its square. Since heavy gas molecules are considered as immobile, the distribution of a light atom jumps $\rho_i$ does not depend on the atom velocity, whereas the mean time between its consecutive collisions is inversely proportional to its velocity, $\tau_u = \tau_0 (u_0/u)$, where $u_0$ is some appropriately fixed value of the velocity. Therefore, the mean value of the jump distance $\bar{\rho}$, which corresponds in the current situation to the mean free path of vapor atoms $\lambda_v$, can be calculated as

$$\lambda_v = \bar{\rho} = \int_0^\infty u\tilde{t}\,\omega(\tilde{t})d\tilde{t} = \int_0^\infty u\tilde{t}\,\tau_u^{-1} \exp(-\tilde{t}/\tau_u)d\tilde{t} = u\tau_u = u_0\tau_0, \tag{1.77}$$

whereas the mean value of their squares, $\left\langle \rho^2 \right\rangle$, is

$$\left\langle \rho^2 \right\rangle = \int_0^\infty (u\tilde{t})^2 \omega(\tilde{t})d\tilde{t} = \int_0^\infty (u\tilde{t})^2 \tau_u^{-1} \exp(-\tilde{t}/\tau_u)d\tilde{t} = 2(u\tau_u)^2 = 2(u_0\tau_0)^2. \tag{1.78}$$

This results in $D_u = D_0\, u/u_0$, i.e. the ratio $D_u/u$ is invariant with respect to the velocity, or

$$\frac{D_u}{u} = \frac{\left\langle \rho^2 \right\rangle}{6u\tau_u} = \frac{\lambda_v}{3} = \frac{D_0}{u_0}. \tag{1.79}$$

Correspondingly, the dimensionless value

$$\Gamma = \frac{4D_u}{uR} = \frac{4\lambda_v}{3R}, \tag{1.80}$$

is also invariant with respect to the velocity.

Eventually, the vapor atoms diffusivity obtained by averaging over the Maxwellian distribution, $w_M(u) = (m_v/2\pi kT)^{3/2}\exp(-m_v u^2/2kT)$, is

$$D = \overline{D}_u = \frac{1}{3}\overline{u}\lambda_v, \tag{1.81}$$

where

$$\overline{u} = (8kT/\pi m_v)^{1/2}. \tag{1. 82}$$

is the mean thermal speed, and Eq. (1.80) can be represented in the form

$$\Gamma = \frac{F_c}{F_{fm}} = \frac{4\overline{D}_u}{\overline{u}R} = \frac{4\lambda_v}{3R} = \frac{4}{3}\mathrm{Kn}_{\mathrm{FS}} \equiv \frac{4}{3}\mathrm{Kn}, \tag{1.83}$$

where the Knudsen number coincides with the definition of Fuchs and Sutugin in Eq. (1.72).

This implies that the uncertainty in determination of the mean free path of vapour molecules (discussed in Section 1.8.2) vanishes at $z \to 0$, and the Meyer formula for the vapour diffusivity, Eq. (1.81), derived from the zeroth order kinetic theory, should be strictly used in this case (rather than Eq. (1.74) or some other options reviewed by Davis [21]). This result will be additionally confirmed by numerical calculations in Section 1.8.4.

Being derived in the same approximation, Eq. (16) naturally coincides with Lorentz's solution of the Boltzmann equation for a dilute mixture of a light gas with a heavy carrier gas (cf. [48]), which yields

$$D = \frac{kT}{3P}\left\langle \frac{u_v}{\sigma_t} \right\rangle \approx \frac{4}{3}\left(\frac{2}{\pi}\right)^{3/2}\left(\frac{kT}{m_v}\right)^{1/2}\frac{1}{n_g d_{gv}^2}, \tag{1.84}$$

where $P = (n_g + n_v)kT \approx n_g kT$ is the gas pressure, $\sigma_t = \pi d_{gv}^2/4$ is the transport cross-section of vapor-gas interactions in the hard sphere approximation, $d_{gv} \approx d_g + d_v = 2(R_g + R_v)$ is the collision diameter, resulting in

$$\lambda_v = \frac{4}{\pi}\frac{1}{n_g d_{gv}^2}. \tag{1.85}$$

If the size $R_v$ of vapour molecules is small in comparison with the size $R$ of the trap particle, $R_v << R$, vapour molecules can be considered as point-wise particles randomly distributed in a sample of volume $V \to \infty$ with the mean concentration $n_u = N_u/V << 1$, i.e. their mutual collisions can be neglected, and thus the condensation rate, $F_u$, can be properly calculated in the steady state approximation as the probability of a collision with the trap per unit time of a vapour molecule (having the speed $u$) multiplied by $n_u$. In its turn, this probability is proportional to the volume sweeping rate, $\beta_u = d\langle V_u \rangle/dt$, by the effective particle of radius $R + R_v \approx R$ migrating with the diffusivity $D_u$, i.e. $F_u = \beta_u n_u$, cf. [10].

From an obvious geometry (scaling) consideration it is clear that the steady-state value of the mean swept volume per unit time depends only on the ratio $\bar{\rho}/R$ (rather than on $\bar{\rho}$ and $R$ separately) and is proportional to $u$, that is confirmed also by numerical calculations. For this reason, the sweeping rate in the dimensionless form can be represented as a function of $\Gamma$, $\beta_u/\beta_u^{(fm)} = \beta_u(\Gamma)/\beta_u^{(fm)}(\Gamma)$, where $\beta_u^{(fm)} = \pi R^2 u$. Since $\Gamma$ and $\beta_u(\Gamma)/\beta_u^{(fm)}(\Gamma)$, do not depend on velocity, $\beta_u(\Gamma)/\beta_u^{(fm)}(\Gamma) = \beta_0(\Gamma)/\beta_0^{(fm)}(\Gamma)$, one obtains $\beta_u(\Gamma) = \beta_0(\Gamma)\beta_u^{(fm)}/\beta_0^{(fm)} = \beta_0(\Gamma)u/u_0$.

As a result, for the total condensation rate $\Phi$ one obtains

$$
\begin{aligned}
F(\Gamma) = \int_0^\infty F_u(\Gamma)du = \int_0^\infty \beta_u(\Gamma)n_u\, du = \int_0^\infty \beta_u(\Gamma)n_v w_M(u)du \\
= \beta_0(\Gamma)n_v \frac{\bar{u}}{u_0} = \pi R^2 \bar{u} n_v \frac{\beta_0(\Gamma)}{\beta_0^{(fm)}(\Gamma)}
\end{aligned}, \qquad (1.86)
$$

or

$$
\frac{F}{F_{fm}} = \frac{F}{F_c}\Gamma = \frac{\beta_u(\Gamma)}{\beta_u^{(fm)}(\Gamma)} = \frac{\beta_0(\Gamma)}{\beta_0^{(fm)}(\Gamma)}. \qquad (1.87)
$$

Therefore, for determination of the total condensation rate one should calculate the sweeping rate $\beta_0(\Gamma)$ for vapor atoms with some fixed velocity $u_0$ and then use Eq. (1.87).

*1.8.2.2. Numerical calculations*

In the transition interval $\mathrm{Kn} = \lambda_v/R \approx 1$, the collision rate for vapor atoms migrating with the velocity $u$ is calculated by evaluation of the mean swept volume per unit time, $\beta_u = d\langle V_u \rangle/dt$, in the numerical approach. In contrast to the random walk approximation used in [10] for heavy vapor molecules with a constant time of their elementary drift and the Maxwellian distribution of their velocities, the jump distance of the effective particle (of radius $R + R_v \approx R$ and fixed velocity $u$) in random directions is calculated in the current approach as $a = u\tilde{t}$, where the time $\tilde{t}$ between collisions of the particle is generated as a random value with the exponential probability density function from Eq. (1.75), $\omega(\tilde{t}) = \tau_u^{-1}\exp(-\tilde{t}/\tau_u)$, where $\tau_u = \lambda_v/u$. The generated data describe the successive positions of the particle center trajectory, which can be further used for calculation of the swept volume. For this calculation the same procedure of random spatial distribution of auxiliary (fictitious) point immobile particles (markers) with a relatively high concentration, $n_* >> \left(\pi R_{12}^2 a_{12}\right)^{-1}$, described in [7], is numerically realized using the Monte Carlo method. Each marker found in the swept volume is counted only once. The number of Monte Carlo markers was increased until the calculated swept volume became invariable with respect to further increase of this number (see below).

The number of jumps $k = t/\tau_u$ was increased (up to $k \approx 10^6$) until the ratio of $\beta_u = d\langle V_u \rangle/dt$ to $\delta V_u/\tau_u = \beta_u^{(fm)}$, where $\delta V_u$ is the mean swept volume during one jump, attained a steady-state value, which in accordance with Eqs. (1.63) and (1.64) had to be equal to $4D_u/uR = \Gamma$ in the limit $\mathrm{Kn} << 1$ and to 1 in the limit $\mathrm{Kn} >> 1$, see Fig. 1.14. Similarly to the previously considered case of heavy vapour molecules, the necessary number of jumps notably increases with the decrease of the Knudsen number; namely, for $\mathrm{Kn} = 10$ the steady state is attained at $k \geq 10^5$, whereas for $\mathrm{Kn} = 0.05$ the minimum number is $k \approx 5 \cdot 10^6$.

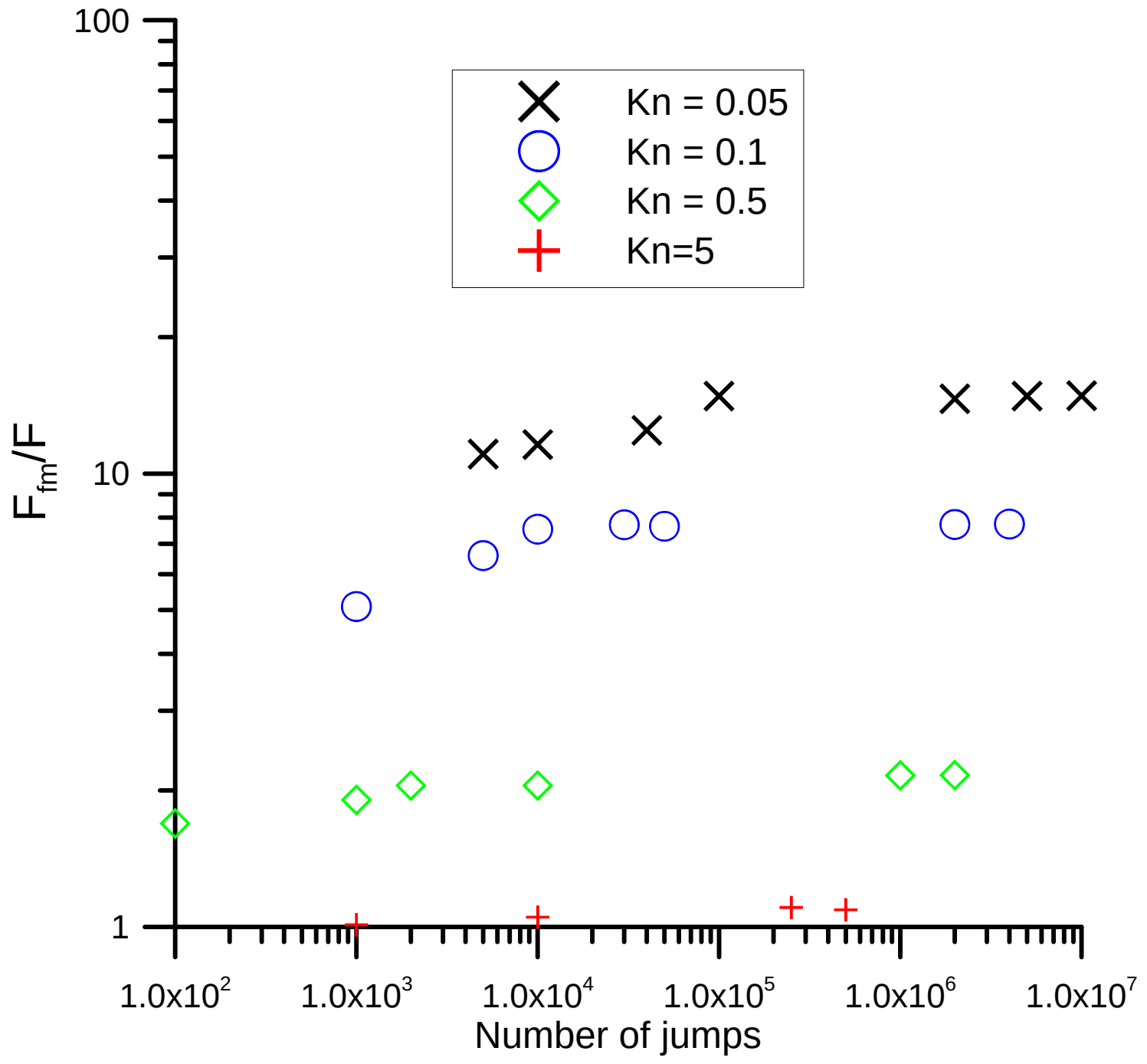

Fig. 1.14. Dependence of $F_{fm}/F = \left(\beta_u/\beta_u^{(fm)}\right)^{-1} = \left(d\langle V_u \rangle/dt \big/ \langle \delta V_u/\tilde{t} \rangle\right)^{-1}$, on the number of jumps $k$ for different values of $\mathrm{Kn}$ .

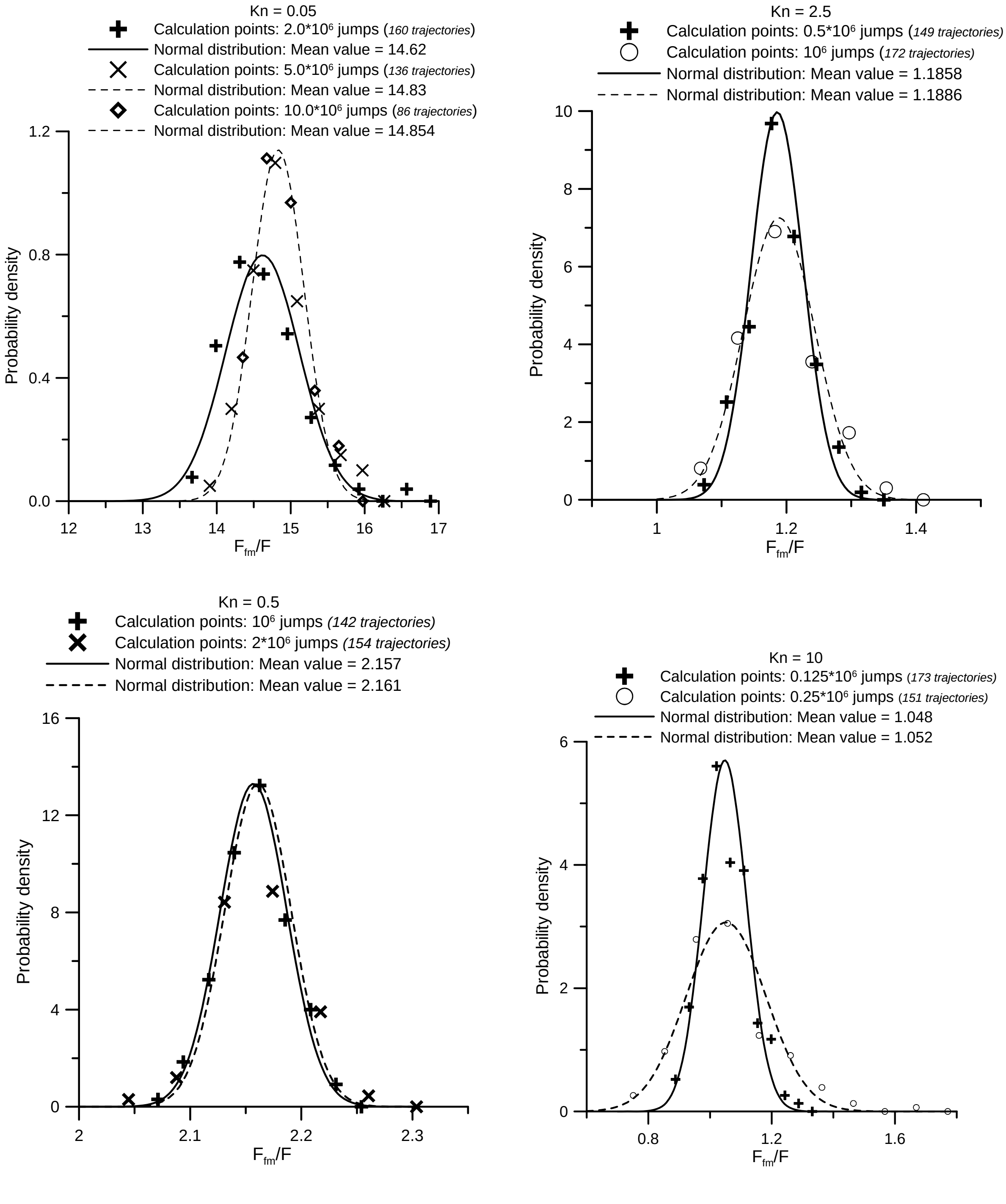

Fig. 1.15. Examples of calculation of the probability density $f(x)$ for $x = F_{fm}/F = \left(\beta_u/\beta_u^{(fm)}\right)^{-1} = \left(d\langle V_u\rangle/dt \big/ \langle \delta V_u/\tilde{t}\rangle\right)^{-1}$ at different $\mathrm{Kn}$; for each number of jumps $k$, 150-200 trajectories are generated resulting in calculation points, which are grouped in intervals of equal width $L$ ($\approx$ 10% of the whole distribution width) and form normal distributions around the mean values (at given $k$).

The number of markers (Monte Carlo points) used in calculations for each trajectory was increased until the calculated mean value of the sweeping rate became invariant with respect to a further increase of this number, Fig. 1.16. It was shown that for the sufficiently large number of Monte Carlo points in the simulation box (normally $10^5$ - $10^6$, depending on the length of a trajectory) a further increase of this number affects only the width of the distribution, rather than the searched mean value.

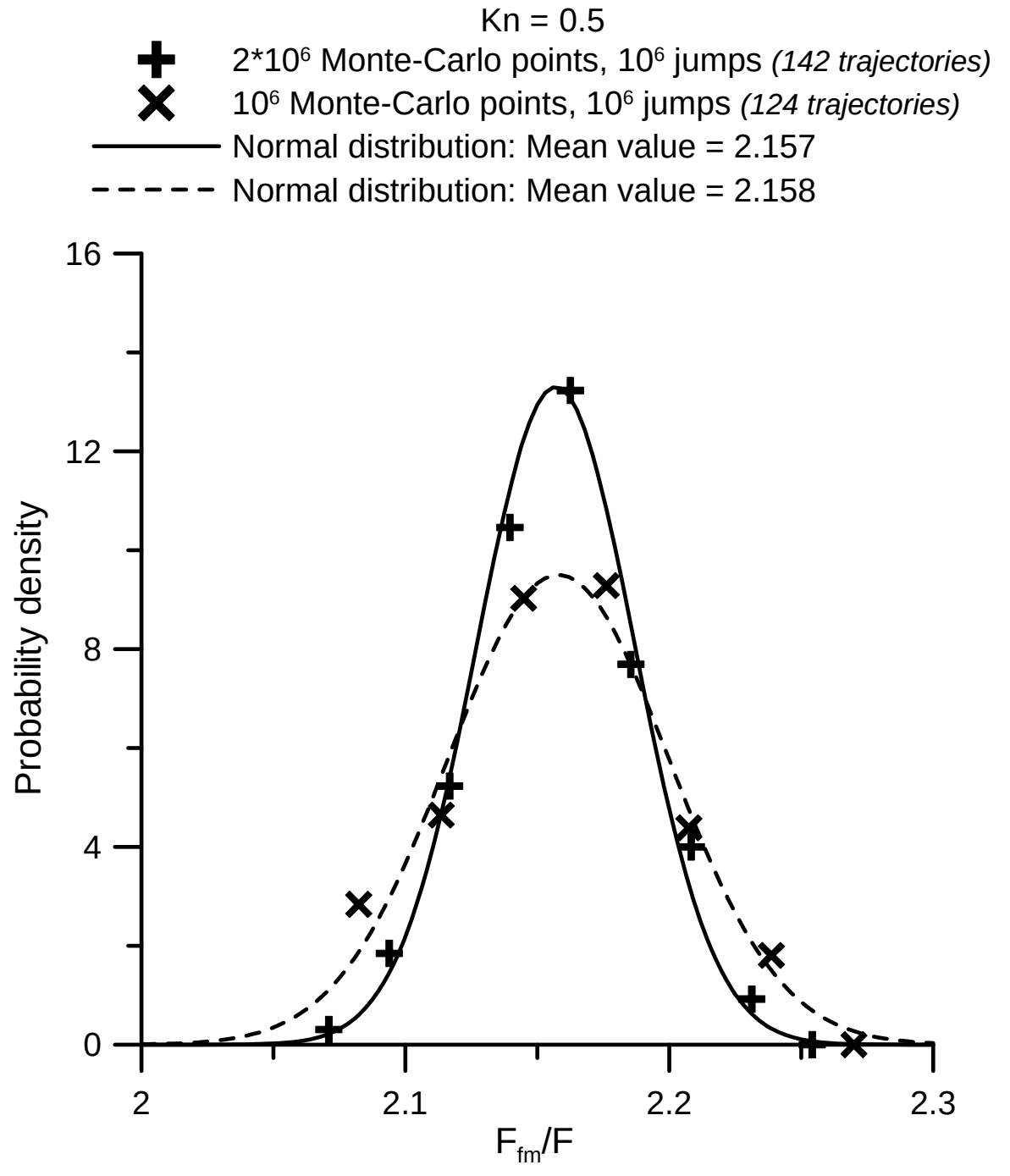

Fig. 1.16. Example of calculated dependence of the distribution function on the number of Monte Carlo points used in calclulations ( $\mathrm{Kn} = 0.5$ ).

The accuracy of the current approach was additionally verified by calculation of the vapour molecules diffusivity, which is subject to Eq. (1.81). As seen from Fig. 1.17, the theoretical value, $D/\bar{u}\lambda_v = 1/3$, is attained for sufficiently long trajectories with the number of jumps $k \geq 10^3$ and for sufficiently large number of trajectories, $N \geq 10^3$.

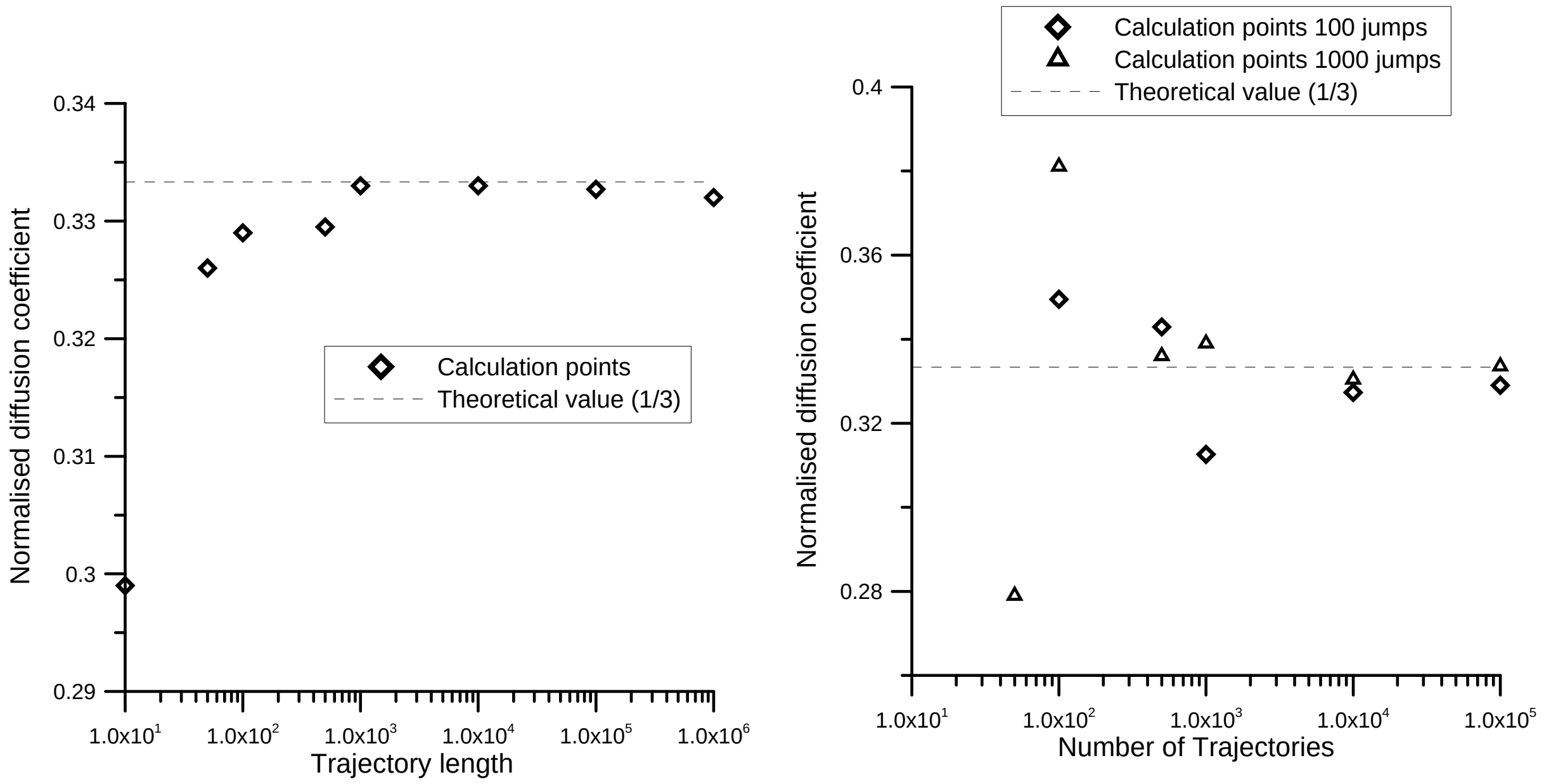

Fig. 1.17. Numerical calculations of the normalised vapour molecule diffusivity, $D/\bar{u}\lambda_v$ .

Results of calculations of the sweeping rate for various $\mathrm{Kn} = 3D/\bar{u}R$ in the transition range from 0.05 to 10 are plotted in Fig. 1.18 along with the interpolation curve

$$\frac{F}{F_{fm}} = \left(\frac{4}{3}\mathrm{Kn}\right)\frac{F}{F_c} = \left(\frac{4}{3}\mathrm{Kn}\right)\frac{1 + A\cdot\mathrm{Kn} + D\cdot\mathrm{Kn}^2 + F\cdot\mathrm{Kn}^3}{1 + B\cdot\mathrm{Kn} + C\cdot\mathrm{Kn}^2 + 2D\cdot\mathrm{Kn}^3 + (4/3)F\cdot\mathrm{Kn}^4}, \tag{1.88}$$

with five parameters found by least squares, $A = 17.065\cdot 10^6$, $B = 0.94A$, $C = 1.98A$, $D = 1.175A$, $F = 1.24A$, which provide accuracy within $\approx 0.4\%$. The interpolation expression, Eq. (1.88), correctly reduces to $4D/\bar{u}R = \Gamma$ in the limit $\mathrm{Kn} << 1$ and to 1 in the limit $\mathrm{Kn} >> 1$, in accordance with Eqs. (1.63) and (1.65).

Somewhat reduced accuracy with the maximum error of $< 2.5\%$ can be attained with the 2-parameter interpolation expression

$$\frac{F}{F_{fm}} = \left(\frac{4}{3}\mathrm{Kn}\right)\frac{F}{F_c} = \left(\frac{4}{3}\mathrm{Kn}\right)\frac{1 + A\cdot\mathrm{Kn}}{1 + B\cdot\mathrm{Kn} + \frac{4}{3}A\cdot\mathrm{Kn}^2}, \tag{1.89}$$

with $A = 6.72$ , $B = 6.1$, which is nevertheless notably higher than that provided by the traditional expressions, Eqs. (1.67), (1.71) and (1.73), with the maximum error of $\leq 5.5\%$, as demonstrated in Fig. 1.19 and in Table 1.4.

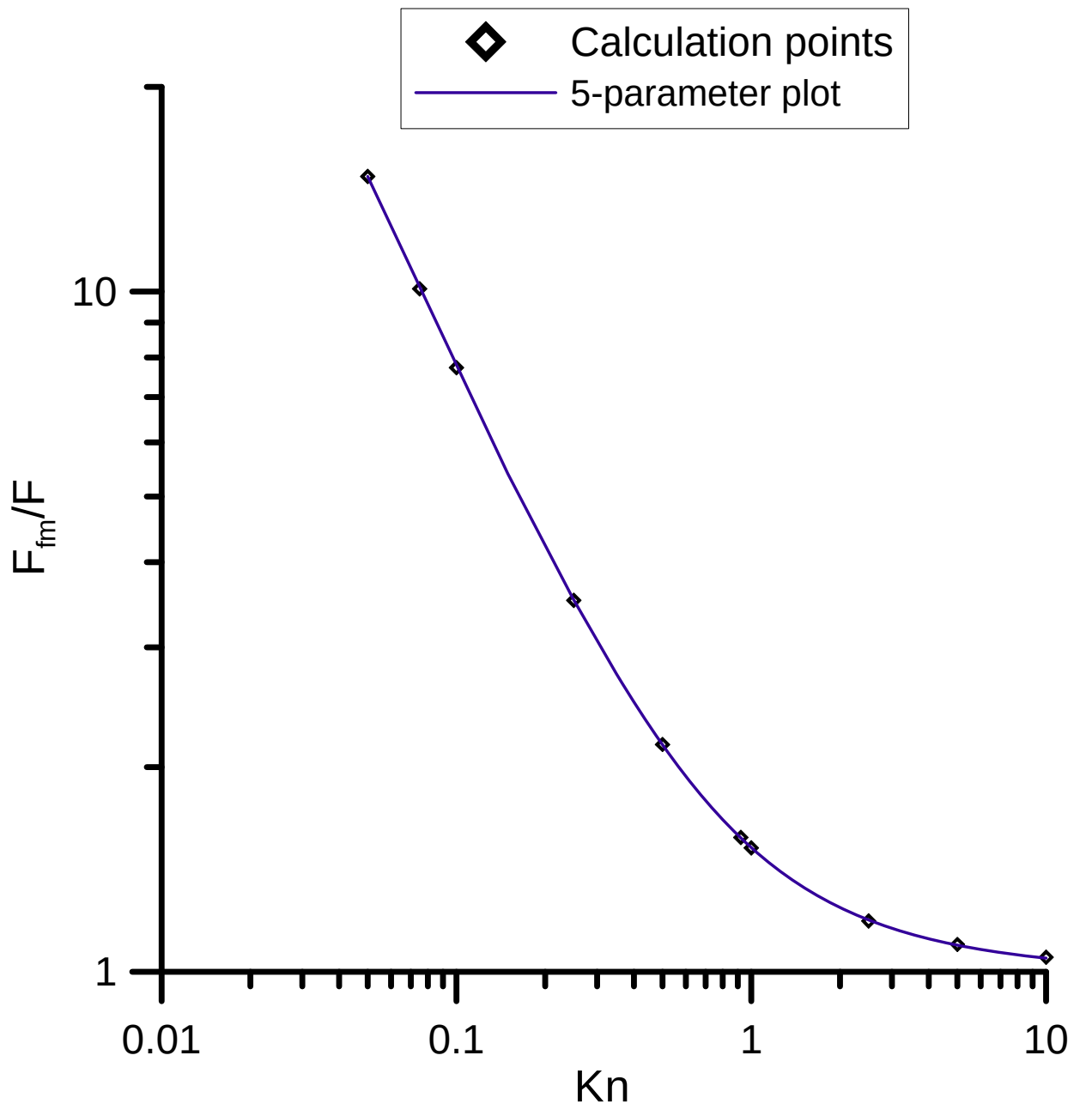

Fig. 1.18. Dependence of $F_{fm}/F = \left(\beta_u/\beta_u^{(fm)}\right)^{-1} = \left(d\langle V_u\rangle/dt/\delta V_u/\tau_u\right)^{-1}$ on $\mathrm{Kn} = 3D/\bar{u}R$: calculation points interpolated by the 5-parameter curve, Eq. (23).

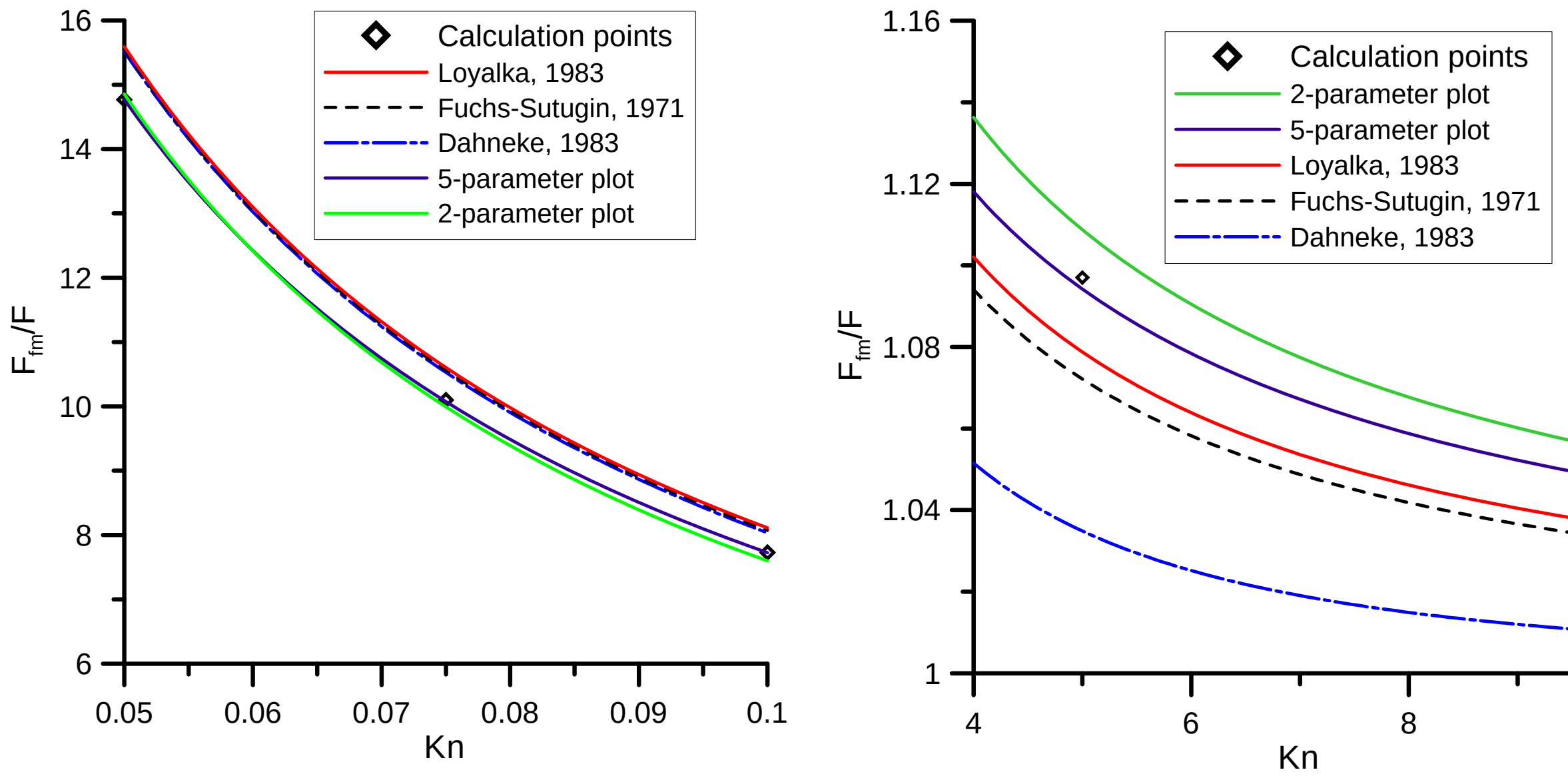

Fig. 1.19. Comparison of different interpolation curves for the condensation rates of light vapour molecules depending on $\mathrm{Kn} = 3D/\bar{u}R$.

Table 1.4. Deviations (in %) of calculation points from different interpolation curves and standard error of the mean (SEM) for calculation points

| Kn | 5-parameter curve, Eq. (1.88) | 2-parameter curve, Eq. (1.89) | Dahneke [26], $\alpha = 1$ Eq. (1.67) | Fuchs-Sutugin [25], Eq. (1.71) | Loyalka [52], Eq. (1.73) | SEM (calculation points) |
|---|---|---|---|---|---|---|
| 0.05 | 0.0123% | 0.65% | 5.1% | 5.33% | 5.59% | 0.78% |
| 0.075 | -0.25% | -1.03 | 4.26% | 4.6% | 4.99% | 0.89% |
| 0.1 | -0.035% | -1.7% | 3.97% | 4.48% | 4.95% | 1% |
| 0.25 | 0.156% | -2.2% | 1.68% | 3.13% | 4.05% | 0.64% |
| 0.5 | -0.122% | -0.14% | -1.44% | 1.37% | 2.6% | 0.67% |
| 0.922 | -0.129% | 1.95% | -4.37% | -0.27% | 1.07% | 0.97% |
| 1.0 | 0.066% | 2.3% | -4.55% | -0.3% | 1.03% | 1.1% |
| 2.5 | 0.377% | 2.43% | -6.23% | -1.1872% | -0.76% | 0.7% |
| 5.0 | -0.256% | 0.89% | -5.66% | -2.26% | -1.66% | 1.05% |
| 10.0 | -0.282% | 0.29% | -3.82% | -1.66% | -1.32% | 1% |

### *1.8.3. Discussion*

As mentioned in Section 1, there does not currently exist a general solution for the condensation rate valid over the full range of Knudsen numbers for arbitrary masses of the diffusing vapour molecules, $m_v$, and the background gas, $m_g$. For this reason, the solution of the problem in the two limiting cases of heavy ($z = m_v / m_g >> 1$) and light ($z << 1$) vapour molecules in a dilute gas mixture, where the problem can be studied more rigorously, attracts a special attention as the first step in search of a more general solution, cf. [23].

In the current paper it was shown that the new approach for calculation of the vapour condensation rate developed for heavy vapour molecules in [10], can be extended with some modifications also to light vapour molecules. However, a further extension of this approach to the general case of arbitrary masses is currently indistinct; in this situation it can be carried out in a simplified manner, similar to that discussed by Davis [23] in relation to the analogous problem of heat fluxes in the Knudsen regime, as follows.

In order to compare the theoretical predictions with experimental data one should eventually insert the value of the binary diffusion coefficient into the calculated condensation rate expression. For instance, in the case of light vapour molecules the diffusion coefficient calculated in Eq. (1.84), should be substituted in Eq. (1.88) or Eq. (1.89). In the general case of vapour molecules with a finite $z$ one can use the expression for the binary diffusion coefficient from the Chapman-Enskog theory [56]

$$D_v = \frac{3}{2}\sqrt{\frac{kT(1+z^{-1})}{2\pi m_g}}\frac{1}{n_g d_{vg}^2}, \qquad (1.90)$$

where $d_{gv} \approx d_g + d_v = 2(R_g + R_v)$ is the collision diameter in the hard spheres approximation, applied in the current approach.

It should be noted that predictions for the binary diffusion coefficient of the Chapman-Enskog theory are accurate on average to about 8% [57]. In particular, comparison of Eq. (1.90) with a more precise expression, Eq. (1.84), in the limit $z << 1$ shows that the value of the numerical coefficient in Eq. (1.90) is underestimated by ≈ 11%.

On the other hand, the difference between the condensation rates calculated in the current approach as functions of $\mathrm{Kn} = 3D/\bar{u}R$ in the two limits $z << 1$ (described by Eq. (1.88)) and $z >> 1$ (described by the expression from [10], similar to Eq. (1.89), but with $A$ = 1.02, $B$ = 1.56), does not exceed 6% (see Fig. 1.20). Therefore, under assumption that the condensation rate smoothly varies within its limiting values at $z \to 0$ and $z \to \infty$, one can apply the same expression (e.g., Eq. (1.89)) in the whole range of $z$ from 0 to ∞ with the error of ≈ 6%. This inaccuracy does not exceed the maximum error of ≈ 8% in calculation of the diffusion coefficient. As a result, one of the formulas derived for the condensation rate in the limits $z << 1$ or $z >> 1$ (e.g. (Eq. (1.89)), can be applied to analysis of experimental measurements after substitution of the general expression for the binary coefficient, Eq. (1.90), with a reasonable accuracy. With a similar accuracy one can also use the traditional correlations of Fuchs-Sutugin, Loylka or Dahneke, since their maximum error (estimated above as ≤ 6%) also does not exceed the uncertainty of the diffusion coefficient.

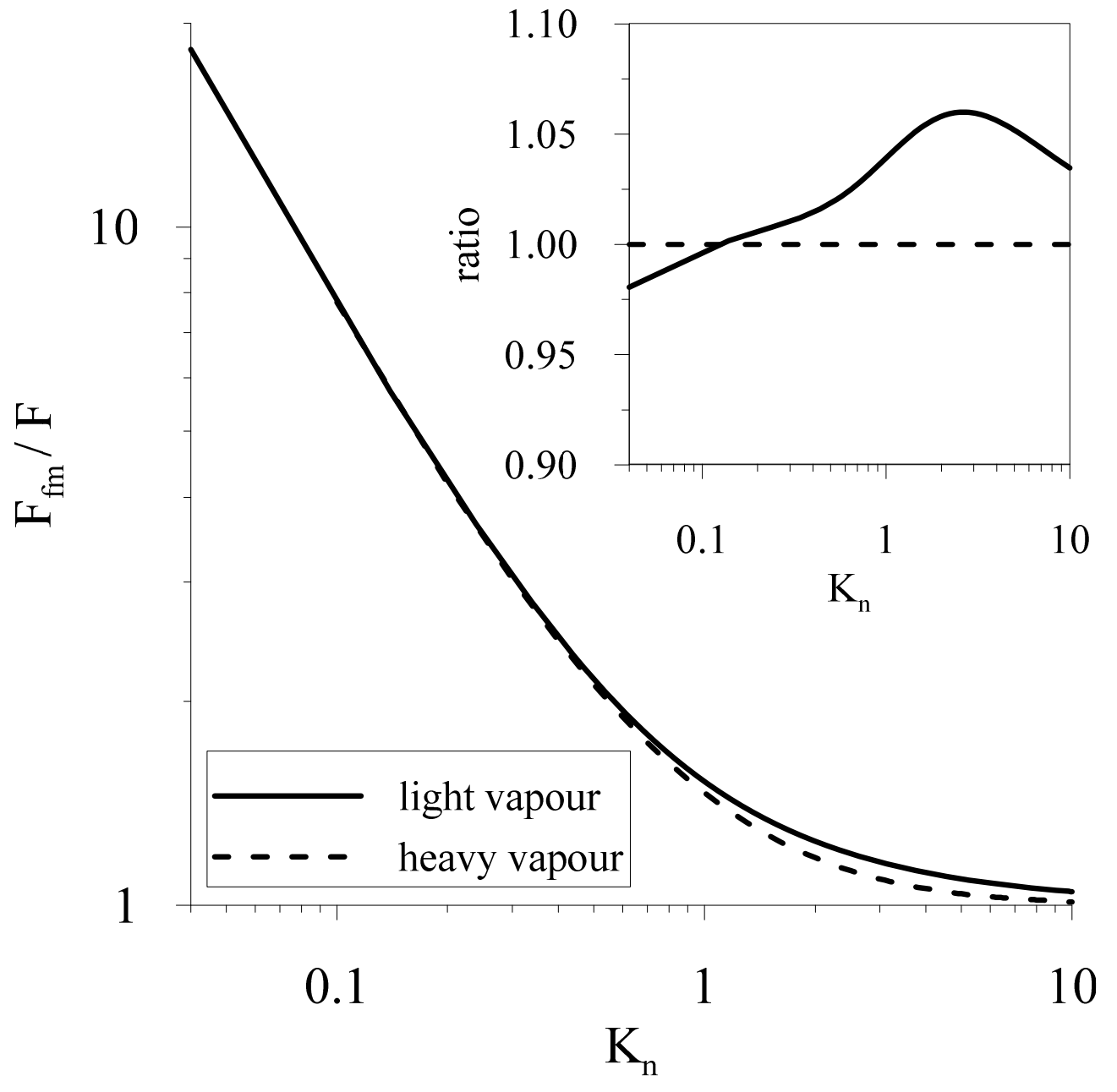

Fig. 1.20. Comparison of the interpolation curves for the condensation rates of light and heavy vapour molecules depending on $\mathrm{Kn} = 3D/\bar{u}R$ and their ratio.

*1.8.4. Conclusions*

The Knudsen aerosol condensation of light vapour molecules, $z = m_v/m_g << 1$, in a dilute gas mixture, $N_v/N_g << 1$, is studied analytically and numerically. In order to overcome uncertainties of the semi-empirical models based on the flux matching theory of Fuchs [4] and to resolve discrepancies between more accurate interpolation formulas of Fuchs and Sutugin [25] and Loyalka [52], derived from various solutions of the Boltzmann equation, a new approach, similar to that recently proposed by the author [10] for consideration of the heavy vapour molecules condensation, $z = m_v/m_g >> 1$, is developed in the present Section.

In this approach the transport of light vapour molecules, which have invariant in time speeds distributed in accord with Maxwell's law and changing direction randomly after each collision (the so called Lorentz gas), can be strictly considered as specific random walks, characterized by the exponential distribution of the elementary displacement (or jump) distances of each walker (rather than the Maxwellian distribution in the previous case, $z >> 1$). In the steady state approximation the condensation rate can be properly related to the mean volume swept per unit time by the effective particle of radius $R + R_v \approx R$ migrating in space with the mean diffusivity of vapour molecules, which can be calculated analytically in the two limiting cases of high and low Knudsen numbers and numerically in the transition mode.

Numerical calculations of the coalescence rate at different values of the Knudsen number, $\mathrm{Kn} = 3D/\bar{u}R$, in the transition range from 0.05 to 10 are carried out with use of the Monte Carlo method by evaluation of the sweeping rate of randomly distributed immobile point markers by the migrating effective particle, developed in [10]. Ten calculated points are approximated by the two analytical interpolation expressions. The accuracy of the approximation (against the calculation points) is determined by the maximum error of $\leq 1\%$ for the interpolation expression with five parameters (determined using the least-squares method). Somewhat reduced accuracy with the maximum error of $< 2.5\%$ can be attained with the two-parameter interpolation expression, more convenient for practical applications; nevertheless, this accuracy is still higher than that provided by the traditional expressions. In particular, the maximum error of 4–6% with respect to the new calculation results is attained using the traditional correlations of Fuchs-Sutugin and Loyalka in a rather wide range of $\mathrm{Kn}$ from 0.05 to 0.25, demonstrating reduction of their accuracy near the border between the transition and continuum modes. Nevertheless, it might be generally concluded that the traditional correlations are confirmed with the reasonable (for practice) accuracy and additionally justified by the independent model (free of simplifications and uncertainties of the traditional approach).

A simplified approach for extension of the derived condensation rate expressions to the general case of arbitrary masses (finite $z$) by use in these expressions of the binary diffusion coefficients from the Chapman-Enskog theory, is analysed. The accuracy of this approach presumably reduces to $\approx$ 8-10% (mainly owing to uncertainties in calculation of the diffusion coefficient), which is generally enough for comparison with experiments and for practical applications.

## 1.9. Extension of the Smoluchowski theory to transitions from dilute to dense regime of Brownian coagulation: triple collisions

The evolution of the size distribution for a system of coagulating Brownian particles over time is described using a kinetic model based on the Smoluchowski equation, Eq. (1.11) [1]. The model assumes binary collisions, whereby the rate of change is second order in particle concentration $n$. For this reason, the Smoluchowski equation cannot describe the coagulation of highly concentrated colloidal or aerosol suspensions (particle volume fraction above ≈ 1%). Such highly concentrated suspensions are well known in emulsions, where the droplet volume fraction may vary from zero to almost one [57] but can also be observed in aerosols. For example, a transition from dilute to dense (or concentrated) particle dynamics may take place during flame aerosol synthesis of nanostructured, fractal-like carbon black, fumed silica [59] or titania [60] particles at industrially relevant conditions. Even though in such processes the particle volume fraction, $f$, is only 0.001 to 0.01%, depending on the process temperature, fractal-like silica agglomerates form and grow to occupy more than 10% of the gas volume during typical reactor residence times in the absence of restructuring or fragmentation [60].

Under these conditions, in order to calculate the coagulation rate of concentrated particulate suspensions, further development of the coagulation theory in the next orders in particle concentration $n$, taking into consideration multiple collisions of Brownian particles, is needed. Direct numerical simulation of particle trajectories by Langevin dynamics (LD) has been used to compute particle motion regardless of concentration. Gutsch et al. [61] investigated the detailed structure evolution of aerosol particles formed by monomer-cluster aggregation by LD simulations. Trzeciak et al. [62] used LD to study the collision frequency function of monodisperse aerosol particles in the Brownian free molecular and continuum modes and found faster coagulation rates for particles larger than the free mean path of the fluid at $f$ >1%. Similarly, Sorensen et al. [63] showed experimentally that soot clusters at high concentrations close to aerogelation grow faster than predicted by classic coagulation theory.

Heine and Pratsinis [64] investigated the growth of spherical (complete coalescence upon collision) and fractal-like particles by Brownian coagulation in the continuum mode by solving the Langevin dynamics equations for each particle trajectory of polydisperse suspensions. By monitoring the LD attainment of the self-preserving size distribution, it was shown that the classic Smoluchowski collision frequency function is accurate for dilute particle volume fractions, $f$, below 0.1%. At higher $f$, coagulation was about 4 and 10 times faster than for the classic theory at $f = 10$ and 20%, respectively. At high particle concentration, an overall coagulation rate was proposed that reduced to the classic one at low concentration. The Langevin dynamics approach was later extended to investigation of coagulation rate of highly concentrated, polydisperse aerosols from the free molecule to the continuum mode by Buesser, Heine and Pratsinis [65].

High concentration significantly increases the coagulation rate of fractal-like particles (agglomerates), consistent with experimental observations. For agglomerates, even initially low particle volume fractions can become effectively large because growing agglomerates occupy far more volume than their equivalent solid mass. As a result, agglomerates may experience a transition from dilute to highly concentrated coagulation dynamics that can lead to gelation. Results showed that the kinetics became more rapid as the system evolves into the cluster dense regime as quantified by an increase in the aggregation rate [64]. A similar approach to investigation of the effect of high concentrations on the aggregation kinetics of colloidal nanoparticles by Langevin dynamics in a broad range of particle volume fractions in the continuum mode was recently presented by Lattuada [66] and Kelkar et al. [67].

Enhanced aggregation rates in dense systems have also been observed with Monte Carlo diffusion-limited cluster–cluster simulations of Sorensen and Chakrabarti [68], both on- and off-lattice, in particulate systems as they evolve from the cluster dilute limit to the cluster dense regime and ultimately the gel point. For instance, Fry et al. [69], Gimel et al. [70, 71] simulated 3d systems with a broad range of monomer volume fractions from 0.0005 to 0.3 and found a crossover in the scaling of mass versus the linear size form fractal dimensions of 1.8 at small size to 2.5 at large.

In order to study the transition from dilute (controlled by binary collisions) to dense (controlled by multiple collisions) regime of coagulation, the Smoluchowski equation was generalized by consideration of triple collisions by Veshchunov and Tarasov [17]. As shown below, following this paper, the transition mode corresponds to a relatively wide interval of the particle volume fraction $f$, from $\approx 0.1\%$ to $\approx 10\%$. Above the upper limit of $\approx 10\%$, multiple collisions among more than three particles should be taken into consideration, which is beyond the scope of the current paper. However, in the regime of triple collisions, solution of the modified Smoluchowski equation is in a rather sound agreement with the results of direct numerical simulation by Langevin dynamics in [64]; this confirms the validity of the new approach to consideration of the transition from dilute to dense regime for spherical particles.

*1.9.1. Model formulation*

In the traditional, "diffusion", approach to analysis of Smoluchowski equation, Eq. (1.11), in the continuum mode, Fick's laws are employed to calculation of particle collision frequency $\beta(R_i, R_j)$ by consideration of a quasi-steady-state concentration profile around colliding particles [1, 2].

In the author's paper [5] it was shown that the traditional diffusion approach is applicable only to the special case of coalescence between large and small particles, $R_1 << \bar{r} << R_2$ (where $\bar{r} \approx n^{-1/3}$ is the mean inter-particle distance), and becomes inappropriate to calculation of the coalescence rate for particles of comparable sizes, $R_1, R_2 << \bar{r}$. In the latter, more general case of comparable size particles, $R_1 \approx R_2 \approx R$, coalescences occur mainly in the kinetic regime (rather than in the diffusion one) characterised by random (homogeneous) spatial distribution of particles. This kinetic regime is realised under the mixing condition, $\tau_d << \tau_c$ (where $\tau_d \approx n^{-2/3}/6D$ is the characteristic time of the particles diffusion redistribution (mixing) on the length scale of the mean inter-particle distance $n^{-1/3}$ after each collision, and $\tau_c^{-1}$ is the collision frequency), which has a clear physical sense in terms of the mean free path $\lambda$ of a particle between its two subsequent collisions, $\lambda \approx (6D\tau_c)^{1/2} >> \bar{r}$ and is valid under the basic "dilution" condition of the theory, $n^{1/3}R \approx R/\bar{r} << 1$.

In the kinetic regime the phenomenological form of the pair-wise kernel $\beta(R_i, R_j)$ in the Smoluchowski kinetic equation, Eq. (1.11), derived for spatially homogeneous systems, is justified, and the original multi-particle problem is rigorously reduced to consideration of two-particle collisions, by calculation of the collision rate between two particles (of radii $R_i$ and $R_j$) randomly migrating (with the diffusivities $D_i$ and $D_j$) in a sample of unit volume.

In its turn, the latter value can be equally calculated as the rate of volume sweeping $d\langle V_{ij}\rangle/dt$ by the effective particle of radius $R_i + R_j$ migrating with the diffusivity $D_i + D_j$, which for spherical particles in the continuum mode takes the form

$$\beta(R_i, R_j) = 4\pi(D_i + D_j)(R_i + R_j) = \frac{2kT}{3\mu}\left(\frac{1}{R_i} + \frac{1}{R_j}\right)(R_i + R_j), \tag{1.91}$$

and fortuitously coincides with the traditional formula derived in the diffusion approximation (as the diffusion flux of particles $j$ into particle $i$), which is valid only for collisions between large and small particles [5]. The swept volume $dV_{ij}$ is calculated during a time step $dt$ that is small enough in comparison with the characteristic time of the particle concentration variation $\tau_c$ , in order to ignore variation in $dt$ of the mean particle concentration $n$, and large enough in comparison with the diffusion relaxation (or mixing) time $\tau_d$, in order to sustain the main assumption of the kinetic regime on random (homogeneous) distribution of coalescing particles. Besides, this time step $dt$ should be large enough in comparison with the characteristic time $\tilde{\tau} \approx 16\left(R_i + R_j\right)^2 / \pi\left(D_i + D_j\right)$, during which the steady state value of the sweeping rate $d\langle V_{ij}\rangle/dt$ is attained, $\tilde{\tau} \le \tau_d << dt << \tau_c$.

For fractal-like particle agglomeration in the continuum mode this expression takes the form (similar to the expression derived by Mountain et al. [39] in the traditional diffusion approach, valid for collisions of small and large particles)

$$\beta(R_i, R_j) = \frac{2kT}{3\mu}\left(\frac{1}{V_i^{1/d_f}} + \frac{1}{V_j^{1/d_f}}\right)\left(V_i^{1/d_f} + V_j^{1/d_f}\right), \tag{1.91a}$$

where $d_f \le 3$ is the fractal dimension of aggregates, $V_i$ is the volume of aggregate $i$, whereas the volume of the newly formed particle "$i + j$" of radius $R_{ij} = \left(R_i^{d_f} + R_j^{d_f}\right)^{1/d_f}$ is $V_{ij} = \left(V_i^{d_f/3} + V_j^{d_f/3}\right)^{3/d_f}$. To account more accurately for the influence of nonspherical particle geometry, a modified kernel can be used

$$\beta(R_i, R_j) = \frac{2kT}{3\mu}\left(\frac{1}{R_{S,i}} + \frac{1}{R_{S,j}}\right)R_{S,ij}, \tag{1.91b}$$

with the Smoluchowski radius for a single particle, $R_{S,i}$, and the combined Smoluchowski radius for the collision of two particles, $R_{S,ij}$, which were recently calculated using Brownian Dynamics approach by Thajudeen and et al. [72]. For simplicity and for direct comparison with the Langevin dynamics simulations of Heine and Pratsinis [64], Eq. (1.91a) will be further used in calculations (Section 1.9.3).

As above explained, the Smoluchowski equation, Eq. (1.11), assumes binary collisions and is derived in the second order of approximation $nR^3 << 1$. In the next, third order, triple collisions of comparable-size particles during their Brownian migration should be taken into consideration. In this approximation collisions which occur among any combination consisting of more than three particles, can be ignored, and the probability $dP_{123}$ of collisions in the above specified $dt$ among three particles randomly migrating in a sample of unit volume is calculated as

$$dP_{123} = P_{1+2,3} dP_{12} + P_{1+3,2} dP_{13} + P_{3+2,1} dP_{32}, \qquad (1.92)$$

where $dP_{ik}$ is the probability of a pair collision between particles $i$ and $j$ (of radii $R_i$ and $R_j$, respectively) in $dt$ calculated, in accordance with Eq. (1.91) or (1.91a), as the volume swept in $dt$ by the effective particle of radius $R_i + R_j$, migrating with the diffusivity $D_i + D_j$,

$$dP_{ik} = \beta(R_i, R_j)dt, \qquad (1.93)$$

and $P_{i+j,k}$ is the probability of a collision between the newly (instantly) formed particle "$i + j$" of radius $R_{ij}$ and the particle $k$ of radius $R_k$ during the considered time step $dt$.

Evaluating the probability $P_{i+j,k}$, one should keep in mind that there are two possibilities of a collision between the particles $k$ and $i + j$: (1) an instantaneous collision (with the probability $P_{i+j,k}(0)$) of the particle $k$ with the particle $i + j$ at the moment of the two particles $i$ and $j$ coalescence, and (2) a collision (with the probability $dP_{i+j,k}$) of the two migrating particles $i + j$ and $k$ in $d\tau$, where $d\tau \le dt$ is the remainder of the time step $dt$ after the collision of the particles $i$ and $j$.

The probability of the second event is calculated as the volume swept by the effective particle of radius $R_{ij} + R_k$ in $d\tau$, i.e. $dP_{i+j,k} = \beta(R_{ij}, R_k)d\tau \le \beta(R_{ij}, R_k)dt$, and thus, should be dropped after substitution in the rhs of Eq. (1.92) as a term of the higher order, $dP_{ij} dP_{i+j,k} = O(dt^2)$, in the first (linear) approximation with respect to the small (on the scale of the particle concentration variation time $\tau_c$) value $dt \to 0$.

The probability of the first event, $P_{i+j,k}(0)$, is equal to the probability that the particle $k$ at the moment of the particles $i$ and $j$ collision is located at the position, where its perimeter overlaps with the perimeter of one of the coalesced particle $i$ and $j$, or, under the above presented assumption of instantaneous coalescence of colliding particles, with the perimeter of the formed particle see Fig. 1.21). This probability is of the zero's order with respect to the time step, and thus should be kept in Eq. (1.92) in the limit $dt \to 0$, since $P_{i+j,k}(0)dP_{ij} = O(dt)$ (see Appendix B).

Therefore, the probability $P_{i+j,k}$ is calculated as

$$P_{i+j,k} = P_{i+j,k}(0) = V(R_i, R_j; R_k), \qquad (1.94)$$

where $V(R_i,R_j;R_k)$ is equal to the sum of two volumes: $V_{ij,k}$ of two overlapping spheres of radii $R_i+R_k$ and $R_j+R_k$, surrounding centers of particles $i$ and $j$ in the moment of their contact (a light grey "dumbbell" in Fig. 1.21), and $\Delta V_{ij,k}$ (cross-hatched area), additionally (instantaneously) swept by the particles $i$ and $j$ in the moment of their coalescence,

$$\begin{aligned}V\left(R_i,R_j;R_k\right) \equiv V_{ij,k}+\Delta V_{ij,k} = \frac{4\pi}{3}\left(\left(R_{ij}+R_k\right)^3+\left(R_i+R_k\right)^3+\left(R_j+R_k\right)^3\right) \\ -u\left(R_{ij}+R_k, R_i+R_k, \frac{\left(R_i+R_j\right)R_j^{d_f}}{R_{ij}^{d_f}}\right) \\ -u\left(R_{ij}+R_k, R_j+R_k, \frac{\left(R_i+R_j\right)R_i^{d_f}}{R_{ij}^{d_f}}\right),\end{aligned} \tag{1.95}$$

where

$$u(x,y,z) \equiv \frac{\pi\left(x+y-z\right)^2\left(\left(x+y+z\right)^2-4\left(x^2-xy+y^2\right)\right)}{12z} \tag{1.96}$$

is the intersection volume of two spheres of radii $x$ and $y$, and $z$ is the inter-center separation, following Chkhartishvili [73], see Appendix B. In particular, for equisize spherical particles of volume $V$

$$V\left(R,R;R\right) = 3\pi\left(8-\sqrt[3]{2}\right)R^3 \equiv q_0\frac{4\pi}{3}R^3, \tag{1.97}$$

where $q_0 \approx 15$ is the numerical factor.

Therefore, the mean number of triple collisions in $dt$ per unit volume among randomly migrating particles, is

$$dn_{123} = \frac{1}{3!}\int_0^\infty\int_0^\infty\int_0^\infty dP_{123}\, n\left(R_1,t\right)n\left(R_2,t\right)n\left(R_3,t\right)dR_1dR_2dR_3, \tag{1.98}$$

and the coagulation rate equation takes the form

$$\begin{aligned}\frac{\partial n(R,t)}{\partial t} = \frac{1}{2}\int_0^\infty\int_0^\infty \beta(R_1,R_2)n(R_1,t)n(R_2,t)\delta(R-R_{12})\, dR_1dR_2 - n(R,t)\int_0^\infty \beta(R,R_1)n(R_1,t)dR_1 \\ +\frac{1}{3!}\int_0^\infty\int_0^\infty\int_0^\infty \beta^{(3)}(R_1,R_2,R_3)n(R_1,t)n(R_2,t)n(R_3,t)\delta(R-R_{123})\, dR_1dR_2dR_3 \\ -\frac{1}{2}n(R,t)\int_0^\infty\int_0^\infty \beta^{(3)}(R,R_1,R_2)n(R_1,t)n(R_2,t)dR_1dR_2\end{aligned} \tag{1.99}$$

where

$$R_{123} = \left( R_1^{d_f} + R_2^{d_f} + R_3^{d_f} \right)^{1/d_f}, \tag{1.100}$$

$$\begin{aligned} \beta^{(3)}(R_i, R_j, R_k) = \beta(R_i, R_j) V(R_i, R_j; R_k) + \beta(R_j, R_k) V(R_j, R_k; R_i) \\ + \beta(R_k, R_i) V(R_k, R_i; R_j) \end{aligned} \tag{1.101}$$

and $V(R_i, R_j; R_k)$ is defined in Eq. (1.95).

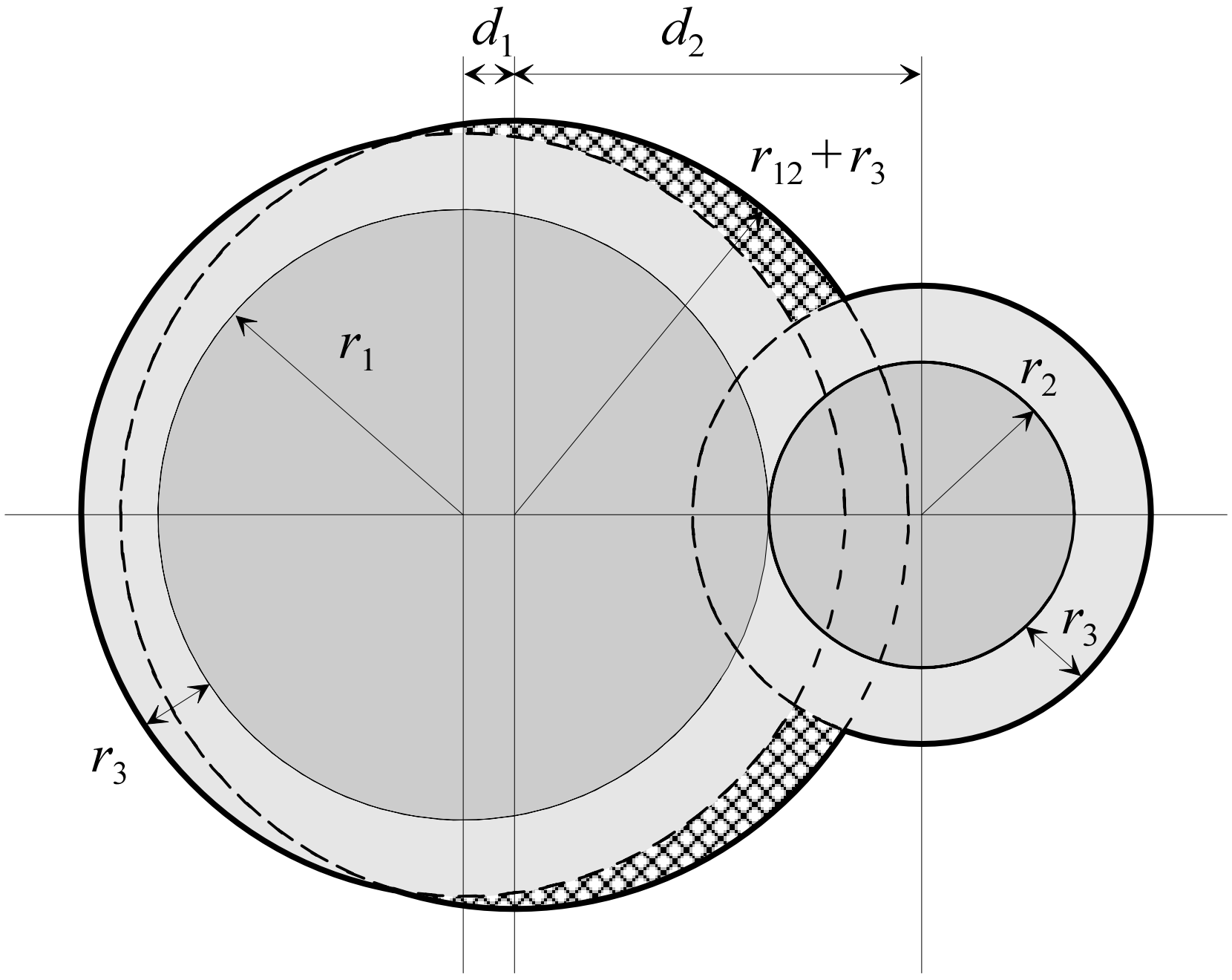

Fig. 1.21. To the derivation of Eq. (1.95): the perimeter of the calculated volume is represented by a thick solid line.

In terms of the discrete variable $i$, the number of monomers in a particle ($i$ = 1 corresponds to monomers), the coagulation rate equation for the concentration of particles containing $i$ monomers, $c_i(t)$, takes the form

$$\begin{aligned} \frac{dc_i}{dt} &= \frac{1}{2}\sum_{j=1}^{\infty}\sum_{k=1}^{\infty}\delta_{i,j+k}\beta_{jk}c_j c_k - c_i\sum_{j=1}^{\infty}\beta_{ij}c_j \\ &+ \frac{1}{6}\sum_{j=1}^{\infty}\sum_{k=1}^{\infty}\sum_{l=1}^{\infty}\delta_{i,j+k+l}\beta_{jkl}^{(3)}c_j c_k c_l - \frac{1}{2}c_i\sum_{j=1}^{\infty}\sum_{k=1}^{\infty}\beta_{ikl}^{(3)}c_j c_k \end{aligned}, \tag{1.102}$$

where $\beta_{ij} \equiv \beta(R_i, R_j)$, $\beta_{jkl}^{(3)} = \beta^{(3)}(R_i, R_j, R_k)$ and $\delta_{ij}$ is the Kronecker delta function.

*1.9.2. Qualitative analysis*

The triple terms in Eq. (1.102) change the particle growth kinetics. Considering the indexes in Eq. (1.102) as continuous variables and integrating the both sides over *i*, one obtains

$$\begin{aligned}\frac{dc(t)}{dt} &= -\frac{1}{2}\int_0^\infty\int_0^\infty \beta(i,j)c(i,t)c(j,t)didj - \frac{1}{3}\int_0^\infty\int_0^\infty\int_0^\infty \beta^{(3)}(i,j,k)c(i,t)c(j,t)c(k,t)didjdk \\ &= -\frac{1}{2}\bar{\beta}c^2(t) - \frac{1}{3}\overline{\beta^{(3)}}c^3(t)\end{aligned} \quad (1.103)$$

where $c(t)=\int_0^\infty c(i,t)di$ is the total number of particles per unit volume, and $\bar{\beta}$, $\overline{\beta^{(3)}}$ are the kernels averaged over the particle size distribution. It is straightforward to demonstrate (see Appendix C) that for homogeneous kernels

$$\beta_{ai,aj} = a^{\lambda}\beta_{ij}, \quad \beta^{(3)}_{ai,aj,ak} = a^{\lambda+3/d_f}\beta^{(3)}_{ijk} \quad (1.104)$$

a self-preserving mass spectrum [76]

$$c_i(t) = \frac{1}{\bar{N}^2(t)}\varphi(x), \quad (1.105)$$

where $x \equiv i/\bar{N}(t)$, is asymptotically attained. Here $\bar{N}$ is the mean number of monomers in the particles, which asymptotically depends on time as

$$\bar{N}(t) \propto t^z, \quad (1.106)$$

where $z = 1/(1-\lambda)$.

However, in contrast to the standard Smoluchowski theory, these conclusions are generally valid only for non-fractal particles ($d_f = 3$). In particular, Eq. (1.106) is not anymore valid, if $d_f \neq 3$ (however, it can be formally used considering *z* as a time dependent value). For this reason, following Heine and Pratsinis (2007b) [64], it is convenient to characterize the enhancement of the coagulation rate by another dimensionless factor

$$\gamma = \frac{2}{\beta_{dilute}}\frac{d}{dt}c^{-1}(t), \quad (1.107)$$

where $\beta_{dilute} = \bar{\beta}(f \to 0)$ is the kernel averaged in the dilute limit; for the kernel Eq. (2) $\beta_{dilute} \approx 1.073\beta_0$ [22].

It is useful to consider the above formulas in the framework of monodisperse approximation, $c(i,t) = c(t)\delta(i - \bar{N}(t))$. Then Eq. (14) reduces to

$$\frac{\partial c(t)}{\partial t} = -\frac{1}{2}\beta_0\left(\bar{R},\bar{R}\right)c^2(t) - \beta_0\left(\bar{R},\right)V\left(\bar{R},\bar{R};\bar{R}\right)c^3(t) = -\frac{1}{2}\beta_0\left(\bar{R},\bar{R}\right)\left(1+2q_0 f\right)c^2(t), \quad (1.108)$$

where $\bar{R}$ is the radius of particle containing $\bar{N}$ monomers, $f = V\left(\bar{R},\bar{R};\bar{R}\right)c(t)$ is the volume fraction, $q_0$ is defined in Eq. (1.97). The first term in the *rhs* of Eq. (1.108) represents the well-known approximation for binary collisions (Friedlander, 2000), whereas the second term (enhancing the concentration change) accounts for the effect of triple collisions. Neglecting this second term and taking into account mass conservation, $c(t)\bar{N}(t) = const$, one easily derives Eq. (1.106). This formula is valid also with triple collisions taken into account but only for the spherical particles ($d_f = 3$), in agreement with the general consideration (Appendix C).

From Eq. (1.107) one derives in the dilute limit for spherical particles ($d_f = 3$)

$$\gamma = \frac{\bar{\beta}(f)}{\beta_{dilute}} + 2qf + O\left(f^2\right), \quad (1.109)$$

where parameter $q \approx 15$ is close to $q_0$ (the difference is due a finite width of the distribution in the dilute limit). As seen dependence of parameter γ is linear at small values of $f$ and almost entirely is due to triple collisions (the second term in the *rhs*). Additional linear contribution is due to dependence of $\bar{\beta}$ on $f$ due to widening of the distribution (which is an indirect effect of the triple collisions); however this contribution is small in comparison with the second term. As for nonlinear terms, they are mainly due to tetrad and higher order collisions and become essential at higher values of $f$.

*1.9.3. Quantitative analysis*

In numerical analysis of Eq. (1.102), the discretization similar to that of sectional method of Gelbard and Seinfeld [75] was used according to which the particle groups (below numerated by the capital letters) are introduced, each group *I* containing particles with the number of monomers within interval $\left(a_I, b_I\right)$ defined as

$$a_{I+1} = \left[sa_I\right] + 1, \quad b_I = a_{I+1} - 1, \quad I = 1,2..., \quad (1.110)$$

where $a_1 = 1$ and $s > 1$ is the progression factor, [...] denoting truncation operation. Within each group, the particle concentrations are approximated as

$$c_i = \frac{c_I}{\Delta_I}, \quad a_I \le i \le b_I, \quad (1.111)$$

where $\Delta_I = a_{I+1} - a_I$ and $c_I$ is the total particle concentration in group *I.* Then Eq. (1.102) is approximated by the group equations

$$\frac{dc_I}{dt}=\sum_{J=1}^{I}\sum_{K=1}^{J}\left(1-\frac{\delta_{JK}}{2}\right)\frac{n_J+n_K}{n_I}\beta_{JK}g_{JK;I}c_Jc_K-c_I\sum_{J=1}^{\infty}\beta_{IJ}c_J$$
$$+\sum_{J=1}^{\infty}\sum_{K=1}^{J}\sum_{L=1}^{\infty}\left(1-\frac{\delta_{JK}}{2}\right)\frac{n_J+n_K+n_L}{n_I}\beta_{JK}V_{JKL}g_{JKL;I}c_Jc_Kc_L \qquad (1.112)$$
$$-c_I\sum_{J=1}^{\infty}\sum_{K=1}^{J}\left(1-\frac{\delta_{JK}}{2}\right)\beta_{IJK}c_Jc_K$$

where $n_I=\left(a_I+b_I\right)/2$, and the group transfer coefficients are defined as

$$g_{JK;I}=\frac{1}{\Delta_J\Delta_K}\sum_{i=a_I}^{b_I}\sum_{k=a_K}^{b_K}\sum_{j=a_J}^{b_J}\delta_{k+j,i},$$
$$g_{JKL;I}=\frac{1}{\Delta_J\Delta_K\Delta_L}\sum_{i=a_I}^{b_I}\sum_{k=a_K}^{b_K}\sum_{j=a_J}^{b_J}\sum_{l=a_L}^{b_L}\delta_{l+k+j,i}, \qquad (1.13)$$

for details see Appendix D.

Results of calculations are compared with the results of direct numerical simulations of particle trajectories by Langevin dynamics [64] for the typical conditions for Brownian coagulation in the continuum mode in air at $T$ = 293 K of spherical particles ($d_f$ = 3), initially monodisperse with $2R_0$ = 1 μm in diameter and with a density of 1 g/cm$^3$. For these conditions the overall collision frequency is evaluated as $\beta_{dilute}$ = 6.4·10$^{-16}$ m$^3$/s, which determines the characteristic time scale, $\tau_{sp}=2/\beta_{dilute}n_0=2V_0/\beta_{dilute}f$ (e.g. $\tau_{sp}\approx 0.2$ s for $f$ = 1 %, however in practice the actual time lag to attain the self-preservation is greater by an order of magnitude, [64]).

The normalized self-preserving size distribution function $\varphi(x)$, Eq. (1.105), calculated for $f$ = 0.3%, 1% and 3% with and without triple collisions taken into consideration, is presented in Fig. 1.22 (for convenience, the relative particle size, $x_R=R/\overline{R}$, is chosen as the ordinate instead of $x=i/\overline{N}$).

From Fig. 1.22 it is seen that the particle size distribution function broadens quickly during growth and attains the self-preserving form for spherical agglomerates, $d_f$ = 3, in accordance with the analysis in Appendix C. The width of the attained self-preserving size distribution function somewhat increases with the increase of the fractional volume $f$ in the transition range, in accordance with LD simulations of Heine and Pratsinis [64], but rather moderately. In particular, it is seen that the majority of particles are concentrated in a relatively narrow size-band (within one order of magnitude, $0.1\overline{R}\le R\le 2\overline{R}$) around the mean size $\overline{R}(t)$, where the particle concentration decreases by ≈ 2–3 orders of magnitude. This allows excluding from consideration, with a sufficient accuracy, the particle sizes outside this narrow band. On the other hand, the remaining sizes (located within this band), being distributed within one order of magnitude, can be considered as comparable. Therefore, only collisions among comparable size particles (distributed around the mean size) can be taken into consideration (despite formally integration over the particle sizes will be extended from $-\infty$ to $+\infty$). This justifies consideration of a homogeneous spatial distribution of particles, which rapidly reinstates in-between particle collisions owing to their diffusion mixing, in accordance with the general assumption of the kinetic approach.

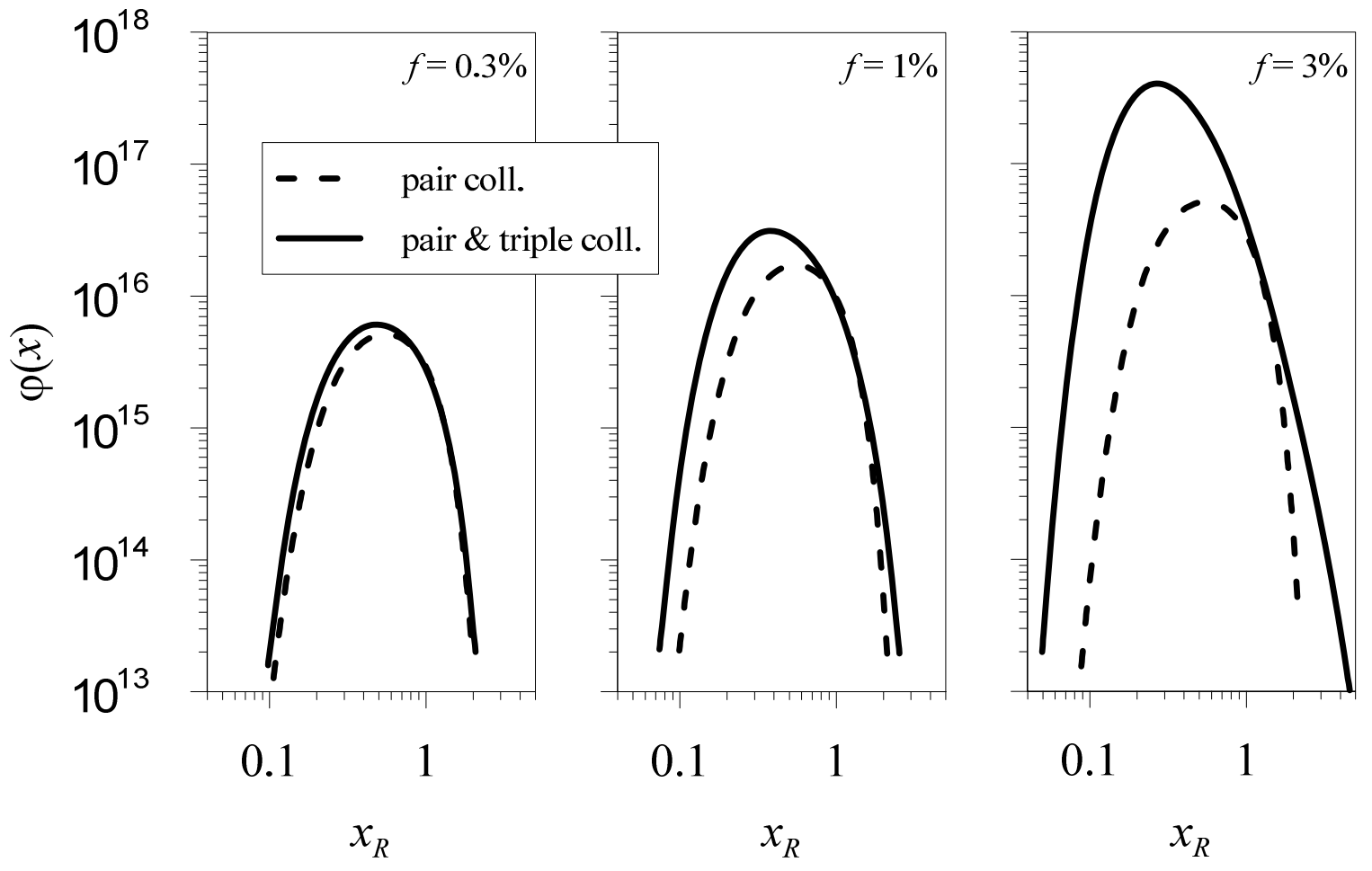

Fig. 1.22. Self-preserving size distribution function calculated with consideration of pair collisions (dashed curves) and pair & triple collisions (solid curves).

The $\gamma$ values calculated at different fractional volumes are plotted in Fig. 1.23, in comparison with the results of Langevin dynamics simulations [64]. A good agreement for the transition from dilute to dense regime at $f \leq 10\%$ reasonably confirms the validity of the new approach. At $f$ higher than $\approx 10\%$, well in the dense regime, the deviation of the current results from LD simulations becomes notable, manifesting the influence of multiple (more than three-particle) collisions.

As shown in the (linear scale) insert of Fig. 1.23, in the transition range ( $f \leq 10\%$) the enhanced collision frequency almost linearly depends on $f$, in a good agreement with the mean field approximation result, Eq. (22). Moreover, this equation provides a good prediction of the slope of the curve; for instance, at $f = 3\%$ the calculated enhancement factor of 1.9 well corresponds to the mean field value of 1.82. These results are additionally illustrated in Fig. 1.24, from which the almost twice acceleration of the kinetics due to triple collisions at $f = 3\%$ is clearly seen.

It should be noted that, in accordance with the analysis presented in Section 3 (and in Appendix C), for fractal agglomerates with $d_f < 3$ the self-preserving size distribution is not attained (asymptotically). Simultaneously the particle volume fraction, $f$, grows and the size distribution function significantly broadens at large times (the width of distribution grows faster than the mean value). In particular, these peculiarities were illustrated by LD analysis [64] of the growth kinetics of agglomerates with $d_f = 1.8$. In these calculations the population was assumed to be initially monodisperse with the monomer radius of 0.11 μm and concentration of $5.4\times10^{16}$ $m^{-3}$, so that the initial volume fraction $f$ was 0.03% corresponding to the characteristic time scale of $\approx 0.06$ s.

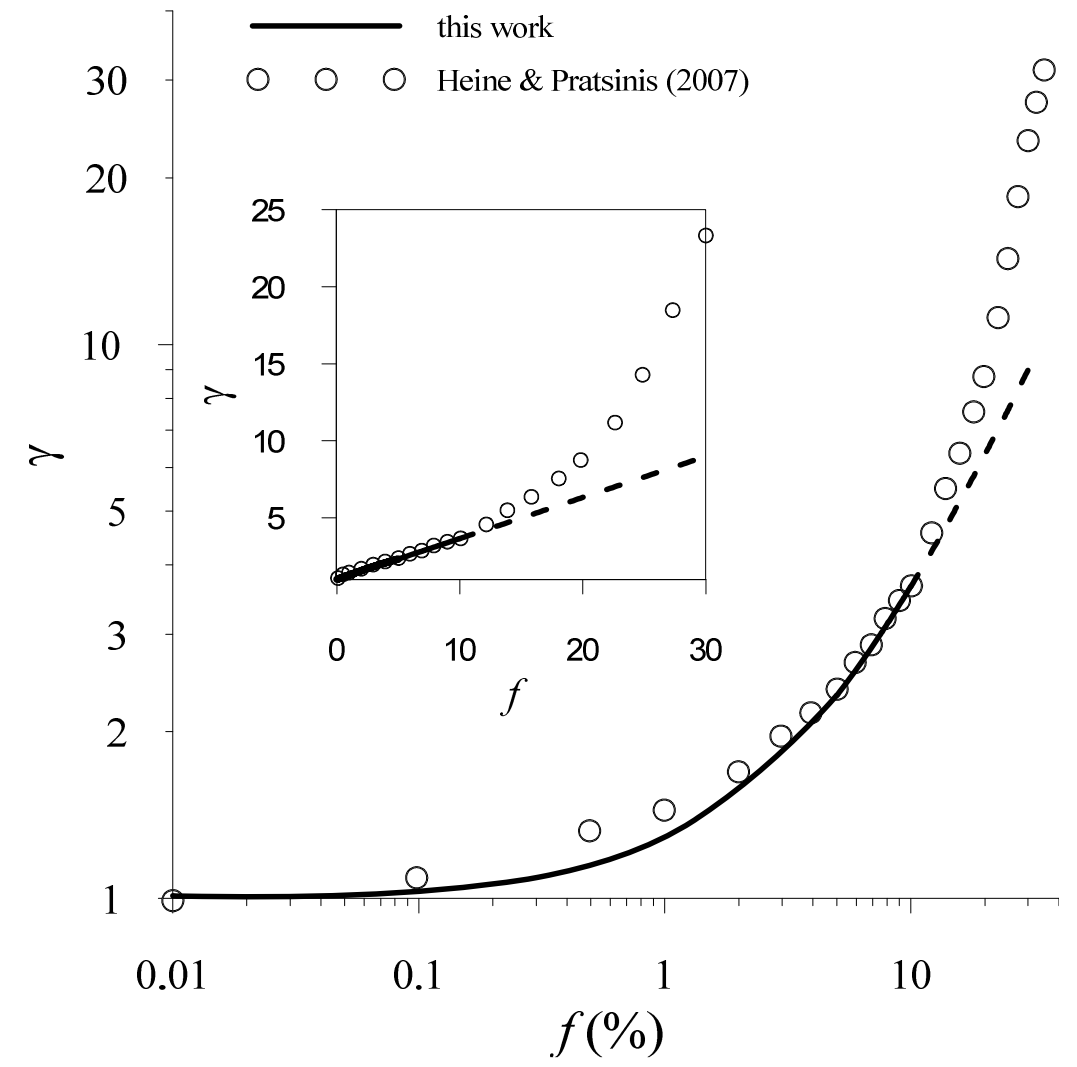

Fig. 1.23. Enhanced collision frequency in simulations of Brownian coagulation of spherical agglomerates, $d_f = 3$, with $\beta_{dilute} = 6.4\times10^{-16}$ m$^3$/s and the initial size $2R_0 = 1$ μm.

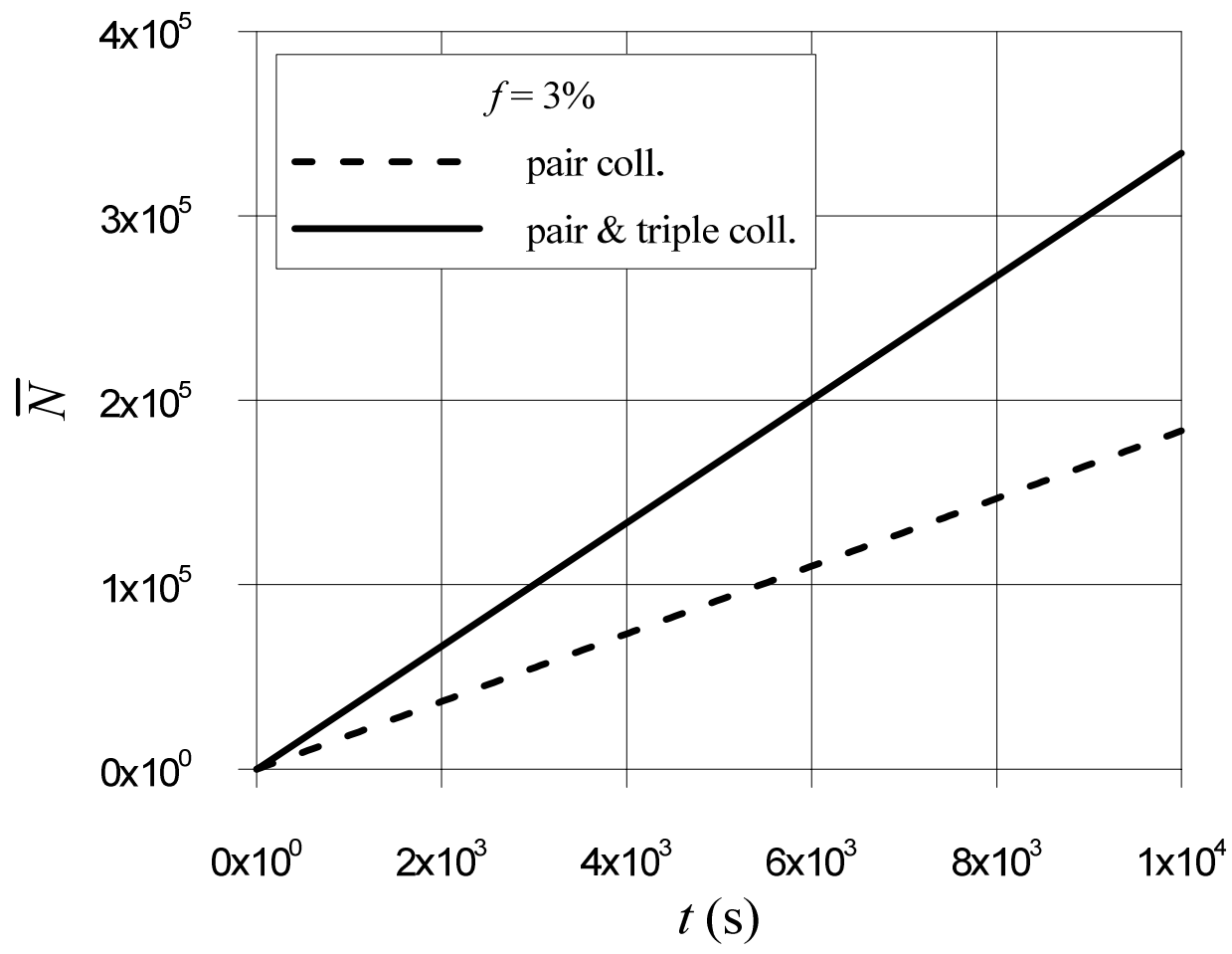

Fig. 1.24. Temporal dependence of the mean number of monomers in spherical agglomerates for $f = 3\%$.

Results of calculations for this regime, using the rate equation, Eq. (1.102), with the kernel for non-spherical, fractal particles, Eq. (1.91a), are presented in Fig. 1.25, where the temporary dependencies are plotted of the volume fraction, $f$, the enhanced collision frequency, $\gamma$, and the geometric standard deviation, $\sigma$, defined as

$$\ln\sigma = \left\langle \ln^2 \left( R/\rho \right) \right\rangle^{1/2}, \quad \ln\rho = \left\langle \ln R \right\rangle. \qquad (1.114)$$

The geometric standard deviation characterizes the relative width of distribution so that it stabilizes with time if the self-preserving limit is attained (see the standard Smoluchowski curve in Fig. 1.25). In the case of the dilute fractal agglomerates the distribution firstly tends to the self-preserving form but then is distorted by the triple collisions resulting in eventual growth of $\sigma$ value at large $f$. For this reason, the kinetic approach, developed for comparable size particles basing on the assumption of a narrow size distribution function, is well justified only in the initial stage of coalescence (i.e. at relatively small $f$), and thus numerical calculations for fractal particles were not attempted at $f \geq 5\%$. Nevertheless, in the initial range of the transition from dilute to dense particle volume fractions, $0.1\% < f < 5\%$, the results of calculations are in a reasonable agreement with LD simulations.

Further modification of the theory for consideration of a broad particle size distribution is foreseen in the near future.

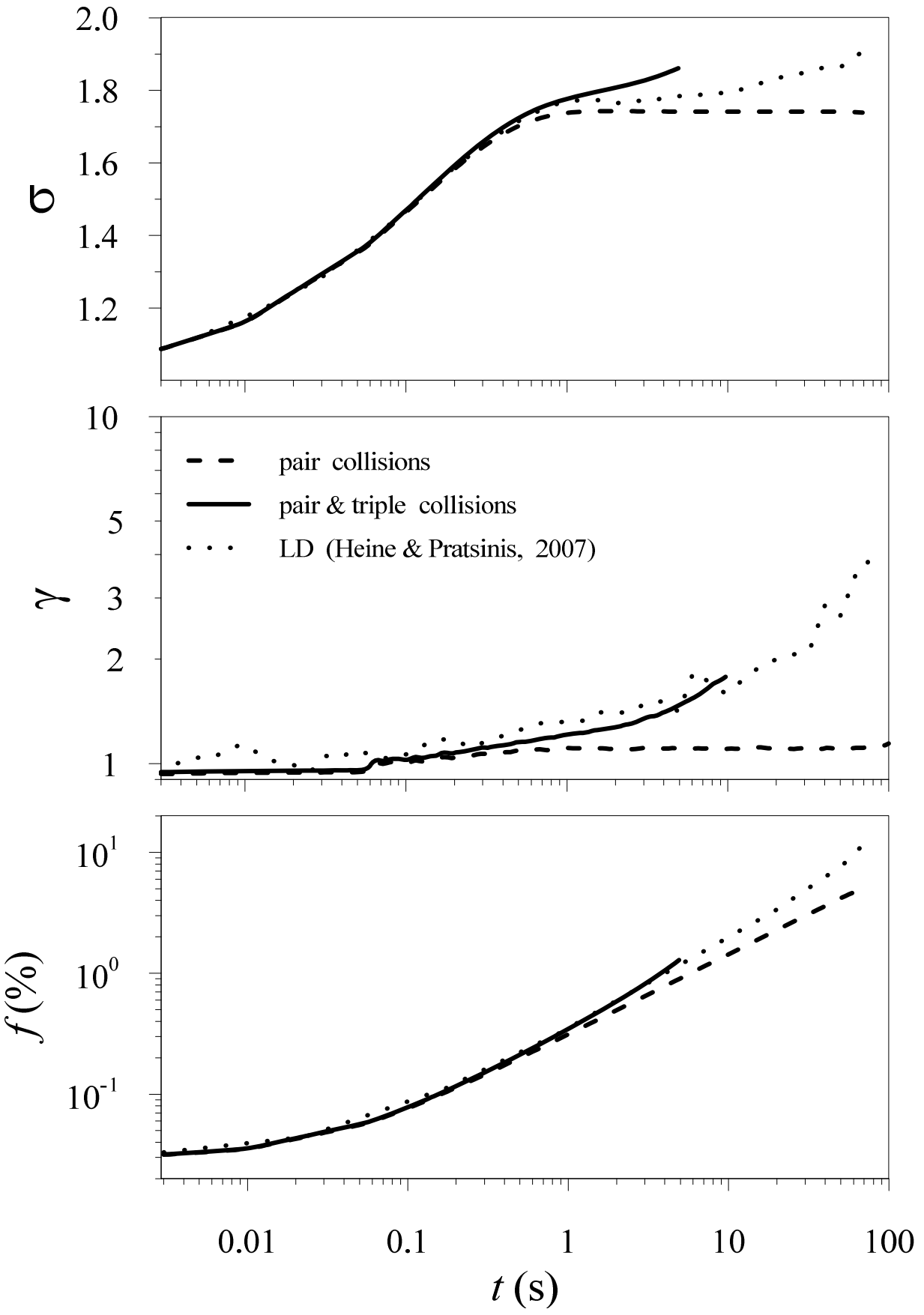

Fig. 1.25. Temporal dependence of the geometric standard deviation, $\sigma$, the enhanced collision frequency, $\gamma$, and the volume fraction, $f$, in simulation of the fractal particles ($d_f = 1.8$) growth kinetics.

### *1.9.4. Conclusions*

The Smoluchowski theory of Brownian coagulation assumes binary collisions, described in the second order in particle concentration $n$, and for this reason cannot describe the coagulation of highly dense (or concentrated) colloidal or aerosol suspensions. In order to study the transition from dilute (controlled by binary collisions) to dense (controlled by multiple collisions) regime of coagulation, the Smoluchowski equation is generalized by consideration of triple collisions. This modification of the classical theory is realized in the kinetic approach, recently proposed by the authors for coagulation of comparable size particles, which cannot be treated within the traditional approach.

After attainment of the self-preserving size distribution function in the system of coagulating spherical particles ($d_f = 3$), the majority of particles are concentrated in a relatively narrow size-band (within one order of magnitude) around the mean size $\overline{R}(t)$, where the particle concentration decreases by $\approx$ 2–3 orders of magnitude. This allows, with a sufficient accuracy, excluding from consideration the particle sizes outside this narrow band and considering the remaining sizes as comparable. This justifies the validity of the new kinetic approach (applied to comparable size particles) and allows comparing the new model predictions with more general results of the direct numerical simulations by Langevin dynamics [64]. A good agreement is attained in a relatively wide range of the fractional volume of spherical particles (up to ≈ 10%), corresponding to the transition from dilute to dense regime of coagulation dynamics, in which multiple collisions among more than three particles can be neglected.

For fractal particles ($d_f < 3$) the particle size distribution function firstly tends to the self-preserving form but then is distorted by the triple collisions and broadens with time. For this reason, numerical calculations for fractal particles were not attempted at large $f$ with a broad size distribution function. Nevertheless, in the initial stage of transition from dilute to dense particle volume fractions, $0.1\% < f < 5\%$, the results of calculations are also in a reasonable agreement with LD simulations.

### 1.10. Discussion

A theoretical basis for the Fuchs semi-empirical approach [4] to calculation of the coagulation kernel in the transition mode has been provided by Sitarski and Seinfeld [76], and by Mork et al. [77], who obtained the coagulation coefficient in the system of equisize particles (of radius $R$ and diffusivity $D$) through the solution of the Fokker-Planck equation for distribution function of Brownian particles. Namely, they calculated the steady rate of absorption of Brownian particles centres by a sink of radius $2R$ and assumed the sink to be fixed in space, and the distribution function of particles centres to be governed by the Fokker-Planck equation (with diffusivity $2D$). In this way, they have reduced the coagulation problem to consideration of condensation (in the large central particle fixed in space) of point masses (representing centres of original particles).

At first glance, such an approach resolves the inconsistency of the Smoluchowski-Chandrasekhar approach to calculation of the collision rate, based on consideration of the diffusion flux of surrounding particles to the central one. Namely, after turning to consideration of point-wise particles, which actually describe positions of the original particles centres, the problem of mutual collisions between particles (of finite size), moving to the central trap, is artificially removed (since collisions of point-wise particles can be ignored). However, as shown in Section 1.3.1, in fact the mutual collisions of the original particles can be neglected only in the case of their small radii in comparison with that of the central particle. Therefore, in the advanced approach of [76] and [77] (based on the Fokker-Planck equation in application to coagulation of equisize particles), in which mutual collisions of surrounding particles were actually neglected (despite they had the same size as the central sink), the inconsistency of the diffusion approach of Smoluchowski-Chandrasekhar was not removed.

Another approach to the Brownian coagulation problem was developed by Nowakowski and Sitarski [78] and by Narsimhan and Ruckenstein [30] through Monte Carlo simulations. Their approach, contrary to [77, 78], was essentially based on an assumption that the motion of only two colliding entities can be considered; however, applicability of this assumption was not grounded (and apparently was just derived from the phenomenological form of the pair-wise kernel in the Smoluchowski kinetic equation, Eq. (1.11)).

As shown in the author's papers [5-7], the coagulation problem for comparable size particles can be really reduced to consideration of two colliding entities only in the case of rapid diffusion mixing of particles between their collisions, i.e. when the system of particles after each coalescence may be considered as spatially homogeneous (see Section 1.3.2). Therefore, the basic assumption of [78, 30] (valid for homogeneous systems) was directly opposite to the Smoluchowski-Chandrasekhar approach, based on the assumption (valid only for collisions between large and small particles) that the collision rate is controlled by the diffusive current of particles, that implies essentially inhomogeneous distribution of particles in space. In situation when the Smoluchowski-Chandrasekhar approach was (and still is) widely accepted for consideration of collisions in the ensemble of comparable size particles (as well as for traditional consideration of diffusion-limited reactions, see Part 2), this question could not be treated as negligible or obvious, and thus required explicit analysis and justification (that was not attempted in [78] or was insufficient in [30], as above explained).

Following the current approach justification that only two colliding entities can be considered owing to rapid diffusion mixing in the ensemble of (comparable size) Brownian particles, the coagulation problem can be properly reduced to consideration of one immobile trap (of radius $R_1 + R_2$) and one migrating point-wise particle, which, in its turn, can be described by the Fokker-Planck equation for distribution function of this particle, $f(\mathbf{r}, \mathbf{v}, t)$. Naturally, this distribution function can be equally applied to formal consideration of an ensemble of point-wise particles of concentration $n(\mathbf{r}, t) = \int f(\mathbf{r}, \mathbf{v}, t) d\mathbf{v}$, however, these particles are fictitious (e.g. markers considered in Sections 1.4.1 – 1.4.3), since they are not related to real Brownian particles from the original multi-particle ensemble. Therefore, the Fokker-Planck approach can be eventually applied to consideration of the Brownian coagulation problem, however, indirectly, after reducing the multi-particle problem to consideration of two particles collision probability (justified under the mixing condition, $\tau_c >> \tau_d$ or $\lambda >> \bar{r}$).

Since the Fokker-Planck equation, used in [76, 77], is generally derived from the Langevin equation, which is also the governing equation of the Monte Carlo approach [78, 30], both methods should bring in similar results. On the other hand, the Langevin equation can be equally reduced at large times $t >> \tau_0$ to the Einstein diffusion equation for the Brownian particle motion (see, e.g. [4, 19]), which can be properly described by the random walk theory applied in the current paper. For this reason, one should expect a rather good coincidence of the semi-analytical predictions of the current approach (especially, in the next, more consistent approximation of the random walk theory, applied in Section 1.5) with results of numerical methods based on the Langevin equation.

More detailed consideration, based on description of each particle motion by the Langevin equation, shows that their relative displacements also obey the Langevin equation for the effective particle only in the case $m_1 b_1 = m_2 b_2$, which, in particular, corresponds to the case of similar particles with $R_1 = R_2$. In this case the collision rate of two original particles can be strictly calculated as the mean volume swept by the effective particle per unit time (as assumed in the current approach), whereas in a more general case of different size particles, $R_1 \neq R_2$, this consideration may be approximate. Nevertheless, calculations of the two-particle coagulation rate in the Langevin approach (Section 1.9) well confirm the conclusion that the current approximation can be used with a reasonable accuracy also for coagulation of different size particles, if the mixing condition is valid for these particles (i.e. when their sizes are comparable).

# Part 2. Diffusion-limited reaction rate theory

## 2.1. Introduction

For many chemical processes, the reaction proceeds from a reaction complex formed by collision of two or more reactants. Each reaction rate coefficient $K$ has a temperature dependency, which is usually given by the Arrhenius equation, $K = K_0 \exp(-E_a/kT)$, where the pre-exponential factor $K_0$ determines the collision frequency of reacting species and the exponential factor determines the number of collisions with energy greater than the activation energy $E_a$ of the complex (i.e. corresponds to the sticking probability of collisions).

Diffusion-limited (or diffusion-controlled) reactions are reactions in which collisions of reactants (determining the pre-exponential factor $K_0$) are controlled by their diffusion migration in suspending solvent (rather than free-molecular collisions typical for molecular reactions in gas mixtures). Diffusion-limited reactions between two different species A and B (A + B $\rightarrow$ C, where C does not affect the reaction) show up in a vast number of applications including not only chemical (see e.g. [79]), but also biological (e.g. [80-82]) and ecological (e.g. [83]) processes that have been studied over many decades. This may apply also to the reaction of point defects, vacancies and interstitials (V + I $\rightarrow$ 0), annihilation in crystals [84] produced by means of high-energy particles or electrons.

A method for calculating the reaction rate of reaction partners migrating by three-dimensional diffusion was developed in [85, 86] by generalization of the Smoluchowski theory for coagulation of colloids [1]. In this method the radius of the activated complex (or the "reaction radius") corresponds to the "influence-sphere radius" in the Smoluchowski theory (roughly equal to the sum of radii of two colliding Brownian particles, $R_{12} \approx R_1 + R_2$), which in the continuum approach is assumed to be large in comparison with elementary drift (or jump) distances $a_1$, $a_2$ of particles migrating by random walks, $R_{12} >> a_1, a_2$. In the opposite limiting case, $R_{12} << a_1, a_2$, the continuum diffusion approach is not anymore valid, therefore, the so-called "free-molecular" (or "ballistic") approximation can be used for colliding Brownian particles [4].

Formulating a reaction-diffusion model, a *d*-dimensional Euclidean space on which A and B particles at initial average densities (number of particles per unit volume) $n_A$ and $n_B$ diffuse freely, is usually considered in the continuum approach (see, e.g. [87-89]). In this approach reactant particles are represented as points or spheres undergoing spatially-continuous Brownian motion, with bimolecular chemical reactions, A + B $\rightarrow$ C, occurring instantly when the particles pass within specified reaction radius $R_{AB}$ between their centres.

The continuum approach was further applied to the diffusion-limited reactions in one (1-d) and two (2-d) dimensions (see, e.g. [90, 91]), the latter case has wide applications also in the membrane biology (see a review in [92]). The diffusion-limited bimolecular reactions between mobile vacancies and interstitials in strongly anisotropic crystals provided the mobile species is constrained to migrate in one plane only, may be also well approximated over a wide range of the reaction by a 2-d second order rate equation [93]. Another example of the 2-d model application is coalescence of intergranular voids on grain faces of irradiated metals or ceramics (e.g. in the practically important case of $UO_2$ nuclear fuel) [94, 95].

In such approach, the same shortcomings of the Brownian coagulation theory that were critically analysed in the author's papers [5-7] (see Part 1), are generally inherited in the diffusion-limited reaction rate models. Namely, the diffusion approach [1, 2] to calculation of the collision rate function (based on assumption that the local collision rate should be equal to the diffusive current of particles) is applicable only to the special case of coalescence between large and small Brownian particles, $R_1 << \bar{r} << R_2$ (where $\bar{r} \approx n^{-1/3}$ is the mean inter-particle distance), and becomes inappropriate to calculation of the coalescence rate for particles of comparable sizes, $R_1, R_2 << \bar{r}$. Correspondingly, the traditional approach to the diffusion-limited reaction rate theory based on a similar assumption that the local reaction rate should be equal to the diffusive current of particles, becomes invalid in the case when the characteristic reaction distance $R_{AB}$ for A↔B complex formation (i.e. the reaction radius), is small in comparison with the mean inter-particles distances, $R_{AB} << \bar{r}_A, \bar{r}_B$, where $\bar{r}_A \approx n_A^{-1/3}$, $\bar{r}_B \approx n_B^{-1/3}$ (see Section 2.2.1).

The new approach developed in [5-7] was generalized in the author's papers [96, 97] to the case of diffusion-limited reaction kinetics. For the base case of continuum mode, $a_A, a_B << R_{AB} << \bar{r}_A, \bar{r}_B$, the reaction rate calculated in the new approach in 3-d (see Section 2.3.1) formally (and in fact, fortuitously) coincides with the traditional result, valid only for reactions with a large reaction radius, $\bar{r}_A << R_{AB} << \bar{r}_B$. However, for the base case $a_A, a_B << R_{AB} << \bar{r}_A, \bar{r}_B$ in 2-d the traditional approach leads to considerable deviations of the reaction decay $n_i(t)$ at large times $t$ from that calculated in the new approach (see Section 2.4), thus explicitly demonstrating inconsistency of the traditional approach.

In the case $R_{AB} << a_A, a_B$, the free molecular (or ballistic) regime is realized. This case can be also considered similarly to the Brownian particles coagulation problem in the corresponding regime, as well as the case of the transition regime, $R_{AB} \approx a_A, a_B$ (Section 2.3.2).

The new approach was further generalized to consideration of reaction kinetics for particles migrating by random walks on discrete lattice sites (with the lattice spacing $a$) in the author's paper [97]. Since the case of large reaction radius, $R_{AB} >> a$, is properly reduced to the continuum media limit, the opposite case, $R_{AB} < a$, with reactions occurred when two particles occupy the same site (see, e.g. [98]), is of the most concern. It will be shown (in Section 2.5) that the traditional approach [91, 98] to consideration of this important case preserves the main deficiencies of the continuum media approach and thus results in erroneous predictions for the reaction kinetics. For this reason, new relationships for the reaction rate constants will be derived either for 3-d (in Section 2.5.1) or for 2-d lattices (in Section 2.5.2). The discrepancy between the new and traditional approach predictions further increases, when more complicated case of catalytically-activated reactions is considered (in Section 2.5.3).

The main outcomes of the new approach to diffusion-limited reaction rate theory are formulated in Section 2.6.

## 2.2. Rate equations

In the approximation $R/\bar{r} << 1$, only pair-wise collisions of particles during their diffusion migration can be taken into consideration, and collisions which occur among any combination consisting of more than two particles, can be ignored.

In the rate theory for a continuous distribution of particles $N(R)dR$, the number of particles of radius $R$ to $R + dR$ per unit volume, under an assumption that collided particles are randomly distributed in space and, upon collision, immediately coalesce to form a new particle, the Smoluchowski coagulation equation takes the form

$$\frac{\partial N(R,t)}{\partial t} = \frac{1}{2}\int_0^\infty\int_0^\infty N(R_1,t)N(R_2,t)\delta\left[R - \left(R_1^3 + R_2^3\right)^{1/3}\right]\beta(R_1,R_2)dR_1dR_2 \\ - N(R,t)\int_0^\infty N(R_1,t)\beta(R,R_1)dR_1, \tag{2.1}$$

where $\beta(R_1,R_2)$ is the collision frequency function. Under the basic condition of the Smoluchowski theory on spatial homogeneity of the particle distribution, $N(R,\mathbf{r},t) = N(R,t)$, the kernel $\beta(R_1,R_2)$ can be defined as a number of collisions in unit time per unit volume between two particles of radii $R_1$ and $R_2$ randomly located in space, which, for this reason, does not depend on time explicitly. For this reason, $\partial N(R,t)/\partial t$ should be calculated from consideration of pair-wise collisions during a relatively short time step $\delta t$ when variation of concentration densities $N(R_1,t)$ and $N(R_2,t)$ can be neglected on the one hand, and $\delta t$ being long enough to attain the steady state value of $\beta(R_1,R_2)$ during this time step, on the other hand.

For the kinetics of an irreversible reaction A + B $\rightarrow$ C (where C does not affect the reaction) in the mean-field approximation, Eq. (2.1) being applied to the two-size ($R_A$ and $R_B$) particle distribution function, is reduced to

$$\frac{dn_A}{dt} = \frac{dn_B}{dt} = -K_{AB}n_A(t)n_B(t), \tag{2.2}$$

where $n_A$ and $n_B$ are the mean concentrations of reacting A and B particles, respectively, and $K_{AB}$ is a rate function (or reaction constant), directly corresponding to the collision frequency function $\beta$ for two particles of different types (A and B), $K_{AB} = \beta(R_1,R_2)\delta(R_1 - R_A)\delta(R_2 - R_B)$. In accordance with the Smoluchowski rate theory, $K_{AB}$ is defined as the frequency of collisions in the unit volume of two particles randomly located in space and for this reason, it should be considered as a value explicitly independent of time. In a self-consistent approach, the reaction rate $dn_A/dt$ should be calculated choosing the time step $dt$ that is short enough to neglect variation of the mean concentrations $n_A$ and $n_B$ in $dt$, and long enough to attain a steady state value of $K_{AB}(t) \approx K_{AB}(\infty) \equiv K_{AB}$. This is important difference from the traditional models for diffusion-limited reaction kinetics (despite they are often termed as the Smoluchowski-type models), where, under assumption that the local reaction rate should be equal to the diffusive current of particles, the "effective" reaction rate is calculated as an explicit time-dependent function $K_{AB}(t)$ (rather than $K_{AB}(\infty)$ in the Smoluchowski theory).

Similarly to analyses of the coagulation problem in Part 1, it will be shown below that this difference is connected with inadequate application of the diffusion approach to calculation of the effective reaction rate (as the diffusive current of particles) for particles with a relatively small reaction radius, $R_{AB} << \bar{r}_A, \bar{r}_B$, that becomes especially critical in the 2-d case. Such an approach being valid in the case of small particles A diffusing into large circular traps B (so called agglomeration), $\bar{r}_A << R_{AB} << \bar{r}_B$ (with time-dependent $K(t)$ properly standing in the growth rate equation), fails in the base case, $R_{AB} << \bar{r}_A, \bar{r}_B$ (corresponding, in particular, to comparable size particles, $R_A \approx R_B << \bar{r}_A, \bar{r}_B$).

*2.2.1. Applicability of the diffusion approach to particles collisions*

The diffusion equation for an ensemble of particles is derived (similarly to consideration of other relaxation processes in weakly inhomogeneous fluids, such as the heat transfer or viscous flow) in the quasi-equilibrium approximation. In this approximation the particles distribution function is considered to be in a local thermodynamic equilibrium, smoothly varying in space and in time following smooth variations of the fluid macroscopic parameters (e.g. temperature, pressure, concentration, velocity). In the case of the mass transfer problem (i.e. the diffusion equation) the varying macroscopic parameter is the number concentration of particles, $n(\mathbf{r},t)$.

Consideration of $n(\vec{r},t)$ as a macroscopic value (i.e. when its thermodynamic fluctuations are small in comparison with its value, $\sqrt{\langle(\delta n)^2\rangle} \approx \sqrt{n} << n$) is valid only if the size of the elementary volume $\delta\widetilde{V} = L^3$, with respect to which $n(\vec{r})$ is defined, is large enough in comparison with the local inter-particle distance, $L >> n^{-1/3}(\mathbf{r})$, that in its turn must exceed the minimum inter-particle distance equal to the particles size, $n^{-1/3}(\mathbf{r}) >> 2R$. For this reason, only heterogeneities of particles spatial distribution on the length scale of $l >> L >> n^{-1/3} >> R$ can be adequately considered in the continuous diffusion approach, under an additional condition for the elementary drift distance, $a << l$ (see Part 1, Section 1.2).

In the case when identical particles (say, of type A with radius $R_A$) are distributed at random throughout a medium of infinite extent with the mean bulk concentration $n_A$ that obeys the dilution condition $n_A R_A^3 << 1$, the particles can be considered as point objects ($R_A << \bar{r}_A$, where $\bar{r}_A \approx n_A^{-1/3}$ is the mean inter-particles distance), which, in accordance with the diffusion equation for an ensemble of point-wise particles, tend to relax with time to a homogeneous spatial distribution (cf. Part 1, Section 1.3.1).

The situation critically changes in the case when a group of B-type traps with a relatively large "influence-sphere", or reaction (with A-particles) radius $R_{AB} >> R_A$ and concentration $n_B$ (obeying $n_B R_{AB}^3 << 1$) appears in the ensemble of A-particles. B-type traps cannot be treated as point objects, if $n_A R_{AB}^3 >> 1$. In this case traps should be considered as macroscopic with respect to A-particles, since the reaction radius $R_{AB}$ is much larger than the mean inter-particle distance $\bar{r}_A \approx n_A^{-1/3}$, and just for this reason additional boundary conditions for diffusion of A-particles emerges on traps surfaces. The induced by these boundary conditions heterogeneities in the spatial distribution of A-particles do not tend to disappear with time, as it was in the previous case (without traps), and the steady state concentration profiles of A-particles around macroscopic trap centres, $n_A(r) = n_A(R_{AB}) + (\bar{n}_A - n_A(R_{AB}))\cdot(1 - R_{AB}/r)$, are attained at $t >> R_{AB}^2/\pi D_A$ [2]. The diffusion flux of A-particles in this concentration profile calculated at the reaction radius, $J_{dif} = 4\pi D_A R_{AB}(\bar{n}_A - n_A(R_{AB})) \approx 4\pi D_A R_{AB} n_A$, if $n_A(R_{AB}) << \bar{n}_A \approx n_A$, determines the accumulation rate of A-particles in a B-trap, and, following consideration in [1, 2], the collision frequency function between A and B particles, taking into consideration migration of traps with the diffusivity $D_B$, eventually takes the form

$$K_{AB}^{(dif)} = 4\pi D_{AB} R_{AB}, \qquad (2.3)$$

where $D_{AB} = D_A + D_B$.

For determination of the applicability range of this result, it should be noted that the characteristic size $l$ of the zone around a large trap in which A-particles concentration varies from the value $n_A(R_{AB}) << n_A$ near the reaction surface to the value of the same order of magnitude as the mean value $n_A$ attained at large distances from the centre, is comparable with $R_{AB}$, i.e. $l \approx R_{AB}$. This value must naturally exceed the mean distance $n_A^{-1/3}(R_{AB})$ between small A-particles in the vicinity of a B-trap surface, $R_{AB} \approx l >> n_A^{-1/3}(R_{AB}) >> n_A^{-1/3}$ (in order to maintain the concentration profile of small particles around the trap), or $n_A R_{AB}^3 >> 1$. This condition logically coincides with the (above mentioned) general requirement to applicability of the diffusion approximation, $l >> n_A^{-1/3}$.

This condition can be confirmed more strictly taking into consideration that the diffusion flux at the reaction surface, $J_{dif} \propto \partial n_A/\partial r|_{r=R_{AB}} \approx \frac{n_A(R_{AB} + \Delta r) - n_A(R_{AB})}{\Delta r}$, can be properly calculated only under assumption $\Delta r << R_{AB}$. n the vicinity of the surface $n_A(R_{AB} + \Delta r) << n_A(2R_{AB}) \approx \bar{n}_A/2$ and thus the mean inter-particle distance in this zone can be evaluated as $\bar{r} \approx n_A^{-1/3}(R_{AB} + \Delta r) >> n_A^{-1/3}(2R_{AB}) = (\bar{n}_A/2)^{-1/3}$. On the other hand, it should be small enough to maintain the concentration profile in this spatial range (where the diffusion flux is calculated), $\bar{r} << \Delta r << R_{AB}$, or $n_A^{1/3} R_{AB} >> 1$.

Therefore, the traditional diffusion approach, that stipulates that the local reaction rate should be equal to the diffusive current of A-particles into the traps (see, e.g. [85, 86]), is valid only for reactions with the large reaction radius, $R_{AB} >> \bar{r}_A \approx n_A^{-1/3}$.

From this analysis it can be seen that the intrinsic reason for steady-state heterogeneities in the small particles spatial distribution is connected with the additional boundary conditions (for these particles diffusion equation) induced by macroscopic (i.e. large scale, $R_{AB} >> \bar{r}_A$) traps. These macroscopic boundary conditions vanish as soon as the reaction radius becomes comparable with the size of small particles ($R_{AB} \approx R_A << \bar{r}_A$), eliminating the driving force for emergence of steady-state spatial heterogeneities.

Indeed, in the opposite case $R_{AB} << \bar{r}_A, \bar{r}_B$, the limit of the point-wise particles restores, which is characterized by the tendency for the system of two type (A and B) particles to a homogeneous spatial distribution (or mixing) owing to their diffusion migration (in the absence of macroscopic boundaries).

*2.2.2. Diffusion mixing condition*

In fact, reactions between point-wise particles induce local heterogeneities in the particles probability density on the length scale of their mean inter-particle distance, which in the case $n_A = n_B = n$ is evaluated as $\bar{r} \approx n^{-1/3}$ (cf. Section 1.2). However, such kind of small-scale heterogeneities of the probability density disappear owing to rapid diffusion relaxation on the length scale of the mean inter-particle distance $\bar{r}$ with the characteristic time $\tau_d \approx \bar{r}^2 / 6D$ (under simplifying assumption $D_A \approx D_B = D$), that is generally much shorter in comparison with the characteristic time $\tau_c \approx (K_{AB}n)^{-1}$ of particles concentration variation, $\tau_d << \tau_c$, as will be explicitly shown below either in the 3-d or in 2-d cases (in Sections 2.3 and 2.5, respectively), and thus do not evolve in heterogeneous distribution of the particles concentration $n(\mathbf{r}, t)$ on a larger time scale. This allows consideration of a random distribution of particles attained during a time step $\tau_d << \delta t << \tau_c$, chosen for calculation of the reaction rate in Eq. (2.2).

In this case (corresponding to the kinetic regime) the spatial distributions of the particle centres $n_{A,B}(\mathbf{r}, t)$ can be considered as homogeneous functions characterized by their mean concentrations $n_{A,B}(t)$, i.e. $n_{A,B}(\vec{r}, t) = n_{A,B}(t)$, slowly varying with time owing to the particles collisions (reactions). Respectively, the collision probability is also a spatially uniform function that can be properly calculated as the frequency of collisions in the unit volume of two particles of different types (A and B) randomly located in space. Nevertheless, this does not prevent extension of the current approach to consideration of long-wavelength fluctuations, which become important at asymptotically large times (see the next Section 2.2.3).

In the opposite case, $D_A >> D_B$ or $D_B >> D_A$, the mixing of slow particles (e.g. B) might be incomplete (if $\tau_d^{(B)} \approx \bar{r}^2 / 6D_B \geq \tau_c$). However, owing to stochastic character of particles movement and collisions, "survived" particles B are still randomly distributed in space, whereas rapidly moving particles A heal up local heterogeneities in particle distribution induced by reactions ("rarefied zones" in locations of two particle reactions) and thus uphold efficient mixing of the reaction system. Therefore, assuming in further analysis $D_A \approx D_B = D$ for simplicity, the mixing condition can be generally represented in the form $\tau_d \approx \bar{r}_A^2 / 6D << \delta t << \tau_c \approx K_{AB}^{-1} n_A^{-1}$.

In a more general case $n_A > n_B$, which at large times ($t >> [K_{AB}(n_A(0) - n_B(0))]^{-1}$) unavoidably turns to $n_A(t) >> n_B(t)$, or $\bar{r}_A(t) << \bar{r}_B(t)$, local homogenization of the system (after disappearance of a two reacting particles A and B) is determined by diffusion mixing of particles A on the length scale of their mean inter-particle distance $\bar{r}_A \approx n_A^{-1/3}$ within the diffusion time $\tau_d \approx \bar{r}_A^2(t)/6D$. On the other hand, to sustain "global" (large-scale) homogeneity of the reaction system, the characteristic time of diffusion mixing of particles A on the scale of the mean distance between particles B, $\bar{r}_B(t)$, should be small in comparison with the characteristic time of the particles B concentration variation, $\tau_c^{(B)} \approx [K_{AB} n_A(t)]^{-1}$. For $K_{AB} = 4\pi D_{AB} R_{AB}$ (see Section 2.3. below), this "global mixing" condition is reduced to $\bar{r}_B(t) << \bar{r}_A(t)\sqrt{\bar{r}_A(t)/R_{AB}}$. Owing to very large value of the ratio $\bar{r}_A(t)/R_{AB}$, this condition fails only at very large times, when $n_B(t)$ becomes very small and $n_A(t)$ practically coincides with its final value $\tilde{n}_A = n_A(0) - n_B(0)$. In the opposite limit $n_A \approx n_B$ the global mixing condition logically converts into the dilution condition, $R_{AB}/\bar{r} << 1$.

Therefore, in further analysis the both cases $n_A = n_B = n$ and $n_A > n_B$ will be considered in the same kinetic approach, with the exception of some special cases for $n_A > n_B$, of very large times (corresponding to $\bar{r}_B(t) \geq \bar{r}_A(t)\sqrt{\bar{r}_A(t)/R_{AB}}$ ) and of $D_A << D_B$, where particles mixing is incomplete and thus the current approach can be applied only approximately.

*2.2.3. Applicability of the reaction rate equation*

As explained above, $K_{AB}$ is defined as the collision frequency of two point-wise particles ( $R_{AB} << \bar{r}_A, \bar{r}_B$ ) of different types (A and B) randomly located in the unit volume. This implies that the size of the unit volume $\delta\tilde{V} = L^3$, with respect to which $K_{AB}$ is defined, is large in comparison with the minimum distance between particles of different type A and B, $L >> R_{AB}$. In this case, if there are $n_A$ particles of type A and $n_B$ particles of type B distributed at random through a sample of the unit volume, the number of collisions between A- and B-particles in the unit time (that defines the reaction rate) reduces to $K_{AB} n_A n_B$.

This definition of the reaction rate can be apparently extended to the case of spatial heterogeneities in distribution of A and B particles, if these heterogeneities are smooth on the length scale of the (adequately defined) unit volume, $l >> L >> R_{AB}$. In this case the number of collisions in $dt$ between A- and B-particles located in the unit volume is calculated as $K_{AB} n_A(\mathbf{r},t) n_B(\mathbf{r},t) dt$, resulting in the local balance equations for the particles numbers

$$\dot{n}_A(\mathbf{r},t) = \dot{n}_B(\mathbf{r},t) = -K_{AB} n_A(\mathbf{r},t) n_B(\mathbf{r},t), \qquad (2.4)$$

where $K_{AB}$ is calculated in the kinetic regime, i.e. under assumption of (locally) homogeneous spatial distribution of particles. For instance, in the continuum limit in 3-d the reaction constant in Eq. (2.4) is calculated as $K_{AB} = 4\pi D_{AB} R_{AB}$ (or $K'_{AB} = 4\pi D_{AB} R_{AB} P_{AB}$, if the sticking probability $P_{AB}$ is smaller than unity, see Eqs. (2.6) and (2.6a) below).

It is important to note that formal extension of Eq. (2.4) to consideration of heterogeneities of a small scale $l \approx R_{AB}$, often performed in the traditional approach (see, e.g. [87-89]), is beyond its applicability range (where $K_{AB}$ is defined, as explained above), and for this reason the obtained in this limit equation loses its original physical sense.

Relaxation of spatial fluctuations in the particles distribution can be taken into consideration by the additional diffusion term in the r.h.s. of Eq. (2.4),

$$\dot{n}_i(\mathbf{r},t) = D_i \Delta n_i(\mathbf{r},t) - K_{AB} n_A(\mathbf{r},t) n_B(\mathbf{r},t), \qquad i = \text{A, B}, \qquad (2.5)$$

since the diffusion term is defined on the length scale of $l >> n^{-1/3} >> R_i$ (as explained above), in a self-consistent manner with the local collision rate definition, $l >> R_{AB} \geq R_i$.

Indeed, as explained in Section 1.2, in the absence of particle collisions (i.e. for point-wise particles) the particle probability density $P(\mathbf{r},t)$ can be considered as a superposition of the probability densities of independently moving particles described by the Einstein-Fokker equation, Eq. (1.5), and thus is a smooth function defined on an arbitrary small length scale, independent on the particle concentration (i.e. including $l << n^{-1/3}$). However, in the opposite case of (finite-size) particle collisions, the probability density is determined by the particle number concentration, $P(\mathbf{r},t) = n(\mathbf{r},t)$, defined on a large length scale, $l >> n^{-1/3}$, and described by the diffusion equation, Eq. (1.3).

This allows extension of the reaction rate theory applicability beyond the mean-field approximation, Eq. (2.2), however, only for fluctuations with long wavelengths, $l >> n^{-1/3}$. Available in the literature [87-89] results of analysis of Eq. (2.5) using instead of $K_{AB}$ an independent "intrinsic" (or "microscopic") rate constant $k$ (entering in the radiative boundary condition for the diffusion flux, $J_{dif}^{(A,B)} = kn_{A,B}\big|_{r=R_{AB}}$, in the traditional approach [85, 86]), demonstrate that effect of renormalization of $k$ by concentration fluctuations, resulting in the effective rate constant $K_{eff} = 4\pi D_{AB}R_{AB}k/(4\pi D_{AB}R_{AB} + k)$ (which reduces to $K_{eff} = 4\pi D_{AB}R_{AB}$ in the limit of high-rate boundary kinetics, $k \to \infty$), occurs on the length scale of the reaction radius, $l \approx R_{AB}$, that is beyond the cut-off limit of Eq. (2.5) for comparable size (or point-wise) particles, $l >> n^{-1/3} >> R_{AB}$. This additionally confirms the above derived conclusion that the results of the traditional approach are grounded only in the case of reactions with a large reaction radius, $\bar{r}_A << R_{AB} << \bar{r}_B$, when short wave-length fluctuations with $\bar{r}_A << l \leq R_{AB}$ in the spatial distribution of A-particles around B-particles can be adequately described by Eq. (2.5). However, in the opposite case, $\bar{r}_A, \bar{r}_B >> R_{AB}$, such short wave-length fluctuations are beyond the cut-off limit of the theory, and thus predictions of the diffusion approach [87-89] fails.

Therefore, the kinetic approach based on Eq. (2.4) (with the reaction constant $K_{AB}$ calculated in the kinetic approach) can be generally used as the first order approximation. In the next order approximation, taking into consideration long wavelength fluctuations, $l >> R_{AB}$, in Eq. (2.5), predictions of the kinetic approach may be violated at large times in the particular case of equal initial concentrations, $n_A(0) = n_B(0)$. In this case the asymptotic ($t \to \infty$) decay, $n_{A,B} \propto (Dt)^{-d/4}$ (where $d < 4$ is the dimension of space) [99, 100], becomes slower as compared with predictions of the mean-field theory, valid for intermediate times ($n_{A,B} \propto (4\pi RDt)^{-1}$ in $d = 3$ and $n_{A,B} \propto \ln(4Dt/R^2)/4\pi Dt$ in $d = 2$, see below).

The "crossover" time from the mean-field behaviour to the fluctuation-induced asymptotic forms can be estimated from comparison of decay laws in these two approximations as $\tilde{t} \propto R^2/D\varepsilon^2$, $\varepsilon = n_{A,B}(0)R^3 << 1$, i.e. is inversely proportional to the square of the initial volume fraction $\varepsilon$ of reactants and, thus, in diluted systems can be very large [101]. At this time the concentration becomes very small, $n_{A,B}(\tilde{t})/n_{A,B}(0) \propto \varepsilon$, i.e. the mean-field approach will correctly describe the reaction kinetics during a large time domain and only a very small amount of active particles will decay via the fluctuation-induced law. In two-dimensional systems the crossover time is smaller and the amount of particles surviving until this time is greater than in three dimensions. For this reason, the crossover from the dependences predicted by the mean-field approximation to the fluctuation-induced asymptotes have been observed in 2-d numerical simulations [100] and also experimentally [102].

Therefore, the reaction kinetics in this case can be calculated by additional consideration of long wavelength fluctuations in Eq. (2.5), e.g. by mapping to a field theory [103, 104] and using the renormalization group methods [105, 106]. However, the reaction rate constant in the master equation of the field theory can be correctly calculated only in the kinetic regime (e.g. $K_{AB} = 4\pi D_{AB} R_{AB} P_{AB}$ in the continuum mode in 3-d, or by more sophisticated expressions in other cases, see Sections 2.3-2.5), rather than taken as the microscopic (intrinsic) rate constant $k$ (cf. [103]). This might be especially important in the case of complete trapping, when the microscopic rate constant $k$ tends to $\infty$, whereas $K_{AB} = 4\pi D_{AB} R_{AB}$ calculated in the current approach is a finite value.

### 2.3. Reaction rate in 3-d case

As explained in Section 2.2.2, in order to calculate the local reaction rate in the kinetic regime, Eq. (2.4), a time step $\delta t$ relatively large in comparison with the diffusion relaxation (or mixing) time should be chosen, $\delta t >> \tau_d \approx n^{-2/3} / 6D$, in order to sustain the main condition of the kinetic regime for random (homogeneous) distribution of reacting particles (where $n = n_A \geq n_B$ and $D_A \approx D_B = D$ are assumed, cf. Section 2.2.2). On the other hand, the time step should be small in comparison with $\tau_c \approx (K_{AB} n)^{-1}$, i.e. $\delta t << \tau_c$, that allows neglecting variation of the mean concentrations $n_A$ and $n_B$ in $\delta t$. Besides, some additional condition for the time step should be valid, $\delta t >> \tilde{\tau}$, in order to attain a steady state value of $K_{AB}(\delta t) \approx const = K_{AB}$, where $\tilde{\tau}$ will be evaluated below.

Following consideration in Part 1 (Section 1.4), let us consider two particles of types A and B located at random through a sample of unit volume. The first ("parent") particle of type A can be surrounded by a sphere with the reaction radius $R_{AB}$. If the second particle centre is located in this exclusion zone, reaction would occur.

As shown in [1, 2], the relative displacements between two particles describing diffusion motions independently of each other and with the diffusion coefficients $D_A$ and $D_B$ also follow the law of diffusion motion with the diffusion coefficient $D_A + D_B$. Therefore, in order to calculate the probability of collisions between the two particles, one can equivalently consider the second particle as immobile whereas the first one migrating with the effective diffusion coefficient $D_{AB} = D_A + D_B \approx 2D$.

In this approximation it is assumed that the effective (mobile) particle jumps to an elementary distance $a_{AB}$ in random directions with a frequency $\nu_{AB} = \tau_0^{-1}$, obeying the relationship for the particle diffusivity from the theory of random walks, $D_{AB} = a_{AB}^2 / 6\tau_0$.

As a result of a jump, the exclusion zone also relocates to the distance $a_{AB}$ and opens the possibility that the second (immobile) particle with its centre located in a zone with the volume $\delta V_0 = \pi R_{AB}^2 a_{AB}$, may be swept out by the mobile particle (cf. Fig. 1.2). Depending upon the ratio between $R_{AB}$ and $a_{AB}$, particles migration can be considered in the continuum mode if $R_{AB} >> a_{AB}$, or in the free molecular mode if $R_{AB} << a_{AB}$, with different results for the collision rate.

*2.3.1. Continuum mode*

In the continuum mode ($a_A, a_B << R_{AB} << \bar{r}_A, \bar{r}_B$) during the time step $\delta t >> \tau_0$ the mobile particle makes many jumps, $k = \delta t / \tau_0 >> 1$, in random directions, however, the total swept zone volume $\delta V$, that determines the probability of the two particles collision in $\delta t$, will be smaller than $k\,\delta V_0 = \delta V_0 \delta t / \tau_0$, owing to strong overlapping of the swept zone segments at $a_{AB} << R_{AB}$. This limit corresponds to the continuum mode of the kinetic regime, characterized by a random spatial distribution of particles (quickly reinstated during the time step). Under this basic condition, the probability to sweep a B-particle in the unit time is reduced to $(\delta V/\delta t) n_B$, if there are $n_B$ B-particles randomly distributed per unit volume. Therefore, the number of collisions $(\delta V/\delta t) n_A n_B$ between A and B particles in the unit time, if there are $n_A$ A-particles randomly distributed per unit volume, will be smaller than $\delta V_0 n_A n_B / \tau_0$.

In order to calculate the volume $\delta V$ swept in $\delta t$, let us uniformly (at random) fill up the space with auxiliary point immobile particles ("markers") of radius $R_* \to 0$ with a relatively high concentration, $n_* >> R_{AB}^{-3}$ (following consideration in Part 1, Section 1.4.1). To facilitate adequate resolution of a fine structure (with the characteristic length of $a_{AB} << R_{AB}$) of the swept zone, the markers concentration $n_*$ should additionally obey the condition that the number of swept markers $\delta N_*^{(0)}$ during one jump must be large, $\delta N_*^{(0)} = \pi R_{AB}^2 a_{AB} n_* >> 1$, or $n_* >> \left(\pi R_{AB}^2 a_{AB}\right)^{-1}$. In this case the swept volume can be calculated as the total number $\delta N_*$ of the swept markers divided by their concentration, $\delta V = \delta N_* / n_*$.

In its turn, for the same reasons (concerning relative displacements of diffusing particles), calculation of the sweeping rate of randomly distributed immobile markers by a large particle of radius $R_{AB}$ migrating with the diffusivity $D_{AB}$ is equivalent to calculation of the condensation rate of the mobile markers migrating with the diffusivity $D_{AB}$ in the immobile trap of radius $R_{AB}$ (see Appendix A).

Owing to $n_* R_{AB}^3 >> 1$, this problem of the (point-wise) markers condensation in the large (macroscopic) trap can be adequately solved in the continuum approach of [1, 2], as above explained in Section 2.2.1. In this approach the total number of swept markers in $\delta t$ is equal to $\delta N_* = 4\pi D_{AB} R_{AB} n_* \delta t \left(1 + 4R_{AB} / \sqrt{\delta t \pi D_{AB}}\right)$ [4], and the swept volume per unit time is equal to $\delta V/\delta t = n_*^{-1}(\delta N_*/\delta t) = 4\pi D_{AB} R_{AB}$, if the time step is sufficiently large, $\delta t >> \tilde{\tau} \approx 16 R_{AB}^2 / \pi D_{AB}$.

The spatial variation of the markers concentration occurs on the length scale $l$ which is comparable with $R_{AB}$ (see Section 2.2.1), i.e. $l \approx R_{AB}$. In accordance with the additional condition of the diffusion equation applicability, $a_{AB} << l$, this result is valid only in the (considered here) case $a_{AB} << R_{AB}$. In this case, the number of collisions $(\delta V/\delta t)n_A n_B$ between A and B particles in the unit time becomes equal to $4\pi(D_A + D_B)R_{AB}n_A n_B$, that results in

$$K_{AB} = 4\pi D_{AB}R_{AB}. \tag{2.6}$$

It is straightforward to see that the first restriction on the time step, $\tau_c >> \delta t >> \tau_d \approx n^{-2/3}/6D$, can be applied if the mixing condition $\tau_c >> \tau_d$, or $n^{1/3}R_{AB} << (3/2\pi)(D/D_{AB}) \approx 3/4\pi$, is valid, that is in agreement with $n^{1/3}R_{AB} << 1$.

The second restriction $\delta t >> \tilde{\tau} \approx 16R_{AB}^2/\pi D_{AB}$, can be applied owing to $\tau_c >> \tilde{\tau}$, or $n^{1/3}R_{AB} << 1/4$, which is practically indistinguishable from the basic condition $n^{1/3}R_{AB} << 1$, within the accuracy of the characteristic times evaluation.

Therefore, the correct expression for the reaction rate, Eq. (2.6), derived in the kinetic regime (by consideration of uniform (random) spatial distribution of reacting particles) for the case of a relatively small reaction radius, $R_{AB} << \bar{r}_A, \bar{r}_B$, coincides with the traditional expression derived in the diffusion regime (by consideration of concentration profiles and diffusive currents of particles) valid in the case of a large reaction radius, $\bar{r}_A << R_{AB} << \bar{r}_B$, yet this coincidence is fortuitous and reflects the internal symmetry in the considered system in 3-d, revealed in Section 1.6.2.

*2.3.2. Finite sticking probability*

This coincidence vanishes in a more general case of sticking probability for A and B particles collisions smaller than unity, $P_{AB} \le 1$, when, for calculation of the reaction rate constant, the collision frequency $K_{AB} = 4\pi D_{AB}R_{AB}$ is multiplied by the probability $P_{AB} = \exp(-E_a/kT)$ of the reaction complex formation

$$K'_{AB} = 4\pi D_{AB}R_{AB}P_{AB}, \tag{2.7}$$

where $E_a$ is the activation energy of the reaction complex formation.

Again, this result is formally similar to predictions of the traditional approach (that is relevant only in the particular case of reactions with a large reaction radius, $\bar{r}_A << R_{AB} << \bar{r}_B$), using in the case of incomplete trapping the radiative boundary condition for the diffusion flux,

$$J_{dif}^{(A,B)} = kn_{A,B}\Big|_{r=R_{AB}}, \tag{2.8}$$

where $k$ is the "intrinsic" (or "microscopic") rate constant at the boundary that, by definition of the boundary kinetics, is independent of the bulk diffusivity $D_{AB}$ and proportional to the boundary area, $\propto R_{AB}^2$. Consequently, the reaction rate constant is calculated [85, 86] as

$$K''_{AB} = \left( \frac{1}{4\pi D_{AB} R_{AB}} + \frac{1}{k} \right)^{-1} = \frac{4\pi D_{AB} R_{AB} k}{4\pi D_{AB} R_{AB} + k}, \qquad (2.9)$$

which, however, coincides with Eq. (2.7) only under an additional assumption $k = 4\pi D_{AB} R_{AB} P_{AB} / (1 - P_{AB})$, that is inconsistent with the (above mentioned) definition of the boundary intrinsic rate constant (which must be independent of the bulk diffusivity $D_{AB}$ and proportional to the boundary area, $\propto R_{AB}^2$).

In fact, derivation of Eq. (2.9) is in direct analogy with the flux matching theory of Fuchs for the coagulation problem. Indeed, the radiative boundary kinetics approach, Eq. (2.8), directly corresponds the free molecular approximation applied to consideration of migrating particles (with the mean free path $a_{AB}$) in a narrow shell (of thickness $\Delta \approx a_{AB} << R_{AB}$) around the reaction sphere (of radius $R_{AB}$), representing flux matches at the shell surface (of radius $r_{sh} = R_{AB} + a_{AB} \approx R_{AB}$) in the harmonic mean approximation of the Fuchs theory (see Section 1.7). For this reason, the intrinsic rate constant can be calculated in the free molecular approximation as $k_0 = R_{AB}^2 \sqrt{8\pi kT \left( m_A^{-1} + m_B^{-1} \right)}$, if particles react after each collision (i.e. $P_{AB} = 1$). This rate is indeed independent of the bulk diffusivity $D_{AB}$ and is proportional to the boundary area, $\propto R_{AB}^2$, as claimed above. In the case of a finite sticking probability, $P_{AB} = \exp(-E_a/kT) \le 1$, the intrinsic rate constant takes the form $k = R_{AB}^2 \sqrt{8\pi kT \left( m_A^{-1} + m_B^{-1} \right)} P_{AB} = k_0 \exp(-E_a/kT) \le 1$, and Eq. (2.9) is reduced to

$$K''_{AB} = \frac{4\pi D_{AB} R_{AB} k_0 \exp(-E_a/kT)}{4\pi D_{AB} R_{AB} + k_0 \exp(-E_a/kT)}. \qquad (2.9a)$$

Comparison of Eq. (2.9a) with the correct (for comparable size particles) Eq. (2.7) shows that the inconsistency of the traditional approach may result in a very strong overestimation of the reaction constant, especially in the case of a large reaction activation energy, $E_a/kT >> 1$ (i.e. small sticking probability, $P_{AB} << 1$), when $k_0 \exp(-E_a/kT) << 4\pi D_{AB} R_{AB}$.

In this case $K''_{AB} \approx k_0 \exp(-E_a/kT)$ and thus for the diffusion-controlled reactions ($a_{AB} << R_{AB}$)

$$\frac{K'_{AB}}{K''_{AB}} \approx \frac{4\pi D_{AB} R_{AB}}{k_0} = \frac{4\pi D_{AB} R_{AB}}{R_{AB}^2 \sqrt{8\pi kT \left( m_A^{-1} + m_B^{-1} \right)}} = \frac{4 D_{AB}}{R_{AB} \bar{c}_{AB}} = \frac{\pi a_{AB}}{4 R_{AB}} \approx \mathrm{Kn_D} << 1, \qquad (2.10)$$

where $\bar{c}_{AB} = \left( 8kT/\pi m_{AB} \right)^{1/2} = \sqrt{\left( \bar{c}_A^2 + \bar{c}_B^2 \right)} = \sqrt{(8kT/\pi)\left( m_A^{-1} + m_B^{-1} \right)}$.

Therefore, the traditional approach can overestimate the reaction rate by several orders of magnitude.

The additional modification of the collision kernel for the cluster-cluster aggregation kinetics, Eqs. (1.61) and (1.62), considered in Section 1.7.2 (Part 1), leads to some quantitative alteration of the obtained results, but does not violate the main qualitative conclusion concerning strong overestimation of the aggregation rate by the traditional approach in the case of the small sticking probability (for comparable size clusters).

It is important to note that incorrect application of the reaction rate theory for reactions of small species (e.g. gas or vapour molecules) with large particles (e.g. aerosols) to the case of reactions between comparable size particles (e.g. between heavy vapour or aerosol particles) often results in an inadequate definition of the bimolecular diffusion-controlled reactions.

Indeed, following the traditional approach, represented by Eq. (2.9) (valid for reactions of small species, e.g. vapour or gas molecules, with large particles, e.g. aerosols), the diffusion-controlled reaction is defined as reactions that occur so quickly that the reaction rate is the rate of transport of the reactant particles through the reaction medium. Indeed, reactions where the activated complex forms easily and the products form rapidly (i.e. $4\pi D_{AB} R_{AB} << k$), are limited by diffusion control, reducing Eq. (2.9) to $K''_{AB} \approx 4\pi D_{AB} R_{AB}$. In the opposite case, $4\pi D_{AB} R_{AB} >> k$, the reaction is said to be chemical controlled, $K''_{AB} \approx k_0 \exp(-E_a/kT)$.

Being extended to bimolecular reactions between comparable size particles (as it is usually done, see, e.g. [79]), this definition becomes erroneous. Indeed, in accordance with Eq. (2.7), in the system of Brownian particles (heavy vapours or aerosols) diffusion migration of (comparable size) reactants in suspending solvent determines the pre-exponential factor $K_0 = 4\pi D_{AB} R_{AB}$ in the Arrhenius equation for the reaction rate constant, $K = K_0 \exp(-E_a/kT)$, where $E_a$ is the activation energy of the reaction complex (corresponding to the sticking probability of collisions), i.e. the reaction rate is always a product of the diffusion and chemical terms, which cannot be reduced only to one step (diffusion or chemical), as it was in the previous case.

In the case of bimolecular reactions in gas mixtures the reaction rate constant has a similar form, $K = K_0 \exp(-E_a/kT)$, however, the pre-exponential factor in this case is determined by the free-molecular collisions, $K_0 = R_{AB}^2 \sqrt{8\pi kT\left(m_A^{-1} + m_B^{-1}\right)}$. This difference from the previous case (with the pre-exponential factor $K_0 = 4\pi D_{AB} R_{AB}$) offers the correct definition of the bimolecular diffusion-controlled reactions as reactions, in which collisions of reactants are controlled by their diffusion migration in suspending solvent (rather than free-molecular collisions, as in the case of reactions between gaseous reactants) and determine the pre-exponential factor in the Arrhenius equation for the reaction rate constant.

## 2.4. Reaction rate in 2-d case

Similarly to the 3-d case, the problem of calculation of the area sweeping rate $\delta S/\delta t$ by an effective particle of radius $R_{AB}$ migrating with the diffusivity $D_{AB} = D_A + D_A \approx 2D$ (where $D_A \approx D_B$ is assumed, cf. Section 2.2.2) in a plane can be properly reduced to consideration of point markers randomly distributed in the plane with the concentration $n_* >> \left(\pi R_{AB}\right)^{-2}$, migrating with the diffusivity $D_{AB}$ into an immobile trap of radius $R_{AB}$ [97]. The markers condensation rate can be calculated using a well-known analogy with the heat-conduction problem in the cylindrical geometry [107]. As a result, the total number of swept markers in $\delta t$ is equal to $\delta N_* \approx 4\pi D_{AB} n_* \delta t / \ln\left(4 D_{AB} \delta t / R_{AB}^2\right)$, if $R_{AB}^2 / 4 D_{AB} << \delta t << \tau_c \approx K_{AB}^{-1} \cdot \min\left[n_A^{-1}, n_B^{-1}\right] \approx K_{AB}^{-1} n_A^{-1}$ (where $n_A \geq n_B$ is assumed) and $\delta t$ obeys the diffusion mixing condition, $\tau_d << \delta t$. Contrary to the 3-d case, the sweeping rate $(\delta S/\delta t) = n_*^{-1}(\delta N_*/\delta t)$ in this case is a function of the time step even for very large $\delta t$, however, this dependence is weak and can be neglected with the logarithmic accuracy.

Indeed, an expression $\ln(xX)$ can be approximated as $\ln(xX) = \ln X + \ln x \approx \ln X$ in the case $X >> x \geq 1$ (and thus $\ln X >> \ln x \geq 0$). Therefore, choosing the time step as $R_{AB}^2/4D_{AB} << \tilde{\tau} << \delta t << \tau_c$, that, under additional condition

$$\tilde{\tau}/\left(R_{AB}^2/4D_{AB}\right) >> \tau_c/\tilde{\tau}, \tag{2.11}$$

can be also represented in the form $0 < \ln(\delta t/\tilde{\tau}) << \ln(\tau_c/\tilde{\tau}) << \ln\left(4D_{AB}\tilde{\tau}/R_{AB}^2\right)$, one obtains $\ln\left(4D_{AB}\delta t/R_{AB}^2\right) = \ln\left(4D_{AB}\tilde{\tau}/R_{AB}^2\right) + \ln(\delta t/\tilde{\tau}) \approx \ln\left(4D_{AB}\tilde{\tau}/R_{AB}^2\right)$. In this approximation the sweeping rate can be calculated as $(\delta S/\delta t) = n_*^{-1}(\delta N_*/\delta t) \approx 4\pi D_{AB}/\ln\left(4D_{AB}\tilde{\tau}/R_{AB}^2\right)$. The number of collisions $(\delta S/\delta t)n_A n_B$ between A and B particles in the unit time becomes equal to $\approx 4\pi D_{AB} n_A n_B/\ln\left(4D_{AB}\tilde{\tau}/R_{AB}^2\right)$, that corresponds to $K_{AB} \approx 4\pi D_{AB}/\ln\left(4D_{AB}\tilde{\tau}/R_{AB}^2\right)$ and thus $\tau_c \approx K_{AB}^{-1} n_A^{-1} \approx \ln\left(4D_{AB}\tilde{\tau}/R_{AB}^2\right)/4\pi D_{AB} n_A$ (if $n_A \geq n_B$ is specified).

Substituting this expression for $\tau_c$ into Eq. (2.11), one obtains $\tilde{\tau} >> \left[\ln\left(4D_{AB}\tilde{\tau}/R_{AB}^2\right)\right]^{1/2} R_{AB}\bar{r}_A/4D_{AB}$, where $\bar{r}_A \approx (\pi n_A)^{-1/2}$; this allows specification $\tilde{\tau} \approx \bar{r}_A^2/4D_{AB}$ (owing to $\left(\bar{r}_A/R_{AB}\right)^2 >> \ln\left[\left(\bar{r}_A/R_{AB}\right)^2\right]$, if $\bar{r}_A/R_{AB} >> 1$ ), that apparently obeys the necessary condition $R_{AB}^2/4D_{AB} << \tilde{\tau} << \tau_c$. Eventually one obtains for the reaction rate in the mean-field approximation

$$K_{AB} \approx 4\pi D_{AB}/\ln\left(\bar{r}_A^2/R_{AB}^2\right), \tag{2.12}$$

that depends on time implicitly (via $\bar{r}_A \approx (\pi n_A(t))^{-1/2}$), rather than explicitly, as obtained in the traditional approach.

In the particular case $n_A = n_B = n$ (or $\bar{r}_A = \bar{r}_B = \bar{r}$), $\tilde{\tau}$ practically coincides with $\tau_d \approx \bar{r}^2/4D_{AB}$, thus, $\delta t$ self-consistently obeys the necessary condition $\tau_d \approx \tilde{\tau} << \delta t$. In this case the reaction rate is reduced to $K_{AB} \approx 4\pi D_{AB}/\ln\left(\bar{r}^2/R_{AB}^2\right) \approx -4\pi D_{AB}/\ln\left(nR_{AB}^2\right)$ (rather than $K_{AB} = 4\pi D_{AB}/\ln\left(4D_{AB}t/R_{AB}^2\right)$ in the traditional approach) and eventually results in solution of the reaction rate equation,

$$\frac{1 + \ln\left(nR_{AB}^2\right)}{n} \approx 4\pi D_{AB} t, \tag{2.13}$$

that at large times, $t >> R_{AB}^2/4\pi D_{AB}$ (before crossover to the asymptotic behaviour at $t \to \infty$, discussed in Section 2.2.3), is close to the traditional solution, $n \approx \ln\left(4D_{AB}t/R_{AB}^2\right)/4\pi D_{AB}t$.

However, in the case $n_A > n_B$ situation critically changes. In this case the initial relationship $n_A(0) > n_B(0)$ at large times turns to $n_A(t) >> n_B(t)$, or $\bar{r}_A(t) << \bar{r}_B(t)$, and the solution of the reaction rate equation (at $t >> [K_{AB}(n_A(0) - n_B(0))]^{-1}$ ) results in the exponential drop of the concentration,

$$n_B(t) \propto \exp(-Ct), \qquad (2.14)$$

where $C \approx 2\pi D_{AB}(n_A(0) - n_B(0))/\ln(\tilde{r}_A / R_{AB})$, and $\tilde{r}_A$ is the final value of $\bar{r}_A(t)$, which variation $\Delta\bar{r}_A(t) = \tilde{r}_A - \bar{r}_A(t)$ at large times, when it approaches to $\tilde{r}_A$, i.e. $\Delta\bar{r}_A(t) << \tilde{r}_A$, is neglected in the expression for $C$ in Eq. (2.14) with the applied logarithmic accuracy, $\ln(\bar{r}_A) = \ln(\tilde{r}_A + \Delta\bar{r}_A) \approx \ln(\tilde{r}_A) + \ln(1 + \Delta\bar{r}_A/\tilde{r}_A) \approx \ln(\tilde{r}_A)$. The obtained solution, Eq. (2.14), is much steeper in the current approach in comparison with that in the traditional approach $n_B(t) \propto \exp(-C_1 t/\ln t)$ ([108-110]), and thus the concentration decay rate $\dot{n}_B$ is strongly underestimated at large times in the traditional approach.

This additionally confirms the importance of the new approach to calculation of the reaction rate in 2-d.

### 2.5. Reactions on discrete lattice

The discrete analogue of the Wiener sausage (mentioned in Part 1) was related to the probability of survival of a Brownian particle by random immobile traps in the Rosenstock approximation [111] or other allied problems, e.g., so called "target annihilation by scavengers" [112]. In the latter problem a single particle A (target) and $N_B = n_B N$ particles B (scavengers) of a finite concentration $n_B$ are randomly located on $N \to \infty$ sites of a 3-dimensional regular lattice. Particle A is immobile, whereas particles B perform independent, homogeneous discrete-time random walks on the lattice sites (including sites occupied by other particles); particle A annihilates as soon as a particle B reaches it.

In fact, the kinetic approach (based on the diffusion mixing condition) allows extending the solution of the target annihilation problem to consideration of many-body effects in the diffusion-limited reaction kinetics. Indeed, since particles B moves independently from each other, the probability of the target annihilation between time $t$ and $t + \delta t$ reduces to the probability of a two-particle (A-B) collision, $w_{AB}(t)\delta t$, multiplied by $N_B$. In the case of mobile particles A of a finite concentration $n_A(0)$, the problem also reduces to consideration of two-particle collisions, if rapid diffusion mixing of particles occurs in-between their mutual collisions. Actually, after each annihilation event (at a moment $t$) when the certain lattice position (where the collision occurred) becomes definitely unoccupied, the random (equiprobable) spatial distribution of particles on lattice sites rapidly reinstates during the mixing time $\tau_d << \delta t$, and a similar to the initial configuration (i.e. random location of particles A and B on lattice sites), but with the new (diminished) particle concentrations, $n_{A,B}(t + \delta t) = n_{A,B}(t) - n_A n_B w_{AB}(t)\delta t$, can be considered in the subsequent time step, if $\delta t << \tau_c$. In the case $\delta t >> \tilde{\tau} \approx 16R_{AB}^2/\pi D_{AB}$, which is generally valid owing to $\tau_c >> \tilde{\tau}$ [7], a steady state value of $w_{AB}(t) \approx w_{AB}(\infty) \equiv w_{AB}$ is attained in the time step $\delta t$, and thus the reaction rate equation takes the form of Eq. (1) with the rate constant $K_{AB} = w_{AB}$, which does not depend on time explicitly (as opposed to condensation of small particles in a large trap, considered in the diffusion approach).

It is important to note that, in contrast to the target annihilation problem where sites can be occupied by several particles, two (or more) point defects (of the same type) cannot occupy the same site. However, under the basic assumption of the reaction rate theory, $n_A, n_B << 1$, "collisions" A-A and B-B (i.e. occupancy of one site by two identical particles) can be generally neglected in calculation of the A-B reaction rate. Indeed, incorporation of these events during the time step $\delta t << \tau_c$, which is used in the derivation of the rate equation and calculation of the reaction constant $K_{AB}$ in the kinetic approach, requires consideration of two simultaneous or successive collisions (A-A and A-B, or B-B and B-A) in unit volume during $\delta t$, with the probabilities $w_{AAB}\delta t \propto n_A^2 n_B \delta t$ and $w_{ABB}\delta t \propto n_B^2 n_A \delta t$, respectively, which can be neglected, owing to $n_A, n_B << 1$, in comparison with the probability of a single pair-wise A-B collision during $\delta t$ in unit volume, $w_{AB}\delta t \propto n_A n_B \delta t$. Therefore, the influence of the forbiddance for identical defects to occupy the same sites can be neglected in calculation of the recombination rate.

Particles migrations by random walks on discrete cubic lattice sites can be considered in two limits, $R_{AB} >> a$ and $R_{AB} < a$. In the case of large reaction radius, $R_{AB} >> a$, the problem is properly reduced to the continuum media limit considered in Section 2.3.1. In the opposite case, the reaction radius $R_{AB}$ is assumed to be small in comparison with the lattice spacing (corresponding to the elementary jump distance, $a = a_A = a_B$), and reactions occur when two particles occupy the same site (see, e.g. [98]). In this case $R_{AB}$ is the minimum length scale of the problem and can be excluded from consideration. This situation is qualitatively different from the above considered (in Section 2.3.2) free molecular regime (for reaction particles suspended in a fluid), in which $R_{AB}$ was also small ($R_{AB} << a_A, a_B$), but non-negligible parameter ($R_{AB} >> \bar{r}_m \approx n_m^{-1/3}$, where $n_m$ is the fluid molecules concentration) that allowed calculating the swept volume for migrating particles.

We start at $t = 0$ with randomly distributed A and B particles on discrete cubic lattice sites, with mean concentrations $n_A$ and $n_B$, respectively; $n_{A,B}a^3 << 1$. Each particle moves by jumps to nearest-neighbouring sites (including sites occupied by other particles of the same type) with the jump frequency $\tau_A^{-1}$ and $\tau_B^{-1}$, respectively; thus all particles perform independent random walks, with the associated diffusion coefficients $D_{A,B} = a^2 / 6\tau_{A,B}$. A reaction between two different species A and B (A + B $\rightarrow$ C, where C does not affect the reaction) occurs as soon as any of the A particles appears on the same lattice site simultaneously with any of the B particles. Again, $n_A \approx n_B = n$ and $D_A \approx D_B = D$, will be considered in further analysis (cf. Section 2.2.2).

Similarly to the above considered continuum limit, reactions between A and B particles induce local heterogeneities in the spatial distribution of these particles probability densities on the length scale of the mean inter-particle distance $\bar{r}_A \approx n^{-1/3} >> a$. However, such kind of heterogeneities quickly disappear owing to rapid diffusion relaxation on the length scale of the mean inter-particle distance $\bar{r}_A$ with the characteristic time $\tau_d \approx \bar{r}_A^2 / 6D$, that is generally much shorter in comparison with the characteristic time $\tau_c \approx (K_{AB} n)^{-1}$ of particles concentration variation, $\tau_d << \tau_c$ (or $n^{1/3} << 6D/K_{AB}$ ). Choosing a time step $\tau_d << \delta t << \tau_c$ for calculation of the reaction rate, this allows considering a random distribution of particles attained in $\delta t$ (owing to $\tau_d << \delta t$) and neglecting variation of the mean concentrations $n_A$ and $n_B$ in $\delta t$ (owing to $\delta t << \tau_c = \min\left[\tau_c^{(A)}, \tau_c^{(B)}\right]$). Besides, some additional condition for the time step should be valid, $\delta t >> \tilde{\tau}$ , in order to attain a steady state value of $K_{AB}(\delta t) \approx const = K_{AB}$ within the time step, cf. Section 2.3.1.

In this case (corresponding to the kinetic regime) the spatial distributions of the particle concentrations $n_{A,B}(\vec{r},t)$ can be considered as homogeneous functions characterized by their mean concentrations $n_{A,B}(t)$, i.e. $n_{A,B}(\vec{r},t) = n_{A,B}(t)$, slowly varying with time owing to particles collisions (reactions). Respectively, the collision probability is also a spatially uniform function that can be properly calculated as the collision frequency of two particles of different types (A and B) randomly located in the unit volume and migrating with diffusivities $D_A$ and $D_B$, similarly to the continuum limit consideration.

In its turn, this problem can be readily reduced to calculation of the collision probability between two particles, randomly located in the unit volume, one of which is immobile (say, particle B) and another (particle A) is mobile, migrating with the effective diffusivity $D_{AB} = D_A + D_B$ .

This approach can be further extended to consideration of spatial heterogeneities in the ensemble of A and B particles, if these heterogeneities are smooth on the length scale of the diffusion equation applicability, $l >> n_{A,B}^{-1/3} >> R_{AB}$ , with the reaction constant calculated in the kinetic regime, similarly to the continuum medium consideration.

However, there is also an important difference with the continuum limit. Indeed, in the continuum limit the probability of the two particles collision in $\delta t$ was calculated as the mean volume swept by the mobile particle (of radius $R_{AB}$ and diffusivity $D_{AB}$). Instead of this, in the discrete lattice limit the collision probability in $\delta t_k$ is determined by the mean number of distinct sites visited by a *k*-step random walk of the mobile particle (so called the range of the random walk, $S_k$), where $k = \delta t_k / \tau_{AB} = \delta t_k 6D_{AB} / a^2 >> 1$ .

*2.5.1. Reaction rate on 3-d discrete lattice*

In the case of a simple 3-d cubic lattice the mean value of $S_k$ can be calculated as [113-116]

$$\overline{S}_k \approx 0.659 \cdot \left[k + 0.729 k^{1/2} + O(1)\right], \qquad (2.15)$$

which for the chosen time step $a^2 / 6D_{AB} << \overline{r}^2 / 6D_{AB} \approx \tau_d << \delta t_k << \tau_c$ , that corresponds to $k >> 1$, can be reduced to

$$\overline{S}_k \approx 0.659 k , \qquad (2.15a)$$

and results in

$$K_{AB} = a^3 \overline{S}_k / \delta t_k \approx 3.96 D_{AB} a \,. \qquad (2.16)$$

For the b.c.c. and f.c.c. lattices the coefficient in Eq. (2.15) is equal approximately to 0.718 and 0.744, respectively [113-115], resulting in

$$K_{AB} = a^3 \overline{S}_k / \delta t_k \approx 4.31 D_{AB} a \qquad (2.16a)$$

for the b.c.c. lattice, and

$$K_{AB} = a^3 \overline{S}_k / \delta t_k \approx 4.47 D_{AB} a \text{ (for f.c.c.)} \qquad (2.16b)$$

for the f.c.c. lattice.

In the case of incomplete sticking of reactant particles, $P_{AB} \le 1$, the reaction constant reduces to

$$K'_{AB} = K_{AB} P_{AB} \approx 3.96 D_{AB} a P_{AB} \,. \qquad (2.17)$$

A formally similar to Eq. (2.16) result was obtained in [91] (following [98]). In that approach the problem was also reduced to consideration of collisions between two particles A and B on discrete lattice sites, however, basing on additional (unjustified) assumptions. Namely, instead of consideration of rapid diffusion mixing of particles (as proposed in the current approach) that allows rigorous reduction of the multi-particle problem to consideration of two-particle collisions and direct calculation of the reaction rate constant, an additional setup (or *Ansatz*) for the reaction rate constant in the multi-particle system was applied in [91] that eventually resulted in a different (apparently erroneous) numerical factor in Eq. (2.16).

Therefore, one can conclude that the currently developed approach can be generalized to consideration of the reaction kinetics on a 3-dimensional lattice, resulting in the new relationship for the reaction rate constant, Eq. (2.16).

This consideration can be naturally reduced to the particular case of reactions between identical particles, A + A $\to$ C, taking place when two A particles appear on the same lattice site simultaneously, by substitution $D_{AB} \to D_{AA} = 2D_A$ in Eqs. (2.16) and (2.17).

*2.5.2. Reaction rate on 2-d discrete lattice*

Reaction rate for particles A and B migrating by random walks on discrete square lattice sites ($n_A \approx n_B = n << a^{-2}$), when the reaction radius is small in comparison with the lattice spacing, $R_{AB} < a$, can be calculated in a similar to 3-d approach (presented in Section 2.5) using the logarithmic approximation (presented in Section 2.4).

As a result, an equation (corresponding to Eq. (2.16) in 3-d case) for the reaction rate constant takes the form

$$K_{AB} = a^2 \overline{S}_k / \delta t_k \,, \qquad (2.18)$$

where $\bar{S}_k = \pi k/\log k + O(k/\log^2 k)$ is the mean number of distinct square lattice sites visited by a $k$-step random walk [116], $k = \delta t_k/\tau_{AB} = \delta t_k 4D_{AB}/a^2$ and, to provide diffusion mixing in $\delta t_k$, the calculation time step is chosen as $a^2/4D_{AB} << n^{-2}/4D_{AB} << \delta t_k << \tau_c \approx K_{AB}^{-1}n^{-1}$. With the logarithmic accuracy one obtains

$$K_{AB} \approx 4\pi D_{AB}/\log(1/na^2), \qquad (2.19)$$

which depends on time implicitly (via $n(t)$).

In the case $n_A > n_B$ at large times this time dependence is weak and can be neglected with the applied logarithmic accuracy

$$K_{AB} \approx 4\pi D_{AB}/\log(1/\tilde{n}a^2), \qquad (2.19a)$$

where $\tilde{n}$ is the final value of $n(t)$, which variation is small, $\Delta n(t) = n(t) - \tilde{n} << \tilde{n}$, and thus, $\log(n(t)a^2)^{-1} = \log((\tilde{n}+\Delta n)a^2)^{-1} \approx \log(\tilde{n}a^2)^{-1} - \log(1+\Delta n/\tilde{n}) \approx \log(\tilde{n}a^2)^{-1}$, since $\log(\tilde{n}a^2)^{-1} >> 1 >> \log(1+\Delta n/\tilde{n})$.

Similarly to the continuum limit in 2-d (considered in Section 2.4), the reaction rate constant differs from that calculated in the traditional approach (with $\log(D_{AB}t/a^2)$ instead of $\log(1/\tilde{n}a^2)$ in the denominator of Eq. (2.19a)) and thus predicts much higher decay rate $\dot{n}_{A,B}$ at large times in comparison with the traditional approach [91, 98].

*2.5.3. Catalytically-activated reactions*

The new approach can be extended to consideration of kinetics of bimolecular, catalytically-activated reactions in 2 or 3 dimensions. The elementary reaction act between reactants takes place only when these meet on a catalytic site (CS); such sites are assumed to be immobile and randomly distributed in space with the mean concentration $n_C$.

We start at $t = 0$ with randomly distributed reactant particles A and B with the mean concentrations $n_A$ and $n_B$, respectively. Each A (B) particle migrates by jumps to nearest-neighbouring sites with the associated diffusivity $D_A$ ($D_B$). Whenever an A particle lands on a catalytic site which is already occupied by a particle B, the two particles may react with a sticking probability $P_{AB} \le 1$. Reacting particles are immediately removed from the system, whereas the corresponding CS remains unaffected. On the other hand, particles never react at non-catalytic sites.

In this case the effective reaction constant reduces to

$$K_{ABC} = K'_{AB}n_C = K_{AB}P_{AB}n_C, \qquad (2.20)$$

where $K'_{AB}$ is derived in Eq. (2.16a) and $n_C$ stands for the probability that a collision occurs on CS (owing to random distribution of CSs in space).

This expression is obtained in the kinetic approach and essentially differs from one obtained in [117] in the traditional diffusion approach, erroneously taking into consideration (following [87-89]) short wavelength fluctuations (on the length scale of the reaction radius, $l \approx R_{AB}$), which are beyond the cut-off limit of the theory (Eq. (2.5)), $l >> n^{-1/3} >> R_{AB}$ (cf. Section 2.3).

### 2.6. Point defect recombination in crystals

The new kinetic approach, based on the "diffusion mixing" condition, to consideration of the reaction kinetics for particles A and B migrating by random walks on discrete lattice sites (with the lattice spacing $a$), and reacting when two particles occupy the same site, i.e. $R_{AB} < a$, was extended in the author's paper [118] to the transition regime, corresponding to $R_{AB}/a \geq 1$, and applied to consideration of the recombination rate of point defects in cubic lattices. In this approach the reaction rate can be reduced to calculation of the mean number of distinct sites visited by the effective particle of radius $R_{AB}$, which is calculated using the numerical algorithm, developed by generalization and further improvement of the original algorithm elaborated by the authors for the continuum limit calculations (Section 1.4.5). The numerical calculations should correctly reproduce the analytical expressions in the two limits, $R_{AB} < a$ and $R_{AB} >> a$, and represent a curve in the intermediate range of the parameter $R_{AB}/a \approx 1-10$, which generally corresponds to Frenkel pair recombination radius in many practical case, e.g. for fcc metals.

The Smoluchowski reaction rate constant from Eq. (2.3), $K_{AB} = 4\pi D_{AB} R_{AB}$, is widely used in the literature in the whole range of the ratio $R_{AB}/a$, including consideration of point defects and impurities in crystals. Correspondingly, the reaction (or recombination) rate constant might be notably overestimated in the limit of point-wise collisions, $R_{AB} \leq a$ (where $K_{AB} \approx 3.96 D_{AB} a$ from Eq. (2.16)), as well in the transition range, $R_{AB}/a \sim 1$.

Indeed, one should expect that in the transition range the reaction rate constant should vary between two limits, Eq. (2.16), if $R_{AB}/a < 1$, and Eq. (2.3), if $R_{AB}/a >> 1$. Calculation of the reaction rate in this transition range can be searched numerically, modifying numerical algorithms developed in the continuum limit [7, 10] (see Part 1). The solution of this problem might be important in many practical cases.

For instance, Frenkel pair recombination radius $R_{iv}$ in fcc copper was evaluated from analysis of the resistivity damage rates, smoothly varying from $4a$ to $3a$ (where $a \approx 0.36$ nm) in the temperature range from 50 to 110 K [119]. Extrapolation to room temperatures of the obtained in [17] correlation for the temperature dependence of the recombination radius yields $R_{iv} \approx 2a$. Similar results were obtained also for other fcc metals [120]. These values of the recombination radius correspond to the transition range, and thus the new approach to calculation of the recombination rate (rather than the traditional expression, Eq. (2.3)) might be important.

In fcc lattices an additional problem of the point defect site positions arise; in particular, interstitial atoms often reside in tetragonal positions with the highest symmetry, which form a simple cubic (sc) lattice for interstitial migrations. For simplicity, only such sc lattices will be currently considered. This consideration can be readily generalized, if defect sites are specified more definitely (e.g. by atomistic calculations).

For the numerical evaluation of the mean number of distinct sites visited by a *k*-step random walk of the effective particle, $\overline{S}_k$, a random migration of a particle of the radius $R_{AB}$ with the fixed jump distance $a$ on the sc lattice and jump frequency $\nu_{AB} = \tau_{AB}^{-1} = 6D_{AB}/a^2$ is numerically generated. The randomly generated data describe the subsequent positions of the particle centre trajectory, which can be further used for calculation of $\overline{S}_k$. Each lattice site visited by the particle is counted only once.

The number of visited distinct sites for each trajectory was calculated using an accelerated numerical algorithm, developed by generalization and further improvement (with respect to the run time, which steeply increases with the trajectory length) of the original algorithm for the continuum limit calculations [7, 10].

For each number of jumps $k$, up to 100-150 random trajectories were generated, which allowed calculating a smooth distribution of the probability density $f(x)$ for $x = S_k/k$, Fig. 2.1. The number of jumps $k = \delta t/\tau_{AB}$ was increased until $\overline{S}_k$ (averaged over the trajectories) attained a steady-state value, which in accordance with the above presented consideration has to converge to the analytically calculated values in the two limits, $R_{AB}/a < 1$ (Eqs. (2)) and $R_{AB}/a >> 1$ (Eq. (4)).

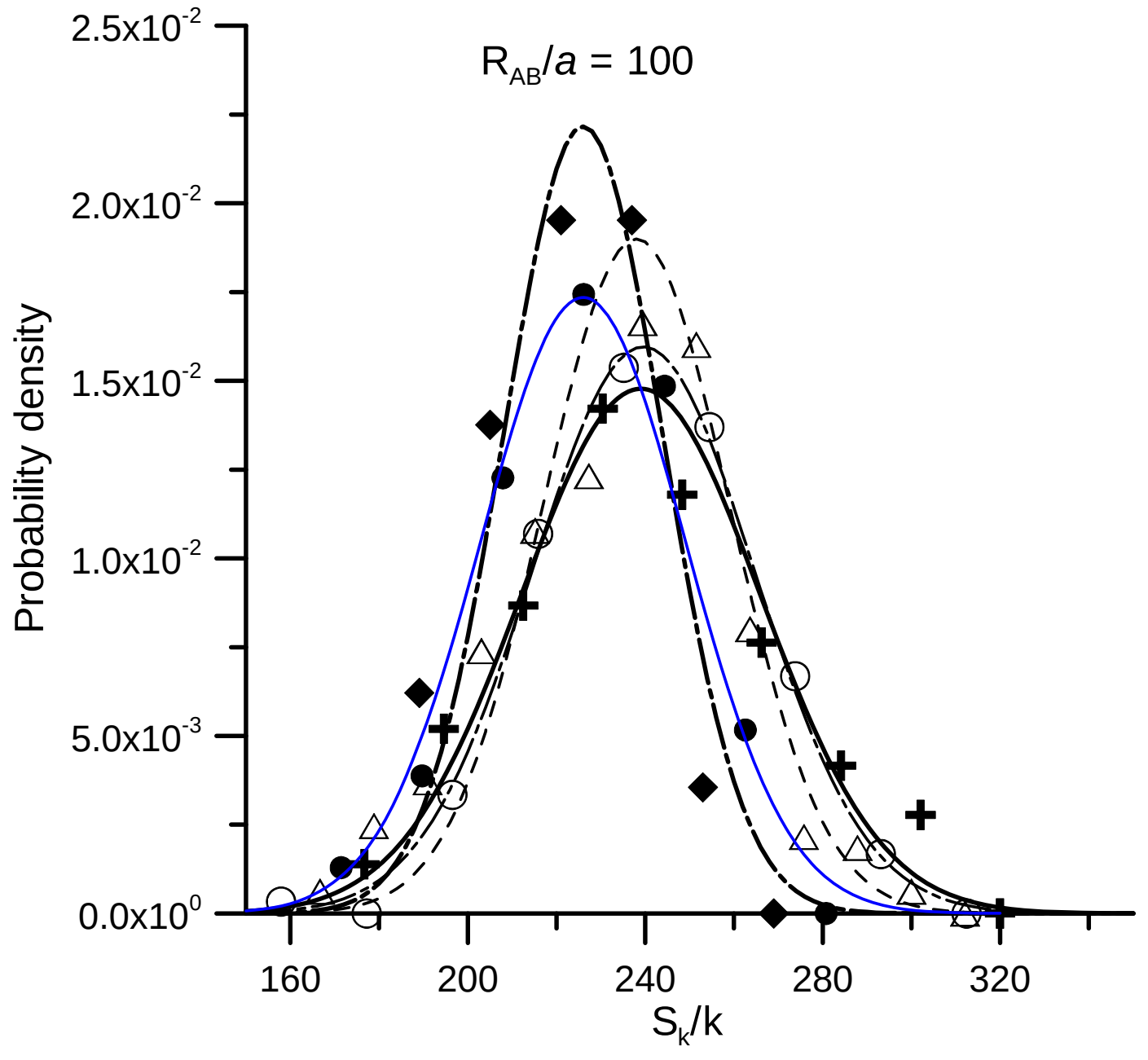

Fig. 2.1. Calculation of the probability density $f(x)$ for $x = S_k/k$ at $R_{AB}/a = 100$; for each number of jumps $k$, 150-200 trajectories are generated resulting in calculation points, which are grouped in intervals of equal width $L$ (≈ 10% of the whole distribution width) and form normal distributions around the mean values (at given $k$). The invariable mean value of $\overline{S_k}/k \approx 226$ is attained at $k \geq 10^7$. Calculation points with the number of elementary jumps:

✚ – $3{\cdot}10^6$; ○ – $4{\cdot}10^6$; △ – $5{\cdot}10^6$; ● – $10^7$; ◆ – $2{\cdot}10^7$.

Similarly to the continuum limit, numerical calculations confirmed that the steady-state value of the mean number of visited distinct sites per unit time depends only on the ratio $R_{AB}/a$ (rather than on $R_{AB}$ and $a$ separately). Besides, the general for the 3-d systems conclusion that the calculated value of $\overline{S_k}/k$ smoothly diminishes and reaches the steady-state limit, invariable with further increase of the number of jumps $k$, is justified. This conclusion is illustrated also in Figs. 2.2 and 2.3, where results of calculations are presented for the two cases $R_{AB}/a<1$ and $R_{AB}/a=10$, respectively.

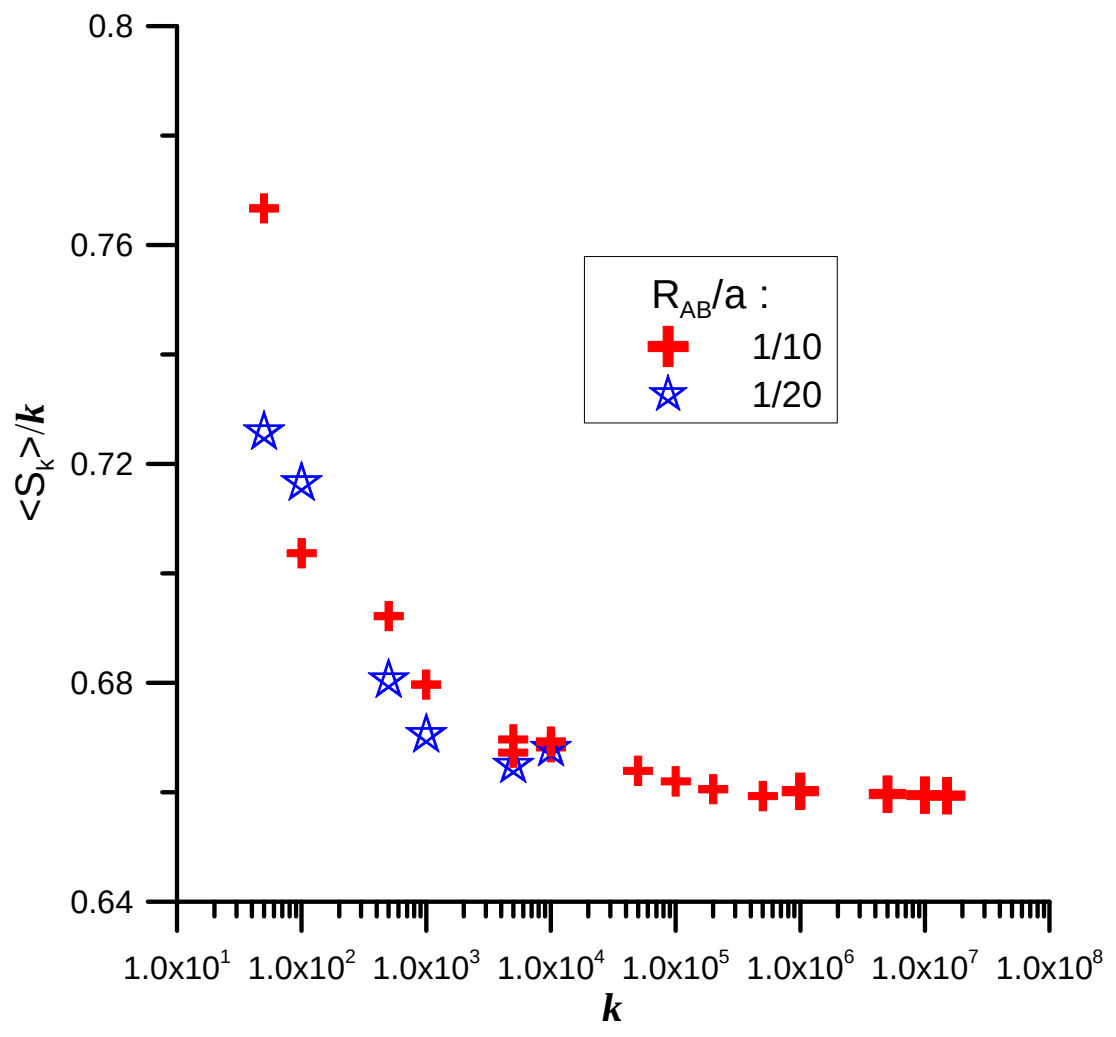

Fig. 2.2. Dependence of the mean number of visited distinct sites per one jump, $\overline{S_k}/k$, on the number of jumps $k$ in the discrete limit, $R_{AB}/a<1$.

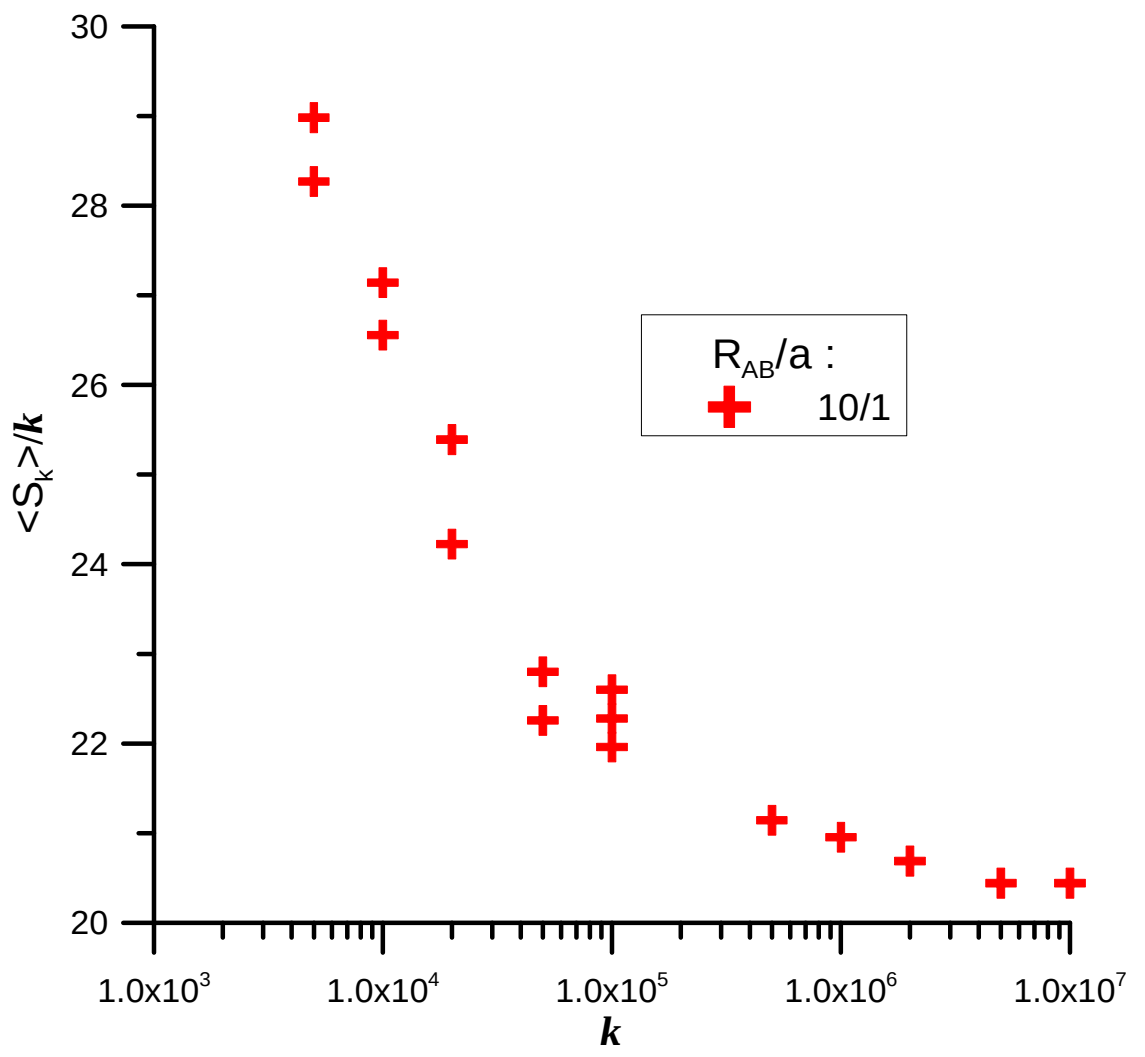

Fig. 2.3. Dependence of the mean number of visited distinct sites per one jump, $\overline{S_k}/k$, on the number of jumps $k$ in the case of $R_{AB}/a=10$.

In particular, in Fig. 2.2 it is seen that the steady-state value, $\overline{S_k}/k \approx 0.66$, numerically calculated for the point-wise collisions ($R_{AB}/a<1$), is in excellent agreement with the analytical prediction of [113, 114], $\overline{S_k}/k=0.659$ (cf. Eq. (2.15a)), which thus should be used for calculation of the reaction rate constant $K_{AB}\approx 3.96 D_{AB}a$ in this limit.

The steady-state values of $\overline{S_k}/k$ calculated in a wide range of the parameter values, $1\le R_{AB}/a\le 100$, are shown in Fig. 4, with more detailed representation of the calculation results in the transition range, $1\le R_{AB}/a\le 4$, in Fig. 2.5.

From Fig. 2.4 it is seen that at very large $R_{AB}/a>>1$ the numerical results fairly converge to the analytical value $\overline{S}_k=2\pi k/3$, or $\overline{S_k}/\delta t_k=2\pi R_{AB}/3a\approx 2.094 R_{AB}/a$, calculated in the continuum limit and corresponding to the Smoluchowski expression for the reaction rate constant, $K_{AB} = 4\pi D_{AB} R_{AB}$. In fact the calculated curve is a stepped rather than a straight line, as seen in Fig. 5; however, the ratio of a step height to the actual value of $\overline{S_k}/k$ smoothly decreases with the growth of the reaction radius $R_{AB}$, and for this reason the calculated curve in Fig. 2.4 looks like a straight line at large $R_{AB}/a >> 1$. Nevertheless, the fine structure of the calculated curve with a step at each value $R_{AB}/a = \sqrt{h^2 + i^2 + j^2}$ (where $h, i, j = 0, 1, 2...$, run through the entire row of integers), becomes important for relatively small reaction radii, corresponding to the transition range, $1 \le R_{AB}/a < 10$, Fig. 2.5. The maximum deviation factor (i.e. the ratio of the real value to that calculated in the continuum limit) of $\approx 3$ is attained at $R_{AB}/a \le 1$ and gradually decreases and tends to 1 at larger $R_{AB}/a$.

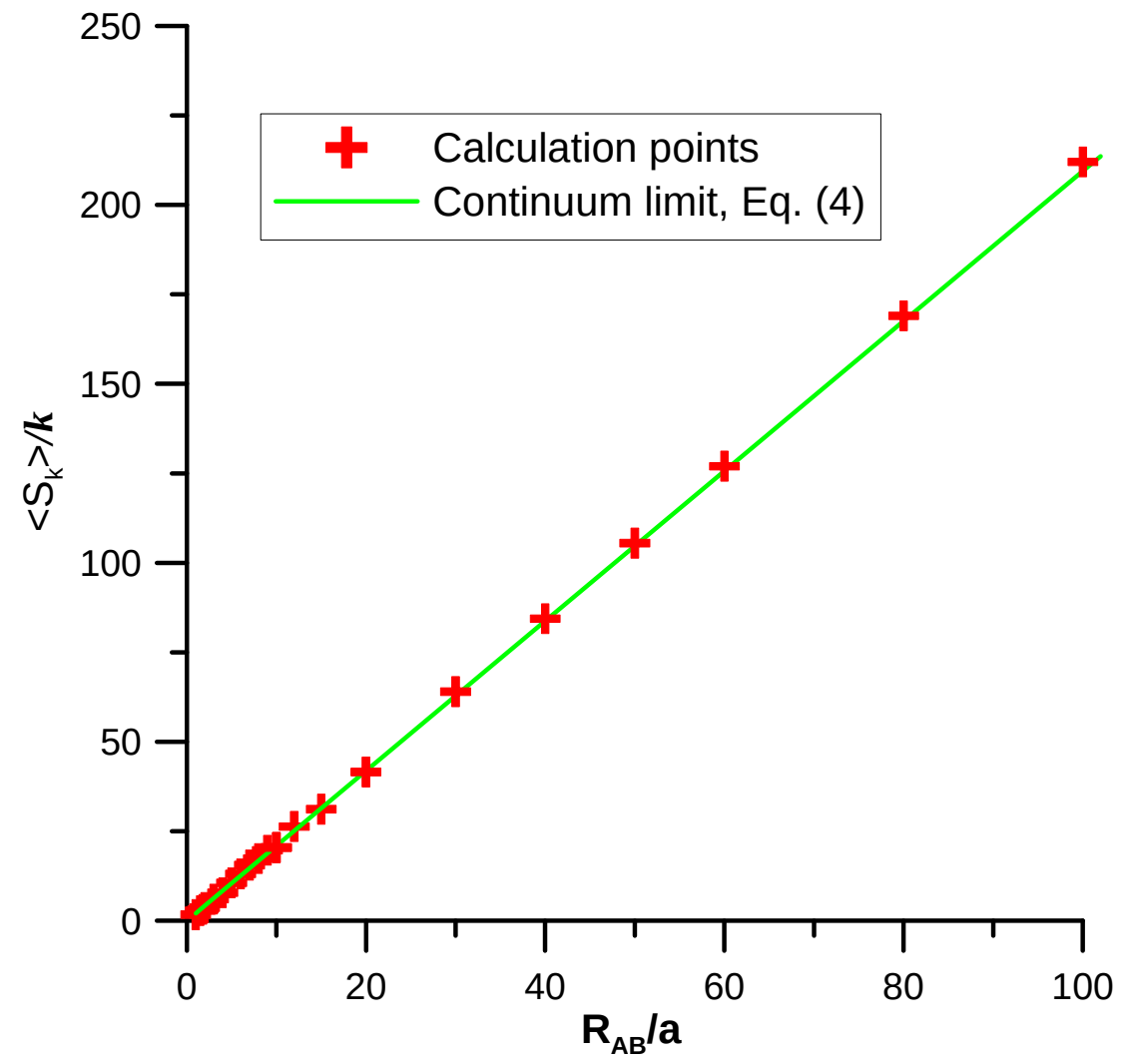

Fig. 2.4. Dependence of the mean number of visited distinct sites per one jump, $\overline{S_k}/k$, on the recombination radius $R_{AB}/a$ in comparison with the theoretical curve calculated in the continuum limit, Eq. (2.3).

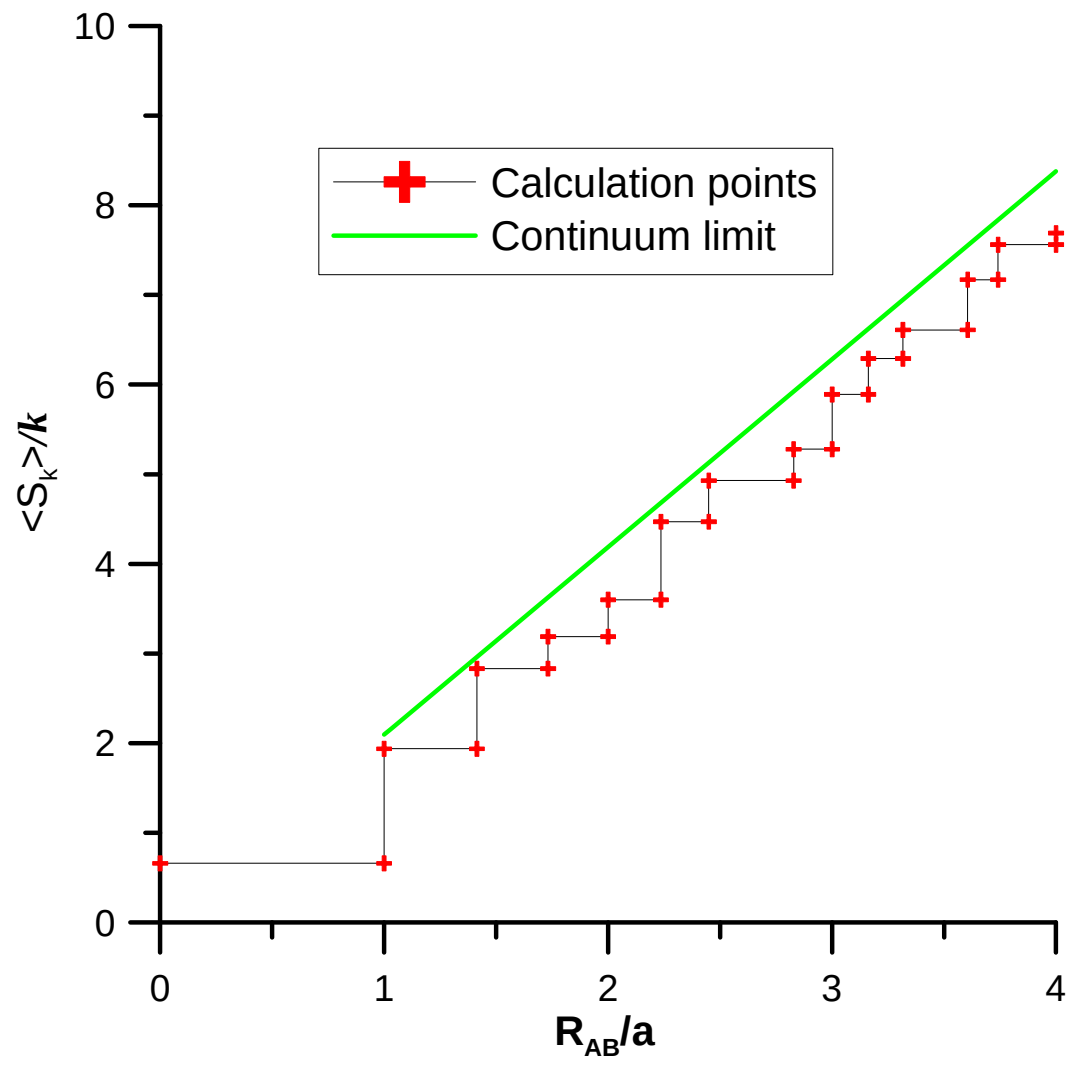

Fig. 2.5. The same as Fig. 2.4 but for the reduced interval of $R_{AB}/a$.

Comparison of the new calculation results for the discrete lattice at large $R_{AB}/a >> 1$ with the previous ones for the continuum model [7, 10] shows that, despite the steady-state values of the collision rate practically coincide in both cases, the number of jumps necessary for the attainment of the steady state increases by several orders of magnitude. Indeed, in the continuum limit the minimum number of necessary jumps fairly agrees with the value $k_{\min} >> (96/\pi)(R_{AB}/a)^2$, derived from the assessment of the analytical solution, $\delta t >> 16 R_{AB}^2/\pi D_{AB}$. For instance, in the case $R_{AB}/a = 10$ the number of jumps in the continuum model [7, 10], $k_{\min} \approx 8\times10^4$, was approximately 2 orders of magnitude smaller than the minimum number of jumps $\approx 5\times10^6$ in the discrete model. In the case $R_{AB}/a = 100$ the latter value further increases and attains $\approx 10^7$, see Fig. 2.1. This value is generally much larger than that used in the molecular dynamics (MD) calculations and thus may induce a serious restriction on applicability of the atomistic calculations to large molecular clusters.

In these cases the current approach might be rather effective, as will be shown elsewhere, for extrapolation of the reaction radii, calculated by MD at relatively short times, to the correct values corresponding to the steady-state values of the reaction constants $K_{AB}$ in the reaction rate equation, Eq. (2.5).

Therefore, the new kinetic approach, based on the "diffusion mixing" condition, to consideration of the reaction kinetics for particles A and B migrating by random walks on discrete lattice sites (with the lattice spacing $a$), and reacting when two particles occupy the same site, i.e. $R_{AB} < a$, was extended to the transition regime, corresponding to $R_{AB}/a \geq 1$, and applied to consideration of the recombination rate of point defects in cubic lattices. In this approach the reaction rate is reduced to calculation of the mean number of distinct sites visited by the effective particle of radius $R_{AB}$, which was calculated using the numerical algorithm, developed by generalization and further improvement of the original algorithm elaborated by the authors for the continuum limit calculations. The numerical calculations correctly reproduced, as foreseen, the analytical expressions in the two limits, $R_{AB} < a$ and $R_{AB} >> a$, and represented a step-wise curve in the intermediate range of the parameter $R_{AB}/a$, which generally corresponds to Frenkel pair recombination radius in many practical case, e.g. for fcc metals.

## 2.7. Conclusions

The new kinetic approach to the diffusion-limited reaction rate theory [96, 97] that is developed on the base of a similar approach to consideration of Brownian coagulation, proposed in the author's papers [5-7] (see Part 1), is overviewed.

The traditional diffusion approach to irreversible reactions A + B $\rightarrow$ C that stipulates that the local reaction rate should be equal to the diffusive current of particles, is critically analysed. In particular, it is shown that the diffusion approach is applicable only to the special case of reactions with a large reaction radius, $\bar{r}_A << R_{AB} << \bar{r}_B$ (where $\bar{r}_A \approx n_A^{-1/3}$, $\bar{r}_B \approx n_B^{-1/3}$ are the mean inter-particle distances), corresponding to small A-particles and large B-traps, and becomes inappropriate to calculation of the reaction rate in the most important (for reaction kinetics) case $R_{AB} << \bar{r}_A, \bar{r}_B$ particularly corresponding to comparable size (or point-wise) particles A and B. Indeed, point-wise particles tend to a homogeneous (in random) spatial distribution owing to their migration and mixing on the scale of the mean inter-particle distance, $l \approx \bar{r}$, with the characteristic diffusion time that is small in comparison with the characteristic reaction time, $\tau_d << \tau_c$. This implies that particles collisions occur in the kinetic regime with the reaction rate calculated as the collision frequency of two particles (A and B) randomly located in the unit volume.

his approach can be further extended to consideration of spatial heterogeneities in the ensemble of comparable size A and B particles, if these heterogeneities are smooth on the length scale of the diffusion equation (for the ensemble of point-wise particles) applicability, $l >> n_{A,B}^{-1/3} >> R_{AB}$, however, with the reaction constant calculated in the kinetic regime.

For the diffusion-controlled reactions, described in the continuum mode of the kinetic regime $a_A, a_B << R_{AB}$, where $a_A, a_B$ are the elementary drift distances of particles migrating by random walks, the calculated reaction rate in 3-d formally (and, in fact, fortuitously) coincides with the expression derived in the traditional approach (that is relevant only in the particular case of reactions with a large reaction radius, $\bar{r}_A << R_{AB} << \bar{r}_B$). This formal coincidence apparently explains a reasonable agreement of predictions of the kinetic equation, derived in the traditional approach, with experimental measurements for 3-d reaction systems.

However, in 2-d geometry, corresponding to reactant particles migration constrained in a plane, the reaction rate calculated in the traditional approach as the diffusive current of A-particles into B-traps naturally (for 2-d) predicts an explicit time-dependence for the reaction rate. On the contrary, in the new approach the original multi-particle problem is reduced (under the mixing condition) to calculation of the area sweeping rate by migrating particles (of radius $R_{AB}$ and diffusivity $D_{AB}$), which depends on time implicitly, via $n_A(t)$ (in the base case $R_{AB} << \bar{r}_A, \bar{r}_B$). As a result, in the case $n_A \neq n_B$ the traditional approach notably underestimates the concentration decay rate $\dot{n}_{A,B}$ at large times in comparison with predictions in the new approach.

Extending consideration of the diffusion-controlled reactions beyond the hard sphere approximation, the inconsistency of the traditional approach can be disclosed explicitly also in 3-d, resulting in overestimation of the reaction rate constant by several orders of magnitude in the case of a large reaction activation energy, $E_a/kT >> 1$. Basing on this analysis, the traditional definition of the diffusion-controlled reactions is criticized and the alternative, more adequate formulation (hardly ever used in the literature) is delineated.

The new approach is further generalized to consideration of reaction kinetics for particles migrating by random walks on discrete lattice sites (with the lattice spacing $a$). Since the case of large reaction radius, $R_{AB} >> a$, is properly reduced to the continuum medium limit, the opposite case, $R_{AB} < a$, with reactions occurred when two particles occupy the same site, was additionally studied. In the new approach the original multi-particle problem is reduced (under the mixing condition) to calculation of the mean number of distinct sites visited by a *k*-step random walk of the mobile particle, which for the simple 3-d and 2-d lattices was evaluated in the literature. As a result, the new relationships for the bimolecular reaction rate constants are derived either for 3-d or for 2-d lattices, whereas the traditional approach preserves the main deficiencies of the continuum medium approach (also in application to catalytically-activated reactions).

Being applied to consideration of the recombination rate of point defects in cubic lattices, the kinetic approach correctly reproduces the analytical expressions in the two limits, $R_{AB} < a$ and $R_{AB} >> a$, and represented a step-wise curve in the intermediate range of the parameter $R_{AB}/a$, which generally corresponds to Frenkel pair recombination radius in many practical case, e.g. for fcc metals.

## Acknowledgements

The author thanks Prof. L.I. Zaichik (IBRAE), Prof. V.V. Lebedev (Landau Institute for Theoretical Physics, Moscow), Dr. Y. Drossinos (JRC, Ispra) and Dr. C.J. Hogan (University of Minnesota, USA) for important critical remarks and valuable discussions. Mr. I.B. Azarov (IBRAE), Dr. V.I. Tarasov (IBRAE) and Mr. P.V. Polovnikov, who are co-authors of the several reviewed papers, are greatly acknowledged for their important and fruitful collaboration. Prof. L.A. Bolshov (IBRAE) is thanked for his interest and support of this work.

## Appendices

### Appendix A

In this Appendix it is shown that the condensation rate of the mobile markers migrating by random walks with the diffusivity $D_{AB}$ in the immobile trap of radius $R_{AB}$ is equivalent to calculation of the sweeping rate of randomly distributed immobile point particles (markers) by a large particle of radius $R_{AB}$ migrating with the diffusivity $D_{AB}$. This assertion is important for derivation of the collision frequency function in the continuum mode of the kinetic regime (Sections 1.4.1 and 2.4). Simultaneously applicability limit of the diffusion approach to calculation of the coagulation (or reaction) rate (revealed in Sections 1.3.1 and 2.3) is additionally confirmed.

Let us consider an ensemble of $N \to \infty$ point particles randomly distributed in a sample of volume $V \to \infty$ with the mean number concentration $n = N/V$ and migrating with the diffusivity $D_{AB}$ into the immobile trap particle of radius $R$ located at $\vec{r} = 0$. The probability for a point particle located at $t = 0$ in the elementary volume $d^3r$ at $\vec{r}$ to reach the trap in $t$ will be designated as $w(\vec{r},t)d^3r$. Therefore, the number of particles located at the distance $r$ and trapped in the time interval between 0 to $t$ is $\int nw(\vec{r},t)d^3r$. Integration of this number over the sample volume determines the total number of point particles trapped in the time interval between 0 to $t$,

$$\Phi_n(t) = \int nw(\vec{r},t)d^3r = 4\pi n\int_R^\infty w(r,t)r^2dr, \qquad (A.1)$$

in accordance with the Einstein-Fokker approach to consideration of particles migration ([121], see also [4]).

Correspondingly, the number of point particles trapped in the time interval between $t$ and $t + \delta t$ is equal to $(d\Phi_n/dt)\delta t = 4\pi n\delta t\int_R^\infty (\partial w(r,t)/\partial t)r^2dr$, that determines the condensation rate of point particles in the trap

$$\nu_n = d\Phi_n/dt = 4\pi n \int_R^{\infty} (\partial w(r,t)/\partial t) r^2 dr . \qquad (A.2)$$

If there is only one point particle randomly located in the sample (of volume $V \to \infty$), it can be found with the probability $V^{-1}d^3r$ in the elementary volume $d^3r$ at each point $\vec{r}$, therefore, the probability for this particle to reach the trap in $t$ can be calculated as

$$\Phi_0(t) = \int_R^{\infty} V^{-1} w(r,t) 4\pi r^2 dr . \qquad (A.3)$$

The probability to reach the trap in $\delta t$ thus becomes equal to

$$(d\Phi_0/dt)\delta t = 4\pi V^{-1} \delta t (\partial/\partial t) \int_R^{\infty} w(r,t) r^2 dr , \qquad (A.4)$$

or, from comparison of Eq. (A.4) with Eq. (A.2),

$$(d\Phi_0/dt)\delta t = (nV)^{-1} (d\Phi_n/dt)\delta t . \qquad (A.5)$$

On the other hand, this latter probability is equal to the probability to sweep in $\delta t$ a sole immobile point particle randomly located in the sample by the trap particle migrating by the same random walks (with the same diffusivity $D_{AB}$). Therefore, the probability of sweeping of the sole point particle in $\delta t$ by the trap particle migrating with the diffusivity $D_{AB}$ is equal to $V^{-1}\delta V$, where $\delta V$ is the volume swept in $\delta t$. Equating this probability to Eq. (A.5), one obtains

$$\delta V/\delta t = n^{-1} d\Phi_n/dt . \qquad (A.6)$$

If there are $N = nV$ immobile point particles randomly distributed in the sample, the total number of swept particles in $\delta t$ is reduced to

$$n\delta V/\delta t = d\Phi_n/dt = \nu_n , \qquad (A.7)$$

with $\nu_n$ from Eq. (A.2).

Therefore, the condensation rate $\nu_n$ of point particles migrating with the diffusivity $D_{AB}$ in the immobile trap particle of radius $R_{AB}$ is equal to the rate of sweeping of immobile point particles by the trap particle of radius $R_{AB}$ migrating by the same random walks (with the diffusivity $D_{AB}$).

## Appendix B

The probability $dP_{123}$ of a collision in the time step $dt << \tau_c$ among three particles randomly migrating in unit volume can be calculated as follows. The full set of events among three particles consists of three options: (a) there are no collisions; (b) there are pair collisions, but no triple collisions; (c) there is a triple collision (instantaneous). As explained in the text, collisions of one of the three particles with a newly formed particle after a moment of the two other particles collision (during the remaining part of the time step $dt$) can be neglected in the first order of approximation $dt/\tau_c << 1$, required for derivation of the collision rate equation, and thus only instantaneous triple collisions should be considered. For this reason, for the probabilities of the three events (a), (b) and (c) one obtains

$$dP_b + dP_b + dP_b = 1. \tag{B.1}$$

Since the pair collisions 12, 23, and 31 are incompatible events, the total probability of the event (a) is calculated as

$$dP_a = 1 - dP_{12} - dP_{23} - dP_{31}, \tag{B.2}$$

where the probability of a pair collision is determined by Eq. (4), $dP_{ij} = \beta(R_i, R_j)dt$.

The probability of a pair collision between particles $i$ and $j$, given no triple collision with a particle $k$ takes place, is equal to the probability $dP_{ij}$ multiplied by the probability that the particle $k$ is outside the volume $V(R_i, R_j; R_k)$, defined in Eq. (6). Similarly to the event (a), three various options are incompatible events so that the total probability of the event (b) is calculated a

$$dP_b = dP_{12} \cdot (1 - V(R_1, R_2; R_3)) + dP_{23} \cdot (1 - V(R_2, R_3; R_1)) + dP_{31} \cdot (1 - V(R_3, R_1; R_2)). \tag{B.3}$$

Substituting Eqs. (B.2) and (B.3) in Eq. (B.1), one obtains

$$\begin{aligned} dP_{123} \equiv dP_c = 1 - dP_a - dP_b &= dP_{12} \cdot V(R_1, R_2; R_3) + dP_{23} \cdot V(R_2, R_3; R_1) + dP_{31} \cdot V(R_3, R_1; R_2) \\ &= (\beta(R_1, R_2)V(R_1, R_2; R_3) + \beta(R_2, R_3)V(R_2, R_3; R_1) + \beta(R_3, R_1)V(R_3, R_1; R_2))dt. \end{aligned} \tag{B.4}$$

where $V(R_i, R_j; R_k)$, presented in Fig. 1.21, can be calculated as the volume $V_{ABC}$ of a figure consisting of three arbitrary intersecting spheres *A*, *B* and *C*,

$$V_{ABC} = V_A + V_B + V_C - V_{A\cap B} - V_{B\cap C} - V_{C\cap A} + V_{A\cap B\cap C}, \tag{B.5}$$

where $V_A$ is the volume of sphere *A*, $V_{A\cap B}$ is the volume of intersection of spheres *A* and *B*, $V_{A\cap B\cap C}$ is the volume of intersection of all three spheres. In the considered particular case the spheres *A*, *B* and *C* have radii $R_A = R_i + R_k$, $R_B = R_j + R_k$ and $R_C = R_{ij} + R_k$, respectively. Taking into account that $R_C > R_A$, $R_B$ and that the center of particle *C* is located between the centers of spheres *A* and *B* one derives from a simple geometrical consideration that the intersection of the spheres *A* and *B* is located entirely within the sphere *C*. This means that $V_{A\cap B\cap C} = V_{A\cap B}$ so that Eq. (B.2) reduces to

$$V_{ABC} = V_A + V_B + V_C - V_{B\cap C} - V_{C\cap A}, \tag{B.6}$$

which coincides with Eq. (1.95), the intersection volume of two spheres being given by Eq. (1.96), e.g. see [73].

## Appendix C

Considering asymptotic behavior of the distribution function in the case of homogeneous kernel

$$\beta(ai,aj)=a^{\lambda}\beta(i,j),$$
$$\beta_3(aj,ak,al)=a^{\lambda+3/d_f}\beta_3(j,k,l)$$

one represents the distribution function in form [74]

$$c(i,t)=\frac{1}{\bar{N}^2(t)}\varphi\left(\frac{i}{\bar{N}(t)}\right), \tag{C.1}$$

where $\bar{N}(t)$ is the mean number of particles in the cluster. Substituting Eq. (C.1) in Eq. (1.103), one derives

$$\frac{d}{dt}\bar{N}(t)=-f_2(x)\bar{N}^{\lambda}(t)-f_3(x)\bar{N}^{\lambda+3/d_f-1}(t), \tag{C.2}$$

where

$$f_2(x)\equiv\frac{\frac{1}{2}\iint\beta(y,z)\delta(x-y-z)\varphi(y)\varphi(z)\,dydz+\varphi(x)\int\beta(x,y)\varphi(y)\,dy}{2\varphi(x)+x\varphi'(x)},$$

$$f_3(x)\equiv\frac{\frac{1}{6}\iiint\beta(y,z,w)\delta(x-y-z-w)\varphi(y)\varphi(z)\varphi(w)\,dydzdw-\frac{1}{2}\varphi(x)\iint\beta(x,y,z)\varphi(y)\varphi(z)\,dydz}{2\varphi(x)+x\varphi'(x)}$$

If only two-particle collisions are taken into consideration ($f_3 = 0$), the variables $x$ and $t$ in Eq. (C.2) separate

$$\frac{1}{\bar{N}^{\lambda}(t)}\frac{d}{dt}\bar{N}(t)=-f_2(x)=const, \tag{C.3}$$

resulting in the self-preserving solution for the particle size spectrum.

If triple collisions are taken into account, the variables in Eq. (C.2) separate only if $d_f=3$. In this case the self-preserving spectrum is attained; however its form (satisfying $f_2(x)+f_3(x)=const$) differs from that for the case of pair collisions, Eq. (C.3). Note that this conclusion is valid also for multiple collisions of higher order, since each new term introduces an additional term in Eq. (C.2), which is asymptotically proportional to $Vc\propto\bar{N}^{\lambda-1+3/d_f}$, similarly to the triple term. However, in the case of fractal particles ($d_f\neq 3$) no self-preserving solution is generally attained.

## Appendix D

The group transfer coefficients are calculated as

$$g_{JK;I}=\frac{1}{\Delta_J\Delta_K}\left(s\left(a_Ib_I;a_Jb_J;b_K\right)-s\left(a_Ib_I;a_Jb_J;b_{K-1}\right)\right)$$
$$g_{JKL;I}=\frac{1}{\Delta_J\Delta_K\Delta_L}\left(u\left(a_Ib_I;a_Jb_J;b_K+a_L,b_K+b_L\right)-u\left(a_Ib_I;a_Jb_J;b_{K-1}+a_L,b_{K-1}+b_L\right)\right)$$

where

$$u\left(a_Ib_I;a_Jb_J;m,n\right)\equiv\sum_{p=m}^{n}s\left(a_Ib_I;a_Jb_J;p\right)$$

and

$$s\left(a_Ib_I;a_Jb_J;n\right)\Big|_{I\geq J}\equiv\sum_{k=1}^{n}\sum_{j=a_J}^{b_J}\sum_{i=1}^{b_L}\Delta\left(k+j-i\right)=\sum_{k=1}^{n}\max\left(0,\min\left(\Delta_J,b_I-a_J-k+1,b_J+k-a_I+1\right)\right)$$
$$=\begin{cases}0, & n<a_I-b_J\\ \dfrac{\left(n-a_I+b_J+1\right)\left(n-a_I+b_J+2\right)}{2}, & a_I-b_J\leq n<a_I-a_J\\ \dfrac{\Delta_J\left(\Delta_J+1\right)}{2}+\Delta_J\left(n-a_I+a_J\right), & a_I-a_J\leq n<b_I-b_J\\ \dfrac{\Delta_J\left(2\Delta_I-\Delta_J+1\right)\left(1-\delta_{IJ}\right)+\left(b_I-a_J+\Delta_J-n\right)\left(n-b_I+b_J\right)}{2}, & b_I-b_J\leq n<b_I-a_J\\ \dfrac{\Delta_J\left(2\Delta_I-\left(\Delta_I+1\right)\delta_{IJ}\right)}{2}, & b_I-a_J\leq n\end{cases}$$

The coefficients satisfy the symmetry condition

$$g_{JK;I}=g_{KJ;I}$$
$$g_{JKL;I}=g_{KJL;I}=g_{LKJ;I}=g_{JLK;I}=g_{LJK;I}=g_{KLJ;I}$$

and the completeness condition

$$\sum_{I=1}^{\infty}g_{JK;I}=\sum_{I=1}^{\infty}g_{JKL;I}=1,$$

the latter providing the particle mass conservation

$$\frac{d}{dt}N_{tot}=\frac{d}{dt}\sum_{I=1}^{\infty}n_Ic_I=0.$$